Universidade de São Paulo
Instituto de Física

# Emaranhamento em teorias quânticas de campos algébricas

Rafael Grossi e Fonseca

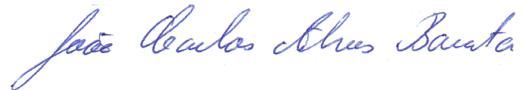

Orientador(a): Prof. Dr. João Carlos Alves Barata

Dissertação de mestrado apresentada ao Instituto de Física da Universidade de São Paulo, como requisito parcial para a obtenção do título de Mestre(a) em Ciências.

Banca Examinadora:
Prof. Dr. João Carlos Alves Barata - Orientador (IF-USP)
Prof. Dr. Hugo Luiz Mariano (IME-USP)
Prof. Dr. Silvio Paolo Sorella (UERJ)

São Paulo
2024

University of São Paulo
Physics Institute

# Entanglement in algebraic quantum field theories

Rafael Grossi e Fonseca

Supervisor: Prof. Dr. João Carlos Alves Barata

Dissertation submitted to the Physics Institute of the University of São Paulo in partial fulfillment of the requirements for the degree of Master of Science.

Examining Committee:
Prof. Dr. João Carlos Alves Barata - Supervisor (IF-USP)
Prof. Dr. Hugo Luiz Mariano (IME-USP)
Prof. Dr. Silvio Paolo Sorella (UERJ)

São Paulo
2024

# Acknowledgements

In respect to the people who are acknowledged, this section is written in Portuguese.

Uma conhecida anedota diz que Fidel Castro possuía o hábito de iniciar seus discursos com uma frase como *"não me delongarei"*, que era seguida de uma verborragia que poderia durar horas, como seu famoso discurso de quatro horas e meia de duração na Assembléia Geral da ONU. Diferentemente d'*el Comandante*, já deixo o leitor avisado que não tenho qualquer pretensão de ser breve aqui. Embora tente ser o mais justo possível, omissões são uma possibilidade e já peço desculpas de antemão caso isso ocorra!

O caminho tortuoso desse Mestrado certamente foi aliviado pelo meu orientador, o professor João Carlos Alves Barata. Ainda recordo-me nas aulas de Métodos de Física Teórica na UFMG quando utilizei suas notas de Física-Matemática pela primeira vez e fui exposto à elegância e à beleza da Matemática de alto nível no contexto de Física. Um outro texto menos conhecido de sua autoria *Como não ser um zumbi científico* foi essencial na minha formação e levo certos pontos deste texto como mantras na minha vida. A sua contribuição para a minha formação foi intensa e de formas que às vezes eu não esperava. Nossas discussões e seminários de grupo me ajudaram a descobrir problemas que eu sequer sabia da existência e me fizeram descobrir a Física-Matemática como o método que me sinto mais à vontade de se fazer Física Teórica (uma forma de ver a disciplina também esclarecida em um de seus textos!). Por sua paciência e a liberdade a mim concedida eu agradeço.

Gostaria de agradecer à minha mãe e ao meu pai por todo o encorajamento e apoio durante essa etapa da minha formação acadêmica. Vocês foram essenciais na minha mudança para São Paulo, sempre me dando suporte emocional e financeiro além de me receberem nas (muitas) idas a Belo Horizonte, quando eu sentia o peso dessa cidade. Destaco também os conselhos que muitas vezes me ajudaram a ver um caminho que eu nem sabia que existia. Obrigado por tudo!

Agradeço também ao meu irmão André, cuja a proximidade etária e acadêmica me abençoou com um porto seguro e a certeza de que sempre haveria alguém que entende as minhas dificuldades, fragilidades e medos além de saber valorizar os momentos de celebração pelas pequenas conquistas. A minha admiração por você é infinita e muitos momentos tomando café em suas visitas à USP funcionaram como momentos de descontração mas também de discussões e geração de novas idéias.

À minha irmã de outra família, Clarisse Reis pela lealdade e carinho em todo esse tempo mesmo que distante. Sua forma peculiar de ver o mundo e sua coragem sempre foram uma inspiração para mim e vou guardar para sempre no peito nossos momentos tomando café, cozinhando e treinando (principalmente braço, já que você claramente não treina perna). Amo você.

Aos meus companheiros de colégio e graduação Daniel Haas e Daniel Miranda que mesmo a um oceano de distância marcaram presença na minha vida. Nossas conversas de domingo de tarde me fizeram gargalhar mesmo em dias sombrios e vocês sempre me inspiraram a querer ser um melhor físico e amigo. Obrigado por me aturarem!

Ao grande amigo Filipe "Uai Física" Menezes com quem pude contar com a amizade desde o início da graduação. Nossas conversas sobre Física e a vida e nosso amor pelo Cruzeiro são momentos que são um tesouro para mim. À Laís Nunes pelo carinho e por sempre me fazer sentir bem em sua casa.

Ao *Casos* que há mais de 10 anos (!) continuam sendo a minha família extendida e grande parte da minha razão para visitar BH. Amo todos vocês mesmo que vocês sejam incapazes de chegar no horário marcado. Agradeço também a Francisco Gallupo, Marcelo Carvalho, Guilherme Carvalho e João Gabriel Carvalho pelo contato e amizade de anos.

Aos amigos que fiz na USP, Arthur "Arthurzito" Meirelles, Henrique Ay Casa Grande, Ivan Romualdo, Luis Felipe Santos, Naim Elias Comar e Rodrigo Ramos que sempre me providenciaram momentos de descontração, idas ao cinema e a shows musicais que acalmaram minhas ansiedades. Agradeço a vocês pela paciência de me ouvir tagarelar sobre a genialidade de David Lynch, sobre o jeito correto de se fazer café ou reclamar sobre o meio acadêmico.

Aos colegas de grupo, Ana Camila Costa, Felipe Dilho Alves, Henrique Ay Casa Grande, Ivan Romualdo, Rodrigo Orselli, Victor Hugo Marques Ramos, Victor Chabu, Vítor Cavalheri e, mais recentemente, (professor) Ricardo Correa da Silva e Thiago Raszeja pelos seminários, discussões e companheirismo no DFMA. Agradeço também a Níckolas Aguiar Alves por ser um exemplo de cientista e amigo e por ter confiado em mim para ajudar a organizar o BHT50.

Agradeço também os amigos que por diversas razões se afastaram ou perderam contato comigo durante esse tempo em São Paulo. Eu entendo as razões e fico contente pelo tempo que nos divertimos juntos. São as memórias boas que ficam e espero que seja o mesmo para vocês.

Ao professor Gustavo Burdman, com quem fiz o curso de Teoria Quântica de Campos I e II e depois foi meu supervisor na monitoria da primeira dessas disciplinas. Embora me sinta estranho de me referir ao senhor na língua portuguesa, saiba que *it comes from the*

*heart*. Além do exemplo de profissionalismo e didática, agradeço pelos conselhos profissionais e conversas com um café. Agradeço também aos membros da secretaria do DFMA, Simone Toyoko Shinomiya e Cecília Cristina Blanco Novello pela paciência e educação além de vestirem a camisa do Departamento. Agradeço a todos os funcionários da lanchonete da Física (nosso querido *Turco*), o seu Valdir da copiadora e o Carvalho da Livraria da Física pelos serviços e boa vontade.

Aos amigos do IFT, em especial João Gomide e Eggon Viana que me fizeram sentir acolhido na Barra Funda. Ressalto ainda as trocas de mensagens catárticas com o João nas noites dos jogos do Cruzeiro.

Aos meus orientadores da UFMG, em especial Gláuber Dorsch e Thaisa Guio que me guiaram no final da minha graduação e me fizeram ter certeza de que queria trabalhar com Teoria Quântica de Campos em alguma de suas muitas formas. Agradeço também ao professor Nelson Braga da UFRJ, que me orientou de maneira exemplar à distância durante a pandemia de 2020. Meu sincero agradecimento ao professor Nelson Yokomizo pelos cursos essenciais na UFMG e pela indicação de fazer o mestrado na USP sob a supervisão do professor Barata.

A Stephanie Fonseca e Lucas Passos pelo carinho desde os tempos da graduação que não diminui. As visitas de vocês a São Paulo fizeram muito bem para a minha alma. Não deixo de mencionar também Leonardo Oliveira e Amanda Campolina, cuja presença na cidade me trouxe segurança no primeiro momento e a Gabriela Militani pelo carinho nas idas a BH.

*"Every sufficiently good analogy is yearning to become a functor."*

John Baez

# Resumo

Existe um recente interesse em aplicar as técnicas de Teoria Quântica de Campos Algébrica (TQCA) a problemas envolvendo emaranhamento em TQC perturbativa. Em particular, a formulação independe de uma escolha de espaço de Hilbert, o que a torna particularmente interessante no contexto de espaços-tempos curvos, e a ênfase na álgebra de observáveis, torna o tratamento das desigualdades de Bell em TQC semelhante ao tratamento na Mecânica Quântica não-relativística. Neste trabalho, apresentamos as estruturas matemáticas necessárias para a formulação de TQCA em termos dos axiomas de Haag-Araki-Kastler (HAK) e discutimos as implicações destes. Além disso, discutimos a abordagem algébrica ao emaranhamento quântico na forma das desigualdades de Bell. Apresentamos uma extensão dessa formulação para espaços-tempos globalmente hiperbólicos gerais usando a chamada abordagem Localmente Covariante à TQC, que estende os axiomas HAK para espaços-tempos gerais utilizando a linguagem da Teoria de Categorias.

**Palavras-chave:** Física-Matemática, Álgebras de von Neumann, Teoria Quântica de Campos Algébrica, Emaranhamento, Teoria de Categorias.

# Abstract

There has been some recent interest in applying the techniques of Algebraic Quantum Field Theory (AQFT) to entanglement problems in perturbative QFT. In particular, the Hilbert space independence of this formulation makes it particularly interesting in the context of curved spacetimes and the emphasis on the algebra of observables makes the treatment of Bell inequalities in QFT resemble such treatment in non-relativistic Quantum Mechanics. In this work, we present the mathematical structures needed for formulating AQFT in terms of the Haag-Araki-Kastler (HAK) axioms and discuss their implications. Moreover, we discuss the algebraic approach to quantum entanglement in the form of Bell inequalities. We provide an extension of this formulation to general globally hyperbolic spacetimes using the so-called Locally Covariant approach to QFT, which extends the HAK axioms to general spacetimes by means of the Category Theory language.

**Keywords:** Mathematical Physics, von Neumann Algebras, Algebraic Quantum Field Theory, Entanglement, Category Theory.

# Contents

# 1 Introduction

*The PCT theorem is a theorem,*
*its not just good advice.*

Edward Witten

## 1.1 On Mathematical Physics

We begin by explaining our understanding of the term *Mathematical Physics*. This designation to Physics has been used throughout History to convey different meanings, with a more recent fad reserving the term to describe the use of advanced techniques to solve partial differential equations or to study the properties of special functions. This terms has even been deemed as a synonym for Fundamental Physics, in more *formal* aspects of theories. That is, many researchers who are called mathematical physicists are those who are many times not interested in applications of theories to models of the real world, for instance.

That is not entirely our understanding of the term. Rather, we reserve this designation not to a specific area of research in Physics, but rather to an *approach* to Theoretical Physics. This approach mimics in some aspects the research conducted in Mathematics, in the sense that one tries to present the working hypotheses in clearest form possible, avoiding any ambiguity in language and trusts the well-founded principles in different areas of Mathematics to make progress. This trust can only be earned once the mathematical physicist becomes well acquainted with the mathematical structures he or she is working with, by learning the correct definitions and the main hypotheses for theorems. This is seen sometimes as pedantic, but we argue that this is needed in order to give a firm structure to new theories or to better understand "established" theories.

The need for rigorous statements is even more critical in areas where experimental evidence is scarce, such as extensions of the Standard Model of Particle Physics or the Cosmology. When flying blind, we turn to Mathematics. However, Mathematics is not just about being to manipulate symbols. Indeed, just because one can *write down* a sentence it doesn't make it true. In fact, carrying on calculations without thinking about the meaning behind certain symbols can lead to dramatic results in Quantum Mechanics, for example (see the very nice paper by Gieres [74], as must read for any serious student of Theoretical Physics).

Indeed, Mathematical Physics is *hard*. It is not as rewarding as other approaches to Theoretical Physics and many times it requires that the researcher becomes well grounded in many areas of Physics and Mathematics, each of which is demanding and a whole area of expertise by itself. Progress is also very slow and the published papers are long and very technical at times.

Why such a specialization be sought by the prospective student then? For one, it is *beautiful*. Physics is built upon the language of Mathematics and it pays reward to be as proficient as possible in such language. Ironically, Physics seems to become *simpler* when one dives into more abstract mathematics, with complex relations emerging as consequences of simple corollaries of very general theorems (see for instance the Reeh-Schlieder Theorem discussed in this Thesis which gives, in a certain sense, rise to Entanglement in Quantum Field Theory). This approach also gives a broader view of Physics, not restricting some of the ideas to a specific area of research. It was the work of mathematical physicists that showed that the relationship between Statistical Mechanics and Quantum Field Theory was more than a mere analogy. It was the work of mathematical physicists that showed that areas of Mathematics such as Topology, at the time regarded as a strictly Pure Math discipline, was needed in Quantum Field Theory and Condensed Matter Physics.

This does not be any means serve as an attack to different approaches to Physics especially to Theory. Rather, our view is that Mathematical Physics is a complementary approach whose value should be recognized and not diminished. A profound understanding of Nature involves tackling problems from different perspectives, and Mathematical Physics does so by deeply exploring and setting in stone the language in which She speaks.

## 1.2 On Non-Perturbative Approaches to QFT

A natural question one might ask is "why bother?". The Standard Model of Particle Physics is arguably one of the most successful scientific models ever and it is built upon the foundations of perturbative QFT, as one learns in the classical textbooks.

There is a caveat to this. It is true that there is a success in describing Nature using ordinary QFT, but every calculation in interesting interacting models is done by means of a *perturbation series*. The terms in this series are interpreted as contributions to a certain process (the famous act of adding Feynman diagrams). A famous argument by Dyson [59] says that these series do not converge for any value of the coupling constant, and subsequent work by Jaffe et al. indeed shows that this is true for many models. Even though the divergence is expected to happen at very high orders in the series, this is still a matter to be thought of.

Algebraic QFT is a *non-perturbative* approach to QFT. The price one pays is that there are fewer models and the treatment of interactions is still in its first steps. It is not the only non-perturbative formulation: topological quantum field theories have been around for a while [140, 10] and they seem to be better suited for studying Quantum Gravity and topological systems in Condensed Matter.

## 1.3 On QFT in Curved Spacetimes

Quantum Field Theory (QFT) incorporates the notion of *locality*, that is, quantum measurements are localized in spacetime and causally separated measurements cannot influence each other being thus not subjected to the usual uncertainty relations of ordinary quantum mechanics, and *covariance*, in the sense that Lorentz observers are indistinguishable, to ordinary Quantum Mechanics.

In the case of a QFT formulated in Minkowski spacetime, we impose covariance to the fields by demanding them to be invariant under the symmetries of the background i.e., the isometries of Minkowski spacetime encoded in the Poincaré Group. However, this procedure fails to more general spacetimes that may not have non-trivial symmetries. In the case of QFT formulated in Minkowski spacetime, we use the symmetries of such background to choose a vacuum state or to define a Wigner-type particle.

### 1.3.1 The energy-momentum tensor

In [136], it is discussed an ambiguity in the definition of a energy-momentum tensor for a quantized field in a curved spacetime, specifically in the renormalization procedure. Roughly speaking, the renormalization of the energy-momentum tensor for the Klein-Gordon field involves taking a quasi-free Hadamard state $\omega$ as a reference state and imposing normal ordering in its GNS representation. This results in an expression for the tensor as a operator-valued distribution.

One can easily see that from the previous discussion, such a procedure is not possible, since it involves a choice of a Hadamard state which is not obvious due to the lack of symmetries in the spacetime.

The solution to this is to impose locality and covariance, that is, the energy-momentum tensor must only depend on the metric locally. Concretely, this solution can be understood in the following way: suppose that there is a prescription to obtain a renormalized energy-momentum tensor $T_{\mu\nu}^{\mathrm{ren}}(x)$ in any curved spacetime. This prescription is local and covariant if for two spacetimes $(M, g)$ and $(M', g')$ and an open subset $U \subset M$ with an isometric diffeomorphism $\kappa : U \to U' \subset M'$ satisfying $\kappa_* g = g'$, we have

$$\alpha'_\kappa(T'^{\text{ren}}_{\mu\nu}(x')) = \kappa_* T^{\text{ren}}_{\mu\nu}(x'), \quad x' \in U', \tag{1.3.1}$$

where $\alpha'_\kappa : \mathcal{A}_{M'}(U') \to \mathcal{A}_M(U)$ is the canonical isomorphism between the local commutation relations algebra of the Klein-Gordon fields in $M$ and $M'$.

This approach has the advantage of allowing an intrinsic definition of an energy-momentum tensor for a globally hyperbolic spacetime that depends solely on the unique construction of a free scalar field on any globally hyperbolic spacetime.

### 1.3.2 Formalization

The formalization of the above procedure and to avoid ambiguities in general when constructing quantities that involve the choice of states in the spacetime is most easily done with the algebraic approach to quantum field theory.

This approach idealized by Haag and Kastler [80] involves working with the local algebra of observables in Minkowski spacetime and, from it, obtaining the states which act upon the observables associating them with a real number. In essence, this corresponds to taking expectation values in the usual approach.

The crucial advantage of this approach is that the algebraic relations of the observables in different theories (even in unitarily inequivalent ones) are *the same*, which allows us to obtain states without ambiguity.

The generalization of this framework to more general spacetimes (globally hyperbolic) is due largely to Fredenhagen, Brunetti and Verch [30] and is our main focus in this work. This is usually called the Locally Covariant QFT (LCQFT) approach.

## 1.4 Bell Inequalities

This Thesis is concerned mainly with the discussion of the violation of Bell's inequalities. It was shown in a seminal work by Summers and Werner [127, 128, 129] and Landau [95] that the vacuum state in a QFT *maximally* violates these inequalities. This shows that this state is maximally correlated and that it is intrinsically different of a classical notion of a vacuum state as the absence of matter.

In the final chapter of this Thesis, we present a formulation of these results in general globally hyperbolic spacetimes, by means of the functorial formulation mentioned in the last section.

## 1.5  Structure of this Thesis

This Thesis is divided in three parts. It was written with the advanced undergraduate student in Physics in mind, however some of its content might be useful for students in Mathematics who have some inclination towards Physics. With this, we start in Part I with the algebraic notions that will become recurrent themes along the Thesis. Some of its content requires some knowledge of basic Functional Analysis, which we thoroughly cover in appendix B.

The main content of Part I is the notion of a $C^*$-algebra. This is an abstraction of the notion of bounded operators acting on some Hilbert space. Because we are free from some Hilbert space, whose construction many times is based on the choice of some set of vectors which are simultaneously eigenstates of some Hamiltonian and a symmetry operator, the problem with the lack of symmetries in curved spacetimes may be remedied. By choosing some normalized functional over the algebra (which we will call a *state*), one can recover the Hilbert space description. We then move to a chapter dedicated to von Neumann algebras, which are a specific type of $C^*$-algebra which are ubiquitous in QFT. In particular, we present a proof of the important Bicommutant Theorem, which shows that a von Neumann algebra is closed in a physically interesting topology. We also briefly discuss the important Tomita-Takesaki Modular Theory, which, in particular, provides a family of automorphisms in the algebra, with interesting physical implications. The last chapter of part I is dedicated to a general formulation of the canonical commutation relations, which is crucial for an algebraic formulation of QFT. The content in this part may seem long and technical, but it pays off in the remainder of the Thesis.

In Part II, we present the algebraic formulation of QFT. In this formulation, the basic objects of interest are not the fields as in ordinary perturbative QFT, but rather the net of algebras (that is, the association of an algebra of observables to a suitable region of spacetime). We then present the Haag-Araki-Kastler axioms which are the physically motivated conditions the net and algebras must obey. We discuss some of the implications of these axioms, which are tight enough to constrain many results about the theory. In particular, we discuss the Reeh-Schlieder theorem, which gives some algebraic conditions a vacuum vector state must obey and implies in physically interesting results. We also discuss how entanglement already seems to manifest in the algebraic formulation, in the form of both the Reeh-Schlieder theorem and the Split property. We also briefly discuss an important result on the structure of the local algebras, which tells us that every algebra which can serve as a *bona fide* algebra of observables for a AQFT must be a von Neumann factor algebra of type $III_1$. Moreover, this algebra is unique up to isomorphisms.

The next chapter in this part deals with the formulation of AQFT in curved spacetimes in the language of Category Theory. This is known in the literature as Locally Covariant Quantum Field Theory (LCQFT) and it provides a very interesting interpretation of the quantum fields and the state space associated with a theory. It also allows one to make very general remarks of QFT without committing to a specific model.

In Part III, we present the formulation of Bell inequalities in the algebraic approach. These inequalities have an historical importance since they were the first strong evidence that entanglement is a intrinsically quantum phenomenon with no classical analogue. We then discuss the violation of these inequalities by the vacuum state in AQFT. Finally, in the last chapter we present a novel formulation of the inequalities in the context of LCQFT. This formulation extends the results to general globally hyperbolic spacetimes and we discuss the implications of such generalization.

# Part I

# Algebras

# 2  $C^*$-Algebras

> *The true foundation of Mathematics is not logic, but structure.*
>
> ———————————————
>
> Nicolas Bourbaki

In this chapter, we introduce the very important concept of $C^*$-*algebras*. They were first introduced by Segal in 1947 [122] already in the context of Quantum Mechanics, building upon foundational work by Gelfand and Naimark [71], which presented an abstract characterization of $C^*$-algebras (that is, without reference to a Hilbert space) already in 1943. Our treatment here is introductory and we follow mainly references [25] and [89]. We assume previous knowledge of basic Functional Analysis, that is, the definition of normed and inner-product spaces as well as some spectral theory. These topics are covered in appendix B and the reader unfamiliar with these topics is encouraged to go through that section before starting this chapter.

## 2.1  Normed Algebras

As mentioned previously, we will be concerned with what is called the *algebra of observables*, which we will take as the basic object of a quantum field theory.

Algebras are vector spaces where a notion of multiplication between vectors is defined. Usually, this product is denoted by the juxtaposition of two elements $a$ and $b$ of an algebra $\mathfrak{A}$, hence $ab \in \mathfrak{A}$. Notice that we denote the elements of an algebra by lowercase letters, leaving the uppercase letter to denote operators in a Hilbert space (which form an example of a $C^*$-algebra, see below). This product is distributive with respect to the vector addition, that is

$$a(b + c) = ab + ac,$$

for all $a, b, c \in \mathfrak{A}$.

A natural example of an algebra is that of the real and complex numbers, where the usual rules of multiplication of numbers is understood. Another good example (which is also a *Lie algebra*) is $\mathbb{R}^3$ with the vector product $\times$.

In this Thesis, we will only be concerned with *associative algebras*, that is, for any triple of elements $a, b, c \in \mathfrak{A}$, we have that

$$a(bc) = (ab)c \equiv abc,$$

whence we can drop the parenthesis whenever clarity is not demanded.

We will be interested in keeping the notion of self-adjointness from the Theory of Operators on Hilbert spaces, which is itself a generalization of the operation of conjugation of complex numbers. This notion is made explicit by defining an *involution*.

**Definition 2.1.1.** A *\*-algebra* or an *involutive complex algebra* or a *$A^*$-algebra* $\mathfrak{A}$ is an algebra over $\mathbb{C}$ together with a map $^* : \mathfrak{A} \to \mathfrak{A}$ called an involution which satisfies the following properties for all $a$, $b \in \mathfrak{A}$ and all $\lambda \in \mathbb{C}$:

1. $(a + b)^* = a^* + b^*$;

2. $(ab)^* = b^* a^*$;

3. $(\lambda a)^* = \overline{\lambda} a^*$;

4. $(a^*)^* = a$.

Furthermore, a subalgebra $\mathcal{S}$ of $\mathfrak{A}$ is called *self-adjoint* if $a \in \mathcal{S}$ implies $a^* \in \mathcal{S}$.

Because we are still interested in keeping the nice topological properties of Banach spaces, we will be interested in algebras which are compatible with a norm. Recall that a norm is a homogeneous, positive-definite functional on a vector space which satisfies the triangle inequality (see appendix B).

**Definition 2.1.2.** A *normed algebra* $\mathfrak{A}$ is a normed vector space whose norm satisfies the *product inequality*

$$||ab|| \leq ||a|| \, ||b||, \tag{2.1.1}$$

for all $a$, $b \in \mathfrak{A}$.

If $\mathfrak{A}$ is unital (that is, there is an element $\mathbf{1}$ such that $a\mathbf{1} = \mathbf{1}a = a$, $\forall a \in \mathfrak{A}$) and $||\mathbf{1}|| = 1$, then $\mathfrak{A}$ is called a *normed unital algebra*.

If $\mathfrak{A}$ is complete in the topology induced by $|| \cdot ||$, then it is called a *Banach algebra*.

If, in addition, $\mathfrak{A}$ is a \*-algebra, and $||a^*|| = ||a||$, then $\mathfrak{A}$ is called a *Banach \*-algebra* or a *$B^*$-algebra*.

## 2.2 The $C^*$ Property

**Definition 2.2.1.** A $B^*$-algebra $\mathfrak{A}$ is called a $C^*$-*algebra* if it satisfies the $C^*$-*property*:

$$||a^*a|| = ||a||\,||a^*|| = ||a||^2, \quad \forall a \in \mathfrak{A}. \tag{2.2.1}$$

A very important example of a $C^*$-algebra is the algebra of bounded operators on a Hilbert space $\mathfrak{B}(\mathcal{H})$ together with the operator norm

$$||A|| = \sup_{\psi \neq 0} \frac{||A\psi||}{||\psi||} = \sup_{||\psi|| = 1} ||A\psi||, \tag{2.2.2}$$

as introduced in section B.6. The involution is given by the conjugate transpose of the matrix corresponding to the operator, which is usually denoted in Physics texts by the "dagger", $\cdot^\dagger$. The $C^*$-property in this case follows from

$$
\begin{aligned}
||A||^2 &= \sup_{\substack{x \in \mathcal{H} \\ ||x|| = 1}} ||Ax||^2 = \sup_{\substack{x \in \mathcal{H} \\ ||x|| = 1}} \langle Ax, Ax \rangle \\
&= \sup_{\substack{x \in \mathcal{H} \\ ||x|| = 1}} \langle x, A^\dagger Ax \rangle \leq \sup_{\substack{x \in \mathcal{H} \\ ||x|| = 1}} ||A^\dagger Ax|| = ||A^\dagger A|| \leq ||A^\dagger||\,||A|| = ||A||^2.
\end{aligned}
$$

Note that if we take the $C^*$-property in equation 2.2.1 together with the product inequality in equation 2.1.1, we automatically get the equality $||A|| = ||A^*||$.

We will be interested in $C^*$-algebras with an identity $\mathbf{1}$ such that $a\mathbf{1} = \mathbf{1}a = a$, $\forall a \in \mathfrak{A}$. These kinds of algebras are called *unital $C^*$-algebras.* A simple exercise left to the reader is to verify that a unital $C^*$-algebra can have, at most, one identity.

The last important definition for understanding $C^*$-algebras is the concept of *ideals*, which we can use to construct algebras and sub-algebras.

**Definition 2.2.2.** A subspace $\mathfrak{B}$ of an algebra $\mathfrak{A}$ is called an *left ideal* if $a \in \mathfrak{A}$ and $b \in \mathfrak{B}$ implies that $ab \in \mathfrak{B}$. Alternatively, if $ba \in \mathfrak{B}$, then $\mathfrak{B}$ is called an *right ideal* of $\mathfrak{A}$. If $\mathfrak{B}$ is both a left and a right ideal, then it is said to be a *two-sided ideal.*

We note that each ideal is automatically an algebra. Furthermore, if $\mathfrak{B}$ is a left or right ideal of an algebra with involution and $\mathfrak{B}$ is self-adjoint, then $\mathfrak{B}$ is automatically a two-sided ideal. Finally, we say that a $C^*$-algebra is *simple* if it has no nontrivial closed two sided-ideals (meaning the only two sided ideals are $\{0\}$ and $\mathfrak{A}$ itself).

## 2.3   Positive Elements

The spectral theory of self-adjoint operators (see appendix B.7) is naturally generalized to the context of Banach algebras and in particular to $C^*$-algebras. This also holds for the classification of operators based on their action in the presence of an inner product. Therefore, we will not bother to repeat the basic definitions here for the $C^*$-algebraic case and instead we will focus on the main differences that occur when uses the more abstract setting.

### 2.3.1   Spectral Mapping Theorem

Let $\mathfrak{B}$ be a unital Banach algebra and let $p(z) = \sum_{j=1}^{n} a_j z^j$ be a polynomial defined at $z \in \mathbb{C}$. Naturally, one can define for the spectrum of $u \in \mathfrak{B}$:

$$p(\sigma(u)) \equiv \{p(\lambda), \lambda \in \sigma(u)\}. \tag{2.3.1}$$

What the Spectral Mapping Theorem shows is that $p(\sigma(u)) = \sigma(p(u))$. In order to prove it, we need the following lemma:

**Lemma 2.3.1.** *Let $\mathfrak{B}$ be a unital Banach algebra and let $u \in \mathfrak{B}$. Then, if $\lambda \in \sigma(u)$, then $(u - \lambda\mathbf{1})q(u)$ is not invertible for any polynomial $q$.*

*Proof.* The proof is by contradiction. Let $p(u) \equiv (u - \lambda\mathbf{1})q(u) \in \mathfrak{B}$. Clearly, $u$ commutes with both $q(u)$ and $p(u)$. We will assume that $p(u)$ is invertible, that is, that there exists $w \in \mathfrak{B}$ such that $wp(u) = p(u)w = \mathbf{1}$.

Define $c \equiv wu - uw$. Now, we multiply $c$ by $p(u)$:

$$p(u)c = u - p(u)uw = u - up(u)w = u - u = 0.$$

Hence, $c = 0$ and $u$ and $w$ commute with each other. Naturally, this implies that $q(u)w = wq(u)$.

From our hypothesis, $w$ satisfies $p(u)w = wp(u) = \mathbf{1}$. Therefore, $(u - \lambda\mathbf{1})q(u)w = \mathbf{1}$ and $w(u - \lambda\mathbf{1})q(u) = \mathbf{1}$. Thus, we have both

$$(u - \lambda\mathbf{1})(q(u)w) = \mathbf{1}$$
$$(q(u)w)(u - \lambda\mathbf{1}) = \mathbf{1}.$$

This means that $(u - \lambda\mathbf{1})$ is invertible with the inverse being given by $(u - \lambda\mathbf{1})^{-1} = q(u)w$, which contradicts the hypothesis that $\lambda \in \sigma(u)$. Hence, $p(u)$ has no inverse. $\blacksquare$

Now, we can present and prove the main theorem of this section.

**Theorem 2.3.1** (Spectral Mapping Theorem). *Let $\mathfrak{B}$ be a unital Banach algebra and $u \in \mathfrak{B}$. Then, for every polynomial $p$ we have*

$$\sigma(p(u)) = p(\sigma(u)) = \{p(\lambda), \lambda \in \sigma(u)\}. \tag{2.3.2}$$

*Proof.* Let $\mu \in \sigma(p(u))$ and let $\{\alpha_j\}$ be the $n$ roots of the polynomial $p(z) - \mu$. Then, we can write $p(z) - \mu = \sum_{j=1} a_j(z - \alpha_j)$, which implies that $p(u) - \mu \mathbf{1} = \sum_{j=1} a_j(u - \alpha_j \mathbf{1})$.

We can say that at least one of the roots $\alpha_j$ is in the spectrum of $u$. If there were not the case, then each factor $(u - \alpha_j \mathbf{1})$ would be invertible, including the product $(u - \alpha_1 \mathbf{1})...(u - \alpha_n \mathbf{1})$, which is a contradiction of the fact that $\mu \in \sigma(p(u))$. Hence, some of the $\alpha_j$'s belong to $\sigma(u)$ and since $p(\alpha_j) = \mu$ then $\sigma(p(u)) \subset \{p(\lambda), \lambda \in \sigma(u)\}$.

Now we prove that $p(\sigma(u)) \subset \sigma(p(u))$. For $\lambda \in \sigma(u)$, it is easy to see that the polynomial $p(z) - p(\lambda)$ has $\lambda$ as a root. Hence, $p(z) - p(\lambda) = (z - \lambda)q(z)$ where $q$ is a polynomial of order $n - 1$. Therefore, $p(u) - p(\lambda)\mathbf{1} = (u - \lambda\mathbf{1})q(u)$ and since $(u - \lambda\mathbf{1})$ is not invertible, then $p(u) - p(\lambda)\mathbf{1}$ can't be either. This means that $p(\lambda) \in \sigma(p(u))$, which establishes that $p(\sigma(u)) \subset \sigma(p(u))$. ∎

### 2.3.2   The Gelfand Homomorphism and the Continuous Functional Calculus

In this section, we present the notion of functional calculus in the context of the theory of operators or, more generally, in the context of $C^*$-algebras. By "functional calculus" we mean to apply the general notion of functions over the field of real or complex numbers to elements of abstract algebras, thus giving sense to expressions such as $f(M)$, where $f$ is a function of, say, real numbers while $M$ is a bounded operator, for example.

We will contend ourselves with the study of polynomials over an element of a $C^*$-algebra, since the *Stone-Weierstrass theorem* guarantees that any continuous function can be arbitrarily approximated by a polynomial function, i.e., the polynomial functions are dense in the continuous functions [117].

The precise way in which we establish the functional calculus is by introducing the so-called *Gelfand homomorphism* $\phi_a$ between a $C^*$-algebra $\mathfrak{A}$ and the algebra of continuous functions defined on the spectrum of an element $a \in \mathfrak{A}$, which is denoted by $C(\sigma(a))$ (the product operation in this vector space is given either by the point-wise product of functions or the convolution between any two functions, since both algebras are isomorphic with the isomorphism being the Fourier transform). Together with the supremum norm $||f||_\infty \equiv \sup_{\lambda \in \sigma(a)} |f(\lambda)|$, this space is indeed a Banach space.

We need the following proposition in order to prove the main theorem of this section:

**Proposition 2.3.1.** *Let $\mathfrak{A}$ be a unital C\*-algebra and $a \in \mathfrak{A}$ a self-adjoint element. Let $p(a) = \sum_{k=0}^{n} \beta_k a^k$ be a polynomial on $a$, in the sense discussed in the previous section (we define $a^0 = \mathbf{1}$). Then, $||p(a)|| = \sup_{\lambda \in \sigma(a)} |p(\lambda)| \equiv ||p||_{\infty}$.*

*Proof.* From the C\*-property, we have that $||p(a)||^2 = ||p(a)^*p(a)||$. On the other hand, using that $a$ is self-adjoint:

$$p(a)^*p(a) = \left(\sum_{k=0}^{n} \beta_k a^k\right)^* \left(\sum_{l=0}^{n} \beta_l a^l\right) = \left(\sum_{k=0}^{n} \overline{\beta_k} a^k\right) \left(\sum_{l=0}^{n} \beta_l a^l\right) = \sum_{k=0}^{n}\sum_{l=0}^{n} \overline{\beta_k}\beta_l a^{k+l} = (\bar{p}p)(a),$$

where $(\bar{p}p)(x) \equiv \overline{p(x)}p(x) = \sum_{k=0}^{n}\sum_{l=0}^{n} \overline{\beta_k}\beta_l x^{k+l}$.

Since $(\bar{p}p)(a)$ is self-adjoint, we can invoke corollary 2.3.2 to write

$$||p(a)^*p(a)|| = ||(\bar{p}p)(a)|| = r((\bar{p}p)(a)) = \sup_{\mu \in \sigma((\bar{p}p)(a))} |\mu| =$$

$$= \sup_{\mu \in \{(\bar{p}p)(\lambda), \lambda \in \sigma(a)\}} |\mu| = \sup_{\lambda \in \sigma(a)} |(\bar{p}p)(\lambda)| =$$

$$= sup_{\lambda \in \sigma(a)} |\overline{p(\lambda)}p(\lambda)| = \sup_{\lambda \in \sigma(a)} |p(\lambda)|^2 = \left(\sup_{\lambda \in \sigma(a)} |p(\lambda)|\right)^2.$$

∎

**Corollary 2.3.1.** *For a self-adjoint element $a \in \mathfrak{A}$, we have*

$$||a^n|| = ||a||^n, \tag{2.3.3}$$

*for all $n \in \mathbb{N} \cup \{0\}$.*

*Proof.* Following the previous proposition and taking $p(a) = a^n$, we have that

$$||a^n|| = \sup_{\lambda \in \sigma(a)} |\lambda^n| = \left(\sup_{\lambda \in \sigma(a)} |\lambda|\right)^n = r(a)^n = ||a||^n.$$

∎

Let $P(\sigma(a))$ be the subspace of $C(\sigma(a))$ consisting of polynomials. Again, the Stone-Weiertrass theorem tells us that $P(\sigma(a))$ is dense in $C(\sigma(a))$. Proposition 2.3.1 tells us that the mapping $\phi_a : P(\sigma(a)) \to \mathfrak{A}$ defined by $\phi(p) = p(a)$ satisfies $||\phi(p)|| = ||p||_{\infty}$. Hence, $\phi_a$ is bounded and the BLT theorem (see appendix) guarantees that we can extend it to the closure of $P(\sigma(a))$ which is given by $C(\sigma(a))$! We denote by $\phi_a(f)$ the limit of a $\phi_a(p)$ such that the polynomials $p$ converge to $f$ in the supremum norm.

**Theorem 2.3.2** (The Gelfand Homomorphism in $C^*$-algebras)**.** *Let $\mathfrak{A}$ be a unital $C^*$-algebra. Let $a \in \mathfrak{A}$ be a self-adjoint element and define the map $\phi \equiv \phi_a : C(\sigma(a)) \to \mathfrak{A}$ as above. Then, the following are true:*

1. *The map $\phi$ is an algebraic $*$-homomorphism:*

$$\phi(\alpha f + \beta g) = \alpha \phi(f) + \beta \phi(g), \ \ \phi(fg) = \phi(f)\phi(g)$$
$$\phi(f)^* = \phi(\overline{f}), \ \ \phi(1) = \mathbf{1} \tag{2.3.4}$$

   *for all $f, g \in C(\sigma(a))$, $\alpha, \beta \in \mathbb{C}$;*

2. *If $f \geq 0$, then $\sigma(\phi(f)) \subset [0, \infty)$;*

3. *If $(f_n)_{n \in \mathbb{N}} \subset C(\sigma(a))$ is a sequence which converges in the norm to $f \in C(\sigma(a))$, then $\phi(f_n)$ converges to $\phi(f)$ in the norm of the algebra. For the converse, if $\phi(f_n)$ converges in the norm of $\mathfrak{A}$, then there exists an $f \in C(\sigma(a))$ such that $\lim_{n \to \infty} \phi(f_n) = \phi(f)$. This means that $\{\phi(f), f \in C(\sigma(a))\}$ is closed in the norm of $\mathfrak{A}$ and hence it is an unital Abelian $C^*$-subalgebra of $\mathfrak{A}$;*

4. *$\sigma(\phi(f)) = \{f(\lambda), \lambda \in \sigma(a)\} \equiv f(\sigma(a))$ for all $f \in C(\sigma(a))$.*

*The $*$-homomorphism $\phi$ is called the Gelfand homomorphism.*

*Proof.* Item 1 can be proved by recognizing that $\phi$ is bounded, hence continuous. The properties are trivially verified for $P(\sigma(a))$ and then extended to $C(\sigma(a))$.

If $f \geq 0$, then $f = g^2$ for some real and continuous function $g$. Since we know $\phi$ is a homomorphism from item 1, then $\phi(f) = \phi(g^2) = \phi(g)^2$. For a real-valued function, $\phi(g)$ is self-adjoint, again from item 1. From the Spectral Mapping Theorem, the spectrum of $\phi(g)^2$ is a subset of the positive line, hence proving item 2.

Convergence of $\phi(f_n)$ to $\phi(f)$ can be verified:

$$||\phi(f_n) - \phi(f)|| = ||\phi(f_n - f)|| = ||f_n - f||_\infty,$$

hence if $||f_n - f||_\infty \to 0$, then $||\phi(f_n) - \phi(f)|| \to 0$. On the other hand, if $\phi(f_n)$ converges in the norm of $\mathfrak{A}$, then $\phi(f_n)$ is a Cauchy sequence in $\mathfrak{A}$, which implies by the equality above that $f_n$ is a Cauchy sequence in the supremum norm. Completeness of $C(\sigma(a))$ implies that this sequence must converge to some $f$, hence $\lim_{n \to \infty} \phi(f_n) = \phi(f)$, thus proving item 3.

Finally, item 4 can be proved by recognizing first that if $\lambda$ is not in the range of $\sigma(a)$ under $f$, then $r \equiv \frac{1}{f - \lambda}$ is continuous. Therefore, $\phi(f)$ is well-defined and we have, from the

properties of item 1:

$$\phi(r)\phi(f-\lambda) = \phi(f-\lambda)\phi(r) = \mathbf{1}.$$

Hence, $\phi(f) - \lambda\mathbf{1}$ is invertible, making $\lambda \in \rho(\phi(f))$, the resolvent set of $\phi(f)$. This establishes that the complement of the range of $f$, i.e.,

$$\mathbb{C} \setminus \{f(\lambda), \lambda \in \sigma(a)\},$$

is a subset of the resolvent set. Therefore, $\sigma(\phi(f)) \subset \{f(\lambda), \lambda \in \sigma(a)\}$.

To show the inclusion in the other way, let $\mu \in \{f(\lambda), \lambda \in \sigma(a)\}$, meaning that $\mu = f(\lambda_0)$ for some $\lambda_0 \in \sigma(a)$. Now, suppose further that $\mu \in \rho(\phi(f))$, implying that $F \equiv \phi(f) - f(\lambda_0)\mathbf{1}$ is invertible. Define $P \equiv \phi(p) - p(\lambda_0)\mathbf{1}$ for some polynomial $p$ such that $||f - p||_\infty < \epsilon$. We have then that $F - P = \phi(f - p) - (f(\lambda_0) - p(\lambda_0))\mathbf{1}$ and

$$||F-P|| \leq ||\phi(f-p)|| + |f(\lambda_0) - p(\lambda_0)| \cdot ||\mathbf{1}|| = ||f-p||_\infty + |f(\lambda_0) - p(\lambda_0)| \leq 2||f-p||_\infty < 2\epsilon.$$

By choosing a small enough $\epsilon$, we can have $||F-P|| \leq ||F^{-1}||^{-1}$, making $P$ invertible in $\mathfrak{A}$, which implies $p(\lambda_0) \notin \rho(\phi(f)) \implies p(\lambda_0) \in \sigma(\phi(f))$. Hence, $\{f(\lambda), \lambda \in \sigma(a)\} \subset \sigma(\phi(f))$, establishing the other inclusion and thus completing the proof for item 4. ∎

### 2.3.3 Spectral Radius

As in the case for operators, the spectrum of an element $u$ of a $C^*$-algebra is given by

$$\sigma(u) = \{\lambda \in \mathbb{C} | \nexists (\lambda\mathbf{1} - u)^{-1}\}. \tag{2.3.5}$$

One can also show that $r(u) \leq ||u||$.

**Definition 2.3.1.** Let $\mathfrak{B}$ be a unital Banach algebra and let $u \in \mathfrak{B}$ with spectrum $\sigma(u)$. The *spectral radius* of $u$ is defined by

$$r(u) \equiv \sup_{\lambda \in \sigma(u)} |\lambda|.$$

The main result we will use in this Thesis is the following theorem due to Beurling:

**Theorem 2.3.3** (Spectral Radius Theorem)**.** *Let $\mathfrak{B}$ be a unital Banach algebra and let $u \in \mathfrak{B}$. Then:*

$$r(u) = \inf_{n \geq 1} ||u^n||^{1/n} = \lim_{n \to \infty} ||u^n||^{1/n}. \tag{2.3.6}$$

*Proof.* See [106] for a proof using the Uniform Boundedness (Banach-Steinhaus) Theorem or [25] and [111] for an alternative proof. ∎

**Corollary 2.3.2.** *If $\mathfrak{A}$ is a unital C\*-algebra and $a \in \mathfrak{A}$ is self-adjoint, then*

$$r(a) = ||a||.$$

*Proof.* Using the $C^*$-property, we can say that a self-adjoint element $a$ satisfies $||a^2|| = ||a||^2$.

Consider the self-adjoint element given by $a^{2n-1}$. Then, we have

$$||a^{2n}|| = ||a^{2n-1}||^2 = ||a^{2n-2}||^{2^2} = ... = ||a||^{2n}.$$

Hence, using equation (2.3.6) and setting $n = 2^m$:

$$r(a) = \lim_{n\to\infty} ||a^n||^{1/n} = \lim_{n\to\infty} ||a^{2^m}||^{1/2^m} = \lim_{n\to\infty} ||a|| = ||a||.$$

∎

One can show that this theorem is valid in the more general case of *normal* elements, i.e., the elements of the algebra satisfying $aa^* = a^*a$. Notice that every self-adjoint element is normal.

### 2.3.4   The Positive Cone

The spectrum of a element of a $C^*$-algebra allows us to define an important class of elements, which are called *positive elements*. The reason for their importance is that they have a natural *order relation*, which is useful when one is interested in making quantitative comparisons. Indeed, these elements will become important in our discussion of Bell's inequalities in $C^*$-algebras.

**Definition 2.3.2.** An element $a$ of a $*$-algebra $\mathfrak{A}$ is defined to be *positive* if it is self-adjoint and its spectrum $\sigma(a)$ is a subset of $\mathbb{R}_+ = [0, \infty)$. The set of all positive elements of $\mathfrak{A}$ is denoted as $\mathfrak{A}_+$.

**Proposition 2.3.2.** *If $a$ and $b$ are self-adjoint and positive elements of a unital C\*-algebra such that $a + b = 0$, then $a = -b = 0$.*

*Proof.* If $\sigma(a) \subset [0, \infty)$, then the Spectral Mapping Theorem guarantees that $\sigma(-a) \subset (-\infty, 0]$. Hence, if $b = -a$, then $\sigma(b) \subset (-\infty, 0]$. If $b$ is positive, this implies that its spectrum is trivial, i.e., $\sigma(b) = \{0\}$. As a consequence, $r(b) = 0$ from corollary 2.3.2 and thus $||b|| = 0$. Hence, $a = -b = 0$. ∎

An important topological property of the positive elements is that they form a *convex cone* named the *positive cone of the C\*-algebra* (hence the name of this section).

A *cone* $C$ is a subset of a vector space over a field $\mathbb{K}$ such that for all $x \in C$, we have that $\alpha x \in C$, for a *positive scalar* $\alpha$. A *convex cone* is a cone closed under linear combinations with positive coefficients, i.e., $\alpha x + \beta y \in C$ for $x, y \in C$ and positive scalar $\alpha, \beta$.

**Theorem 2.3.4.** *The set $\mathfrak{A}_+$ defined above is a convex cone which is closed in the topology of $\mathfrak{A}$ and satisfies $\mathfrak{A}_+ \cap (-\mathfrak{A}_+) = \{0\}$.*

*Proof.* See [25]. ∎

From this theorem, it is easy to see that for $a, b, z \in \mathfrak{A}_+$, we have that $a + b \in \mathfrak{A}_+$ and if $-z^*z \in \mathfrak{A}_+$, then $z = 0$.

We now prove an important theorem which provides an alternative definition for a positive element. First, we state without proof the following important lemma:

**Lemma 2.3.2.** *If $a$ and $b$ are self-adjoint, positive elements of a unital C\*-algebra $\mathfrak{A}$ such that*

$$ab = ba,$$

*then $ab$ is also self-adjoint and positive.*

*Proof.* See [111]. ∎

**Theorem 2.3.5.** *If $a$ is a self-adjoint, positive element of a unital C\*-algebra $\mathfrak{A}$, then the following statements are equivalent:*

1. *$\sigma(a) \subset [0, ||a||]$,*

2. *$\left\| \mathbf{1} - \frac{a}{||a||} \right\| \leq 1$,*

3. *There exists a unique self-adjoint, positive element $b \in \mathfrak{A}$ such that $b^2 = a$.*

*One usually denotes $b \equiv \sqrt{a}$ and, appropriately, calls it the square-root of $a$.*

*Proof.* We first show that $(1) \implies (2)$. This follows from the Spectral Mapping Theorem:

$$\sigma \left( \mathbf{1} - \frac{a}{||a||} \right) = \left\{ \mathbf{1} - \frac{\lambda}{||a||}, \lambda \in \sigma(a) \right\} \subset \left\{ \mathbf{1} - \frac{\lambda}{||a||}, \lambda \in [0, ||a||] = [0, 1] \right\}.$$

Hence, from corollary 2.3.2:

$$\left|\left|\mathbf{1} - \frac{a}{||a||}\right|\right| = r\left(\mathbf{1} - \frac{a}{||a||}\right) \leq 1.$$

Next, we show that $(2) \implies (3)$. One can show that such $b$ is explicitly given by

$$b = ||a||^{1/2}\left[\sum_{n=0}^{\infty} c_n \left(\mathbf{1} - \frac{a}{||a||}\right)^n\right], \tag{2.3.7}$$

where $c_0 = 1$, $c_1 = -\frac{1}{2}$ and $c_n = -\frac{(2n-3)!!}{(2n)!!}$ (see [25], theorem 2.2.10).

Finally, we show $(3) \implies (1)$. Since $b$ is self-adjoint, then the $C^*$-property implies that $||b||^2 = ||b^2|| = ||a||$ and again we invoke the Spectral Mapping Theorem. ∎

In light of theorem 2.3.5, we will use the two definitions of positive elements (making reference to the spectrum or to the element $b \in \mathfrak{A}$) interchangeably.

The following corollary and proposition are important in the context of Bell's inequalities, since they justify the use of $C^*$-algebras to treat more general problems as a two-level system, as we will see in Part III.

**Corollary 2.3.3.** *Let $\mathfrak{A}$ be a unital $C^*$-algebra and $b \in \mathfrak{A}$ self-adjoint satisfying $||b|| \leq 1$. Then, there exists an self-adjoint, positive element $y \in \mathfrak{A}$ such that $y^2 = \mathbf{1} - b$. Such element will be denoted as $\sqrt{\mathbf{1} - b}$.*

*Proof.* Let $w = \mathbf{1} - b$. By the triangle inequality, $||\mathbf{1} - w|| = ||b|| \leq 1$. Hence, from theorem 2.3.5, there exists a unique self-adjoint and positive element $y$ such that $y^2 = w = \mathbf{1} - b$. ∎

**Proposition 2.3.3.** *Every self-adjoint element $a$ of a unital $C^*$-algebra $\mathfrak{A}$ can be written as $a_+ - a_-$, where $a_{\pm}$ are themselves self-adjoint and positive. We also have that $a_{\pm}$ commute with $a$ and satisfy $a_+ a_- = a_- a_+ = 0$.*

*Proof.* Consider the real-valued functions $f_{\pm}(\lambda) \equiv \frac{1}{2}(|\lambda| \pm \lambda)$. It is easy to see that both functions are continuous, positive and satisfy $f_+ f_- = 0$ and $\lambda = f_+(\lambda) - f_-(\lambda)$. Using the Gelfand homomorphism $\phi_a$, we define

$$a_{\pm} \equiv \phi_a(f_{\pm}). \tag{2.3.8}$$

The Gelfand Homomorphism Theorem guarantees that these elements have the desired properties. ∎

The final theorem of this section gives us an important criteria do characterize and construct positive elements.

**Theorem 2.3.6.** *Every element of the form $x^*x$ in a unital $C^*$-algebra $\mathfrak{A}$ is positive.*

*Proof.* Let $v = x^*x$. This element is easily seen to be self-adjoint. From proposition 2.3.3, we can write $v = v_+ - v_-$ where $v_\pm$ are self-adjoint and positive satisfying the properties in that proposition.

Let $w = xv_-$. Then, we have that $-w^*w = -v_-x^*xv_- = -v_-(v_+ - v_-)v_- = (v_-)^3$. Since $v_-$ is positive, then $(v_-)^3$ is also positive. Hence, $-w^*w$ is positive, which implies that $w = 0$, implying that $xv_- = 0$. Multiplying this expression by $x^*$ on the left, we have that $0 = x^*xv_- = (v_+ - v_-)v_- = -(v_-)^2$. Since $v_-$ is self-adjoint, the $C^*$-property implies that $||v_-||^2 = ||(v_-)^2|| = 0$. Therefore, $x^*x = v_+$, which is positive by construction. ■

Combining the previous theorem with theorem 2.3.5, we conclude that a self-adjoint element $a \in \mathfrak{A}$ is positive if, and only if, there exists an $x \in \mathfrak{A}$ such that $x^*x = a$.

The importance of this construction is that we can impose an *order relation* between the elements of a $C^*$-algebra, which will prove useful once we start working with inequalities in $C^*$-algebras (in particular, the $C^*$-algebraic form of Bell's inequalities).

Recall that a *partial order* in a set $X$ is a relation $\preceq$ between any two elements $a$ and $b$ such that we have

1. Reflexivity: $a \preceq a$, for all $a \in X$;

2. Transitivity: if $a \preceq b$ and $b \preceq c$, then $a \preceq c$;

3. Antissymetry: if $a \preceq b$ and $b \preceq a$, then $a = b$.

This notion is, of course, a generalization of the order relation between real numbers. A partial order which extends to all the elements of the set is called a *total order*.

Positive elements of a $C^*$-algebra yield an order relation in the following way: if $a$ and $b$ are self-adjoint elements of $\mathfrak{A}$, then $a \leq b$ if $b - a \in \mathfrak{A}_+$. Naturally, a positive element is such that $a \geq 0$. Notice that this is a natural generalization of the ordering of real numbers, in the sense that the difference of any two successive real numbers is a positive real number.

For an involutive algebra, we define a *congruence transformation* to be

$$\mathfrak{A} \ni a \mapsto c^*ac \in \mathfrak{A}, \tag{2.3.9}$$

for some fixed $c \in \mathfrak{A}$. An important feature of the partial order relation defined above is that it is invariant under congruence transformations:

**Proposition 2.3.4.** *Let $\mathfrak{A}$ be a unital C\*-algebra and let $a$ and $b$ be self-adjoint elements such that $a \geq b$. Then,*

$$c^* a c \geq c^* b c, \forall c \in \mathfrak{A}.$$

*Proof.* By definition, if $a \geq b$, then $a - b$ is positive. Hence, from theorem 2.3.5, there is an element $d \in \mathfrak{A}$ such that $a - b = d^* d$. Hence,

$$c^*(a - b)c = c^*(d^* d)c = (dc)^*(dc) \in \mathfrak{A}_+.$$

Therefore, $c^*(a - b)c \geq 0 \implies c^* a c \geq c^* b c.$     ∎

Finally, some of the important properties implied by the partial order introduced in this section can be stated in the following proposition:

**Proposition 2.3.5.** *Let $\mathfrak{A}$ be a unital C\*-algebra and let $a$ and $b$ be self-adjoint elements. Then,*

1. *$a \geq 0 \implies \|a\|\mathbf{1} \geq a \geq 0$;*

2. *$a \geq b \geq 0 \implies \|a\| \geq \|b\|$;*

3. *$a \geq 0 \implies a\|a\| \geq a^2 \geq 0$.*

*Proof.* For item 1, we recall from theorem 2.3.5 that if $a \geq 0$ then $\sigma(a) \subset [0, \|a\|]$. The Spectral Mapping Theorem then tells us that

$$\sigma(\|a\|\mathbf{1} - a) = \{\|a\| - \lambda, \lambda \in \sigma(a)\} \subset [0, \|a\|].$$

Hence, $\|a\|\mathbf{1} - a \geq 0$, which proves item 1.

For item 2, we use item 1 to infer that $\|a\|\mathbf{1} \geq a \geq b \geq 0 \implies \|a\|\mathbf{1} \geq b$. The Spectral Mapping Theorem then tells us that

$$\sigma(\|a\|\mathbf{1} - b) \subset [0, \infty],$$

which is possible if, and only if $\|a\| \geq \sup\{\lambda, \lambda \in \sigma(b)\} = \|b\|$.

Finally, for item 3, we invoke the Spectral Mapping Theorem again to write for $a \geq 0$

$$\sigma\left(\left(a - \frac{\|a\|}{2}\mathbf{1}\right)^2\right) = \left\{\left(\lambda - \frac{\|a\|}{2}\mathbf{1}\right)^2, \lambda \in \sigma(a)\right\} \subset \left\{\left(\lambda - \frac{\|a\|}{2}\mathbf{1}\right)^2, \lambda \in [0, \|a\|]\right\} \subset \left[0, \frac{\|a\|^2}{4}\right].$$

Hence, the spectral radius is given by $r\left(\left(a - \frac{||a||}{2}\mathbf{1}\right)^2\right) \le \frac{||a||^2}{4}$, implying that $\left|\left|\left(a - \frac{||a||}{2}\mathbf{1}\right)^2\right|\right| \le \frac{||a||^2}{4}$. From item 1, $\frac{||a||^2}{4}\mathbf{1} \ge \left(a - \frac{||a||}{2}\mathbf{1}\right)^2$. By expanding the right hand side, we find $0 \ge a^2 - a||a||$, hence proving item 3. ∎

## 2.4   Representations and States

In practice, we are interested in using the features of the abstract elements of such algebras to describe the properties of more concrete elements, such as operators acting on a Hilbert space. This is the motivating point to introduce the notion of a *representation*.

In Quantum Physics, one is often interested in attributing numerical values to observables which can be measured in an experimental setting. Since Quantum Physics has an intrinsic uncertainty, the best one can often do is obtain expected values. Such values will be incorporated in the theory by means of linear functionals from which take positive values on the positive elements of the algebra. These functionals are dubbed *states*.

### 2.4.1   Representations

**Definition 2.4.1.** A *∗-morphism* between two ∗-algebras $\mathfrak{A}$ and $\mathfrak{B}$ is a mapping $\pi : \mathfrak{A} \ni a \to \pi(a) \in \mathfrak{B}$ such that

1. $\pi(\alpha a + \beta b) = \alpha\pi(a) + \beta\pi(b)$, for all $\alpha, \beta \in \mathbb{C}$;

2. $\pi(ab) = \pi(a)\pi(b)$;

3. $\pi(a^*) = \pi(a)^*$.

A ∗-morphism is called a ∗-*monomorphism* if it is injective (one-to-one), a ∗-*epimorphism* if it is surjective (onto) and a ∗-*isomorphism* if it is both injective (monic) and surjective (epic)[1].

We note that a ∗-morphism is a ∗-isomorphism if, and only if

$$\ker \pi \equiv \{a \in \mathfrak{A}, \pi(a) = 0\} = \{0\}. \tag{2.4.1}$$

Note that properties 1 and 2 are those of any structure-preserving morphism between algebras. Property 3 is exclusive to algebras with a involutive structure.

An important property common to all ∗-morphisms is that they are automatically continuous. This is illustrated by the following lemma:

---

[1]   The adjectives *monic* and *epic* are commonly used to describe morphisms in Category Theory. We already employ them here since we will make extensive use of Category Theory latter in this Thesis.

**Lemma 2.4.1.** *Let* $\mathfrak{A}$ *and* $\mathfrak{B}$ *be two C\*-algebras and* $\pi : \mathfrak{A} \to \mathfrak{B}$ *a* $*$*-morphism. It follows that*

1. *$a \geq 0 \implies \pi(a) \geq 0$, that is, $\pi$ is* positivity preserving*;*

2. *$||\pi(a)|| \leq ||a||, \forall a \in \mathfrak{A}$, that is, $\pi$ is bounded, hence continuous.*

*As a consequence, the range of $\pi$ is itself a C\*-algebra.*

*Proof.* To prove 1, we recall that from theorem 2.3.5, we know that if $a \geq 0$, then there exists some $b \in \mathfrak{A}$ such that $b^*b = a$. Hence, properties 1 and 2 of definition 2.4.1 imply that

$$\pi(a) = \pi(b^*b) = \pi(b)^*\pi(b) \geq 0.$$

From proposition 2.3.5, we have that

$$||\pi(a)||^4 = ||\pi(a^*a)||^2 \leq ||\pi(a^*a)||\,||a^*a|| = ||\pi(a)||^2||a||^2,$$

which implies that $||\pi(a)|| \leq ||a||$, thus proving item 2.  ∎

With these concepts in mind, we now introduce the proper definition of a representation in the context of C\*-algebras.

**Definition 2.4.2.** A *representation of a C\*-algebra* $\mathfrak{A}$ is a pair $(\mathcal{H}, \pi)$ consisting of a complex Hilbert space $\mathcal{H}$, deemed the *representation space*, and a $*$-morphism $\pi : \mathfrak{A} \to \mathfrak{B}(\mathcal{H})$. If $\pi$ is a $*$-isomorphism, i.e., $\ker \pi = \{0\}$, then $(\mathcal{H}, \pi)$ is said to be a *faithful representation*.

We mention *en passant* that given a representation $(\mathcal{H}, \pi)$, one can obtain a faithful representaion by taking the quotient $\mathfrak{A}_\pi \equiv \mathfrak{A}/\ker \pi$. Since faithful representations are the most important ones, we give a criteria for faithfulness:

**Proposition 2.4.1.** *Let* $(\mathcal{H}, \pi)$ *be a representation of the C\*-algebra* $\mathfrak{A}$*. This representation is faithful if, and only if, it satisfies the following equivalent conditions:*

1. $\ker \pi = \{0\}$*;*

2. *$||\pi(a)|| = ||a||, \forall a \in \mathfrak{A}$;*

3. *$\pi(a) > 0, \forall a > 0$.*

*Proof.* See [25], proposition 2.3.3.  ∎

A very important concept which is important in the context of QFT is that of a *cyclic representation*. This representation involves a specific vector $\Omega$ in the representation space which is cyclic, meaning that the set

$$\{A\Omega, A \in \mathcal{M} \subset \mathfrak{B}(\mathcal{H})\} \tag{2.4.2}$$

is dense in $\mathcal{H}$.

Acting with an operator on a cyclic vector allows one to obtain a new vector of the Hilbert space with arbitrary precision. In fact, the celebrated *Reeh-Schlieder Theorem* [112, 4] states that the vacuum state is cyclic for the algebra of observables (more precisely, for the representation of such algebra in a Hilbert space) of a QFT. Physically, this means that any state (in the Physics sense) can be constructed from the vacuum state by the action of local operators [142]. The use of the word "local" here is important since the Hilbert space where $\Omega$ lives may not be the full Hilbert space of the theory, thus opening the possibility of superselection sectors.

**Definition 2.4.3.** A *cyclic representation of a C\*-algebra* $\mathfrak{A}$ is a triple $(\mathcal{H}, \pi, \Omega)$ consisting of a representation $(\mathcal{H}, \pi)$ together with a vector $\Omega \in \mathcal{H}$ which is cyclic for $\pi$.

## 2.4.2   States

As mentioned before, we will be interested in functionals over the $C^*$-algebra. In particular, we will be interested in *states*, which are positive and normalized functionals, as defined below.

We should emphasize the difference between the concept of a "state" in the Physics literature, which generally refers to a vector of a Hilbert space[2]. This vector is then associated with a "state" which the physical system of interest finds itself in. In our case, "state" will have the same meaning as an "expectation value". In fact, the Riesz Representation Theorem B.5.1 makes this connection more explicit for a broader class of functionals.

Let us denote the space of linear functionals over a $C^*$-algebra $\mathfrak{A}$ (that is, the *dual space*) by $\mathfrak{A}^*$. We can define a norm in this space by

$$||f|| = \sup_{a \in \mathfrak{A}, ||a||=1} \{|f(a)|\}, \tag{2.4.3}$$

for $f \in \mathfrak{A}^*$.

---

[2]   Some texts call the vector a "state vector", a notation we find appropriate.

**Definition 2.4.4.** A linear functional $\omega$ over a $C^*$-algebra (or more generally, a $*$-algebra) is defined to be *positive* if

$$\omega(a^*a) \geq 0, \forall a \in \mathfrak{A}. \tag{2.4.4}$$

A positive linear functional $\omega$ over $\mathfrak{A}$ with

$$||\omega|| = 1 \tag{2.4.5}$$

is called a *state*. If

$$\omega(a^*a) > 0, \forall a \neq 0, \tag{2.4.6}$$

then $\omega$ is said to be a *faithful state*.

The set of all states over $\mathfrak{A}$ is denoted by $E_{\mathfrak{A}}$.

One can show that positivity entails continuity ([25], proposition 2.3.11). Since theorem 2.3.6 tells us that every positive element of a $C^*$-algebra is of the form $a^*a$, then an equivalent definiton of a positive functional is that of it being positive on positive elements.

An important tool for dealing with positive elements is a form of the Cauchy-Schwarz inequality for positive states:

**Lemma 2.4.2** (Cauchy-Schwarz Inequality). *Let $\omega$ be a positive linear functional over the $*$-algebra $\mathfrak{A}$. It follows that*

1. $\omega(a^*b) = \overline{\omega(b^*a)}$;

2. $|\omega(a^*b)|^2 \leq \omega(a^*a)\omega(b^*b)$, *for all $a, b \in \mathfrak{A}$.*

*Proof.* The proof follows from the fact that positivity of $\omega$ implies

$$\omega((\lambda a + b)^*(\lambda a + b)) \geq 0.$$

Since the functional is linear, we can write

$$|\lambda|^2 \omega(a^*a) + \overline{\lambda}\omega(a^*b) + \lambda\omega(b^*a) + \omega(b^*b) \geq 0.$$

For this equation to be positive, we need both items 1 and 2. ∎

In the same spirit of the ordering by positive elements in the $C^*$-algebra, there is a natural ordering for the dual space induced by the positive states. We say that $\omega_1 \geq \omega_2$ if $\omega_1 - \omega_2$ is positive. In this case, we also say that $\omega_1$ *majorizes* $\omega_2$.

Not all linear combinations of states result in a new state. This is due to the normalization condition (2.4.5).

**Definition 2.4.5.** Let $\omega_1$ and $\omega_2$ be two states over a $C^*$-algebra $\mathfrak{A}$. A *convex combination* of these states is such that

$$\omega = \lambda\omega_1 + (1-\lambda)\omega_2, \qquad (2.4.7)$$

where $\lambda \in (0,1)$.

It is easy to see that $\omega$ defined in equation (2.4.7) is also a state. This state has the property of majorizing both $\lambda\omega_1$ and $(1-\lambda)\omega_2$ and also multiples of both $\omega_1$ and $\omega_2$. The state $\omega$ is usually referred to as a *mixed state*, and its connection with Quantum and Statistical Mechanics is direct [16].

However, one can obtain states which are not obtained from convex combinations of pairs of states. This yields the natural definition of a *pure state*:

**Definition 2.4.6.** A state $\omega$ overa a $C^*$-algebra $\mathfrak{A}$ is said to be *pure* if the only positive linear functionals majorized by $\omega$ are of the form $\lambda\omega$ with $\lambda \in [0,1]$. Hence, $\omega$ cannot be written as a convex combination of other states.

The set of all pure states over $\mathfrak{A}$ is denoted by $P_{\mathfrak{A}}$.

One can define at least two topologies over the sets $E_{\mathfrak{A}}$ and $P_{\mathfrak{A}}$. These will become important in the context of the Bicommutant Theorem for Von Neumann algebras, and we leave this discussion for then.

To finish this section, we prove that there are always non-trivial states over a unital $C^*$-algebra[3].

**Proposition 2.4.2.** *Let $\mathfrak{A}$ be a unital $C^*$-algebra. Then, for all $a \in \mathfrak{A}$, there exists a state $\omega_a$ over $\mathfrak{A}$ such that*

$$\omega_a(a^*a) = ||a||^2. \qquad (2.4.8)$$

*Proof.* Let $a \in \mathfrak{A}$ and define the linear subset $V_1 \equiv \{\alpha\mathbf{1} + \beta a^*a, \alpha, \beta \in \mathbb{C}\}$. It is easy to see that this subset is normal, meaning its elements comute with their adjoints.

We define the linear functional $\phi_a : V_1 \to \mathbb{C}$ by

$$\phi_a(\alpha\mathbf{1} + \beta a^*a) \equiv \alpha + \beta||a^*a|| = \alpha + \beta||a||^2,$$

where the last equality follows from the $C^*$-property.

---

[3]  The proof for $C^*$-algebras that have no unit is a bit more involved, and the interested reader is referred to [25].

Corollary 2.3.2 tells us that

$$||\alpha\mathbf{1} + \beta a^*a|| = r(\alpha\mathbf{1} + \beta a^*a) = \sup\{|\alpha\mathbf{1} + \beta\lambda|, \lambda \in \sigma(a^*a)\} \geq |\alpha\mathbf{1} + \beta||a||^2|.$$

Hence:

$$|\phi_a(\alpha\mathbf{1} + \beta a^*)| \leq |\alpha + \beta||a^*a||| \leq ||\alpha\mathbf{1} + \beta a^*a||,$$

which establishes that $\phi_a \leq 1$ and, therefore, bounded. By definition, $\phi_a(\mathbf{1}) = 1$, which implies that $||\phi_a|| = 1 = \phi_a(\mathbf{1})$.

From the Hahn-Banach Theorem, there exists a bounded extension for $\phi_a$ given by $\omega_a : \mathfrak{A} \to \mathbb{C}$ such that $||\omega_a|| = ||\phi_a|| = 1$. This implies that $\omega_a(\mathbf{1}) = 1$, and thus $\omega_a$ is continuous satisfying $||\omega_a|| = \omega_a(\mathbf{1}) = 1$. $\omega_a$ is positive if and only if it is continuous. Therefore, it is a state over $\mathfrak{A}$. Since $\omega_a$ is an extension of $\phi_a$, we have that $\omega_a(a^*a) = \phi_a(a^*a) = ||a||^2$, thus proving the proposition. ∎

## 2.5   The GNS Construction

In this section, we will develop the tools to prove the following crucial theorem about $C^*$-algebras.

**Theorem 2.5.1.** *Let $\mathfrak{A}$ be a $C^*$-algebra. It follows that $\mathfrak{A}$ is isomorphic to a norm-closed self-adjoint algebra of bounded operators on a Hilbert space.*

This theorem has already been alluded to when we discussed the use of the algebraic approach in Quantum Field Theory in Curved Spacetimes. This theorem implies that we can make general statements about a theory without committing ourselves to a specific model associated to a Hilbert space and to states defined therein. Then, we can use this theorem to obtain a concrete Hilbert space where we can perform calculations.

The interesting part about this theorem is that in spite being a theorem that guarantees the existence of an isomorphism, the proof for it involves constructing a *cyclic representation* of the algebra in a Hilbert space. This is the main content of what is called the *GNS construction*, which we now present. The name is after Gelfand, Naimark and Segal. The proof is rather long but extremely important (we follow the exposition on both [25, 89]).

**Theorem 2.5.2** (GNS Construction Theorem). *Let $\omega$ be a state over a $C^*$-algebra $\mathfrak{A}$. Then, one can construct a Hilbert space $\mathcal{H}_\omega$ and a $*$-representation $\pi_\omega$ of $\mathfrak{A}$ by bounded operators on $\mathcal{H}_\omega$ such that $\pi_\omega(a^*) = \pi_\omega(a)^\dagger$.*

*If $\mathfrak{A}$ is unital, then there exists a cyclic vector $\Omega$ for the representation $\pi_\omega$ such that*

$$\omega(a) = \langle \Omega, \pi_\omega(a)\Omega \rangle_{\mathcal{H}_\omega}. \tag{2.5.1}$$

*The triple $(\mathcal{H}_\omega, \pi_\omega, \Omega)$ is referred to as the GNS triple associated to $(\mathfrak{A}, \omega)$ or the cyclic representation of $(\mathfrak{A}, \omega)$. This representation is faithful if $\omega$ is a faithful state and unique up to unitary equivalence.*

*Proof.* We start by defining a inner product on $\mathfrak{A}$, making it a inner product space. The next step would then be to show completeness.

From the state $\omega$, one can naively define

$$\langle a, b \rangle \equiv \omega(a^*b), \tag{2.5.2}$$

for all $a, b \in \mathfrak{A}$. However, this product allows for the existence of elements $n \in \mathfrak{A}$ such that $\langle n, n \rangle = 0 \not\Rightarrow n = 0$. Hence, this sesquilinear form is not positive-definite, hence it can't be a inner product over $\mathfrak{A}$.

On the other hand, we can construct a quotient space from $\mathfrak{A}$ such equation [2.5.2](#) indeed defines an inner product and we make such space into a Hilbert space by adding completeness.

Denote the set

$$\mathcal{N} \equiv \{n \in \mathfrak{A}, \omega(n^*n) = 0\}. \tag{2.5.3}$$

Let $\mathcal{N}_1 \equiv \{n \in \mathfrak{A}, \omega(b^*n) = 0, \forall b \in \mathfrak{A}\}$. We show that $\mathcal{N} = \mathcal{N}_1$. From the Cauchy-Schwarz inequality

$$|\omega(b^*n)|^2 \leq \omega(b^*b)\omega(n^*n).$$

Hence, if $n \in \mathcal{N}$, then $\omega(b^*n) = 0$ for all $b \in \mathcal{N}$. From this, we conclude that $\mathcal{N} \subset \mathcal{N}_1$. On the other hand, let $n' \in \mathcal{N}_1$. Then, $\omega(b^*n') = 0$ for all $b$. In particular, choosing $b = n'$, we have that $\omega((n')^*n') = 0$, and hence $n' \in \mathcal{N}$, implying $\mathcal{N}_1 \subset \mathcal{N}$. Both inclusions imply that $\mathcal{N} = \mathcal{N}_1$.

Now, we show that $\mathcal{N}$ is a *closed* linear subspace of $\mathfrak{A}$. This will allow us to eventually take the quotient of $\mathfrak{A}$ by $\mathcal{N}$, thus allowing us to get rid of the problem of non-positivity definetess.

Let $m, n \in \mathcal{N}$ and $\alpha, \beta \in \mathbb{C}$ . Then, for every $b \in \mathfrak{A}$, we have that $\omega(b^*m) = \omega(b^*n) = 0$. Hence,

$$\omega(b^*(\alpha m + \beta n)) = \alpha\omega(b^*m) + \beta\omega(b^*n) = 0.$$

Therefore, $\alpha m + \beta n \in \mathcal{N}$. Let $(n_i)_{i \in \mathbb{N}}$ be a sequence in $\mathcal{N}$ that converges to $n \in \mathfrak{A}$. Since $\omega$ is continuous, we have

$$\omega(b^*n) = \lim_{i \to \infty} \omega(b^*n_i) = 0,$$

proving that $n \in \mathcal{N}$. Thus, $\mathcal{N}$ is closed.

Now, we can construct the quotient space $\mathfrak{A}/\mathcal{N}$. This space is comprised by the equivalence classes

$$[a] = \{a + n, n \in \mathcal{N}\}.$$

Naturally, the zero vector is given by the equivalence class $[0] = \{n, n \in \mathcal{N}\} = \mathcal{N}$. We thus take the inner product (2.5.2) to be

$$\langle [a], [b] \rangle = \omega(a^*b). \tag{2.5.4}$$

This expression is well-defined in the sense that it is not dependent on the representative of the equivalence class. This can be seen by changing $a \to a + n$:

$$\omega((a + n)^*b) = \omega(a^*b) + \omega(n^*b) = \omega(a^*b).$$

Equation (2.5.4) is a *bona fide* inner product since $\langle [a], [a] \rangle = \omega(a^*a) = 0$ if, and only if $[a] \in \mathcal{N} \implies [a] = [0]$.

Completeness of $\mathfrak{A}/\mathcal{N}$ follows from the canonical completion of metric spaces (see, for instance, [93], chapter 1). Hence, the Hilbert space is given by $\mathcal{H}_\omega = \widetilde{\mathfrak{A}/\mathcal{N}}$.

Next, we construct the representation $\pi_\omega$. From the canonical completion, we can consider $\mathfrak{A}/\mathcal{N}$ to be dense in $\widetilde{\mathfrak{A}/\mathcal{N}}$. Define

$$\pi_\omega(a)[z] = [az], \; z \in \mathfrak{A}. \tag{2.5.5}$$

This definition is well-defined because $\mathcal{N}$ is a left ideal of $\mathfrak{A}$: for $n \in \mathcal{N}$ and $a, b \in \mathfrak{A}$ we have

$$\omega(b^*(an)) = \omega((a^*b)^*n) = 0,$$

following the positivity of the state and the Cauchy-Schwarz inequality. So, equation (2.5.5) is well-defined since $a(z + n) = az + an$ and since $an \in \mathcal{N}$ (because it is a left-ideal), it follows that $[a(z + n)] = [az]$. It is easy to see that operator defined on (2.5.5) is linear and an algebraic morphism.

What we need to show now is that indeed $\pi_\omega(a)$ is a bounded operator on the Hilbert space we constructed. Consider $[z] \in \mathfrak{A}/\mathcal{N}$. Then

$$||\pi_\omega(a)[z]||^2 = ||[az]||^2 = \langle [az], [az] \rangle = \omega((az)^*(az)) =$$
$$= \omega(z^*(a^*a)z) = \frac{\omega(z^*(a^*a)z)}{\omega(z^*z)}\omega(z^*z) = \frac{\omega(z^*(a^*a)z)}{\omega(z^*z)}||[z]||^2.$$

On the other hand, the expression

$$\phi(a) \equiv \frac{\omega(z^*az)}{\omega(z^*z)} \tag{2.5.6}$$

defines a state on $\mathfrak{A}$, since it is positive

$$\phi(c^*c) = \frac{\omega(z^*(c^*c)z)}{\omega(z^*z)} = \frac{\omega((cz)^*(cz))}{\omega(z^*z)} \geq 0,$$

and $\phi(\infty) = \frac{\omega(z^*z)}{\omega(z^*z)} = 1$.

Thus, we have that $||\phi|| = 1$ and that $|\phi(c)| \leq ||\phi|| \, ||c|| \leq ||c||$, for all $c \in \mathfrak{A}$. In light of this, we have

$$||\pi_\omega(a)[z]||^2 = \phi(a^*a)||[z]||^2 \leq ||\phi|| \, ||a^*a|| \, ||[z]||^2 = ||a^*a|| \, ||[z]||^2 = ||a||^2 \, ||[z]||^2,$$

which implies that $||\pi_\omega(a)|| \leq ||a||$, thus making $\pi_\omega(a)$ a bounded operator on the dense subspace $\mathfrak{A}/\mathcal{N}$. From the BLT theorem B.2.5, we can extend $\pi_\omega(a)$ to all the Hilbert space. It is easy to that $\pi_\omega$ is linear and an algebraic morphism. What is left to show is that it is indeed a $*$-morphism.

For $[a], [b] \in \mathfrak{A}/\mathcal{N}$, we have

$$\langle [x], \pi_\omega(a^*)[y] \rangle = \langle [x], [a^*y] \rangle = \omega(x^*a^*y) = \omega((ax)^*y) =$$
$$= \langle [ax], [y] \rangle = \langle \pi_\omega(a)[x], [y] \rangle = \langle [x], \pi_\omega(a)^\dagger[y] \rangle.$$

Hence, $\pi_\omega(a^*) = \pi_\omega(a)^\dagger$.

If $\mathfrak{A}$ is unital, define

$$\Omega \equiv [\mathbf{1}]. \tag{2.5.7}$$

Now:

$$\langle \Omega, \pi_\omega(a)\Omega \rangle = \langle [\mathbf{1}], \pi_\omega(a)\mathbf{1} \rangle = \langle [\mathbf{1}], [a\mathbf{1}] \rangle = \omega(\mathbf{1}a) = \omega(a).$$

We also have that $\{\pi_\omega(a)\Omega, a \in \mathfrak{A}\} = \{[a], a \in \mathfrak{A}\} = \mathfrak{A}/\mathcal{N}$, implying that $\Omega$ is a cyclic vector.

We now show that the representation $(\mathcal{H}_\omega, \pi_\omega)$ is faithful. In order to do so, we invoke the first item of proposition 2.4.1. Assuming that $\omega$ is faithful and $\pi_\omega(a) = 0$, we have, for all $b \in \mathfrak{A}$:

$$0 = \langle \pi_\omega(a)\pi_\omega(b)\Omega, \pi_\omega(a)\pi_\omega(b)\Omega \rangle = \omega((ab)^*ab),$$

by the definition of the scalar product in the representation space. Since the state is faithful, we have that $ab = 0$ for all $b$. In particular, $aa^* = 0$, which implies that $a = 0$. Hence, $\ker \pi_\omega = \{0\}$ and the representation is faithful by proposition 2.4.1.

The last step in the proof is to show uniqueness.

Let $(\mathcal{H}'_\omega, \pi'_\omega, \Omega')$ be a second GNS triple associated to the pair $(\mathfrak{A}, \omega)$. Define an operator $U$ such that

$$U\pi_\omega(a)\Omega = \pi'_\omega(a)\Omega'. \tag{2.5.8}$$

This expression is well-defined since it preserves the scalar product:

$$\langle U\pi_\omega(a)\Omega, U\pi_\omega(b)\Omega \rangle = \langle \pi'_\omega(a)\Omega, \pi'_\omega(b)\Omega \rangle$$
$$= \omega(a^*b) = \langle \pi_\omega(a)\Omega, \pi_\omega(b)\Omega \rangle.$$

By taking $a = \mathbf{1}$ in equation (2.5.8), we get that

$$U\Omega = \Omega'. \tag{2.5.9}$$

From that same expression, we obtain

$$\langle U^*U\pi_\omega(a)\Omega, \pi_\omega(b)\Omega \rangle = \langle \pi_\omega(a)\Omega, \pi_\omega(b)\Omega \rangle$$

Which implies that $U^*U = \mathbf{1}$. Hence, $U$ is unitary. By the definition in equation (2.5.8), $U$ is only defined on a dense subset of $\mathcal{H}_\omega$. However, one can extend it to the hole Hilbert space by invoking the BLT theorem and this extension is easily seen to be unitary.

Finally, by inserting an identity in

$$\langle U\pi_\omega(a)\mathbf{1}\Omega, U\pi_\omega(b)\Omega \rangle = \langle U\pi_\omega(a)U^{-1}U\Omega, U\pi_\omega(b)\Omega \rangle = \pi'_\omega(A)\Omega'$$

and using expression (2.5.9), we get

$$\pi'_\omega(a) = U^{-1}\pi_\omega(a)U, \tag{2.5.10}$$

thus proving that the two cyclic representations can differ at most by a unitary transformation. ∎

Sometimes, this construction is also referred as the *canonical representation of $\mathfrak{A}$ associated with $\omega$*, to emphasize its importance.

An interesting consequence of this construction is connected to *automorphisms* of a $C^*$-algebra. A *$*$-automorphism* $\tau$ of a $\mathfrak{A}$ is a $*$-isomorphism of $\mathfrak{A}$ to itself. Since $\tau$ is an isomorphism, it is invertible and it follows that each $*$-automorphism of $\mathfrak{A}$ is norm preserving, i.e.:

$$||\tau(a)|| = ||a||, \ \forall a \in \mathfrak{A}. \tag{2.5.11}$$

**Corollary 2.5.1.** *Let $\omega$ be a state over $\mathfrak{A}$. Suppose there is a $*$-automorphism $\tau$ which leaves $\omega$ invariant, i.e.:*

$$\omega(\tau(a)) = \omega(a), \ \forall a \in \mathfrak{A}. \tag{2.5.12}$$

*Then, there exists a unique unitary operator $U_\omega$ on the Hilbert space of the GNS representation $(\mathcal{H}_\omega, \pi_\omega, \Omega)$ such that*

$$U_\omega \pi_\omega(a) U_\omega^{-1} = \pi_\omega(\tau(a)), \ \forall a \in \mathfrak{A}$$

*and that leaves the cyclic vector invarinat:*

$$U_\omega \Omega = \Omega.$$

*Proof.* This corollary follows trivially from the uniqueness of the GNS representation applied to $(\mathcal{H}_\omega, \pi_\omega \circ \tau, \Omega)$ with $(\pi_\omega \circ \tau)(a) = \pi_\omega(\tau(a))$. ∎

Another important consequence of the GNS construction is related to the *irreducibility* of the representation which infers that $\omega$ is a pure state. A set of operators $\mathcal{M} \subset \mathfrak{B}(\mathcal{H})$ is said to be *irreducible* if the only closed subspaces of $\mathcal{H}$ which are invariant under $\mathcal{M}$ are the trivial subspaces $\{0\}$ and $\mathcal{H}$. A representation $(\mathcal{H}, \pi)$ of a $C^*$-algebra is said to be irreducible if the set $\pi(\mathfrak{A})$ is irreducible on $\mathcal{H}$. Naturally, a set of operators/representation which is not irreducible is said to be reducible.

**Proposition 2.5.1.** *Let $\mathcal{M} \subset \mathfrak{B}(\mathcal{H})$ be a self-adjoint set of bounded operators. The following conditions are equivalent:*

1. *$\mathcal{M}$ is irreducible;*

2. *The commutant of $\mathcal{M}$, written as $\mathcal{M}'$ and consisting of the set of all operators that commute with all operators in $\mathcal{M}$ consists only of multiples of the identity operator;*

3. *Every non-zero vector $\psi \in \mathcal{H}$ is cyclic for $\mathcal{M}$ in $\mathcal{H}$ unless $\mathcal{M} = \{0\}$ and $\mathcal{H} = \mathbb{C}$.*

*Proof.* See [25], proposition 2.3.8. ∎

**Theorem 2.5.3.** *Let $\omega$ be a state over a $C^*$-algebra $\mathfrak{A}$ and $(\mathcal{H}_\omega, \pi_\omega, \Omega)$ its GNS representation. The following conditions are equivalent:*

1. *$(\mathcal{H}_\omega, \pi_\omega)$ is irreducible;*

2. *$\omega$ is a pure state.*

*Proof.* We first prove that $1 \implies 2$. We prove the contrapositive, i.e., the negation of 2 implies the negation of 1. Assuming thus that 2 is false, then there are other elements that majorize $\omega$ which are not of the form $\lambda\omega$. Hence, there is a positive functional $\rho$ such that

$$\rho(a^*a) \leq \omega(a^*a),$$

for all $a \in \mathfrak{A}$. By means of the Cauchy-Schwarz inequality, we have:

$$|\rho(b^*a)|^2 \leq \rho(b^*b)\rho(a^*a) \leq \omega(b^*b)\omega(a^*a)$$
$$= ||\pi_\omega(b)\Omega||^2 \, ||\pi_\omega(a)\Omega||^2.$$

Hence, we can define the functional $(\pi_\omega(b)\Omega, \pi_\omega(a)\Omega) \mapsto \rho(b^*a)$ on $\mathcal{H}_\omega \otimes \mathcal{H}_\omega$. Since each copy of the Hilbert space is dense in the tensor product, this functional is densely defined on $\mathcal{H}_\omega \otimes \mathcal{H}_\omega$. It is also easily seen to be bounded and sesquilinear. From Riesz's Theorem, there is a unique bounded operator $T$ on $\mathcal{H}_\omega$ such that

$$\langle \pi_\omega(b)\Omega, T\pi_\omega(a)\Omega \rangle = \rho(b^*a).$$

It follows trivially that since $\rho$ is not a multiple of $\omega$ than $T$ can't be a multiple of the identity. Furthermore, positivity of $\rho$ implies that

$$0 \leq \rho(a^*a) = \langle \pi_\omega(a)\Omega, T\pi_\omega(a)\Omega \rangle$$
$$\leq \omega(a^*a) = \langle \pi_\omega(a)\Omega, \pi_\omega(a)\Omega \rangle.$$

From this, it follows that $0 \leq T \leq \mathbf{1}$. On the other hand,

$$\langle \pi_\omega(b)\Omega, T\pi_\omega(c)\pi_\omega(a)\Omega \rangle = \rho(b^*ca) = \rho((c^*b)^*a)$$
$$= \langle \pi_\omega(b)\Omega, \pi_\omega(c)T\pi_\omega(a)\Omega \rangle,$$

implying that $T$ is in the commutant $\pi_\omega(\mathfrak{A})'$ of $\pi_\omega(\mathfrak{A})$. From proposition 2.5.1, it follows that the representation is reducible, and 1 is false.

Now, we show that $2 \implies 1$. The proof is again by the contrapositive, hence we assume 1 is false. Hence, again by proposition 2.5.1, the commutant $\pi_\omega(\mathfrak{A})'$ contains elements which are not multiples of the identity. If $T \in \pi_\omega(\mathfrak{A})'$, then $T^* \in \pi_\omega(\mathfrak{A})'$ and so are the combinations $T + T^*$ and $(T - T^*)/i$. Hence, there is selfadjoint element $S \in \pi_\omega(\mathfrak{A})'$ which is not a multiple of the identity. There is then a spectral projector $P$ with $0 < P < \mathbf{1}$ contained in the commutant.

We now consider the functional given by

$$\rho(a) = \langle P\Omega, \pi_\omega(a)\Omega \rangle.$$

We verify that it is positive:

$$\rho(a^*a) = \langle P\Omega, \pi_\omega(a^*a)\Omega \rangle = \langle P\pi_\omega(a)\Omega, P\pi_\omega(a)\Omega \rangle \geq 0,$$

where we have also used the self-adjointness and idempotency of the projector $P$.

Now, $\omega$ majorizes $\rho$:

$$\omega(a^*a) - \rho(a^*a) = \langle \pi_\omega(a)\Omega, (\mathbf{1} - P)\pi_\omega(a)\Omega \rangle \geq 0.$$

Since $\rho$ is not a multiple of $\omega$, condition 2 must be false. ∎

## 2.6  Matrix Algebras

In this section, we present a concrete application of the constructions mentioned in this chapter, based on [15]. For this, we will work with matrices. Recall that for finite dimensional Hilbert spaces, the space of bounded operators $\mathfrak{B}(\mathcal{H})$ coincides with the space of finite-dimensional square matrices (as elucidated in section B.2.3.3). We start with general constructions for the $n \times n$ dimensional complex matrices, $\mathrm{Mat}(n, \mathbb{C})$ and then explicitly apply the GNS construction to the case of $\mathrm{Mat}(2, \mathbb{C})$.

We first note that $\mathrm{Mat}(n, \mathbb{C})$ is a $C^*$-algebra with the involution operation given by $(M^*)_{ij} = \overline{M_{ji}}$ or, as it is usually denoted in Physics textbooks as the *conjugate transpose*, $M^\dagger$. The inner product is given by trace operation:

$$\langle A, B \rangle_2 \equiv \mathrm{Tr}(A^*B). \tag{2.6.1}$$

That this is a scalar product which makes $\mathrm{Mat}(n, \mathbb{C})$ into a Hilbert space can be easily verified (see, for instance, [111] or [93]).

### 2.6.1  Density Matrices

The positive elements of $\mathrm{Mat}(n, \mathbb{C})$ are given by the matrices which have non-negative eigenvalues. The self-adjoint, positive matrices $\rho$ with unit trace are called *density matrices*. The set of density matrices is easily seen to be a convex set, in the same spirit as the positive elements of a $C^*$-algebra (the astute reader may have already expect this).

From the cyclic property of the trace, $\mathrm{Tr}(AB) = \mathrm{Tr}(BA)$, one can easily see that for a density matrix $\rho$ and a unitary matrix $U$, the matrix $U^*\rho U$ is also unitary.

The *Spectral Theorem for Self-Adjoint Matrices*[4] yields the following representation for a denstity matrix:

$$\rho = \sum_{k=1}^{n} \varrho_k P_k, \tag{2.6.2}$$

where $\varrho_k$ are the eigenvalues of $\rho$ (whose sum is 1 following positivity and the unit trace of $\rho$) and $P_k$ are the spectral projectors in the one-dimensional subspaces spanned by the eigenvectors of $\rho$. Each of these projectors satisfy $P_k = P_k^*$, they are mutually orthogonal $P_k P_l = \delta_{kl} P_k$ and $\sum_{k=1}^{n} P_k = \mathbf{1}$.

## 2.6.2   States

**Proposition 2.6.1.** *Let $\rho$ be a density matrix in $Mat(n, \mathbb{C})$. It follows that every state $\omega$ on $Mat(n, \mathbb{C})$ is given by*

$$\omega(A) = Tr(\rho A). \tag{2.6.3}$$

*This is a one-to-one identification, and we can thus identify the set of states on $Mat(n, \mathbb{C})$ with the set of density matrices on $Mat(n, \mathbb{C})$.*

*Proof.* Consider the vector space $\mathrm{Mat}(n, \mathbb{C})$ with the inner product given by equation (2.6.1). The norm induced by this inner product is given by

$$||A||_2 = \sqrt{\mathrm{Tr}(A^*A)}. \tag{2.6.4}$$

Because $\mathrm{Mat}(n, \mathbb{C})$ is a finite dimensional vector space, all norms are equivalent (see appendix B). Hence, the norm given by (2.6.4) is equivalent to the operator norm on $\mathfrak{B}(\mathcal{H})$ since both spaces are isomorphic to each other. This equivalence means that continuity on one norm implies continuity on the other.

Let $\omega$ be a state on $\mathrm{Mat}(n, \mathbb{C})$. By the Riesz Representation Theorem, we can guarantee that $\omega$ is of the form

$$\omega(A) = \langle \rho, A \rangle_2 = \mathrm{Tr}(\rho^* A),$$

for some matrix $\rho \in \mathrm{Mat}(n, \mathbb{C})$. We now show that $\rho$ is a density matrix.

The normalization condition of the state $\omega(\mathbf{1}) = 1$ implies that $\mathrm{Tr}(\rho^*) = 1$. Positivity of the state, on the other hand, implies that $\mathrm{Tr}(\rho^* A^* B) = \overline{\mathrm{Tr}(\rho^* B^* A)}$ and for any matrix $M$ we have that $\overline{\mathrm{Tr}(M)} = \mathrm{Tr}(M^*)$. Hence

$$\mathrm{Tr}(\rho^* A^* B) = \mathrm{Tr}(A^* B \rho) = \mathrm{Tr}(\rho A^* B),$$

---

[4]   See any book on Linear Algebra on finite dimensional vector spaces.

where we used the cyclic property of the trace. For $A = \mathbf{1}$, we have that $\text{Tr}(\rho^* B) = \text{Tr}(\rho B)$, from where we easily see that $\rho = \rho^*$. Thus, $\rho$ is self-adjoint and has unit trace. All we need to show is that it is also a positive matrix.

Let $\rho = \sum_{k=1}^{n} \varrho_k P_k$ be the spectral decomposition of the matrix $\rho$. From the positivity of the state, we have

$$0 \leq \omega(A^* A) = \text{Tr}(\rho A^* A) = \sum_{k=1}^{n} \varrho_k \text{Tr}(P_k A^* A).$$

Since this is valid for any $A \in \text{Mat}(n, \mathbb{C})$, we pick $A$ to be a projection matrix $P_j$, from which we conclude that $\text{Tr}(P_k A^* A) = \delta_{jk}$, which implies that the eigenvalues $\varrho_k$ are positive. Hence, $\rho$ is a density matrix.

The last step in the theorem is to show that the representation in (2.6.3) is unique. In order to do this, suppose there are two distinct density matrices $\rho$ and $\rho'$ such that $\text{Tr}(\rho A) = \text{Tr}(\rho' A)$, for all $A$. Then, we have that $0 = \text{Tr}((\rho - \rho')A) = \langle (\rho - \rho'), A \rangle_2$, which implies that $\rho = \rho'$, proving uniqueness. ∎

This form of the states gives rise to the probabilistic interpretation of states in the context of Quantum Physics. Indeed, if we consider a self-adjoint matrix $A$ with spectral decomposition $\sum_{k=1}^{n} \alpha_k P_k$, we have that for a density matrix $\rho$:

$$\text{Tr}(\rho A) \equiv \omega_\rho(A) = \sum_{k=1}^{n} \alpha_k \text{Tr}(\rho P_k) \equiv \sum_{k=1}^{n} \alpha_k \mathcal{P}_k^{\rho, A},$$

where we have defined $\mathcal{P}_k^{\rho, A} = \text{Tr}(\rho P_k)$. The properties of the projection operators imply that $\mathcal{P}_k^{\rho, A} \geq 0$ and that $\sum_{k=1}^{n} \mathcal{P}_k^{\rho, A} = 1$, which are the expected properties of a probability distribution on the spectrum (composed of eigenvalues) of the matrix $A$. Hence the expression

$$\omega_\rho(A) = \sum_{k=1}^{n} \alpha_k \mathcal{P}_k^{\rho, A} \tag{2.6.5}$$

gives rise to the interpretation of the states $\omega_\rho$ as an average on the spectrum of the matrix $A$, which is precisely the definition we give to an expectation value in Quantum Mechanics!

We can also describe the pure and mixed states in the context of matrices [16]. The only pure states on $\mathfrak{B}(\mathcal{H}) = \text{Mat}(n, \mathbb{C})$ are of the form

$$\omega(A) = \text{Tr}(PA), \tag{2.6.6}$$

where $P$ is an orthogonal projector on a one-dimensional subspace.

### 2.6.3   The GNS Construction in $\text{Mat}(n, \mathbb{C})$

We now present the realization of the GNS construction in terms of matrices.

The ingredients we need are an algebra, which we already have, and a state over this algebra. As we saw in the previous subsection, the set of states in $\text{Mat}(n, \mathbb{C})$ coincides with the set of density matrices. Since a density matrix $\rho$ is, by definition, self-adjoint, there is a unitary transformation realized by a matrix $V$ which leaves it in diagonal form with its eigenvalues on the principal diagonal:

$$V^* \rho V = D_\rho = \begin{pmatrix} \varrho_1 & & \\ & \ddots & \\ & & \varrho_k \end{pmatrix}. \tag{2.6.7}$$

Next, we define the inner product by

$$\langle A, B \rangle_\rho = \omega_\rho(A^* B) = \text{Tr}(\rho A^* B), \tag{2.6.8}$$

which is just the inner product in the GNS construction, equation (2.5.2). The set $\mathcal{N}$ as defined in the construction is given by

$$\begin{aligned} \mathcal{N} &= \{ N \in \text{Mat}(n, \mathbb{C}), N\rho^{1/2} = 0 \} \\ &= \{ N \in \text{Mat}(n, \mathbb{C}), \ker(N) \supset \text{ran}(\rho^{1/2}) \}, \end{aligned} \tag{2.6.9}$$

where $\rho^{1/2} = V D_\rho^{1/2} V^*$ is obtained by writing the matrix $D_\rho$ with the square-roots of the eigenvalues, i.e.:

$$D_\rho^{1/2} = \begin{pmatrix} \sqrt{\varrho_1} & & \\ & \ddots & \\ & & \sqrt{\varrho_k} \end{pmatrix}. \tag{2.6.10}$$

The fact that we need to consider the square-roots of the density matrix comes from recognizing that $\langle A, A \rangle_\rho = \text{Tr}((A\rho^{1/2})^* A\rho^{1/2})$. It is easy to show that $\rho^{1/2} \rho^{1/2} = \rho$.

We now define the orthogonal projection operator over $\text{ran}(\rho^{1/2})$, $P_\rho$. Hence, $\text{ran}(P_\rho) = \text{ran}(\rho^{1/2})$. If $u = v + w$ with $v \in \text{ran}(\rho^{1/2})$ and $w \in (\text{ran}(\rho^{1/2}))^\perp$, then $P_\rho u = v$. From the definition, we see that $P_\rho \rho^{1/2} = \rho^{1/2}$, and $\rho^{1/2}$ being self-adjoint implies that $\rho^{1/2} = \rho^{1/2} P_\rho$. A useful consequence of this fact is that $P_\rho \rho P_\rho = \rho$.

This definition of the projection operator allows us to recast the set $\mathcal{N}$ in a convenient equivalent way:

$$\mathcal{N} = \{ N \in \text{Mat}(n, \mathbb{C}), N P_\rho = 0 \}.$$

In order to define the quotient space $\mathrm{Mat}(n, \mathbb{C})/\mathcal{N}$, we notice that the equivalence classes are given by $[A] = \{A + N, N \in \mathcal{N}\}$ and $A \sim B \iff A - V \in \mathcal{N}$. An equivalent way to state the equivalence relation is that $A \sim B \iff AP_\rho = BP_\rho$. In fact, if $A \sim B$, then $(A - B)P_\rho = 0$, because $A - B = N \in \mathcal{N}$. The converse is trivially seen to be true. Hence

$$\mathrm{Mat}(n, \mathbb{C})/\mathcal{N} = \{AP_\rho, A \in \mathrm{Mat}(n, \mathbb{C})\}.$$

The scalar product can then be defined from equation (2.6.8) as

$$\langle AP_\rho, BP_\rho \rangle_\rho = \mathrm{Tr}(\rho A^* B) = \omega_\rho(A^* B),$$

by a straightforward computation. That $\mathrm{Mat}(n, \mathbb{C})/\mathcal{N}$ is a Hilbert space with this scalar product follows from the fact that $\mathrm{Mat}(n, \mathbb{C})$ is a finite-dimensional Hilbert space under the operator norm[5] and that projection operators are bounded.

Next, we define a representation $\pi_\rho$ of $\mathrm{Mat}(n, \mathbb{C})$ on $\mathrm{Mat}(n, \mathbb{C})/\mathcal{N}$ by

$$\pi_\rho(A)BP_\rho = (AB)P_\rho, \tag{2.6.11}$$

for $A, B \in \mathrm{Mat}(n, \mathbb{C})$. That $P_\rho$ is a cyclic vector for $\pi_\rho$ can be seen from

$$\{\pi_\rho(A)P_\rho, A \in \mathrm{Mat}(n, \mathbb{C})\} = \{AP_\rho, A \in \mathrm{Mat}(n, \mathbb{C})\} = \mathrm{Mat}(n, \mathbb{C})/\mathcal{N}.$$

Finally, we define $\Omega_\rho \equiv \mathbf{1}P_\rho$, which yields

$$\langle \Omega_\rho, \pi_\rho(A)\Omega_\rho \rangle_\rho = \mathrm{Tr}(\rho A) = \omega_\rho(A).$$

Hence, we have a representation of the state $\omega_\rho$ in terms of a vector $\Omega_\rho \in \mathrm{Mat}(n, \mathbb{C})/\mathcal{N}$.

---

[5]   Recall that for finite dimensions all norms are equivalent, see appendix B.

# 3 von Neumann Algebras

*In Mathematics, you don't understand things. You just get used to them.*

John von Neumann

In this chapter, we introduce the notion of a specific algebra of bounded operators on a Hilbert space, which is known as a von Neumann algebra. They were introduced and extensively studied in the seminal papers by Murray and von Neumann in the 1930's [107]. In particular, they classified a specific subset of von Neumann algebras which are called *factors*, to be defined below. These factors are classified as being of the (not so creative) types I, II or III. Type I factors can be further subdivided depending on the dimension of the Hilbert space, hence being either type $I_d$, for a countable dimension $d$ or type $I_\infty$. Type III factors can also be subdivided by means of the Tomita-Takesaki modular theory, to be introduced in the next chapter.

Factors can be understood as the building blocks for more general von Neumann algebras, using a construction called *direct integration*. Hence, it makes sense for one to study more deeply these algebras. On another hand, factors have shown themselves to be deeply connected with the universal structure of the algebras in QFT [69, 34]. A particular type of factor deemed *type* $III_1$ is believed to be isomorphic to any consistent algebraic QFT.

## 3.1 A Physicist's Point of View

The classification of factors due to Murray and von Neumann is a challenging subject. Because we have relegated the discussion of fields to the second part of this Thesis, we shall encourage the reader by presenting already some intuition and results before delving into the theorems and more mathematically precise statements. For a nice review of the subject tailored to physicists, we recommend [142] and for a pedagogical exposition of type $III_1$ factors in the context of QFT we recommend [145].

A von Neumann algebra is a subalgebra of the algebra of bounded operators acting on a Hilbert space. They are defined via an algebraic relation which implies the important *Bicommutant Theorem*: this algebra is closed under different locally convex topology. This is a useful fact because, among these topologies, there is the *topology generated by the vector states*: we say that a sequence of operators $(A_n)_{n\in\mathbb{N}}$ converges to $A$ in this topology if for any pair of vectors $\psi, \varphi \in \mathcal{H}$

$$\lim_{n\to\infty} |\langle\psi, A\varphi\rangle - \langle\psi, A_n\varphi\rangle| = 0. \tag{3.1.1}$$

In a more mathematical approach, this is called *weak convergence*, and it is related to the *weak topology*, generated by the vectors on the Hilbert space. We will define this properly below, but we already discuss the physical importance of this property of von Neumann algebras.

Since what we effectively measure in a lab are expectation values of observables, we would like to have a notion of convergence which reflects the limitations of our measurements. What is interesting is that in a von Neumann algebra, weak convergence *is equivalent* to convergence in other topologies. Hence, we do not loose information by considering this topology instead of others. This is a motivation to consider von Neumann algebras as the algebra of observables in a theory.

Moreover, this kind of convergence allows us to give a notion of entropy in QFT. As we will discuss in the next chapter and towards the end of this Thesis, one can define the notion of *relative entropy* which has the properties that one would expect of a entropy measure (monotonicity, positivity). For local algebras, such entropy is given by

$$\mathcal{S}_{\omega,\omega'}(\mathcal{U}) = \langle\Omega, \ln\Delta_{\omega,\omega'}\Omega\rangle, \tag{3.1.2}$$

where $\mathcal{U}$ is the region of spacetime where the algebra is defined, $\Omega$ is the vacuum state and the operator $\Delta_{\omega,\omega'}$ is called the *relative modular operator*, to be discussed in the context of the Tomita-Takesaki modular theory, in the next chapter. Naturally, one would expect a convergence of such functional when one considers a net of algebras indexed by different regions of spacetime, which one could interpret as measurements of the relative entropy for states on algebras at different times.

With the motivation for von Neumann being (hopefully) clear, we will be interested in the algebras which are called *factors*. For an algebra $\mathfrak{M}$, we define its *commutant* $\mathfrak{M}'$ as the set of operators which commute with those in $\mathfrak{M}$. The *center* of the algebra is then defined as

$$Z(\mathfrak{M}) = \mathfrak{M} \cap \mathfrak{M}'. \tag{3.1.3}$$

For an algebra with a unit, the center contains at least the multiples of the identity. If these are the only elements of $Z(\mathfrak{M})$, then we say that the center is *trivial*. A factor is a von Neumann algebra with trivial center. These algebras are, in a sense, the fundamental building blocks of more general algebras. This was one of the motivations behind the work of Murray and von Neumann in classifying them [107].

Both Murray and von Neumann were motivated by mathematical and physical problems to present their classification of factors. From the physical standpoint, a problem they addressed was the division of a quantum mechanical system into two subsystems [145]. In the simplest case, one factorizes the Hilbert space $\mathcal{H}$ as a tensor product of the two subsystems:

$$\mathcal{H} = \mathcal{H}_1 \otimes \mathcal{H}_2. \tag{3.1.4}$$

The observables of one subsystem are then ascribed to the algebra $\mathfrak{M} = \mathcal{B}(\mathcal{H}_1) \otimes \mathbf{1}$ while the observables of the other subsystem are ascribed to the commutant, which can be written naturally as $\mathfrak{M}' = \mathbf{1} \otimes \mathcal{B}(\mathcal{H}_2)$. Hence, we obtain the factorization given by

$$\mathcal{B}(\mathcal{H}) = \mathcal{B}(\mathcal{H}_1) \otimes \mathcal{B}(\mathcal{H}_2). \tag{3.1.5}$$

Notice that in this case, $Z(\mathfrak{M}) = \mathfrak{M} \cap \mathcal{M}' = \mathbf{1}$. This kind of factorization of the algebra (based on the tensor product of the underlying Hilbert space) is what is called a *type I case*, in the terminology of Murray and von Neumann. One then calls $\mathfrak{M}$ a *type I factor*.

The main characteristic of type I factors is that it contains *minimal projections*, meaning that for some vector $\psi \in \mathcal{H}_1$, one can construct a projection onto the one-dimensional subspace spanned by such vector. Such projection is constructed from the projection given by $E_\psi = |\psi\rangle \langle \psi|$, in physicist's notation and then extended to the algebra by

$$\mathfrak{M} \ni E = E_\psi \otimes \mathbf{1}. \tag{3.1.6}$$

It is important to notice that one can always define such projection operator, but it may or may not be contained in the von Neumann algebra, dependingo on its type.

These factors are the ones that arise in the Quantum Mechanics with finite degrees of freedom. It was found in the studies of the foundations of QFT that the algebras that describe infinite degrees of freedom cannot be of type I, as it is illustrated by the so-called *Fermi* gedankenexperiment, discussed in the next chapter.

The type which von Neumman was apparently more interested in was what is called *type II factor*. This factor is characterized by having no minimal projections, but every non-zero projection $E$ has a *subprojection $F$*, denoted as $F < E$ (the order relation will be discussed bellow) which is finite in the sense that for a partial isometry $W$, if $WW^* = F$ and $W^*W = F'$, then $F = F'$. For a long time, these factors were believed to be of more mathematically than physically interesting, but more recent work has shown its importance in approaches to quantum gravity [40].

The last factor in the Murray-von Neumann classification is the so-called *type III factor*. These are the factor which arise in QFT, and they are characterized by again the lack of minimal projections and for any projection $E \in \mathfrak{M}$, there exists a $W \in \mathfrak{M}$ (which is *not* a partial isometry) such that

$$\begin{aligned} W^*W &= \mathbf{1}, \\ WW^* &= E, \end{aligned} \tag{3.1.7}$$

that is, $W$ maps $\mathfrak{M}$ isometrically in $E\mathcal{H}$.

One can further classify type III factors by means of the Tomita-Takesaki modular theory, which was proposed by Alain Connes [48]. This classification produces a continuum of type $III_\lambda$ factors with $\lambda \in [0, 1]$. In particular, the type $III_1$ is the most interesting from the physical standpoint for the following reasons:

1. The spectrum of the modular operator is the whole real line;

2. Up to unitary equivalence, there is only one type $III_1$ *hyperfinite* (generated by an increasing family of finite-dimensional subalgebras) factor.

The second of these reasons can already be clarified: the uniqueness fixes the general structure of the local algebras that can occur in general QFTs [34].

We now proceed into making the statements presented in this section precise. We follow the expositions in [25, 89, 118, 130].

## 3.2 Elementary Definitions

We begin with the algebraic definition of a von Neumann algebra. We will consider a separable Hilbert space $\mathcal{H}$ and a subalgebra $\mathfrak{M}$ of $\mathcal{B}(\mathcal{H})$.

**Definition 3.2.1.** The *commutant* of $\mathfrak{M}$ is the subset $\mathfrak{M}'$ of $\mathcal{B}(\mathcal{H})$ which commutes with all elements of $\mathfrak{M}$, that is

$$\mathfrak{M}' = \{B \in \mathcal{B}(\mathcal{H}), AB = BA, \forall A \in \mathfrak{M}\}. \tag{3.2.1}$$

The commutant $\mathfrak{M}'$ is trivially non-empty since it contains at least the identity. One can also show that $\mathfrak{M}'$ is a Banach subalgebra of $\mathcal{B}(\mathcal{H})$: if $(B_j)_{j \in \mathbb{N}}$ is a sequence in $\mathfrak{M}'$ which converges to some element $B \in \mathcal{B}(\mathcal{H})$ in the uniform topology, then for each $A \in \mathfrak{M}$ we have that $BA - AB = (B - B_j)A - A(B - B_j)$, for all $j \in \mathbb{N}$. Hence, $||BA - AB|| \leq 2||B - B_j||\,||A|| \to 0$ for large $j$. This establishes that $BA - AB = 0$, which implies that $B \in \mathfrak{M}'$.

Since one can regard the commutant as a subalgebra itself of $\mathcal{B}(\mathcal{H})$, then we can take its own commutant. This gives rise to the notion of a *double commutant* or *bicommutant*: $\mathfrak{M}'' \equiv (\mathfrak{M}')'$. From the definition, it is easy to see that $\mathfrak{M} \subset \mathfrak{M}''$.

**Definition 3.2.2.** A *von Neumann algebra* is a $*$-subalgebra $\mathfrak{M}$ of $\mathcal{B}(\mathcal{H})$ satisfying

$$\mathfrak{M} = \mathfrak{M}''. \tag{3.2.2}$$

It is important to notice that we can continue taking the commutant of the commutants which still generates a subalgebra of $\mathcal{B}(\mathcal{H})$. For instance, $(\mathfrak{M}')''$ is by definition $((\mathfrak{M}')')'$ which in turn is easily seen to coincide with $(\mathfrak{M}'')'$, which is denoted as $\mathfrak{M}'''$ (the *triple commutant*). In general, $(\mathfrak{M}^{(n)})^{(m)} = \mathfrak{M}^{(n+m)}$.

Now, one can easily see that $\mathfrak{M}' = \mathfrak{M}'''$ by the following argument: since $\mathfrak{M}$ is at least a subset of $\mathfrak{M}''$, then we can take the commutant on both sides of the expression $\mathfrak{M} \subset \mathfrak{M}'$ to get $\mathfrak{M}''' \subset \mathfrak{M}'$. On the other hand, $\mathfrak{M}' \subset (\mathfrak{M}')'' \equiv \mathfrak{M}'''$, establishing that $\mathfrak{M} = \mathfrak{M}'''$. An iteration of this idea leads us to the following equalities:

$$\begin{aligned}
\mathfrak{M} \subset \mathfrak{M}'' &= \mathfrak{M}^{(4)} = \mathfrak{M}^{(6)} = ... \\
\mathfrak{M}' = \mathfrak{M}''' &= \mathfrak{M}^{(5)} = \mathfrak{M}^{(7)} = ...
\end{aligned} \tag{3.2.3}$$

Now we present the important definition of a *factor*. As we stated in the previous section, this is the most important type of von Neumann algebras since all von Neumann algebras can be constructed as a direct integral of these algebras. The work of von Neumann and Murray on the classification of factors will be presented in a couple of sections.

**Definition 3.2.3.** A von Neumann algebra $\mathfrak{M}$ is said to be a *factor* if

$$\mathfrak{M} \cap \mathfrak{M}' = \mathbb{C}\mathbf{1}, \tag{3.2.4}$$

that is, the intersection between the algebra and its commutant (sometimes denoted by $\mathcal{Z}(\mathfrak{M})$ and referred to as the *center*) comprises of multiples of the identity (which is sometimes rephrased as saying that the center is trivial).

Of course that $\mathcal{B}(\mathcal{H})$ is itself a von Neumann algebra and it is easily seen to also be a factor. Trivially, the set $\mathbb{C}\mathbf{1}$ is also a factor.

Every subset $\mathcal{N} \subset \mathcal{B}(\mathcal{H})$ is contained in at least one von Neumann algebra, $\mathcal{B}(\mathcal{H})$ itself. We can then define $\mathfrak{M}[\mathcal{N}]$ as the von Neumann algebra *generated* by the set $\mathcal{N}$. This algebra is constructed by taking the intersection of all the von Neumann algebras which contain $\mathcal{N}$, and in a sense $\mathfrak{M}[\mathcal{N}]$ is the "smallest" von Neumann algebra which contains $\mathcal{N}$.

## 3.3 Locally Convex Topologies

There are some reasons to consider more general and weaker topologies on $\mathcal{B}(\mathcal{H})$ than the operator topology, defined by the operator norm

$$||A|| = \sup_{\psi \in \mathcal{H}} \frac{||A\psi||_{\mathcal{H}}}{||\psi||_{\mathcal{H}}}.$$

Some of these reasons include the fact that $\mathcal{B}(\mathcal{H})$ is non-separable in this topology, which stops one from using some interesting results from Functional Analysis such as the use of a Schauder basis [93]. Another motivation is that one may be interested in defining a topology for a set of operators where the condition of positive-definetness of the norm is not desired. Such is the case for the space of distributions, for example.

In the following section, we provide, for completeness, a introductory discussion about locally convex spaces in a more general form. This section can thus be skimmed over by the reader more interested in the topologies relevant to the bicommutant theorem.

### 3.3.1 Locally Convex Spaces

A generalization of the notion of a topology induced by a norm is that of a *locally convex topology*. The origin of the name stems from one of the two equivalent ways of thinking about this topology. We present both of them here for completeness, but the latter is the one we find more useful for our discussion (the reader may wish then to skip to definition 3.3.2 if they are not interested in more general definitions).

Let us begin with some basic definitions. We will assume previous knowledge of the basic definitions of topology and topological vector spaces (TVS) (see appendix A).

**Definition 3.3.1.** A *locally convex vector space* (LCVS) is a TVS whose local base consists of balanced, convex sets.

We use an equivalent characterization for an LCVS using a *family of separating seminorms*. One can show that such a family generates a locally convex topology and vice versa [118].

**Definition 3.3.2.** A *seminorm* on a vector space $X$ is a function $p : X \to \mathbb{R}$ such that

1. $p(x + y) < p(x) + p(y)$, for all $x, y \in X$;

2. $p(\lambda x) = |\lambda| p(x)$ for all $x \in X$ and $\lambda \in \mathbb{K}$.

The astute reader may wish to compare this definition with the definition of a norm (contained in definition B.1.1). One clearly sees that a seminorm is a norm if it also satisfies $p(x) \neq 0$ if $x \neq 0$. In order to get a notion of distances, convergences and, more generally, of a "topology", we require the following definiton:

**Definition 3.3.3.** A family $\mathscr{P}$ of seminorms on $X$ is said to be *separating* if to each $x \neq 0$ corresponds at least one $p \in \mathscr{P}$ with $p(x) \neq 0$.

The term "separating" is indicative of the fact that the family $\mathscr{P}$ allows one to distinguish between two elements. Recall that in a metric space, if the distance between two elements is zero, then they must be the same. Dropping the requirement that $p(x) = 0 \iff x = 0$ results in one not being able to distinguish two elements by the same definition. However, with a whole separating family, we can *separate* the points (hence, a vague notion of distance).

The two definitions are then shown to be equivalent by the following theorem which we shall not prove.

**Theorem 3.3.1.** *With each separating family of seminorms on $X$, we can associate a locally convex topology $\tau$ on $X$ and vice versa: every locally convex topology is generated by some family of separating seminorms.*

*Proof.* See [118]. ■

### 3.3.2 Topologies on $\mathcal{B}(\mathcal{H})$

In this section, we enumerate some of the locally convex topologies one can define on $\mathcal{B}(\mathcal{H})$. These topologies are "weaker" or "finer" than the uniform topology, in the sense that uniform convergence implies convergence in all these topologies, but not the converse. We begin by the weakest or "coarsest" topology, which recieves the unflattering name of *weak topology*, while the second topology we will present receives the uninventive name of *strong* topology. The weak topology turns out to be the most important locally convex topology we will consider, and the reader in a hurry can feel free to just become acquainted with this definition. The strong topology will be used here to present one of the most common formulations of the Bicommutant theorem.

**Definition 3.3.4.** The *weak topology* on $\mathcal{B}(\mathcal{H})$ is the topology induced by the family of seminorms $\{p_{\chi,\psi}, \chi, \psi \in \mathcal{H}\}$ defined by

$$p_{\chi,\psi}(X) = |\langle \chi, X\psi \rangle|, \qquad (3.3.1)$$

for all operators $X \in \mathcal{B}(\mathcal{H})$. We say that a sequence $(X_i)_{i \in \mathbb{N}}$ (or more generally, a *net*) in $\mathcal{B}(\mathcal{H})$ converges to $X$ in the weak topology if

$$|\langle \chi, X_i\psi \rangle - \langle \chi, X\psi \rangle| \to 0. \qquad (3.3.2)$$

We then say that $X_i$ converges weakly to $X$.

Again, this is the type of convergence which is referred to in the Physics literature as "convergence defined by the states", with the "states" being the vectors on the underlying Hilbert space. From a physical standpoint, this is the expected topology to be relevant in experiments since it is the one that concerns expectation values, which are essentially those we measure in an experiment.

A related topology is given by the *weak\* topology*[1]. This topology is defined on general dual spaces of normed spaces, and we say that $(f_n) \subset X^*$ converges *-weakly to $f$ if for all $x \in X$, we have that $f_n(x) \to f(x)$, in the usual topology of the underlying field. This topology turns out to be very important in defining a notion of convergence for divergent series (the so-called *regular methods*, see [93], chapter 4.10) or the very important *Alaoglu's theorem* (see [52], chapter V).

Following our presentation, we introduce the following topology.

**Definition 3.3.5.** The *strong topology* on $\mathcal{B}(\mathcal{H})$ is the topology induced by the family of seminorms $\{p_\psi, \psi \in \mathcal{H}\}$ defined by

$$p_\psi(X) = ||X\psi||, \qquad (3.3.3)$$

for all operators $X \in \mathcal{B}(\mathcal{H})$. We say that a net $(X_i)$ in $\mathcal{B}(\mathcal{H})$ converges to $X$ in the strong topology if

$$||X_i\psi - X\psi|| \to 0, \qquad (3.3.4)$$

for all $\psi \in \mathcal{H}$.

---

[1]   Pronounced as "weak star".

It is clear from the Cauchy-Schwarz inequality that strong convergence implies weak convergence: suppose we have that $\lim_n ||(A_n - A)x|| = 0$. Then

$$|\langle y, (A_n - A)x \rangle| \leq ||y|| \, ||(A_n - A)x|| \to 0.$$

Since $y \in \mathcal{H}$ is arbitrary, then we have that $A_n$ converges weakly to $A$.

We can characterize the strong and weak topologies in terms of a local base (definition A.1.6). For the weak topology is generated by the sets of the form

$$\mathscr{U}_w = \{\mathfrak{A} \subset \mathcal{B}(\mathcal{H}), |\langle x, Ay \rangle - \langle x, A'y \rangle| < r\}, \tag{3.3.5}$$

for all $A$, $A' \in \mathfrak{A}$ and some $r > 0$ and $x, y \in \mathcal{H}$.

The strong topology on the other hand can be characterized by

$$\mathscr{U}_s = \{\mathfrak{A} \subset \mathcal{B}(\mathcal{H}), ||Ax - A'x|| < r\}, \tag{3.3.6}$$

for all $A$, $A' \in \mathfrak{A}$ and some $r > 0$ and $x \in \mathcal{H}$.

## 3.4 The Bicommutant Theorem

We now present the first main result of the theory of von Neumann algebras in the form of the Bicommutant (sometimes called the Double Commutant) theorem. We present a simpler statement and proof of the theorem as can be found in [130] or [5]. We stress that this theorem is valid for other locally convex topologies such as the $\sigma$-weak and $\sigma$-strong topologies. For a more general version of this theorem, see [25], theorem 2.4.11.

We start with the following proposition which is important by itself.

**Proposition 3.4.1.** *If $\mathcal{N} \subset \mathcal{B}(\mathcal{H})$, then $\mathcal{N}'$ is weakly and strongly closed.*

*Proof.* We start by showing that $\mathcal{N}'$ is weakly closed. Let $(A')_{n \in \mathbb{N}} \subset \mathcal{N}'$ be a sequence in $\mathcal{N}'$ which converges weakly to $A' \in \mathcal{B}(\mathcal{H})$, meaning

$$\lim_{n \to \infty} |\langle x, (A'_n - A')y \rangle| = 0,$$

for all $x, y \in \mathcal{H}$. To prove that $A' \in \mathcal{N}'$, we first consider some $B \in \mathcal{N}$ and the commutation relation $A'B - BA'$. Since $A_n \in \mathcal{N}'$ for all $n \in \mathbb{N}$, we can then write

$$BA' - A'B = B(A' - A'_n) - (A' - A'_n)B,$$

which in turn implies that for all $x, y \in \mathcal{H}$:

$$|\langle x, (BA' - A'B)y\rangle| = |\langle x, (B(A' - A'_n) - (A' - A'_n)B)y\rangle| =$$
$$|\langle B^*x, (A' - A'_n)y\rangle - \langle x, (A' - A'_n)By\rangle| \to 0,$$

where the limit comes from the definition of weak convergence, i.e., the functional goes to zero for all elements of $\mathcal{H}$, including $B^*x$ and $By$.

Hence, we have that $\langle x, (BA' - A'B)y\rangle = 0$, since it is not dependent on $n$. This implies that $BA' = A'B \implies A' \in \mathcal{N}'$, since $B$ is arbitrary. Hence, $\mathcal{N}'$ is weakly closed.

Now, we show that $\mathcal{N}'$ is strongly closed. In analogy for what we did for the weak topology, we assume there is a sequence $(A')_{n \in \mathbb{N}}$ in $\mathcal{N}'$ which converges strongly to $A' \in \mathcal{B}(\mathcal{H})$, meaning

$$\lim_{n \to \infty} ||(A'_n - A')x|| = 0,$$

for all $x \in \mathcal{H}$. Again considering $B \in \mathcal{N}$ and $A'B - BA'$, we can write

$$||(BA' - A'B)x|| = ||B(A' - A'_n)x - (A' - A'_n)Bx|| \leq ||B(A' - A'_n)x|| + ||(A' - A'_n)Bx||$$
$$\leq ||B|| \, ||(A' - A'_n)x|| \to 0,$$

where we have used the triangle inequality and the identity $||Cy|| \leq ||C|| \, ||y||$ for bounded operators (see appendix B). With this, we show that $A' \in \mathcal{N}'$ and hence $\mathcal{N}'$ is strongly closed. ∎

Before showing the main theorem of this section, we should consider an important definition which will be one of the hypotheses of the theorem. We say that an algebra $\mathfrak{A} \subset \mathcal{B}(\mathcal{H})$ is *non-degenerate* if the linear space spanned by range of the algebra is dense in $\mathcal{H}$, that is $\overline{\text{span}(\mathcal{AH})} = \mathcal{H}$. The proof of this theorem is based on [15].

**Theorem 3.4.1** (The Bicommutant Theorem). *Let $\mathfrak{M} \subset \mathcal{B}(\mathcal{H})$ be a self-adjoint and non-degenerate algebra. Then, the following statements are equivalent:*

1. *$\mathfrak{M} = \mathfrak{M}''$;*

2. *$\mathfrak{M}$ is closed in the weak topology;*

3. *$\mathfrak{M}$ is closed in the strong topology.*

*Proof.* The implications (1) $\implies$ (2) and (1) $\implies$ (3) follows from proposition 3.4.1 by taking $\mathcal{N} = \mathfrak{M}'$ and applying the proposition to the bicommutant. We also have that (2) $\implies$ (3) by noting that since strong convergence implies weak convergence, if we have a sequence $(A_n)_{n \in \mathbb{N}} \subset \mathfrak{M}$ which converges *strongly* to $A \in \mathcal{H}$, then $(A_n)_{n \in \mathbb{N}}$ converges *weakly* to $A$. If $\mathfrak{M}$ is weakly closed, then $A \in \mathfrak{M}$ and $\mathfrak{M}$ is also strongly closed. We now show that $(3) \implies (1)$.

We thus assume that $\mathfrak{M}$ is closed in the strong topology. We now show that this implies that $\mathfrak{M}'' \subset \mathfrak{M}$. This can be done by showing that for each $A'' \in \mathfrak{M}''$ there is an open set (in the strong topology) $\mathscr{U}$ which contains $A''$ and satisfies $\mathscr{U} \cap \mathfrak{M} \neq \varnothing$. It suffices to consider the elements $\mathscr{U}$ which form a local basis to $\mathfrak{M}$.

Under this consideration, the statement $\mathfrak{M}'' \subset \mathfrak{M}$ is equivalent to show that for each $A'' \in \mathfrak{M}''$, $N \in \mathbb{N}$, $r_j > 0$ and $x_j \in \mathcal{H}$ for $j \in \{1, ..., N\}$ there is an element $A \in \mathfrak{M}$ such that

$$\|A'' x_j - A x_j\| < r_j,$$

for all $j$ considered. We have just followed the characterization laid out in equation (3.3.6). We will prove this statement for $N = 1$ and then for $N > 1$.

For the first case, we just need to show that if $A'' \in \mathfrak{M}''$, then there is some $A \in \mathfrak{M}$ such that $\|A'' x - A x\| < r$, for some $r > 0$ and all $x \in \mathcal{H}$. Since $\mathfrak{M}$ is an algebra, we have trivially that $\operatorname{span}(\mathfrak{M}x) = \mathfrak{M}x$, which implies (also trivially) that $\overline{\operatorname{span}(\mathfrak{M}x)} = \overline{\mathfrak{M}x}$. Since $\mathfrak{M}$ is non-degenerate, $\overline{\operatorname{span}(\mathfrak{M}x)}$ coincides with $\mathcal{H}$, hence $\overline{(\mathfrak{M}x)}$ is a closed linear subspace of $\mathcal{H}$. Now, we notice that $\mathfrak{M}\overline{\mathfrak{M}\psi} \subset \overline{\mathfrak{M}\psi}$, since if we take $\varphi \in \overline{\operatorname{span}(\mathfrak{M}\psi)}$ then there is some $C \in \mathfrak{M}$ such that $\|\varphi - C\psi\| < \epsilon$, for all $\epsilon > 0$. But that also implies that for all $B \in \mathfrak{M}$ we have that $\|B\varphi - BC\psi\| < \|B\|\epsilon$, which implies that $B\varphi \in \overline{\mathfrak{M}\psi}$, from which we conclude that $\mathfrak{M}\overline{\mathfrak{M}\psi} \subset \overline{\mathfrak{M}\psi}$.

Now, we denote by $P$ the orthogonal projector over $\overline{\mathfrak{M}\psi}$ e we show that it is in the commutant, i.e., $P \in \mathfrak{M}'$. The fact stated above that $\mathfrak{M}\overline{\mathfrak{M}\psi} \subset \overline{\mathfrak{M}\psi}$ implies that for all $A \in \mathfrak{M}$ we have that $AP = PAP$. By taking the adjoint on both sides we have that $PA^* = PA^*P$, where we used the fact that orthogonal projectors are self-adjoint. But since $\mathfrak{M}$ itself is self-adjoint, we have that $PA = PAP$, which in turn implies that $AP = PA$. Hence, $P \in \mathfrak{M}'$.

With this in hand, consider $A'' \in \mathfrak{M}''$. By the same argument as above, we have that $A''P = PA''$, which in turn implies that $A''\overline{\mathfrak{M}\psi} \subset \overline{\mathfrak{M}\psi}$. We would like to prove that $\psi \in \overline{\mathfrak{M}\psi}$. For that, we observe that for all $B \in \mathfrak{M}$ we have trivially that $B\psi \in \overline{\mathfrak{M}\psi}$. Hence, $PB\psi = B\psi$, which implies $(P - \mathbf{1})B\psi = 0$. Since $P \in \mathfrak{M}'$ as we found above, then $B(P - \mathbf{1})\psi = 0$, which implies $\mathfrak{M}(P - \mathbf{1})\psi = \{0\}$, since we consider all $B \in \mathfrak{M}$. Non-degeneracy and self-adjointness of $\mathfrak{M}$ entails that $(P - \mathbf{1})\psi = 0$, which in turn implies that $\psi = P\psi \in \overline{\mathfrak{M}\psi}$, as we wanted to show.

Now, we consider the case for $N > 1$. This can be obtained from the $N = 1$ by considering $N$ copies of the Hilbert space and von Neumann algebra. Denote by $r = \min\{r_1, ..., r_N\}$ and it suffices to prove there is an $A \in \mathfrak{M}$ such that $||A''\psi_j - A\psi_j|| < r$, for all $j = 1, ..., N$.

The copies are obtained as

$$\mathcal{H}^N \equiv \underbrace{\mathcal{H} \oplus ... \oplus \mathcal{H}}_{N \text{ times}},$$

$$\mathfrak{M}^N \equiv \underbrace{\mathfrak{M} \oplus ... \oplus \mathfrak{M}}_{N \text{ times}}.$$

Now it remains to prove that $\mathfrak{M}^N$ is non-degenerate and strongly closed. Non-degeneracy follows from the observation that for $\Psi \equiv \psi_1 \oplus ... \oplus \psi_N \in \mathcal{H}^N$, then $\mathfrak{M}^N\Psi = A\psi_1 \oplus ... \oplus A\psi_N$, for $A \in \mathfrak{M}$. Now, since $\mathfrak{M}$ is non-degenerate, we have that $\mathfrak{M}^N\Psi = 0$ if, and only if, we have $\psi_j = 0$ for each $j \in \{1, ..., N\}$. Hence, $\mathfrak{M}^N$ is non-degenerate.

To prove strong closure, we consider a strongly convergent sequence (or even a net) $A_n \oplus ... \oplus A_n \in \mathfrak{M}^N$. Then, for all $\psi_1 \oplus ... \oplus \psi_N \in \mathcal{H}^N$ we have that $A_n\psi_1 \oplus ... \oplus A_n\psi_N \in \mathcal{H}^N$ is convergent. Hence, each term $A_n\psi_j$ will converge to some $A\psi_j$ since each copy of $\mathfrak{M}$ is itself strongly closed. Therefore, $\mathfrak{M}^N$ is strongly closed.

Finally, considering each element of $(\mathfrak{M}^N)''$ as $(A^N)'' \equiv \underbrace{A'' \oplus ... \oplus A''}_{N \text{ times}}$, the $N = 1$ case tells us that there is an element $A^N \equiv \underbrace{A \oplus ... \oplus A}_{N \text{ times}}$ such that $||(A^N)''\Psi - A^N\Psi|| < r$. Hence, for each $j \in \{1, ..., N\}$ we have that $||A''\psi_j - A\psi_j|| < r \leq r_j$, which completes the proof. ∎

**Corollary 3.4.1.** *Let $\mathfrak{M}$ be a non-degenerate $*$-algebra of operators on a Hilbert space $\mathcal{H}$. It follows that $\mathfrak{M}$ is dense in $\mathfrak{M}''$ in the weak and strong topologies.*

*Proof.* See [25], corollary 2.4.15 (which includes other locally convex topologies). This result is usually referred to as the *von Neumann density theorem.* ∎

## 3.5 Comparison Theory of Projections

As we have seen in the previous chapter and in appendix B, given a von Neumann algebra (or more generally, a $C^*$-algebra) $\mathfrak{M}$, one can write any element $A \in \mathfrak{M}$ as a combination of two self-adjoint operators, namely

$$A = \frac{A + A^*}{2} + i\frac{A - A^*}{2i}, \tag{3.5.1}$$

which can be easily verified. Moreover, the Spectral Theorem for self-adjoint operators tells us that any self-adjoint operator $A$ can be written in terms of a spectral measure

$$A = \int_{\sigma(A)} \lambda dP_\lambda, \tag{3.5.2}$$

with $\sigma(A)$ being the spectrum and each $P_\lambda$ is an orthogonal projection in $\mathcal{B}(\mathcal{H})$. We recall that an orthogonal projector $P$ is a self-adjoint operator which satisfies $P^2 = P$. A direct consequence of this statement is that any self-adjoint operator can be arbitrarily approximated by linear combinations of its spectral projections, by partitioning the spectrum and approximating the integral formally by Riemann sums of the form

$$A \simeq \sum_i \lambda_i P(\Delta_i),$$

where $\Delta_i \subset \sigma(A)$ and $\lambda_i \in \Delta_i$ are chosen points within each partition interval. Therefore, it follows that a von Neumann algebra is generated by taking the weak closure of the algebra spanned by its projections. Hence, one is led to study the structure of a von Neumann algebra by analyzing its projections.

### 3.5.1 Lattice of projections

We begin by defining an ordering for projections. We start with the very simple definition of a lattice.

**Definition 3.5.1.** A *lattice* is a non-empty set $L$ with two binary functions denoted by $\wedge$ (read as "and") and $\vee$ (read as "or") satisfying for all $a, b, c \in L$:

1. Idempotence: $a \wedge a = a$ and $a \vee a = a$;

2. Commutativity: $a \wedge b = b \wedge a$ and $a \vee b = b \vee a$;

3. Associativity: $a \wedge (b \wedge c) = (a \wedge b) \wedge c$ and $a \vee (b \vee c) = (a \vee b) \vee c$;

4. Absorption: $a \wedge (a \vee b) = a$ and $a \vee (a \wedge b) = a$.

We discussed the notion of a partial order in the context of positive elements in a $C^*$-algebra (see section 2.3). From a partial order, one can construct a lattice. To see this, we state and prove the following simple lemma.

**Lemma 3.5.1.** *Let $L$ be a lattice. Then, two elements $x, y \in L$ satisfy*

$$x = x \wedge y \iff y = x \vee y.$$

*Proof.* We first assume that there are elements $x$ and $y$ in $L$ such that $x = x \wedge y$. Then, it follows immediately from commutativity and distributivity that

$$x \vee y = (x \wedge y) \vee y = y.$$

Assuming now the converse, i.e., $y = x \vee y$ it follows from the same properties that

$$x \wedge y = x \wedge (x \vee y) = x,$$

thus proving the lemma. ∎

Hence, by defining $x \leq y$ if, and only if, $x = x \wedge y$ or $y = x \vee y$, one gets a partial ordering for a lattice. In the case of projections, the $\vee$ and $\wedge$ operations are defined by

$$E \vee F = E + F - EF, \quad E \wedge F = EF. \tag{3.5.3}$$

We leave to the reader to verify that the set of projections with the above operations indeed forms a lattice.

Since the projection operators are self-adjoint, they inherit the ordering by positive elements, discussed in section 2.3.4, that is: let $E$ and $F$ be two projection operators in $B(\mathcal{H})$, then we say that $E \leq F$ if $E - F \in B(\mathcal{H})_+$. In fact, we have the more general statement:

**Proposition 3.5.1.** *Let $E$ and $F$ be two projection operators on $\mathcal{H}$ onto closed subspaces $Y$ and $Z$, respectively. The following conditions are then equivalent:*

1. *$Y \subseteq Z$;*

2. *$FE = E$;*

3. *$EF = E$;*

4. *$||Ex|| \leq ||Fx||, \forall x \in \mathcal{H}$;*

5. *$E \leq F$.*

*In this case, $E$ is said to be a* subprojection *of $F$. Furthermore, if $E$ is a subprojection of $F$ and $E \neq F$, then it is said to be a* proper subprojection *and we denote by $E < F$.*

*Proof.* We first show that $1 \implies 2$. Since $Y \subseteq Z$, then $Ex \in Y$ is also in $Z$, for all $x \in \mathcal{H}$. Hence, since $Z$ is invariant under $F$, we have that $FEx = Ex$.

Now, we show that $2 \implies 3$. Since $FE = E$ and both $E$ and $F$ are self-adjoint, we have that $EF = (FE)^* = E^* = E$.

We proceed to show that $3 \implies 4$. If $EF = E$, then for all $x \in \mathcal{H}$ we have that $||Ex|| = ||EFx|| \leq ||Fx||$ because $||E|| \leq 1$.

Using the defining properties of an orthogonal projection, we see that $\langle Ex, x \rangle = \langle E^2 x, x \rangle = \langle Ex, Ex \rangle = ||Ex||^2$, and similarly for $F$. Hence, we have that $4 \implies 5$.

Finally, we show that $5 \implies 1$. This is obtained by considering that if $E \leq F$, then for each $y \in Y$ we have that $||y||^2 = \langle Ey, y \rangle \leq \langle Fy, y \rangle = ||Fy||^2 \leq ||y||^2$. Hence, $||Fy|| = ||y||$ which implies that $y \in Z$. ∎

Together with the partial order relation above, the set of projection operators on a Hilbert space has the structure of a lattice.

### 3.5.2   Equivalence relation

For completeness and clarity, we state what is meant by an equivalence relation.

**Definition 3.5.2.** An *equivalence relation* $\sim$ is a relation between two elements in a non-empty set $A$ satisfying for all $a, b, c \in A$

1. Reflexivity: $a \sim a$;

2. Symmetry: $a \sim b \implies b \sim a$;

3. Transitivity: if $a \sim b$ and $b \sim c$, then $a \sim c$.

**Definition 3.5.3.** Let $\mathfrak{M}$ be a von Neumann algebra and $E$ and $F$ two projection operators in $\mathfrak{M}$. We say that $E$ and $F$ are *equivalent* (*relative to* $\mathfrak{M}$) when there is a $V \in \mathfrak{M}$ such that $E = V^*V$ and $F = VV^*$. In this case, we write $E \sim F$. The operator $V$ is called a *partial isometry with initial projection $E$ and final projection $F$*.

It is very easy to see that this defines an equivalence relation between projections[2]. We leave to the reader to check this statement.

**Proposition 3.5.2.** *Let $E \in \mathfrak{M}$ be a projection and define*

$$E\mathfrak{M}E = \{EAE : A \in \mathfrak{M}\}. \tag{3.5.4}$$

*Then, $E\mathfrak{M}E$ is closed under products and under the involution operation. Furthermore, it is also weakly closed, and hence a von Neumann algebra on $E\mathcal{H}$.*

---

[2]   The notation for this equivalence relation varies in the literature, and sometimes is referred to as *relative dimension* (i.e., two projections have the same dimension relative to the algebra if the operator $V$ satisfying the conditions in the definition exists). We will use the standard notation for equivalence relations and hope that it is clear from the context that we are referring to the equivalence relation between projections in a von Neumann algebra.

*Proof.* It is straightforward to see that $E\mathfrak{M}E$ is a linear subspace of $\mathfrak{M}$. For $A, B \in \mathfrak{M}$, we have that $EAB \in \mathfrak{M}$ since $\mathfrak{M}$ is an algebra, hence $E\mathfrak{M}E$ is closed under products using that $E^2 = E$. Self-adjointness of $E$ implies that $E\mathfrak{M}E$ is closed under involution.

Now, weak closure follows from considering a sequence $(EA_nE)_{n\in\mathbb{N}}$ which converges weakly to some element $B \in \mathfrak{M}$, that is

$$\langle EA_nEx, y \rangle \to \langle Bx, y \rangle,$$

for all $x, y \in \mathcal{H}$. Now, because $E$ is self-adjoint, we can write this limit as

$$\langle EA_nx, Ey \rangle \to \langle Bx, y \rangle.$$

Since this convergence holds for all $x, y \in \mathcal{H}$, including $Ex, Ey \in E\mathcal{H}$, it follows that $B$ acts as $E\overline{B}E$ for some $\overline{B} \in \mathfrak{M}$. Hence, $B = E\overline{B}E \in E\mathfrak{M}E$, which implies that $EA_nE$ converges weakly to an element of $E\mathfrak{M}E$, proving it is a weakly closed subspace and hence a von Neumann algebra when restricted to $E\mathcal{H}$. ∎

### 3.5.3  Classification of Factors

With the previous definitions, we can give a classification for the possible projections in a von Neumann algebra. From these, we can classify the Murray-von Neumann classification of factors as advertised.

**Definition 3.5.4.** Let $\mathfrak{M}$ be a von Neumann algebra. A projection $E \in \mathfrak{M}$ is said to be

1. *minimal* if $\mathfrak{M}$ contains no proper subprojection of $E$;

2. *abelian* if $E\mathfrak{M}E$ is an abelian von Neumann algebra;

3. *infinite* if there exists a projection $E_0 \in \mathfrak{M}$ such that $E_0 < E$ (proper subprojection) and $E_0 \sim E$;

4. *finite* if it is not infinite.

Naturally, we have that a minimal projection is also abelian and finite. In fact, in the case of factors a projection is minimal if and only if it is abelian (see [90], proposition 6.4.3). Intuitively, one can see that a minimal projector must have a one-dimensional range, justifying the name. Notice that we have that being a minimal implies being abelian which in turn implies being finite, in all cases.

For general von Neumann algebras, we can extend this classification (and consequently the Murray-von Neumann classification) by considering *central projections*, that is, projections which are contained in the center $Z(\mathfrak{M})$. However, since we are considering only factors in our discussion, the center is trivial and the only central projection is the identity. For more details, see [90], chapter 6.

We are finally in position to give the Murray-von Neumann classification of factors.

**Definition 3.5.5** (Murray-von Neumann Classification)**.** A factor $\mathfrak{M}$ is said to be of

1. type I if it contains a minimal (abelian) projection;

2. type II if it contains a finite projection but no minimal (abelian) projection;

3. type III if it is neither type I nor type II.

A further refinement for type I and type II factors can already be presented. In fact, Murray and von Neumann had already subdivided the type I factors according to the dimension of the underlying Hilbert space. In more technical terms, for each cardinal number $n$, there is a unique (up to isomorphism) type $\mathrm{I}_n$ factor, namely $B(\mathcal{H}) \otimes \mathfrak{A}$, where $\mathcal{H}$ has dimension $n$ (which can be infinite) and $\mathfrak{A}$ is some Abelian von Neumann algebra. Hence, for Quantum Mechanics, where we work in Hilbert spaces of *finite* dimension, we are concerned with type I factors which are nothing more than the algebra of bounded operators on that space. For these algebras, expressions such as "$P_\psi = |\psi\rangle \langle\psi|$" which are routinely used in textbooks on Quantum Mechanics represent the minimal projections, since they are constructed from the vectors living in the Hilbert space.

The existence of minimal projections alongside the spectral theorem allows one to factor the algebra according to the factorization of the underlying Hilbert space. In the type I case, this can be seen by writing the algebra over $\mathcal{H} = \mathcal{H}_1 \otimes \mathcal{H}_2$ as $B(\mathcal{H}) = B(\mathcal{H}_1) \otimes B(\mathcal{H}_2)$.

We can subdivide type II factors according to the fineteness of the identity (which is trivially a projection). If the identity is finite, then we speak of a type $\mathrm{II}_1$ factor and if the identity is infinite, we speak of a type $\mathrm{II}_\infty$ factor. Even though von Neumann was mostly interested originally in type II factors, they seemed to be more relevant in Mathematics than in Physics. Recent work, on the other hand, has shown that these algebras are important in the context of Quantum Gravity [40].

Finally, one can refine type III algebras as type $\mathrm{III}_\lambda$ where $\lambda \in [0, 1]$. This is done using the Tomita-Takesaki Modular Theory, to be discussed in the next section.

An equivalent classification which was in fact the one proposed by Murray and von Neumann in [107] takes into consideration the *dimension function*.

**Definition 3.5.6.** Let $\mathfrak{M}$ be a factor. Define the map $d : \mathcal{P}(\mathfrak{M}) \to [0, \infty]$, where $\mathcal{P}(\mathfrak{M})$ denotes the set of projectors in $\mathfrak{M}$. The map $d$ is called (*von Neumann's*) *dimension function* if it satisfies for all $E, F \in \mathcal{P}(\mathfrak{M})$

1. $d(E) = 0$ if, and only if, $E = 0$;

2. If $E$ is orthogonal to $F$, then $d(E + F) = d(E) + d(F)$;

3. $d(E) = d(F)$ if, and only if, $E \sim F$;

4. $d(E) \leq d(F)$ if, and only if, $E \leq F$;

5. $d(E) < \infty$ if, and only if, $E$ is a finite projection;

6. $d(E) + d(F) = d(E \vee F) + d(E \wedge F)$.

From this, we can define a classification for the factors:

1. if $\text{Ran}(d) = \{0, 1, ..., n\}$ then $\mathfrak{M}$ is type $\text{I}_n$;

2. if $\text{Ran}(d) = \{0, 1, ..., \infty\}$ then $\mathfrak{M}$ is type $\text{I}_\infty$;

3. if $\text{Ran}(d) = [0, 1]$ then $\mathfrak{M}$ is type $\text{II}_1$;

4. if $\text{Ran}(d) = [0, \infty]$ then $\mathfrak{M}$ is type $\text{II}_\infty$;

5. if $\text{Ran}(d) = \{0, \infty\}$ then $\mathfrak{M}$ is type III;

### 3.5.4 Trace and density matrices

From the definitions, we can divide the factors in *finite factors*, which include the type $\text{I}_n$ and $\text{II}_1$ (because they only contain finite projections) and *infinite* factors, which include the type $\text{II}_\infty$ and type III (the latter containing only infinite projections). The distinction between finite and infinite factors is characterized by the existence of a *tracial state*.

**Definition 3.5.7.** A *faithful normalized trace* on a von Neumann algebra $\mathfrak{M}$ is a state Tr on $\mathfrak{M}$ such that

1. Tr is *tracial*: $\text{Tr}(AB) = \text{Tr}(BA)$, for all $A, B \in \mathfrak{M}$;

2. $\rho$ is *faithful*: $\text{Tr}(A^*A) = 0$ implies that $A = 0$.

Notice that these are the properties one would expect from an abstract notion of a trace (especially condition 1, which is sometimes known as the "cyclic property"). From these state, we can define *trace states* as

$$\omega_{\text{Tr}}(A) = \text{Tr}(\rho A), \tag{3.5.5}$$

where $\rho$ is a density matrix. For type I and II factors one can define a density matrix from the minimal projectors as in the formal expression $\rho = |\psi\rangle\langle\psi|$. The existance of a trace is guaranteed by the following theorem.

**Theorem 3.5.1.** *A von Neumann factor $\mathfrak{M}$ is finite if, and only if, there is a faithful normal tracial state $\rho$ on $\mathfrak{M}$.*

*Proof.* See [25]. ∎

This theorem is also revealing in the sense that it shows that type III factors do not have a trace. In fact, because they only contain infinite projections, *they also have no density matrices or pure states* (recall our discussion about the pure states in a matrix algebra, preceding equation 2.6.6). This also implies that a notion of a von Neumann entropy is not well defined in type III algebras. This is revealing in the case of Quantum Field Theory, whose algebra of observables is universally a type III algebra: it can be argued [45, 119] that the absence of pure states contributes as a further obstacle for an "ignorance" interpretation of Quantum Physics and one needs a more sensible definition in order to attribute a notion of entropy to a QFT system.

Another consequence of this discussion is the following. If one is presented with a quantum system living in a Hilbert space $\mathcal{H}_1$, the observables one has access to are described by $B(\mathcal{H}_1)$, so long one has some natural scale, for instance[3]. Naturally, we can define density matrices in this system. However, suppose now we are presented with a second quantum system which is light-years away from us, described by a Hilbert space $\mathcal{H}_2$ [124]. Naturally, causality requirements (which will become clearer when we present the Haag-Araki-Kastler axioms) imply that we can only access the observables of the full Hilbert space $\mathcal{H}_1 \otimes \mathcal{H}_2$ of the form

$$B(\mathcal{H}_2) = B(\mathcal{H}_1) \otimes \mathbf{1}_{\mathcal{H}_2}.$$

---

[3] This is to avoid complicated issues with unbounded operators. Hence, instead of working with a full-fledged position operator which is unbounded, we can work with a bounded version of it either by exponentiating or by confining our system to a box or a line.

Now here is the caveat: the question whether or not one has access to density matrices of the full system becomes rather non-trivial. In the case of *finite dimension*, we are working in type I algebras where we can define a full density matrix for the system by an expression like $\rho \otimes \mathbf{1}_{\mathcal{H}_2}$. However, if $\mathcal{H}_2$ is *infinite dimensional*, one can no longer define a sensible density matrix for the composite system, even if $\mathcal{H}_1$ is finite dimensional! From a physical standpoint, one can still work with $\rho \otimes \mathbf{1}_{\mathcal{H}_2}$ as an "effective" density matrix since one only has access to the observables in $B(\mathcal{H}_1)$. This fact opens questions about the renormalizability of certain observables in a QFT [124].

## 3.6 Tomita-Takesaki Modular Theory

In this section, we will present some elements of Modular Theory, first presented in two unpublished papers of Tomita in 1967 and then recast in modern form and notation by Takesaki in 1970 [131]. The importance of the original papers were only realized after Takesaki's work was published, in an episode that illustrates one of the goals in Mathematical Physics: to clarify and make precise mathematical statements of relevance in Physics. Hence, this theory is better known nowadays as Tomita-Takesaki theory. We will use both names interchangebly.

The importance of this theory to Quantum Field Theory cannot be overstated. In fact, one might claim that QFT was revolutionized due to this theory [24]. A reason for that is, as we mentioned elsewhere, the Tomita-Takesaki theory allows one to classify the type III factors, which are the most relevant in QFT. A very important result [34] indicates that all algebras which describe an AQFT are indeed isomorphic to the *unique* type III$_1$ hyperfinite factor.

Our treatment here is introductory and mainly based on an approach due to Rieffel and van Daele [113] which is the more relevant approach to Physics. For a more general form, we recommend the original monograph by Takesaki [131] and a more modern approach by the same author [132]. We also recommend [90].

### 3.6.1 Basic structure

In the context of a von Neumann algebra $\mathfrak{M}$ acting on a Hilbert space $\mathcal{H}$, we can always refer to the Modular Theory of the algebra if we have a vector $\Omega \in \mathcal{H}$ which is *cyclic* and *separating* for $\mathfrak{M}$. We have defined what we mean by a cyclic vector in section 2.4.1 and its existence is guaranteed by the GNS construction.

**Definition 3.6.1.** Let $\mathcal{H}$ be a Hilbert space and $\mathfrak{M}$ a von Neumann algebra on $\mathcal{H}$. A subset $\mathcal{R} \subseteq \mathcal{H}$ is *separating for* $\mathfrak{M}$ if $A\Omega = 0$ implies $A = 0$ for all $A \in \mathfrak{M}$ and $\Omega \in \mathcal{R}$. An equivalent statement is that the semi-norm $p$ on $\mathfrak{M}$ defined by $p(A) = ||A\Omega||$ is in fact a norm.

**Proposition 3.6.1.** *Let $\mathfrak{M}$ be a von Neumann algebra on $\mathcal{H}$ and $\mathcal{R} \subseteq \mathcal{H}$ a subset. Then, $\mathcal{R}$ is cyclic for $\mathfrak{M}$ if and only if it is separating for the commutant $\mathfrak{M}'$.*

*Proof.* We first assume that $\mathcal{R}$ is cyclic for $\mathfrak{M}$. We then choose some $A' \in \mathfrak{M}$ such that $A'\mathcal{R} = \{0\}$. Then for $B \in \mathfrak{M}$ and $\Omega \in \mathcal{R}$, we have that $A'B\Omega = BA'\Omega = 0$. Hence, $A'[\mathfrak{M}\mathcal{R}] = 0$, where $[\mathfrak{M}\mathcal{R}]$ denotes the space generated by the elements of $\mathfrak{M}$ acting on the vectors of $\mathcal{R}$. This implies that $A' = 0$ which in turn implies that $\mathcal{R}$ is separating for $\mathfrak{M}'$.

Now we assume the converse. Set $P' = [\mathfrak{M}\mathcal{R}]$. Then $P'$ is a projection in $\mathfrak{M}'$ and $(\mathbf{1} - P')\mathcal{R} = \{0\}$. Hence, $\mathbf{1} - P' = 0$ and $[\mathfrak{M}\mathcal{R}] = \mathcal{H}$. Since $\mathcal{H}$ is dense in itself, we conclude that $\mathcal{R}$ is cyclic for $\mathfrak{M}$. ∎

The starting point of Modular Theory is the pair $(\mathfrak{M}, \Omega)$, comprising of a von Neumann algebra $\mathfrak{M}$ on a Hilbert space $\mathcal{H}$ and a vector $\Omega \in \mathcal{H}$ which is cylic and separating for $\mathfrak{M}$. This pair is usually referred to as being in the *standard form*.

Let $(\mathfrak{M}, \Omega)$ be in standard form as defined above. We define operators $S_0$ and $F_0$ on $\mathcal{H}$ as

$$\begin{aligned} S_0 A\Omega &= A^*\Omega; \\ F_0 A'\Omega &= A'^*\Omega, \end{aligned} \qquad (3.6.1)$$

for $A \in \mathfrak{M}$ and $A' \in \mathfrak{M}'$. These operators are well-defined in the dense domains $D(S_0) = \mathfrak{M}\Omega$ and $D(F_0) = \mathfrak{M}'\Omega$.

**Proposition 3.6.2.** *The operators $S_0$ and $F_0$ defined above are closable and*

$$S_0^* = \bar{F}_0 \equiv F, \quad F_0^* = \bar{S}_0 \equiv S, \qquad (3.6.2)$$

*where the bar denotes the closure.*

*Proof.* For $A \in \mathfrak{M}$ and $A' \in \mathfrak{M}'$, we can compute from the definition of $S_0$

$$\langle A'\Omega, S_0 A\Omega \rangle = \langle A'\Omega, A^*\Omega \rangle = \langle A\Omega, A'^*\Omega \rangle = \langle A\Omega, F_0 A'\Omega \rangle.$$

Thus, $F_0 \subseteq S_0^*$, which implies that $S_0^*$ is densely defined. We conclude that $S_0$ is closable. The argument for $F_0$ is completely analogous.

Now we show that $S_0^*$ is the closure of $F_0$. We choose $x \in D(S_0^*)$ and set $y = S_0^* x$. For $A \in \mathfrak{M}$, we have

$$\langle A\Omega, y \rangle = \langle A\Omega, S_0^* x \rangle = \langle x, S_0 A\Omega \rangle = \langle x, A^*\Omega \rangle.$$

∎

Therefore, it is more convenient to work with the extensions $S$ and $F$ defined in equation 3.6.2. In particular, we can write $S$ in its unique polar decomposition.

**Definition 3.6.2.** In the polar decomposition

$$S = J\Delta^{1/2}, \tag{3.6.3}$$

we call the positive, self-adjoint operator $\Delta$ the *modular operator* and the anti-unitary operator $J$ the *modular conjugation* or *modular involution* associated with the pair $(\mathfrak{M}, \Omega)$.

Now, since $\Delta$ is a self-adjoint operator, we can apply the spectral or functional calculus to it by means of the spectral theorem for self-adjoint operators. Recall that this theorem states that the modular operator can be decomposed as

$$\Delta = \int_{\mathbb{R}_+} \lambda \, dP_\lambda,$$

where $P_\lambda$ are the spectral projectors and positivity of $\Delta$ entails that the integral is over the positive real line. We are then led to define the family of *unitary* operators indexed by $t \in \mathbb{R}$

$$\Delta^{it} \equiv \int_{\mathbb{R}} \lambda^{it} \, dP_\lambda. \tag{3.6.4}$$

This family of operators forms a *one-parameter strongly continuous unitary group*, meaning that

$$\lim_{t \to t_0} \Delta^{it} = \Delta^{it_0}, \tag{3.6.5}$$

as can be seen from the definition in equation 3.6.4.

We mention *en passant* that one can apply Stone's Theorem [82] to write

$$\Delta^{it} = e^{i\tau H}, \tag{3.6.6}$$

where $H$ is a possibly unbounded operator with the domain suitably defined. In fact, one can think of $H$ as being the Hamiltonian associated with the modular operator. It turns out that this Hamiltonian is unbounded by below. We will comment further the physical implication of this statement.

**Proposition 3.6.3.** *The modular operator and conjugation together with the operators $S$ and $F$ satisfy the following relations:*

1. $\Delta = FS$;             3. $F = J\Delta^{-1/2}$;             5. $J^2 = \mathbf{1}$;

2. $\Delta^{-1} = SF$;          4. $J = J^*$;               6. $\Delta^{-1/2} = J\Delta^{1/2}J$.

*Proof.* See [25], proposition 2.5.11. ∎

We now present the main theorem in Modular Theory, which is referred to as the *Tomita-Takesaki theorem*. The proof for this theorem is somewhat intricate and can be found in [25] (theorem 2.5.14), in [130] (theorem 2.3.3) or in [90] (theorem 9.2.9). For a more recent short proof for the special case of bounded operators, see [124].

**Theorem 3.6.1** (Tomita-Takesaki Theorem)**.** *Let $\mathfrak{M}$ be a von Neumann algebra with a cyclic and separating vector $\Omega$. Then $J\Omega = \Omega = \Delta\Omega$ and the following equalities hold*

$$\begin{aligned} J\mathfrak{M}J &= \mathfrak{M}'; \\ \Delta^{it}\mathfrak{M}\Delta^{-it} &= \mathfrak{M}, \end{aligned} \tag{3.6.7}$$

*for all $t \in \mathbb{R}$.*

## 3.6.2  Modular automorphism group and the KMS condition

We now discuss some important physical consequences of the Tomita-Takesaki theorem. Because of the invariance of the algebra under the action of the modular operator, we can define a one-parameter automorphism group called the *modular automorphism group of $\mathfrak{M}$ relative to $\Omega$*:

$$\sigma_t(A) = \Delta^{it}A\Delta^{-it}, \tag{3.6.8}$$

with $A \in \mathfrak{M}$ and $t \in \mathbb{R}$. This is an example of what is called a *$C^*$-dynamical system*, which comprises of a $C^*$-algebra $\mathfrak{A}$, a (locally compact) group $G$ and a strongly continuous homomorphism of $G$ into the set of $*$-automorphisms of $\mathfrak{A}$, sometimes denoted as $Aut(\mathfrak{A})$.

One of the difficult matter in working in AQFT is defining physically sensible states. In Modular Theory, this is aided by the fact that we can define a faithful normal state on $\mathfrak{M}$ induced by $\Omega$

$$\omega(A) = \frac{1}{||\Omega||^2}\langle\Omega, A\Omega\rangle, \tag{3.6.9}$$

which yields a family of states due to its invariance under $\{\sigma_t\}$, i.e., $\omega(\sigma_t(A)) = \omega(A)$ for all $A \in \mathfrak{M}$ and $t \in \mathbb{R}$. Indeed, this is the starting point of the famous result by Bisognano and Wichmann which now bears their name [21]:

**Theorem 3.6.2** (Bisognano and Wichmann Theorem). *Let $\mathfrak{M}$ be a von Neumann algebra with a cyclic and separating vector $\Omega$. Let $\{U(t)\}$ be a continuous unitary group satisfying $U(t)\mathfrak{M}U(-t) \subset \mathfrak{M}$ for all $t \geq 0$. Then, any two of the following conditions implies the third.*

1. *$U(t) = e^{itH}$, for a positive operator $H$;*

2. *$U(t)\Omega = \Omega$, for all $t \in \mathbb{R}$;*

3. *$\Delta^{it}U(s)\Delta-it = U(e^{-2\pi t}s)$ and $JU(s)J = U(-s)$, for all $s, t \in \mathbb{R}$.*

This theorem allowed Bisognano and Wichmann to characterize the modular group associated with a von Neumann algebra and a vacuum vector living in what is called a *Rindler wedge*. Their result shows that the modular group implements Lorentz boosts leaving the wedge invariant.

Another direct application of the Tomita-Takesaki theorem to QFT is that the modular automorphism group satisfies what is called the *Kubo-Martin-Schwinger (KMS) condition* [26, 53]. The notion of a *KMS state* was first introduced by Kubo [94], Martin and Schwinger [104] in order to describe a vacuum state in thermal equilibrium. The condition these states must satisfy was formulated by Haag, Hugenholtz and Winnink [79].

The condition can be stated as the following: let $\mathfrak{M}$ be a von Neumann algebra with $\{\alpha_t : t \in \mathbb{R}\}$ being a ($\sigma$-weakly) continuous one-parameter group of automorphisms of $\mathfrak{M}$. Then, we say that a state $\phi$ on $\mathfrak{M}$ satisfies the KMS condition with respect to $\{\alpha_t\}$ if for any $A, B \in \mathfrak{M}$ there exists a complex function $F_{A,B}$ with

$$F_{A,B}(t) = \phi(\alpha_t(A)B), \quad F_{A,B}(t + i\beta) = \phi(B\alpha_t(A)), \qquad (3.6.10)$$

where $\beta \in (0, \infty)$ is the inverse temperature. The function $F_{A,B}$ is required to be analytic in $\{z \in \mathbb{C} : 0 < \mathrm{Im}\{z\} < \beta\}$ and continuous on the closure of this set. It follows that $\phi(\alpha_{i\beta}(A)B) = \phi(BA)$ for all $A$ and $B$ in $\mathfrak{M}$. It also follows from the previous discussion on the Tomita-Takesaki theorem that such states are $\alpha$-invariant, meaning that $\phi(\alpha_t(A)) = \phi(A)$.

An example of a state which satisfies this condition is given by the Gibbs equilibrium state

$$\phi_G(A) = \frac{\mathrm{Tr}_{\mathcal{H}}(e^{-\beta H}A)}{\mathrm{Tr}_{\mathcal{H}}(e^{-\beta H})},$$

where $H$ is a self-adjoint operator and $e^{-\beta H}$ is of trace-class. In fact, any faithful normal state satisfies the KMS-condition at $\beta = 1$. This is sometimes referred as the *modular condition*. We have the following genereal theorem which we state without proof.

**Theorem 3.6.3.** *Let $\mathfrak{M}$ be a von Neumann algebra with a cyclic and separating vector $\Omega$. Then, the induced state $\omega$, defined by $\omega(A) = \langle \Omega, A\Omega \rangle$ with $A \in \mathfrak{M}$ satisfies the modular condition with respect to the modular automorphism group $\{\sigma_t : t \in \mathbb{R}\}$ associated with the pair $(\mathfrak{M}, \Omega)$.*

*Proof.* See [26]. ∎

## 3.7 Classification of type III factors

To finish this chapter, we present one of the most important applications of the Tomita-Takesaki Modular Theory to Quantum Field Theory, which is the classification of type III factors. This classification was first proposed by Connes [48]. Unbeknownst to him, this classification would turn out to be important with the discovery by Haagerup that type III$_1$ hyperfinite von Neumann algebras are unique up to an isomorphism [81]. Later, it was Fredenhagen who realized that all local algebras describing a QFT are isomorphic to the unique hyperfinite type III$_1$ factor [69].

### 3.7.1 The crossed product

We begin with the notion of a *crossed product* between an von Neumann algebra and a group $G$. This can be seen as a particular case of a *semi-direct product* between a group and a non empty set which we now present.

**Definition 3.7.1.** Let $M$ be a non-empty set and $G$ a group. A function $\alpha : G \times M \to M$ is said to be an *action of $G$ from the left over $M$* if it satisfies

1. For all $g \in G$, the function $\alpha(g, \cdot) : M \to M$ is a bijection;

2. If $e$ is the neutral element of $G$, then $\alpha(e, x) = x$, for all $x \in M$;

3. For all $g, h \in G$ and $x \in M$, we have that

$$\alpha(g, \alpha(h, x)) = \alpha(gh, x).$$

An action $\beta$ *from the right* over $M$ is defined is defined in an analogous way, changing condition 3 to

3'. $\beta(g, \beta(h, x)) = \beta(hg, x).$

If $G$ is an Abelian group, then both notions coincide.

In the case where $M = \mathfrak{M}$ is a von Neumann algebra, we say that the automorphic representation of $G$ on $\mathfrak{M}$ is *unitarily implemented* if there is a representation $g \mapsto U(g)$ by unitary operators on the underlying Hilbert space such that

$$\alpha_g(A) = U(g)AU(g)^*, \quad \forall A \in \mathfrak{M}. \tag{3.7.1}$$

Now, one can naturally regard the group $G$ as being $(\mathbb{R}, +)$, the addition group of real numbers and consider the action $\alpha_g(.)$ as being given by the modular automorphism group $\sigma_t(.)$, for $t \in \mathbb{R}$. Hence, the *crossed product of a von Neumann algebra $\mathfrak{M}$ relative to the modular automorphism group of any faithful normal state $\omega$* is just the semi-direct product

$$\mathfrak{N} = \mathbb{R} \rtimes_{\sigma^\omega} \mathfrak{M}. \tag{3.7.2}$$

It turns out that in the case that $\mathfrak{M}$ is a type III factor, the resulting algebra $\mathfrak{N}$ is a type $\mathrm{II}_\infty$ von Neumann factor. Moreover, the algebra $\mathfrak{M}$ is isomorphic to $\mathbb{R} \rtimes_\theta \mathfrak{N}$, where $\theta$ is in some sense the dual of $\sigma^\omega$ on $\mathfrak{N}$. For a proof of this statement, see chapter XII of [132], theorem 1.1. This observation is the starting point for considering type $\mathrm{II}_\infty$ algebras in the context of Quantum Gravity [99, 144].

### 3.7.2 Modular spectrum

It is interesting to analyze the spectrum of the modular operator $\Delta$ since it gives a measure of the periodicity of the modular automorphism group. Intuitively, the smaller the spectrum $\mathrm{sp}(\Delta)$ the closer the automorphism group $(\sigma_t)_{t \in \mathbb{R}}$ is to the identity in the algebra. In fact, in the extreme case where the spectrum reduces to the singlet set, the automorphsims coincide with the identity, for all $t \in \mathbb{R}$. In the other extreme, where $\mathrm{sp}(\Delta) = \mathbb{R}_+$, the group becomes *ergodic*, meaning that if regard the $C^*$-algebra dynamical system with the dynamics given by the action of the modular automorphism group than a point "visits" all parts of the system in a random sense. In other words, there are no fixed points.

**Definition 3.7.2.** The *modular spectrum* of $\mathfrak{M}$ is given by

$$S(\mathfrak{M}) \equiv \bigcap_\omega \mathrm{sp}(\Delta_\omega), \tag{3.7.3}$$

where $\omega$ runs over all faithful normal states over $\mathfrak{M}$ and $\Delta_\omega$ are the corresponding modular operators.

We are now in position to give a classification of type III factors. Originally, Connes used a more general classification by introducing *weights*. It will suffice for us the following definition which can be shown to be equivalent.

**Theorem 3.7.1.** *Let $\mathfrak{M}$ be a factor on a Hilbert space $\mathcal{H}$. Then, we have the following:*

1. *If $\mathfrak{M}$ is finite (i.e., type $I_n$ or $II_1$), then $S(\mathfrak{M}) = \{1\}$;*

2. *If $\mathfrak{M}$ is of type $I_\infty$ or $II_\infty$, then $S(\mathfrak{M}) = \{0, 1\}$;*

3. *If $\mathfrak{M}$ is of type III, then $S(\mathfrak{M}) \ni 0$ and we have the following classification:*

    a) *$\mathfrak{M}$ is of type $III_\lambda$ with $0 < \lambda < 1$ if $S(\mathfrak{M}) = \{0\} \cup \{\lambda^n : n \in \mathbb{Z}\}$;*

    b) *$\mathfrak{M}$ is of type $III_0$ if $S(\mathfrak{M}) = \{0, 1\}$;*

    c) *$\mathfrak{M}$ is of type $III_1$ if $S(\mathfrak{M}) = \mathbb{R}_+$.*

*Proof.* See [130], page 111 or [132], page 122. ∎

From a first glance, it is not clear how different the physical content of a theory would be if the algebra were of type $III_\alpha$ or type $III_\beta$ with $\alpha \neq \beta$. Even though we have not introduced yet the axioms which an algebra of observables must satisfy in order for it to describe a *bona fide* Quantum Field Theory, we can already indicate the importance of the type $III_1$, in particular.

We first note that, irrespective of the factor type, the spectra of the modular operators for all faithful normal states coincide in many physical situations. This implies that it suffices to compute the spectrum of a single, usually convenient, modular operator in order to determine the type of $\mathfrak{M}$. In fact, normal states play an even more significant role in the physics of type III factors. This is encapsulated in a result by Connes and Størmer [51] which we briefly comment, following [132], chapter XII.

First, we need the notion of an *orbit of an action* and its generated *orbit space*.

**Definition 3.7.3.** Let $G$ be a group and $\alpha : G \times M$ be the action of $G$ over a non-empty set $M$. For $m \in M$, we define the *orbit of the action* as being the set

$$\mathrm{Orb}_\alpha(m) \equiv \{\alpha_g(m) : g \in G\} \subset M. \tag{3.7.4}$$

The corresponding orbit space is comprised of the equivalence classes of elements of $M$ under the following equivalence relation: $m \sim_\alpha n$ if $\mathrm{Orb}_\alpha(m) = \mathrm{Orb}_\alpha(n)$.

In the case where $M = \mathfrak{M}$ is a von Neumann algebra, the action of the group has a subset of operators as its range, hence the composition $\omega \circ \alpha$ with a state $\omega$ is well-defined. In fact, we define the *orbit of a state $\omega$* as being the set $\{\omega \circ \alpha \; \alpha \in Aut(\mathfrak{M})\}$ and the *state orbit space* is taken as the space generated by the norm closure of the orbit of a state. The closure is taken for topological convenience.

**Definition 3.7.4.** Let $\mathfrak{M}$ be a von Neumann algebra and $\mathfrak{M}_*$ the space of normal states over $\mathfrak{M}$. We define the *diameter* of the state orbit space as

$$d(\mathfrak{M}) = \sup_{\omega_1, \omega_2 \in \mathfrak{M}_*} \inf_{\alpha \in Aut(\mathfrak{M})} ||(\omega_1 \circ \alpha) - \omega_2||. \tag{3.7.5}$$

Equivalently, let $[\omega]$ define the equivalence class defined above. Then, the diameter of the state orbit space can be written in terms of the induced metric

$$\overline{d}([\omega_1], [\omega_2]) = \inf_{\omega'_i \in [\omega_i]} ||\omega'_1 - \omega'_2||. \tag{3.7.6}$$

An intuition for this definition [83] is that it gives a measure of "how noncommutative" the algebra $\mathfrak{M}$ is, with type $III_1$ factors being the most noncommutative, as a consequence of the following theorem.

**Theorem 3.7.2.** *If $\mathfrak{M}$ is a type $III_\lambda$ factor with $0 \leq \lambda \leq 1$, then*

$$d(\mathfrak{M}) = 2\frac{1 - \lambda^{1/2}}{1 + \lambda^{1/2}}.$$

*Proof.* See [132], page 427, theorem 5.12. ∎

Hence, for type $III_1$ factors, the diameter of the state orbit space is zero, which means that the orbit of any normal state is norm dense in the state space. An analysis by Connes [49] leads to the conclusion that one can't use the invariance under automorphisms as a criterion to distinguish between two states over a type $III_1$ algebra. From a physical standpoint, this means that two unitarily equivalent states are "equally mixed".

# 4 Canonical Commutation Relations

*The career of a young theoretical physicist consists of treating the harmonic oscillator in ever-increasing levels of abstraction.*

Sidney Coleman

In this chapter, we present the formulation of the canonical commutation relations (CCR) in a representation independent way compatible with our algebraic approach to Quantum Physics. A similar construction exists for the canonical anti-commuting relations (CAR), but we specialize to the CCR since it will be mostly used in this Thesis. The interested reader may find reference [26] useful, and that is the main reference for this chapter.

## 4.1 The Permutation Group

We begin by making some remarks about the simple but important concept of a permutation group. We assume the reader is familiar with the basic definition of a group: a set which is equipped with an operation which is associative, there is a neutral element 1 and for each element $a$ in the set there is an inverse $a^{-1}$ such that

$$aa^{-1} = a^{-1}a = 1. \tag{4.1.1}$$

Now, consider a non-empty and finite set $C$ and let $\text{Perm}(C)$ be the set of all bijections of $C$ into itself. Naturally, $\text{Perm}(C)$ is a group where the group operation is the composition of functions. This operation is associative, the neutral element is given by the identity function and the inverse of a function $f$ is its corresponding inverse function $f^{-1}$ (which exists since we are considering bijections). The group $\text{Perm}(C)$ is called the *permutation group of the set $C$*.

We mention *en passant* the important *Cayley's theorem* which reflects the importance of permutation groups in Group Theory.

**Theorem 4.1.1** (Cayley's Theorem)**.** *Let $G$ be a group. Then, $G$ is isomorphic to a subgroup of $Perm(G)$.*

The proof for this theorem can be found in any standard book on Group Theory, but we recommend the book by Weyl [139] which is an excellent reference for Group Theory in the context of Quantum Mechanics. We also strongly recommend [15][1].

We consider in particular the set $\{1, ..., n\}$ and the permutation group $S_n = \text{Perm}(\{1, ..., n\})$. Then, we can represent each element as a $2 \times n$ matrix:

$$\pi \equiv \begin{pmatrix} 1 & 2 & ... & n \\ \pi(1) & \pi(2) & ... & \pi(n) \end{pmatrix}. \tag{4.1.2}$$

For instance, if we consider the set $\{1, 2, 3\}$, then the permutation group $S_3$ is comprised of $3! = 6$ elements

$$\begin{aligned} \pi_1 &= \begin{pmatrix} 1 & 2 & 3 \\ 1 & 2 & 3 \end{pmatrix}, & \pi_2 &= \begin{pmatrix} 1 & 2 & 3 \\ 2 & 1 & 3 \end{pmatrix}, & \pi_3 &= \begin{pmatrix} 1 & 2 & 3 \\ 1 & 3 & 2 \end{pmatrix}, \\ \pi_4 &= \begin{pmatrix} 1 & 2 & 3 \\ 3 & 2 & 1 \end{pmatrix}, & \pi_5 &= \begin{pmatrix} 1 & 2 & 3 \\ 3 & 1 & 2 \end{pmatrix}, & \pi_6 &= \begin{pmatrix} 1 & 2 & 3 \\ 2 & 3 & 1 \end{pmatrix}. \end{aligned} \tag{4.1.3}$$

We first notice that $\pi_1$ is the group identity. Secondly, while $\pi_1$, $\pi_5$ and $\pi_6$ are *cyclic* permutations (meaning that the order 123 when read from left to right does not change), the elements $\pi_2$, $\pi_3$ and $\pi_4$ are not cyclical. When a group has permutations such as these, we speak of *symmetric* permutations for cyclic permutations and *anti-symmetric* permutations for the non-cyclic permutations.

## 4.2   Identical Particles

We now proceed to the construction of the space of $n$ discrete quantum mechanical particles, which receives the name of *Fock space*. We will then be able to define a general notion of creation and annihilation operators, familiar in the algebraic study of the simple harmonic oscillator in an introductory course in Quantum Mechanics, and then we will construct the CCR in a concrete way. We then proceed to generate the CCR algebra which is representation independent.

First, some considerations about symmetry. In ordinary Quantum Mechanics, the physical states a particle can occupy are described by vector in a Hilbert space. Typically, for $n$ particles with configuration space given by $\mathbb{R}^m$, one considers the Hilbert space as being $L^2(\mathbb{R}^{nm})$ of complex square-integrable functions defined on $\mathbb{R}^{nm}$. If $n$ is not fixed, then the total Hilbert space which gives all possible quantum configurations for the particles is given by the direct sum

$$\mathfrak{F} = \bigoplus_{n=0}^{\infty} L^2(\mathbb{R}^{nm}), \tag{4.2.1}$$

where we define $L^2(\mathbb{R}^0) = \mathbb{C}$. The vectors in this space are then sequences $\Psi = (\psi_n)_{n \geq 0}$ with $\psi^0 \in \mathbb{C}$ and $\psi_n \in L^2(\mathbb{R}^{nm})$ for $n \geq 1$. We will constantly work with these vectors in component form, i.e., $\Psi = (\psi_1, \psi_2, ...)$. An alternative way to represent these vectors is using the direct sum symbol: $\Psi = (\psi_1 \oplus \psi_2 \oplus ...)$. This notation serves to remind us that the direct sum operation endows the resulting cartesian product with a vector space structure. However, we will use the "cartesian product notation" for clarity, but one should be reminded that the resulting space is even a normed space, with the norm given by

$$||\Psi|| = |\psi_0| + \sum_{n=1}^{\infty} \int dx_1...dx_n |\psi_n(x_1, ..., x_n)|^2, \tag{4.2.2}$$

that is, the sum of the norms in each subspace which comprised $\mathfrak{F}$. This is a general procedure for defining the norms on the direct sum of normed spaces[2].

---
[1]   Sadly, only available in Portuguese.

Now here enters the probabilistic interpretation of Quantum Mechanics. If $\Psi$ is normalized, then we can define the probability density

$$d\rho(x_1, ..., x_m) = |\psi_n(x_1, ..., x_m)|^2 dx_1...dx_m, \tag{4.2.3}$$

which gives the probability of finding the $n$th particle in a neighborhood around the point $(x_1, ..., x_m)$ in the configuration space. In Quantum Statistics, on the other hand, particles are *indistinguishable*, hence the probability density should be invariant under the permutation of the coordinates of each particle. This permutation naturally defines a unitary representation of the permutation group, defined in the previous section. The symmetry will then be assured if $\Psi$ transforms correctly.

By inspecting equation (4.2.3), we see that the functions $\psi_n$ can change at most by a relative sign in order to maintain the symmetry. Hence, we will be concerned with two representations of the permutation group:

1. The representation where the components $\psi_n$ are *symmetric* under interchange of coordinates, or *bosonic representation*;

2. The representation where the components $\psi_n$ are *anti-symmetric* under interchange of coordinates, or *fermionic representation*.

---

2   The same works for defining inner products. For the tensor product, one changes the sum of the norms and inner products by their product.

Functions which transform under the bosonic representation are said to describe *bosons* while those transforming in the fermionic representation are said to describe *fermions*. Respectively, they are said to satisfy *Bose-Einstein* and *Fermi-Dirac statistics*, in a beautiful connection illustrate by the *spin-statistics theorem*[3].

We are then interested in studying the subspaces $\mathfrak{F}_\pm$ of $\mathfrak{F}$ which describe bosons (+ sign) and which describe fermions (− sign). We should note that if we have particles with more structure (such as spin or color), then we need to generalize this construction. We will not do this here and we will work with particles with no internal structure.

## 4.3  The Fock Space

Assuming that the states of each particle form a complex Hilbert space $\mathcal{H}$, we denote the $n$-fold tensor product of $\mathcal{H}$ with itself by

$$\underbrace{\mathcal{H} \otimes \mathcal{H} \otimes ... \otimes \mathcal{H}}_{n\,\text{times}} \equiv \mathcal{H}^{\otimes n}. \tag{4.3.1}$$

We can then introduce the *Fock space* by

$$\mathfrak{F}(\mathcal{H}) = \bigoplus_{n=0}^{\infty} \mathcal{H}^{\otimes n}, \tag{4.3.2}$$

where $\mathcal{H}^{\otimes 0} = \mathbb{C}$. The construction presented in the last section was an example of a Fock space.

The subspaces describing bosons and fermions can be obtained by the following construction. Define the operators $P_\pm$ on $\mathcal{H}^{\otimes n}$ by

$$\begin{aligned}
P_+(\psi_1 \otimes ... \otimes \psi_n) &= \frac{1}{n!} \sum_\pi \psi_{\pi_1} \otimes ... \otimes \psi_{\pi_n}; \\
P_-(\psi_1 \otimes ... \otimes \psi_n) &= \frac{1}{n!} \sum_\pi \epsilon_\pi \psi_{\pi_1} \otimes ... \otimes \psi_{\pi_n},
\end{aligned} \tag{4.3.3}$$

where the summation is over all permutations $\pi_i$ in the permutation group and the symbol $\epsilon_\pi$ is 1 for a symmetric permutation and −1 for an anti-symmetric permutation.

**Proposition 4.3.1.** *The operators $P_\pm$ defined above are bounded, self-adjoint operators on $\mathcal{H}^{\otimes n}$. They also satisfy $P_\pm^2 = P_\pm$ and hence they are orthogonal projections.*

---

[3]  See, for instance, Streater and Wightman [125] chapter 4-4.

*Proof.* From the triangle inequality, we find that

$$||P_+(\psi_1 \otimes ... \otimes \psi_n)|| \leq |(n!)^{-1}| \sum_\pi ||\psi_{\pi_1} \otimes ... \otimes \psi_{\pi_n}|| = |(n!)^{-1}| \sum_\pi \prod_{i=1}^n ||\psi_i||,$$

and similar for $P_-$ showing clearly that they are bounded.

To show that $P_\pm$ are self-adjoint, it is convenient to introduce the notation

$$P_+ = \frac{1}{n!} \sum_\pi \mathcal{P}_n(\pi), \qquad P_- = \frac{1}{n!} \sum_\pi \epsilon_\pi \mathcal{P}_n(\pi), \tag{4.3.4}$$

where $\mathcal{P}_n$ is a representation of the permutation group, acting on $\mathcal{H}^{\otimes n}$ as

$$\mathcal{P}_n(\pi) \left( \sum_{k=1}^\ell \alpha_k \psi_1^k \otimes ... \otimes \psi_n^k \right) = \sum_{k=1}^\ell \alpha_k \psi_{\pi(1)}^k \otimes ... \otimes \psi_{\pi(n)}^k,$$

with $\alpha_k \in \mathbb{C}$. It is easy to see that this is indeed a representation and by an explicit calculation taking the norm on both sides of the above equation that it is an isometry, hence a unitary representation. This means that it satisfies $\mathcal{P}_n(\pi)^* = \mathcal{P}_n(\pi)^{-1} = \mathcal{P}_n(\pi^{-1})$. Hence

$$P_+^* = \frac{1}{n!} \sum_\pi \mathcal{P}_n(\pi^{-1}) = \frac{1}{n!} \sum_\pi \mathcal{P}_n(\pi) = P_+,$$

because we sum over all permutations (including the inverses). Thus, $P_+$ is self-adjoint.

Finally, we show that $P_\pm^2 = P_\pm$. The following identities are useful

$$\begin{aligned} P_+ \mathcal{P}_n(\pi) = \mathcal{P}_n(\pi) P_+ = P_+, \\ P_- \mathcal{P}_n(\pi) = \mathcal{P}_n(\pi) P_- = \epsilon_\pi P_-. \end{aligned} \tag{4.3.5}$$

This can be proved for $P_+$ by

$$P_+ \mathcal{P}_n(\pi') = \frac{1}{n!} \sum_\pi \mathcal{P}_n(\pi) \mathcal{P}_n(\pi') = \frac{1}{n!} \sum_\pi \mathcal{P}_n(\pi \pi') = \frac{1}{n!} \sum_\pi \mathcal{P}_n(\pi'') = P_+,$$

where we used that $\pi \mapsto \pi \pi' \equiv \pi''$ is a bijection for each $\pi'$. The proof for $P_-$ is analogous. We then compute

$$P_+^2 = \frac{1}{n!} \sum_\pi P_+ \mathcal{P}_n(\pi) = \frac{1}{n!} \sum_\pi P_+ = P_+,$$

since there are exactly $n!$ elements in the permutation group and there are no elements of the group being summed over in the second equality. Again, the proof for $P_-$ is completely analogous. ∎

The *Bose-Fock* and the *Fermi-Fock spaces* (or symmetric and anti-symmetric spaces, respectively) are then defined from the ranges of the projection operators:

$$\mathfrak{F}_\pm(\mathcal{H}) = \bigoplus_{n=0}^\infty \mathcal{H}_\pm^{\otimes n} \tag{4.3.6}$$

where $\mathcal{H}_\pm^{\otimes n} = \mathrm{Ran}(P\pm)$.

### 4.3.1 Creation and annihilation operators

It will be convenient for our purposes to define a *number operator* on $\mathfrak{F}(\mathcal{H})$:

$$N\Psi = (n\psi_{(n)})_{n\geq 0},$$
$$D(N) = \{\Psi \in \mathfrak{F}(\mathcal{H}) : \sum_{n=0}^{\infty} n^2 ||\psi^{(n)}||^2 < \infty\}, \tag{4.3.7}$$

where we have recognized that $N$ is an *unbounded operator* and therefore we should be careful in specifying the domain on which it acts.

Now, the $C^*$-algebras describing bosons and fermions can be defined by means of the annihilation and creation operators. For each $f \in \mathcal{H}$, they can be defined on $\mathcal{H}^{\otimes n}$ respectively as

$$a(f)(\psi_1 \otimes ... \otimes \psi_n) = n^{1/2}\langle f, \psi_1 \rangle \psi_2 \otimes \psi_3 \otimes ... \otimes \psi_n,$$
$$a^*(f)(\psi_1 \otimes ... \otimes \psi_n) = (n+1)^{1/2} f \otimes \psi_1 \otimes ... \otimes \psi_n. \tag{4.3.8}$$

For the $\psi_0 \in \mathbb{C}$ component of the vectors, we define

$$a(f)\psi_0 = 0, \ \ a^*(f)\psi_0 = f. \tag{4.3.9}$$

Now, one can compute

$$||a(f)\psi_{(n)}|| \leq n^{1/2}||f|| \, ||\psi_{(n)}||, \ \ ||a^*(f)\psi_{(n)}|| \leq (n+1)^{1/2}||f|| \, ||\psi_{(n)}||, \tag{4.3.10}$$

where we have used the Cauchy-Schwarz inequality. We consider the extension of these operators to the domain $D(N^{1/2})$.

Moreover, these operators satisfy

$$\langle a^*(f)\varphi, \psi \rangle = \langle \varphi, a(f)\psi \rangle. \tag{4.3.11}$$

This can be verified by the straightforward computation

$$\langle a^*(f)\varphi, \psi \rangle = \langle a^*(f)\varphi_1 \otimes ... \otimes \varphi_n, \psi_1 \otimes ... \otimes \psi_n \rangle = \sqrt{n+1}\langle f \otimes \varphi_1 \otimes ... \otimes \varphi_n, \psi_1 \otimes ... \otimes \psi_n \rangle$$
$$= \sqrt{n+1}\langle f, \psi_1 \rangle\langle \varphi_1 \otimes ... \otimes \varphi_n, \psi_2 \otimes ... \otimes \psi_n \rangle \rightarrow \sqrt{n}\langle f, \psi_1 \rangle\langle \varphi_1 \otimes ... \otimes \varphi_n, \psi_2 \otimes ... \otimes \psi_n \rangle$$
$$= \langle \varphi, a(f)\psi \rangle,$$

where we have taken the limit $n \rightarrow \infty$, as per the definition of the Fock space.

Now we are interested in defining the creation and annihilation operators in the Bose and Fermi Fock spaces defined in equation (4.3.6):

$$a_\pm(f) = P_\pm a(f)P_\pm,$$
$$a^*_\pm(f) = P_\pm a^*(f)P_\pm. \tag{4.3.12}$$

It is easy to verify that these operators satisfy the same expression as in (4.3.11). Moreover, $a(f)$ leaves the subspaces $\mathfrak{F}_\pm(\mathcal{H})$ invariant since its action in each $\mathcal{H}^{\otimes n}$ is to take the inner product of $f$ with the first entry $\psi_1$. However, the action of $P_\pm$ defined in equation (4.3.3) shuffles all the entries by considering the sum over all permutations and we consider all possible $n$. Hence, we can simply write

$$a_\pm(f) = a(f)P_\pm, \tag{4.3.13}$$

and by taking the adjoint on both sides of the equation and using the self-adjointness of the projection operators, we get

$$a_\pm^*(f) = P_\pm a^*(f). \tag{4.3.14}$$

The construction of particle states follows from applying the creation operator $a^*(f)$ to the vacuum state, which we write as $\Omega = (1, 0, 0, ...) \in \mathfrak{F}(\mathcal{H})$. We interpret the vacuum state as a zero particle state and the vectors

$$\psi_\pm(f) \equiv a_\pm^*(f)\Omega \tag{4.3.15}$$

as representing a particle in a physical state described by the vector $f \in \mathcal{H}$. The multiple particle states then follow from successive application of these operators with a suitable normalization factor:

$$\psi_\pm(f_1, ..., f_n) \equiv \frac{1}{\sqrt{n!}} a_\pm^*(f_1)...a_\pm^*(f_n)\Omega = P_\pm(f_1 \otimes ... \otimes f_n). \tag{4.3.16}$$

In a similar interpretation, the annihilation operators $a(f)$ "destroy" a particle with physical state described by $f$. The *Pauli Exclusion Principle* follows directly from this observation by considering that if $f_i = f_j$ for some pair $i, j$ with $1 \le i < j \le n$, then

$$\psi_n(f_1, ..., f_n) = P_-(f_1 \otimes ... \otimes f_n) = 0,$$

by anti-symmetry. The Principle can then be represented by the operator equation

$$a_-^*(f)a_-^*(f) = 0, \tag{4.3.17}$$

which has the physical interpretation that it is impossible to create two fermions occupying the same physical state.

In fact, one can compute the more general form of the commutation and anti-commutation relations using the definitions and properties shown here. We thus obtain

$$\begin{aligned} [a_+(f), a_+(g)] &= 0 = [a_+^*(f), a_+^*(g)], \\ [a_+(f), a_+^*(g)] &= \langle f, g \rangle \mathbf{1}, \end{aligned} \tag{4.3.18}$$

which are the *canonical commutation relations* or the *CCR* and

$$\{a_-(f), a_-(g)\} = 0 = \{a_-^*(f), a_-^*(g)\},$$
$$\{a_-(f), a_-^*(g)\} = \langle f, g \rangle \mathbf{1}, \tag{4.3.19}$$

which are the *canonical anti-commutation relations* or the *CAR*.

### 4.3.2 The CCR relations

The most notable difference between bosons and fermions (both from a theoretical and experimental viewpoints) is that the former does not obey the Pauli Exclusion Principle. This means that we can have an arbitrarily large number of bosons occupying some physical state, which is reflected on the unboundedness of the annihilation operator: if we consider $\psi_{(n)}$ as the $n$-fold tensor product of $f \in \mathcal{H}$ with itself, then equation (4.3.10) implies that

$$||a(f)\psi_{(n)}|| = n^{1/2}||f|| \, ||\psi_{(n)}||,$$

which is clearly unbounded. This unboundedness in not present for fermions, and it motivates us to look for bounded functions of the creation and annihilation operators. This is done by defining the operators

$$\Phi(f) \equiv \frac{a(f) + a^*(f)}{\sqrt{2}},$$
$$\Pi(f) \equiv \Phi(if) = \frac{a(f) - a^*(f)}{i\sqrt{2}}. \tag{4.3.20}$$

Naturally, one can invert these expressions to obtain $a(f)$ and $a^*(f)$ in terms of $\Phi(f)$ and $\Pi(f)$, which are the operators we will work with. Notice that we are essentially mimicking the quantization of the simple harmonica oscillator in Quantum Mechanics by means of creation and annihilation (sometimes called *ladder*) operators.

**Proposition 4.3.2.** *Let $\mathcal{H}$ be a complex Hilbert space, $\mathfrak{F}_+(\mathcal{H})$ be the Bose-Fock space and $\Phi(f)$ defined as above, with $a(f)$ and $a^*(f)$ satisfying the CCR. It follows that*

1. *For each $f \in \mathcal{H}$, $\Phi(f)$ is essentially selfadjoint on $\mathfrak{F}_+(\mathcal{H})$ and if $||f_n - f|| \to 0$, then $||\Phi(f_n) - \Phi(f)|| \to 0$ for all $\psi \in D(N^{1/2})$;*

2. *If $\Omega = (1, 0, 0, ...)$, then the linear span of the set $\{\Phi(f_1)\Phi(f_2)...\Phi(f_n)\Omega\}$ is dense in $\mathfrak{F}_+(\mathcal{H})$;*

3. *For each $\psi \in D(N)$ and $f, g \in \mathcal{H}$, we have*

$$(\Phi(f)\Phi(g) - \Phi(g)\Phi(f))\psi = i \operatorname{Im}\langle f, g \rangle \psi. \tag{4.3.21}$$

*Proof.* See [26], proposition 5.2.3.                                            ∎

Because of the first item, we will consider the self-adjoint closure of $\Phi(f)$ and use the same notation to denote the self-adjoint operator that results from this extension. This extension will generate the unitary groups that result in the CCR algebra, which we will be interested in.

**Theorem 4.3.1.** *Consider the self-adjoint operator $\Phi(f)$ defined above. Define the unitary operator*

$$W(f) = \exp(i\Phi(f)). \tag{4.3.22}$$

*It follows that*

1. *For all $f, g \in \mathcal{H}$, we have*

$$W(f)D(\Phi(g)) = D(\Phi(g)),$$
$$W(f)\Phi(g)W(f)^* = \Phi(g) - \operatorname{Im}\langle f, g\rangle \mathbf{1};$$
$$W(f)W(g) = e^{-i\operatorname{Im}\langle f, g\rangle/2}W(f+g);$$

2. *If $||f_n - f|| \to 0$, then $||W(f_n)\psi - W(f)\psi|| \to 0$ (strong convergence);*

3. *For all $f \in \mathcal{H} \setminus \{0\}$*

$$||W(f) - \mathbf{1}|| = 2.$$

*Proof.* See [26], proposition 5.2.4.                                            ∎

Item 1 of theorem 4.3.1 introduced the so-called *Weyl operators* and the *Weyl form of the CCR*:

$$W(f)W(g) = e^{-i\operatorname{Im}\langle f, g\rangle/2}W(f+g) = e^{-i\operatorname{Im}\langle f, g\rangle}W(g)W(f). \tag{4.3.23}$$

## 4.4   The CCR Algebra

In this section, we analyze the abstract $C^*$-algebras generated by elements satisfying the Weyl form of the CCR. In order to do this, we will consider a vector space $H$ equipped with a *nondegenerate symplectic bilinear form* $\sigma : H \times H \to \mathbb{R}$. This form is linear on both entries, and it satisfies

1. Anti-symmetry: $\sigma(f, g) = -\sigma(g, f)$ and

2. Non-degeneracy: $\sigma(f, g) = 0, \forall f \in H \implies g = 0$.

If $H = \mathcal{H}$ is a complex (pre) Hilbert space, then we can recover the form of the CCR described previously by choosing

$$\sigma(f, g) = \text{Im}\langle f, g\rangle. \tag{4.4.1}$$

In fact, our choice to work with a symplectic form is more general in the sense that if $H$ is a *real* vector space and is equipped with such form and there exists an operator $J$ on $H$ satisfying

$$\sigma(Jf, g) = -\sigma(f, Jg), \quad J^2 = -1, \tag{4.4.2}$$

then we can make $H$ into a *complex* inner product space by defining

$$\begin{aligned}(\lambda_1 + i\lambda_2)f &= \lambda_1 f + \lambda_2 Jf, \\ \langle f, g\rangle &= \sigma(f, Jg) + i\sigma(f, g),\end{aligned} \tag{4.4.3}$$

with $f, g \in H$ and $\lambda_1, \lambda_2 \in \mathbb{R}$. One can construct the operator $J$ by a Gram-Schmidt type procedure, where we look for elements $\{x_i, y_i\}$ in $H$ which span a dense subspace and satisfy

$$\begin{aligned}\sigma(x_i, x_j) &= \sigma(y_i, y_j) = 0, \\ \sigma(x_i, y_j) &= \delta_{ij}.\end{aligned}$$

We then define $J$ by $Jx_i = y_i$ and $Jy_i = -x_i$, and extend to the whole space by continuity.

**Theorem 4.4.1.** *Let $H$ be a real vector space equipped with a nondegenerate symplectic bilinear form $\sigma$. Let $\mathfrak{A}_i$ with $i = 1, 2$ be two $C^*$-algebras generated by nonzero elements $W_i(f)$, $f \in H$ satisfying*

1. *$W_i(-f) = W_i(f)^*$,*

2. *$W_i(f)W_i(g) = e^{-i\sigma(f,g)/2}W_i(f+g)$ for all $f, g \in H$.*

*It follows then that there exists a unique $*$-isomorphism $\alpha : \mathfrak{A}_1 \to \mathfrak{A}_2$ such that*

$$\alpha(W_1(f)) = W_2(f), \tag{4.4.4}$$

*for all $f \in H$. Thus, the $C^*$-algebra generated by the Weyl operators $W(f)$ is unique up to an isomorphism. Furthermore*

1. *$W(f)$ is unitary for all $f \in H$, $W(0) = \mathbf{1}$ and $||W(f) - \mathbf{1}|| = 2$ for all nonzero $f \in H$;*

*2. If $T$ is a real linear invertible opeartor on $H$ such that*

$$\sigma(Tf, Tg) = \sigma(f, g),$$

*for all $f, g \in H$, then there is a unique $*$-automorphism $\gamma$ of $\mathfrak{A}(H)$ such that*

$$\gamma(W(f)) = W(Tf). \tag{4.4.5}$$

*These are the so-called* Bogoliubov transformations.

*Proof.* See [26], theorem 5.2.8. ∎

## 4.5 States and Representations

We will be interested in working with the so-called *regular representations* of the CCR algebra, which correspond to those where the generators $t \mapsto W(tf)$ exist for $t \in \mathbb{R}$ and $W(f)$ are the Weyl operators defined in equation (4.3.22). In more technical terms, a representation is said to be regular if the one-parameter unitary groups $t \mapsto \pi(W(tf))$ are strongly continuous. Similarly, a state $\omega$ over the CCR algebra is said to be regular if the cyclic representation associated $(\mathcal{H}_\omega, \pi_\omega, \Omega_\omega)$ is regular.

A first example of a regular state is the *Fock state*, which we denote by $\omega_F$. This state is defined by the vacuum state $\Omega = (1, 0, 0, ...) \in \mathfrak{F}_+(\mathcal{H})$. One then computes

$$\omega_F(W(f)) = \langle \Omega, W(f)\Omega \rangle = e^{-\|f\|^2/4}. \tag{4.5.1}$$

In order to obtain this, we recall the *Baker-Campbell-Hausdorff formula* (BCH)[4]

$$\exp(X)\exp(Y) = \exp\left(X + Y + \frac{1}{2}[X, Y] + \frac{1}{12}[X, [X, Y]] + ...\right).$$

Our Weyl operators can be written in the form

$$W(f) = \exp(i\Phi(f)) = \exp\left(\frac{i}{\sqrt{2}}a(f) + \frac{i}{\sqrt{2}}a^*(f)\right),$$

with the creation and annihilation operators obeying the commutation relations

$$[a(f), a^*(g)] = \langle f, g \rangle \mathbf{1}.$$

---

[4]   Technically, we can only *formally* employ the BCH relation since it is valid for *bounded operators*. However, there are ways to make this expression valid [15], with the result being the same. Hence, for the sake of the argument, we will allow us a moment of sloppiness.

Since the commutation relation results in the identity with some numerical factor, it commutes with all other elements, hence we just need to consider the BCH formula up to the first commutator in equation (4.5). Hence, we can write

$$W(f) = \exp(i\Phi(f)) = \exp\left(\frac{i}{\sqrt{2}}a^*(f)\right)\exp\left(\frac{i}{\sqrt{2}}a(f)\right)\exp\left(-\frac{1}{4}[a(f), a^*(f)]\right)$$

$$= \exp\left(\frac{i}{\sqrt{2}}a^*(f)\right)\exp\left(\frac{i}{\sqrt{2}}a(f)\right)\exp\left(-\frac{1}{4}||f||^2\right).$$

From this, we recall that the exponentials are defined as

$$\exp\left(\frac{i}{\sqrt{2}}a(f)\right) = \mathbf{1} + \sum_{i=1}^{\infty}\left(\frac{i}{\sqrt{2}}a(f)\right)^n,$$

and similar for $a^*(f)$. Hence, the action on the vacuum vector is simply

$$\left(\exp\left(\frac{i}{\sqrt{2}}a(f)\right)\right)\Omega = \left(\mathbf{1} + \sum_{i=1}^{\infty}\left(\frac{i}{\sqrt{2}}a(f)\right)^n\right)\Omega = \Omega,$$

since $a(f)\Omega = 0$. Using the property in equation (4.3.11), we compute

$$\langle\Omega, W(f)\Omega\rangle = \exp\left(-\frac{1}{4}||f||^2\right)\left\langle\Omega, \exp\left(\frac{i}{\sqrt{2}}a^*(f)\right)\exp\left(\frac{i}{\sqrt{2}}a(f)\right)\Omega\right\rangle$$

$$= \exp\left(-\frac{1}{4}||f||^2\right)\left\langle\exp\left(\frac{i}{\sqrt{2}}a(f)\right)\Omega, \exp\left(\frac{i}{\sqrt{2}}a(f)\right)\Omega\right\rangle$$

$$= \exp\left(-\frac{1}{4}||f||^2\right)\langle\Omega, \Omega\rangle = \exp\left(-\frac{1}{4}||f||^2\right).$$

In fact, the Fock representation turns out to be the most important for our purposes because of the *Stone-von Neumann Uniqueness Theorem*. In order to present this theorem, we mention that one is concerned with normal states with respect to the Fock representation. These are states of the form $\omega(A) = \text{Tr}(\rho A)$, which is equivalent to the GNS representation being a direct sum of copies of the representation of the CCR on $\mathcal{H}$. Normality is an important criterion since they correspond to convex combinations of finite-particle vector states. Hence, they are interpreted as states with a finite number of particles.

**Theorem 4.5.1 (Stone-von Neumann).** *Let $\mathfrak{A}(\mathcal{H})$ be the CCR algebra over a finite-dimensional Hilbert space $\mathcal{H}$. It follows that each regular state $\omega$ over $\mathfrak{A}(\mathcal{H})$ is normal with respect to the Fock representation. Hence, any regular representation of $\mathfrak{A}(\mathcal{H})$ is a multiple of the Fock representation.*

*Proof.* See [26], corollary 5.2.15 for one of the (many) proofs for this theorem. ∎

Even though the Stone-von Neumann states that the finite-dimensional representations of the CCR are equivalent to the Fock representation, this is not true in the infinite-dimensional case. One of the first observations that led to the algebraic approach to QFT was the importance of non-unitarily equivalent representations of the CCR. We will see more of this in the forecoming chapters.

# Part II

# Fields

# 5  Algebraic Quantum Field Theory

*AQFT proceeds by isolating some structural assumptions that hold in most known QFT models. It formalizes these structural assumptions, and then uses "abstract but efficient nonsense" to derive consequences of these assumptions.*

Hans Halvorson

In this chapter, we present the formulation of Algebraic Quantum Field Theory in terms of the Haag-Kastler-Araki (HAK) axioms. We show that these axioms are strong enough to constrain certain aspects of the theory and give rise to certain interesting properties such as the one encoded by the Reeh-Schlieder theorem. We also comment on the "universal structure" of the local algebras, which is a set of results that show that all local algebras of consistent QFTs are determined by a unique type of von Neumann factor. We finish by presenting the so-called *Fermi paradox*, which rises from attempting to treat Quantum Field Theory in the same footing as non-relativistic Quantum Mechanics.

## 5.1  Introduction

The program of attempting to make QFT rigorous began around the 1950s, mainly with the efforts of Wightman et al. Their approach which is nowadays understood in terms of the *Wightman axioms* emphasized the mathematical structure of the quantum fields, regarding them as *operator-valued distributions*. They also emphasized the construction of the *Wightman functions*, which is a mathematically precise formulation of the correlation functions or *n*-point functions from standard QFT, in the language of distributions. We will not work in this formulation in this Thesis, referring the interested reader to the original monograph by Streater and Wightman [125], which is, in our opinion, still the best exposition of these ideas together with the standard proofs of foundational theorems in QFT such as the Edge of the Wedge theorem, Haag's theorem, the PCT theorem and the spin-statistics theorem (the last two give the iconic title of the book).

By the 60's, Haag, Kastler, Araki and many others shifted the conceptual focus of the axiomatization program to the *local algebras of observables* (or, stated more precisely, to the *net* of algebras, meaning the mapping $\mathcal{O} \mapsto \mathcal{A}(\mathcal{O})$ of an algebra to a open set in spacetime). This program has received the name of *algebraic quantum field theory* or AQFT for short, and it relies heavily on the concepts of $C^*$ and von Neuman algebras developed in the previous chapters.

A motivation behind the development of the AQFT framework can be traced to what is called the *van Hove model* of a neutral scalar with an external source [133], where it was first realized that unitarily inequivalent representations of the CCR algebras are indeed important. Moreover, such work inspired Haag to develop what is now called *Haag's theorem*, which stated the inexistence of the interaction picture, in general. A nice review of the subject can be found in [68].

One can state that the main goal of AQFT is to provide very general and model-independent statements which any bona fide quantum field theory should satisfy. In theory, any quantum field theory should have a equivalent algebraic formulation, with this equivalence being verified by the agreement in the observables. It should be noted that AQFT, like the Wightman program, is a *non-perturbative* formulation of QFT.

In the past decades, many statements about free theories have been casted in this language with some rigorous proofs being made for low dimensional models. QFT in curved spacetimes has also benefited from the Hilbert space-independent formulation of AQFT, with algebraic methods becoming the modern language for QFT in curved spacetimes. In even more recent years, there has been a better understanding on how to handle interactions in the context of $C^*$-algebras [36]. A summary of these recent developments can be found in [28]. The classic monographs of Haag [78] and of Araki [4] remain the standard references for an introduction to AQFT.

The mathematical pre-requisites include the topics worked in the fist part of this Thesis. We should now be able to state the axioms of AQFT and present its main results needed for an understanding of entanglement and Bell's inequalities in the Part III.

## 5.2   The Haag-Araki-Kastler Axioms

The use of $C^*$-algebras to describe Quantum Physics was first championed by Segal in 1947 [122]. In that work, he pointed out that many questions of physical interest such as the determination of spectral values for observables can be addressed without reference to a specific Hilbert space, one the algebra of observables is chosen to be a $C^*$-algebra. Remarkably, this feature would be used to argue in favor of algebraic methods in QFT in curved spacetimes in the 1980's [136].

Segal was in fact interested in circumventing the difficulties associated with the existence of non-equivalent representations of the commutation relations. Building upon these ideas, Haag and Kastler proposed in 1964 the first version of the axioms which would give birth to AQFT [80]. This set of axioms is commonly referred as the *Haag-Kastler axioms*, but due to the many contributions of Araki to the field of research, some authors call these the *Haag-Araki-Kastler axioms*. We will adhere to the latter convention, henceforth abbreviating to HAK axioms.

Originally, Haag and Kastler presented the following version of the axioms: let $\mathcal{O}$ be a "region" of Minkowski spacetime $\mathbb{M}$ to which we associate a $C^*$-algebra $\mathfrak{A}(\mathcal{O})$, that is

$$\mathcal{O} \mapsto \mathfrak{A}(\mathcal{O}). \tag{5.2.1}$$

Then, we require

1. The regions $\mathcal{O} \subset \mathbb{M}$ for which the mapping 5.2.1 is defined are open sets with compact closures.[1]

2. **Isotony:** If $\mathcal{O}_1 \subset \mathcal{O}_2$ then $\mathfrak{A}(\mathcal{O}_1) \subset \mathfrak{A}(\mathcal{O}_2)$ and we assume that either $\mathfrak{A}(\mathcal{O}_1)$ and $\mathfrak{A}(\mathcal{O}_2)$ have a common unit element or that neither of them has a unit.[2]

3. **Einstein Causality**[3]**:** If $\mathcal{O}_1$ and $\mathcal{O}_2$ are spacelike seaparated from each other[4] then $\mathfrak{A}(\mathcal{O}_1)$ and $\mathfrak{A}(\mathcal{O}_2)$ commute with each other, i.e.,

$$[\mathfrak{A}(\mathcal{O}_1), \mathfrak{A}(\mathcal{O}_2)] = \{0\}. \tag{5.2.2}$$

---

[1]   In more modern terminology, this requirement is usually referred to as *causal convexity* and it physically represents 4-dimensional regions with finite extension. That is, the algebras are associated to bounded, open and causally convex regions, the latter meaning that every causal curve whose endpoints are in the region is completely contained in that region.

[2]   Of course, if the latter prevails, one can obtain the first situtation by a formal adjunction of a unit to the algebras.

[3]   Originally called *local commutativity*. Nowadays, this is sometimes called *microcausality*.

[4]   We shall review causal structures in the next chapters when we introduce the notion of curved spacetimes.

4. **Algebra of Quasilocal Observables:** The set-theoretic union of all $\mathfrak{A}(\mathcal{O})$ becomes a normed $*$-algebra due to the isotony assumption. We take its completion and use the uniqueness of the norm to make the union a $C^*$-algebra denoted by $\mathfrak{A}$. This algebra is called the quasilocal algebra of observables and it contains the uniform limits of all bounded observables describing measurements performable in finite regions of spacetimes. Since we consider uniform limits, we do not change the local character of the observables (hence the name quasilocal). This algebra is also deemed *primitive*, meaning that it admits a faithful irreducible representation (by pure states).

5. **Poincaré[5] Covariance:** The (identity connected component of the) Poincaré group $\mathscr{P}$ is represented algebraic automorphisms $\alpha(P) : \mathfrak{A}(\mathcal{O}) \to \mathfrak{A}(P\mathcal{O})$, $P \in \mathscr{P}$ and $P\mathcal{O}$ is the image of the region $\mathcal{O}$ under the Poincaré transformation $P$. Additionally, we require for the identity $\mathrm{id}_{\mathscr{P}} \in \mathscr{P}$ that $\alpha(\mathrm{id}_{\mathscr{P}}) = \mathrm{id}_{\mathfrak{A}(\mathcal{O})}$ and that the automorphism preserves composition of transformations, i.e. $\alpha(P_1) \circ \alpha(P_2) = \alpha(P_1 \circ P_2)$ for any $P_1, P_2 \in \mathscr{P}$.

In more modern treatments of the HAK axioms it is common to include an extra axiom for the existence of dynamics. Sometimes, this axiom is called the *time-slice axiom*, and its stated as

6. **Existence of Dynamics or Time-Slice Axiom:** If $\mathcal{O}_1 \subset \mathcal{O}_2$ and $\mathcal{O}_1$ contains a Cauchy surface[6] of $\mathcal{O}_2$, then

$$\mathfrak{A}(\mathcal{O}_2) = \mathfrak{A}(\mathcal{O}_1). \tag{5.2.3}$$

This axiom is frequently weakened. An example of how such weakening occurs is by associating to each causal inclusion $\mathcal{O}_1 \hookrightarrow \mathcal{O}_2$ an isomorphism between the corresponding algebras. However, some authors completely omit this axiom since the axiom about Poincaré covariance already carries dynamical information by defining the automorphisms of the algebra [55]. This omission is valid only when working in Minkowski spacetime.

An axiom which is also convenient to include is that of *weak additivity*. It is particularly important in the context of the Reeh-Schlieder theorem and the Tomita-Takesaki modular theory, so we defer its presentation to section 5.3.2.

---

5   Originally, Haag and Kastler referred to this axiom as Lorentz covariance and they talked about the representations of the inhomogeneous Lorentz group, which is nowadays simply called the Poincaré group.

6   A Cauchy surface for $\mathcal{O}_2$ is a subset met exactly once by every inextendible timelike curve contained in $\mathcal{O}_2$. In the next chapter, we will review this concept.

### 5.2.1 Some remarks and interpretations

The first natural interpretation of the axioms presented is that the self-adjoint elements of the local algebra of observables $\mathfrak{A}(\mathcal{O})$ are the physical observables which can be measured within the region $\mathcal{O}$. Naturally, these values which can be measured lie in the spectrum of the elements of the algebra.

For the local regions, one in general treats only *double-cones*, which are the sets of all points lying on smooth timelike curves between two points $p, q \in \mathbb{M}$. An elementary example is given by

$$\mathcal{O} = \{(t, \mathbf{x}) \in \mathbb{R}^4 : |t| + |\mathbf{x}| < \ell_0\}, \tag{5.2.4}$$

for some length-scale $\ell_0$. By scalings, boosts and translations, every double-cone in $\mathbb{M}$ can be obtained from this example.

We also comment about the Causality axiom. Notice that if we are considering Fermi field operators in causally separated regions, then the observables should *anti-commute*, not commute. However, this is not inconsistent with the axiom as long as we observables as elements of the algebra, rather than field operators (which are regarded as being *unobservable*).

We should also note that we will be interested in two types of objects: *states* and *operations*. We have already discussed the algebraic definition of a state, which is employed to characterize a statistical ensemble of physical systems. In the notation due to von Neumann, a state can be a mixed state if it is composed of pure states (which themselves cannot be decomposed further). An operation, on the other hand is a physical apparatus which may act on the systems of an ensemble during a limited amount of time, yielding a transformation from an initial state to a final state. They are represented by linear transformations of the dual $\mathfrak{A}^*$ which maps positive functionals into themselves and preserves the norm. If the operation transforms a pure state into a new pure state, it is called a *pure operation*. For a general state $\varphi$, the transformation by a pure operation $a$ is given by $\varphi \to \varphi_a$ with

$$\varphi_a(b) = \varphi(a^* b a), \tag{5.2.5}$$

for some $b \in \mathfrak{A}$.

Many examples of theories described by Lagrangians are compatible with the HAK axioms and can be found in the literature. Some of the original work of Araki was in constructing algebras for free fields [1, 2, 3]. A very important extension was achieved by Dimock who was able to construct the algebras of observables for scalar fields in arbitrary manifolds [55]. In a more recent development, Buchholz and Fredenhagen were able to use the $C^*$-algebraic paradigm to include interacting models for bosons [36] and Brunetti et. al.

extended the formalism to include fermions [29]. The inclusion of electromagnetism as an example of a gauge theory in this paradigm can be found in [33]. More examples including conformal theories can be found in chapter 2 of [86].

## 5.3 The Reeh-Schlieder Theorem

In this section we present the important theorem due to Reeh and Schlieder, first presented in [112].

### 5.3.1 The spectrum condition

For the Poincaré Covariance axiom, we will particularly be interested in representations of the algebra where the Poincaré automorphisms are unitarily implemented, meaning that

$$\pi(\alpha(P)a) = U(P)\pi(a)U(P)^{-1}, \tag{5.3.1}$$

where $U(P)$ corresponds to a representation of the Poincaré group by bounded operators acting on the Hilbert space associated with the representation $\pi$. In this case, we say that *the representation is Poincaré covariant*. Correspondingly, a state is said to be Poincaré covariant if

$$\omega(\alpha(P)a) = \omega(a), \tag{5.3.2}$$

for all $a \in \mathfrak{A}$ and $P \in \mathscr{P}$. An alternative notation that will be used later uses the notion of a *pullback*

$$\alpha(P)^*\omega = \omega. \tag{5.3.3}$$

A convenient fact illustrated by the following theorem tells us that the GNS representation associated to the state $\omega$ is Poincaré covariant.

**Theorem 5.3.1.** *Let $\alpha$ be an automorphism of a unital $*$-algebra $\mathfrak{A}$. If a state $\omega$ on $\mathfrak{A}$ is invariant under $\alpha$, i.e., satisfying equation* (5.3.3), *then $\alpha$ is unitarily implemented in the GNS representation of $\omega$ by a unitary that leaves $\Omega_\omega$ invariant. Moreover, any group of automorphisms which leaves $\omega$ invariant is unitarily represented in $\mathcal{H}_\omega$.*

*Proof.* One considers the left ideal (see definition 2.2.2) of $\mathfrak{A}$ given by

$$\mathcal{I}_\omega = \{a \in \mathfrak{A} : \omega(a^*a) = 0\}.$$

Next, the observation that $\omega(\alpha(a)^*\alpha(a)) = (\alpha^*\omega)(a^*a) = \omega(a^*a)$ shows that $\mathcal{I}_\omega$ is invariant under the action of $\alpha$. Hence, we can define the map $U : \mathfrak{A}/\mathcal{I}_\omega \to \mathfrak{A}/\mathcal{I}_\omega$ by

$$U[a] = [\alpha(a)],$$

where the square brackets denote the equivalence class determined by the ideal. By considering $\alpha^{-1}$, we can show this map is invertible and it fixes $\Omega_\omega = [\mathbf{1}]$, which is trivially left invariant under the action of $U$. Now,

$$\langle U[a]\Omega_\omega, U[b]\Omega_\omega \rangle = \omega(\alpha(a)^*\alpha(a)) = \omega(a^*a) = \langle [a]\Omega_\omega, [b]\Omega_\omega \rangle.$$

Hence, $U$ defines an invertible isometry and it can be extended uniquely to an unitary operator acting on $\mathcal{H}_\omega$. This operator implements $\alpha$:

$$\pi_\omega(\alpha(a))[b] = [\alpha(a)b] = [\alpha(a a \alpha^{-1}(b))] = U[a\alpha^{-1}(b)] = U\pi_\omega(a)[\alpha^{-1}b] = U\pi_\omega(a)U^{-1}[b].$$

Now, if we consider a second automorphism $\beta$ which leaves $\omega$ invariant and $V$ its implementation as described above, we have

$$UV[a] = [\alpha(\beta(a))] = [(\alpha \circ \beta)(a)],$$

which shows that $UV$ implements $\alpha \circ \beta$, and hence any automorphism can be implemented by a unitary operator obtained by successive compositions. ∎

The automorphism can, in particular, represent a translation, which is part of the Poincaré group. Some physically relevant states such as thermal states are translation invariant but not Lorentz invariant, since they have a definite rest frame[7]. Even so, not all Poincaré invariant ou translation invariant states are physically acceptable, so we need an extra criterion to determine the physical states.

Let $\omega$ be a translation invariant state whose unitary implementation of the translation group is strongly continuous, i.e., $\mathbb{R}^4 \ni x \mapsto U(x)\psi \in \mathcal{H}_\omega$ is continuous. This defines a strongly continuous one-parameter unitary group, and we can apply Stone's theorem and its generalization[8] [82] to obtain

$$U(x) = \exp(iP_\mu x^\mu), \tag{5.3.4}$$

where the summation over the repeated indices is implied and $\mu \in \{0, 1, 2, 3\}$ (hence, there are four generators, one for each spacetime direction) and $P_\mu$ is the momentum *operator*.

Now, to any (Borel[9]) subset $\delta \subset \mathbb{R}^4$, we assign a projection operator $E(\delta)$, whose interpretation is that it tells whether or not the result of the value of the four-momentum measured by the generator $P^\mu$ would lie in $\delta$. This assignment $\delta \mapsto E(\delta)$ is called a *projection-valued measure* or *PVM* and it allows us to write

$$U(x) = \int \exp(ip_\mu x^\mu)dE(p), \tag{5.3.5}$$

---

[7] This observation leads to the so-called *Unruh effect*, see [136] for a standard introduction.

[8] Referred to as the Stone-Naimark-Ambrose-Gelfand or *SNAG* theorem.

[9] See [117] or any good book on Real Analysis or Measure Theory.

where the lower-case $p$ denotes the values of the four-momentum.

With this, we can now define the condition for a physically acceptable state.

**Definition 5.3.1.** The state $\omega$ is said to satisfy the *spectrum condition* if the support of the measure $E$ lies in the *closed forward cone*, i.e.,

$$\operatorname{supp} E \subset \overline{V^+} = \{p \in \mathbb{R}^4 : p^\mu p_\mu \geq 0, p^0 \geq 0\}. \tag{5.3.6}$$

This condition is sometimes referred to as saying that the joint spectrum of the momentum operators $P^\mu$ lies in $\overline{V^+}$, and it reflects the choice for states of positive-definite energy.

An important consequence of this condition is that one can analytically-continue the definition of $U(x)$ to include complex vectors $x \in \mathbb{R}^4 + iV^+$, where $V^+ = \operatorname{int}(\overline{V^+})$. It also allows us to give a precise definition of the vacuum state.

**Definition 5.3.2.** A *vacuum state* is a translation invariant state $\omega_0$ which obeys the spectrum condition and whose GNS vector $\Omega_{\omega_0}$ is the unique translation invariant vector up to scalar multiples in $\mathcal{H}_{\omega_0}$. The corresponding GNS representation is called the vacuum representation.

A very important that can be shown to hold in physically interesting theories is called *Haag's duality*. It was shown to hold in physically interesting theories by Araki [2] and Eckmann and Osterwalder [61] (see [70] for a modern discussion). This was the basis for the celebrated works by Bisognano and Wichmann, where they showed, among other things, that this duality holds for general scalar fields [21, 22].

**Definition 5.3.3.** We say that the vacuum state satisfies *Haag's duality* if its GNS representation $\pi_0$ satisfies

$$\pi_0(\mathfrak{A}(\mathcal{O}'))' = \pi_0(\mathfrak{A}(\mathcal{O}))'', \tag{5.3.7}$$

where $\mathcal{O}'$ denotes the causal complement of the spacetime region $\mathcal{O}$.

This axiom will become particularly important in the context of violations of Bell's inequalities.

## 5.3.2 Weak additivity

We now present an additional axiom for the HAK which will allow us to prove the Reeh-Schlieder theorem in Minkowski spacetime. This theorem "came as a surprise" as stated in [125] and it shows how different QFT can be from ordinary Quantum Mechanics.

The axiom is given as

7. **Weak Additivity:** For any causally convex open region $\mathcal{O}$, $\pi(\mathfrak{A}(\mathbb{M}))$ is contained in the weak closure of the $*$-algebra generated by the algebras $\pi(\mathfrak{A}(\mathcal{O} + x))$ as $x$ runs over $\mathbb{R}^4$.

Because of the Bicommutant Theorem, we can state the condition as saying that $\pi(\mathfrak{A}(\mathbb{M})) \subset \pi(\mathfrak{B}(\mathcal{O}))''$, where $\mathfrak{B}(\mathcal{O})$ is the algebra generated described above.

This essentially says that the algebra of observables over all Minkowski spacetime can be built as limits of algebraic combinations of translates of observables in a given region $\mathcal{O}$. One should expect this from a translation-invariant theory such as QFT in Minkowski spacetime. However, assuming this axiom results in the Reeh-Schlieder theorem. Our proof follows closely [4].

**Theorem 5.3.2** (**Reeh-Schlieder Theorem**)**.** *Let $\mathcal{O}$ be a causally convex bounded open region in $\mathbb{M}$ and $\mathfrak{A}(\mathcal{O})$ its local algebra. Let $\omega$ be a state over $\mathfrak{A}(\mathcal{O})$ which obeys the spectrum condition and $\Omega$ its GNS vector. Then, $\Omega$ is cyclic and separating for $\mathfrak{A}(\mathcal{O})$.*

*Proof.* The proof for $\Omega$ being cyclic follows by contradiction. Hence, we suppose that the set of vectors of the form $A\Omega$ for $A \in \pi(\mathfrak{A}(\mathcal{O}))$ which is *not* dense in $\mathcal{H}$. This means that there is a non-zero vector $\psi \in \mathcal{H}$ such that $\langle \psi, A\Omega \rangle = 0$, for all $A \in \pi(\mathfrak{A}(\mathcal{O}))$.

Now consider a subset $\mathcal{O}_1$ of $\mathcal{O}$ such that $\overline{\mathcal{O}_1} \subset \mathcal{O}$, where the overline denotes the causal closure of $\mathcal{O}_1$. Then, for any sufficiently small $x \in \mathbb{R}^4$, we have that

$$\mathcal{O}_1 + x \subset \mathcal{O} \tag{5.3.8}$$

and because we are assuming translation invariance, we can say that

$$U(x)Q_iU(x)^{-1} \in \pi(\mathfrak{A}(\mathcal{O})), \tag{5.3.9}$$

where $U(x)$ is the unitary implementation of the translation group and $Q_i$ for $i \in \{1, ..., n\}$ is a collection of operators in $\pi(\mathfrak{A}(\mathcal{O}))$. Now, the condition in equation (5.3.8) implies that for a sufficient small collection of $x_i$, we hae

$$\langle \psi, U(x_1)Q_1U(x_2 - x_1)Q_2U(x_3 - x_2)...U(x_n - x_{n-1})Q_n\Omega \rangle,$$

where we have already used that $U(x_i)U(x_j) = U(x_i + x_j)$ and $U(x)^{-1} = U(-x)$.

Now, the analytic continuation of $U(x)$ allows us to define the function $F : (y_1, ..., y_n) \mapsto \langle \psi, U(y_1)Q_1U(y_2)...U(y_n)Q_n\Omega \rangle$ as an analytic function in $(\mathbb{R}^4 + iV^+)^n \subset \mathbb{C}^{4n}$. Now, because this function is continuous on $\mathbb{R}^{4n}$ and it vanishes in some neighborhood of the origin, it must vanish identically in the domain of analyticity, which includes its boundary values[10]. Hence, denoting by $\pi(\mathfrak{B}(\mathcal{O}))$ the algebra generated by $\pi(\mathfrak{A}(\mathcal{O}_1 + x))$ for $x \in \mathbb{R}^4$, we have that

$$\langle \psi, \pi(\mathfrak{B}(\mathcal{O}))\Omega \rangle = 0.$$

On the other hand, by weak additivity the weak closure of $\pi(\mathfrak{B}(\mathcal{O}))$ is $\pi(\mathfrak{A}(\mathbb{M}))$. Furthermore, because $\pi(\mathfrak{A}(\mathcal{O}))\Omega$ is dense due to the GNS construction, we have that $\pi(\mathfrak{B}(\mathcal{O}))\Omega$ is also dense. These conditions imply that $\psi = 0$, a contradiction. Hence, $\Omega$ is cyclic.

To prove that $\Omega$ is separating, we suppose that $A\Omega = 0$ for some $A \in \pi(\mathfrak{A}(\mathcal{O}_1))$. Now we choose some subset $\mathcal{O}_2$ which is causally disjoint from $\mathcal{O}_1$. Using Einstein causality, we note that for each $\psi \in \mathcal{H}$ and all $B \in \pi(\mathfrak{A}(\mathcal{O}_2))$, we have

$$\langle A^*\psi, B\Omega \rangle = \langle \psi, AB\Omega \rangle = \langle \psi, BA\Omega \rangle = 0.$$

Therefore, $A^*\psi$ is orthogonal to a dense set of vectors, and then must vanish by the first part of the theorem. Since $\psi$ is arbitrary, this implies that $A^* = 0$ which in turn implies that $A = 0$. Hence, $\Omega$ is separating. ∎

For an alternate proof which does not assume weak additivity but is formulated in terms of Wightman functions, see [142]. That proof is based on [125]. Notice the crucial assumption of translation invariance in equation (5.3.9), which poses some technical difficulty when generalizing the Reeh-Schlieder theorem to curved spacetimes.

We first recall that a cyclic and separating vector is the starting part for the Tomita-Takesaki Modular Theory, which will eventually allow us to classify the von Neumann algebra of observables. However, the physical consequences of the theorem can already be appreciated.

Essentially, the theorem says that *entanglement* is a deep physical property of physical states in QFTs. This is because the cyclicity property allows one in principle to construct any physical state with an arbitrary precision by acting with elements of the algebra on the vacuum vector state. Even more, because the vacuum is separating and because of proposition 3.6.1, the same is true for the commutant of the algebra, which is located at a causally disjoint region of spacetime, in light of Einstein causality. In fact, the vacuum state is cyclic and separating both for the algebra and its commutant, which starts to resemble some idea of entanglement.

Some corollaries can be derived from the Reeh-Schlieder theorem.

---

[10]   One can see this in two ways: either by proving that any analytic function in some domain $D$ which vanishes in a straight line contained in $D$ vanishes identically (see for instance [27], chapter 2, section 28) or as a consequence of the *edge of the wedge theorem*, which is important on its own (see for instance [125], where it is used in proving the analytic continuation of Wightman functions).

**Corollary 5.3.1.** *A binary test, i.e., a measurement with two outcomes (say, 0 or 1) have a nonzero probability of yielding 1 in the vacuum state.*

*Proof.* We can represent a binary test by a orthogonal projector operator $P$. The result 0 is taken to be the kernel of such operator. Hence, suppose $P \in \pi(\mathfrak{A}(\mathcal{O}))$ is a orthogonal projector with vanishing expectation value in the vacuum state (i.e., zero probability of yielding 1). Then

$$||P\Omega||^2 = \langle P\Omega, P\Omega \rangle = \langle \Omega, P^*P\Omega \rangle = \langle \Omega, P\Omega \rangle = 0,$$

where we have used the self-adjointness and idempotence of $P$. Therefore, $P\Omega = 0$ which implies that $P = 0$ by the Reeh-Schlieder theorem. ∎

This corollary tacitly assumes that the binary tests are non-trivial (i.e., $P$ is not the identity operator) and that they are "sharply defined".

As a prelude to the formulation of Bell's inequalities in AQFT, we comment that Redhead was able to show that the Reeh-Schlieder property in QFTs entails in non-local correlations of the vacuum state [110, 109]. This is illustrated by the following corollary.

**Corollary 5.3.2.** *For every pair of local regions $\mathcal{O}_1$ and $\mathcal{O}_2$, there are vacuum correlations between the algebras $\pi(\mathfrak{A}(\mathcal{O}_1))$ and $\pi(\mathfrak{A}(\mathcal{O}_2))$ which act on a Hilbert space of dimension $dim\,\mathcal{H} \geq 1$.*

*Proof.* The proof is by contradiction. For this, we assume that the correlation between two observables can be decomposed as a product of two states, namely

$$\langle \Omega, \pi(a_1)\pi(a_2)\Omega \rangle = \omega_1(a_1)\omega_2(a_2), \tag{5.3.10}$$

for all $a_1, a_2 \in \mathfrak{A}(\mathcal{O})$ and for some $\omega_1, \omega_2 \in \mathfrak{A}(\mathcal{O})^*$. In particular, by choosing $\pi(a_1) = \mathbf{1}$ and repeating the process for $a_2$, we can write

$$\langle \Omega, \pi(a_1)\pi(a_2)\Omega \rangle = \langle \Omega, \pi(a_1)\Omega \rangle \langle \Omega, \pi(a_2)\Omega \rangle. \tag{5.3.11}$$

By fixing $a_2$ and letting $a_1$ vary as we translate between regions in $\mathfrak{A}(\mathcal{O}_1)$, as constructed in the proof for the Reeh-Schlieder theorem, we find

$$\pi(a_2)\Omega = \langle \Omega, \pi(a_2)\Omega \rangle \Omega,$$

which contradicts the Reeh-Schlieder theorem, except when $\dim \mathcal{H} = 1$. ∎

The proof for this corollary already uses the notion of a product of states as representing the absence of correlations, in equation (5.3.10). We will develop these notions further in Part III of this Thesis.

### 5.3.3   The cluster property

The result obtained in equation (5.3.11) (by assuming the correlations are written as a product of states) is an example of a general feature of QFTs called *cluster property* ou *cluster decomposition principle* [138]. Essentially, this says that the vacuum expectation value of pairs of observables can be factorized as

$$\langle \Omega, \pi(a_1)\pi(\alpha(x)a_2)\Omega \rangle \to \langle \Omega, \pi(a_1)\Omega \rangle \langle \Omega, \pi(a_2)\Omega \rangle, \tag{5.3.12}$$

for action of the symmetry group $\alpha$. In the context of AQFT, this decomposition follows from the fact that the weak-closure of the algebra generated by the vacuum GNS representation is a von Neumann factor (see [4], theorem 4.6).

We mention *en passant* that in the case where the theory has a *mass gap*, meaning that the vacuum energy is an isolated singularity in the energy-momentum spectrum, than a stronger form of the cluster property holds.

**Theorem 5.3.3.** *Let $\omega$ be a vacuum state and $E$ the PVM associated with the unitary implementation of the translation group (see definition 5.3.1). Suppose that the support for $E$ lies in $\{p \in \mathbb{R}^4 : p^0 \geq m\}$, for some real number $m$.*

*Then, for two regions $\mathcal{O}_1$ and $\mathcal{O}_2$ satisfying $\mathcal{O}_2 + te \subset \mathcal{O}_1'$ with $|t| < T$ and $e = (1, 0, 0, 0)$ the following inequality holds*

$$|\omega(a_1 a_2) - \omega(a_1)\omega(a_2)| \leq Gk(mT/2), \tag{5.3.13}$$

*where $k$ is a spacetime-valued function defined by $k(x) = \frac{1}{2}((\pi x)^{-1/2} + (\pi x)^{-1})e^{-x}$ and $G = (\omega(a_1 a_1^*)\omega(a_2^* a_2))^{1/2} + (\omega(a_1^* a_1)\omega(a_2 a_2^*))^{1/2}$.*

*Proof.* See [4], theorem 4.7.                                                                      ■

What this theorem tells us is that in spacelike directions, as $x \to \infty$, one has the cluster property at an exponentially rapid convergence rate, provided the theory has a mass-gap. Hence, to a very good approximation, the vacuum is a product state for observables located in two spacelike separated regions.

## 5.4   The Universal Structure of the Local Algebras

We now present one of the most fundamental results in AQFT, namely the characterization of the local algebras. This theme has been echoed throughout this Thesis and now we have sufficient tools to understand this result. Of course a full understanding of the original results and proofs in [69, 34] and beyond would require a deeper understanding of the Tomita-Takesaki Modular Theory which we only discussed aspects of. Nevertheless, we hope to leave the reader acquainted with some of these results and we hope to encourage them in learning more about Modular Theory and von Neumann algebras.

The first observation is that the local algebras for a QFT are factors. This seems surprising at first, since we stated that more complex von Neumann algebras can be built out of factors by a direct integration. Why should the local algebra be the simplest building block? An argument for this is the following: the axiom of Einstein causality implies that for two causally disjoint regions $\mathcal{O}_1$ and $\mathcal{O}_2$, we have that the GNS representations of the local algebras for some state $\omega$ satisfy $\pi_\omega(\mathfrak{A}(\mathcal{O}_1)) \subset \pi_\omega(\mathfrak{A}(\mathcal{O}_2))'$ and vice-versa. Hence, if we assume that observers on $\mathcal{O}_1$ can't access non-trivial observables in $\mathcal{O}_2$, then one should expect that

$$\pi_\omega(\mathfrak{A}(\mathcal{O}_1)) \cap \pi_\omega(\mathfrak{A}(\mathcal{O}_2)) = \mathbb{C}\mathbf{1},$$

which is exactly the definition of a factor.

The next characterization is that the local algebras are *locally quasi-equivalent* to some vacuum (or translation invariant) representation. To make this precise, we present the following definition.

**Definition 5.4.1.** Let $\mathcal{O} \mapsto \mathfrak{M}_1(\mathcal{O})$ and $\mathcal{O} \mapsto \mathfrak{M}_2(\mathcal{O})$ be two nets of von Neumann algebras acting on a Hilbert space $\mathcal{H}$. We say that $\mathfrak{M}_1$ and $\mathfrak{M}_2$ are *locally quasi-equivalent* if for each double cone $\mathcal{D}$ there is an isomorphism $i_\mathcal{D} : \mathfrak{M}_1(\mathcal{D}) \to \mathfrak{M}_2(\mathcal{D})$.

We will take this as an axiom even though there are some arguments that this is true, in what is called the *Doplicher-Haag-Roberts* or *DHR criterion* [57, 58] or the *Bucholz-Fredenhagen criterion* [35].

Now, if we assume weak additivity and that the vacuum state satisfies the spectrum condition, then the theory satisfies what is sometimes referred to as *property B*.

**Definition 5.4.2.** Let $\mathcal{O} \to \mathfrak{M}(\mathcal{O})$ be a net of von Neumann algebras on a Hilbert space $\mathcal{H}$. Then, we say that the net *satisfies property B* if for any two double cones $\mathcal{D}_1$ and $\mathcal{D}_2$ such that $\overline{\mathcal{O}_1} \subseteq \mathcal{O}_2$ and a nonzero projection $E \in \mathfrak{M}(\mathcal{O}_2)$ we have that $E \sim \mathbf{1}$, i.e., there is an isometry $V \in \mathfrak{M}(\mathcal{O}_2)$ such that $VV^* = E$.

Notice that by definition, type III factors satisfy property B.

**Proposition 5.4.1.** *Let $\mathcal{O} \mapsto \mathfrak{M}(\mathcal{O})$ be a net of von Neumann algebras satisfying Einstein causality and weak additivity with a vacuum state (which by definition satisfies the spectrum condition). Then, the net satisfies property B.*

*Proof.* See chapter 6 of [91] or [23] for the original proof. ∎

From this proposition and the characterization of property B, it follows that the local algebras should be of type III, provided they satisfy the physically motivated axioms of Einstein causality and weak additivity (which is required for the Reeh-Schlieder property) and there is a vacuum state.

One could envision though that the algebras might be of infinite type, say type $\text{I}_\infty$. However, there are some good reasons to believe that these algebras are *hyperfinite*.

**Definition 5.4.3.** Let $\mathfrak{M}$ be a von Neumann algebra. Then, $\mathfrak{M}$ is said to be *hyperfinite* if there is a family $(\mathfrak{M}_\lambda)_{\lambda \in \Lambda}$, with $\Lambda$ being some index set, of *finite dimensional* von Neumann algebras such that $\mathfrak{M} = \left( \bigcup_{\lambda \in \Lambda} \mathfrak{M}_\lambda \right)''$.

The meaning of this definition is quite clear: a hyperfinite factor is one that can be obtained by unions of matrix algebras. From a physical standpoint, it is reasonable to require hyperfiniteness in order to make the connection between the QFT and laboratory procedures. This happens because in a real world experiment, one has a finite number of tasks that can be performed on a system which results in some finite quantity which is ascribed to the system.

It is not difficult to see that every type I factor is hyperfinite. Hence, one can use the intuition from Quantum Mechanics to understand that applying, say, a Hamiltonian to some eigenstate corresponds to the measurement of the energy of that state. Even though mathematically such operator may be unbounded, one obtains a finite result for the energy measurement. If we proceed to consider QM as the non-relativistic limit of QFT, then such connection between the algebras (i.e., hyperfiniteness) should be expected. To see the explicit construction of some of these factors, we recommend section VI of [142].

Finally, we can use an argument by Longo in [100] to show that the algebras must be of type $\text{III}_1$. There are more sophisticated ways of showing this (see for instance [69, 34]), but this is the quickest way and it serves our purposes. An interesting feature of type $\text{III}_1$ factors is that they are unique up to an isomorphism, as shown by Haagerup in [81].

The argument goes as follows: we would like our axiomatic scheme to eventually describe fields with $n$-point functions having a well-behaved scaling limit at short distances. As so, we consider unbounded operators smeared by test-functions $f \mapsto \Phi(f)$ [125] defined on some dense domain $\mathcal{D} \subset \mathcal{H}$ of the Hilbert space and with the Wightman functions given by

$$W_n(f_1, ..., f_n) = \langle \Omega, \Phi(f_1)...\Phi(f_n)\Omega \rangle.$$

The scaling limit is given by scaling maps on the test functions $(\beta_{p,\lambda} f)(x) = f((x-p)/\lambda)$ and some positive monotone function $\nu$ which scales the Wightman functions appropriately:

$$W_n^{s.l.}(f_1, ..., f_n) = \lim_{\lambda \to 0^+} \nu(\lambda)^n W_n(\beta_{p,\lambda} f_1, ..., \beta_{p,\lambda} f_n).$$

The functions $W_n^{s.l.}$ are supposed to themselves be satisfy the Wightman axioms [125]. Going further, we say that the fields satisfy an *asymptotic scale invariance* if there is in some sense a field in the limit such that

$$\langle \Omega, \Phi(f)\Omega \rangle = 0.$$

For a theory which satisfies the HAK axioms (including weak additivity) *and* satisfies asymptotic scale invariance, than the local algebras are given by the unique hyperfinite type $\mathrm{III}_1$ factor, as shown in [34]. Hence, for scale-invariant theories (i.e., massless theories), this result is valid. For general theories, the argument by Longo is that for a continuous change in the mass parameter of the theory $m \in [0, \infty)$, the change on the algebras is isomorphic since free fields are described by representations of the CCR and the corresponding massive theories are connected to one another by a Bogoliubov transformation. Therefore, all theories are described by the unique hyperfinite type $\mathrm{III}_1$ factor. See [88] for further details on these arguments. The scaling argument can be put in a more algebraic form as an inductive limit based on techniques from renormalization [32].

## 5.5 The Split Property

As we described in the previous sections, the act of measurement is represented by the expectation value of some observable which is placed in some region of spacetime. The axiom of Einstein's causality then tells us that for spacelike separated regions, the algebras must commute element-wise. This reflects the idea of independence of measurements conducted in causally separated regions.

However, we would like to go further and impose that one can actually *prepare* the experiments in different regions without influence from another. This enforces the independence of the regions and avoids further issues of superluminal communication, since an experimenter can in principle prepare two experiments simultaneously in two regions distant from one another. The preparation is encapsulated in the construction of the underlying Hilbert space with its vectors representing physical states which the system may access. The measurement of expectation values of the observables then tell us properties of such system (such as energy, total angular momentum or polarization, for instance).

In Quantum Mechanics, this is achieved by means of constructing the *total* Hilbert space as the tensor product of the Hilbert space localized in each region. That is, given two local systems localized at $\mathcal{O}_1$ and $\mathcal{O}_2$ with each Hilbert space given by $\mathcal{H}_1$ and $\mathcal{H}_2$, respectively, then the total Hilbert space of the system is given by $\mathcal{H} = \mathcal{H}_1 \otimes \mathcal{H}_2$ and each physical state is described by $\psi = \psi_1 \otimes \psi_2$ with $\psi_1 \in \mathcal{H}_1$ and $\psi_2 \in \mathcal{H}_2$. Since the observables of QM are members of a type I von Neumann algebra, as discussed, we factorize the algebra of bounded operators following the tensor product: $B(\mathcal{H}) = B(\mathcal{H}_1) \otimes B(\mathcal{H}_2)$. In this particular case then, Einstein causality follows from the element-wise commutation of each sub-algebra. We would like to generalize this idea to QFT (i.e., algebras other than type I which *don't* factorize according to the tensor product) using an idea which goes by the name of *split property*[11]. The idea goes as follows.

**Definition 5.5.1.** Let $\mathfrak{M}_1$ and $\mathfrak{M}_2$ be two von Neumann algebras (of any type) acting on a Hilbert space $\mathcal{H}$. Then, we say that $\mathfrak{M}_1$ and $\mathfrak{M}_2$ form a *split inclusion* if their is a type I von Neumann factor $\mathfrak{N}$ such that

$$\mathfrak{M}_1 \subset \mathfrak{N} \subset \mathfrak{M}_2.$$

A net $\mathcal{O} \mapsto \mathfrak{M}(\mathcal{O})$ of von Neumann is then said to have the *split property* if $\overline{\mathcal{O}_1} \subset \mathcal{O}_2$ implies that $\mathfrak{M}(\mathcal{O}_1) \subset \mathfrak{M}(\mathcal{O}_2)$ is a split inclusion.

A net having the split property entails some consequences for the states over these algebras. We shall discuss these now.

A consequence of a factor $\mathfrak{M}$ being type I is that there is a unitary $U : \mathcal{H} \to \mathcal{H}_1 \otimes \mathcal{H}_2$ for some Hilbert spaces $\mathcal{H}_i$ such that $\mathfrak{M} = U^*(B(\mathcal{H}_1) \otimes \mathbf{1}_{\mathcal{H}_2})U$, as we have stated before. Hence, we can have that $\mathfrak{N} = U^*(B(\mathcal{H}_1) \otimes \mathbf{1}_{\mathcal{H}_2})U$ with the commutant given by $\mathfrak{N}' = U^*(\mathbf{1}_{\mathcal{H}_1} \otimes B(\mathcal{H}_2))U$. It follows from the split inclusion that

$$\mathfrak{M}_1 \subset U^*(B(\mathcal{H}_1) \otimes \mathbf{1}_{\mathcal{H}_2}), \;\; \mathfrak{M}_2' \subset U^*(\mathbf{1}_{\mathcal{H}_1} \otimes B(\mathcal{H}_2))U.$$

From this analysis, we can describe a joint state over $\mathfrak{M}_1 \otimes \mathfrak{M}_2$ of two states $\omega_1$ and $\omega_2$ over $\mathfrak{M}_1$ and $\mathfrak{M}_2'$ respectively. If these states can be expressed in terms of density matrices,

---

[11] Some authors call this property the *funnel property*. In [34], this property is shown to be derivable by a more fundamental property called *nuclearity*, which essentially puts bounds on the number of degrees of freedom. For our purposes, it is sufficient to consider the split property as being fundamental.

i.e.,

$$\omega_1(A) = \text{Tr}((\rho_1 \otimes \mathbf{1}_{\mathcal{H}_2})UAU^*), \quad \omega_2(B) = \text{Tr}((\mathbf{1}_{\mathcal{H}_1} \otimes \rho_2)UBU^*), \tag{5.5.1}$$

with $A \in \mathfrak{M}_1$ and $B \in \mathfrak{M}'_2$, then the joint state is given by

$$\omega(C) = \text{Tr}((\rho_1 \otimes \rho_2)UCU^*), \tag{5.5.2}$$

with $C \in \mathfrak{M}_1 \otimes \mathfrak{M}'_2$. It is straightforward to see that this state can be written as a product state:

$$\omega(AB) = \omega_1(A)\omega_2(B),$$

for $A \in \mathfrak{M}_1$ and $B \in \mathfrak{M}'_2$. In fact, this construction is possible whenever the states can be written in terms of density matrices [67].

A consequence of this construction is that if we have two causally separated regions $\mathcal{O}_1$ and $\mathcal{O}_3$, then we can have observers in each region preparing independently their experiments, whose observables are part of the larger algebra

$$\mathfrak{M}(\mathcal{O}_1)\overline{\otimes}\mathfrak{M}(\mathcal{O}_3), \tag{5.5.3}$$

where the overline denotes the topological closure of the algebra. The reason we can do this is that in a connected spacetime we can certainly find a neighborhood $\mathcal{O}_2$ of $\overline{\mathcal{O}_1}$ which is contained in the causal complement of $\mathcal{O}_3$. Then, $\mathfrak{M}(\mathcal{O}_1) \subset \mathfrak{M}(\mathcal{O}_2)$ is split and $\mathfrak{M}(\mathcal{O}_3) \subset \mathfrak{M}(\mathcal{O}_2)'$.

This property has been shown to hold for massive scalar theories, Dirac fields and lower dimensional models [96] and a formulation for general globally hyperbolic spacetimes has been achieved in [66]. A lesson learned from this models is that the split property is closely related to the existence of well-behaved thermal states and to quantum energy inequalities. This can be understood intuitively by attributing an energy cost to disentangling the degrees of freedom between the two regions. Should the number of degrees of freedom increase with the energy scale, it then becomes impossible to disentangle the states with finite energy unless the algebras are located in regions far from each other. This again echoes with the theme of entanglement being part of the structure of QFT.

## 5.6 Type-III$_1$ Factors and Fermi's Paradox

To finish this chapter, we present an example based on [145] which demonstrates the difference between QM and QFT from the point of view of von Neumann algebras. This example is itself based on a *gedankenexperiment* conducted by Fermi in [65] and the apparent paradox was first observed by Hegerfeldt in [85]. The observation that this "paradox" lied in the use of the incorrect type of factor to describe the system was observed by Buchholz and Yngvason in [37].

The *gedankenexperiment* can be formulated as follows: suppose we have two atoms, which we label as $A$ and $B$ separated by a distance $R$. At time $t = 0$, atom $B$ finds itself in an excited state while atom $A$ is in the ground state. By considering the system to be isolated, there is a non-zero probability that atom $B$ will decay and emit a photon which is then absorbed by atom $A$. We are interested in computing the probability for such an event. As figure 1 makes it clear, the limit in the speed of light $c$ implies that atom $A$ will remain in its ground state for $t < R/c$. Hence, what we assume in this thought experiment is that radiation is made of quanta (i.e., Quantum Mechanics) which travels at finite speed (i.e., Special Relativity).

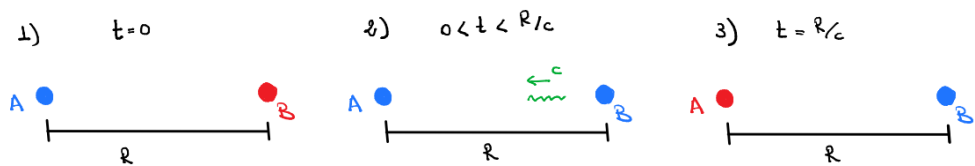

Figure 1 – An illustration of the *gedankenexperiment*. At time $t = 0$, atom $A$ is in its ground state (represented by the blue color) while atom $B$ is in an excited state (red color). At some time $0 < t < R/c$, atom $B$ emits a photon (in green) which is absorbed by atom $A$ at time $t = R/c$, which is the required time of travel in our idealization.

The "paradox" arises when one tries to analyze this *gedankenexperiment* from a type I framework, that is, by treating this a problem of Quantum Mechanics. This is done by first constructing the Hilbert spaces of each component of the experiment: the space $\mathcal{H}_A$ is spanned by the eigenstates of the atom $A$, similar to $\mathcal{H}_B$ and for the photon in $\mathcal{H}_\gamma$. We assume these spaces are complete and we take the (topologically closed) tensor product to describe the full Hilbert space

$$\mathcal{H} = \mathcal{H}_A \otimes \mathcal{H}_B \otimes \mathcal{H}_\gamma. \tag{5.6.1}$$

The corresponding algebra, in a type I framework, is then given by

$$B(\mathcal{H}) = B(\mathcal{H}_A) \otimes B(\mathcal{H}_B) \otimes B(\mathcal{H}_\gamma), \tag{5.6.2}$$

and the state of the system at time $t = 0$ is simply

$$\omega_0 = \omega_A \otimes \omega_B \otimes \omega_\gamma, \tag{5.6.3}$$

where $\omega_A$ is the ground state of atom $A$, $\omega_B$ is the excited state of atom $B$ and $\omega_\gamma$ is the state with no photons (at time $t = 0$, atom $B$ has not decayed yet and hence there are no photons).

Now, suppose that time evolution is governed by an Hamiltonian $H$, which induces a one-parameter group of automorphisms of the total algebra in equation (5.6.2), in the spirit of Stone's theorem. This induces a time evolution for the state $\omega_0$ in terms of the Heisenberg picture

$$\omega_t(E) = \omega_0 \left( e^{iHt} E e^{-iHt} \right), \tag{5.6.4}$$

where $E \in B(\mathcal{H})$ is some observable of interest. In particular, we would like to compute the probability $\mathcal{P}(t)$ of finding atom $A$ in an excited state. This means that want to project the total state into the subspace which describes the excited states of $A$. Since we are working in a type I framework, we have minimal projections and thus we can safely speak of projections into vector states. By choosing $E$ to be of the form

$$E = (\mathbf{1}_A - P_A) \otimes \mathbf{1}_B \otimes \mathbf{1}_\gamma, \tag{5.6.5}$$

with $P_A$ being the projection operator[12] on the ground state of atom $A$, the probability is then given by $\mathcal{P}(t) = \omega_t(E)$. This probability is expected to satisfy $\mathcal{P}(t) = 0$ for all $0 < t < R/c$.

However, we are always assuming a condition on the Hamiltonian which describes the system: it is bounded by below, meaning its spectrum is semibounded. This means that we can consider $\sigma(H) \subset [0, +\infty)$ and we can define an extension of the time-evolution operator by

$$e^{-iHt} \to e^{-iH(t+iy)}. \tag{5.6.6}$$

This extends implies an analytic continuation of the probability function $\mathcal{P}(t)$ to $\mathcal{P}(z)$ with $z = t + iy \in \mathbb{C}$. This function is continuous in $\operatorname{Im} z \leq 0$ and holomorphic in $\operatorname{Im} z < 0$. From the *Schwarz reflection principle*, a function satisfying these conditions has an extension to the lower part of the complex plane given by

$$\mathcal{P}(\bar{z}) = \overline{\mathcal{P}(z)}.$$

But if the function is continuous and it vanishes in a segment of the real line (where $0 < t < R/c$), then the function must be identically zero! From this analysis, we can draw two conclusions: either atom $A$ *never* absorbs the photon or it does so *instantly* (which is the case of superluminal propagation, or simply stating that $P(t)$ never vanishes in the real line). Ergo, it seems that we have a paradox. This is the original argument in [85].

---

[12] If one is more comfortable using the Dirac notation, we can express this projector as $P_A = \mathbf{1} - |\psi\rangle \langle \psi|$, where $\psi$ is the ground state of atom $A$.

But it was pointed out in [37] that there is an alternative: the hypothesis that the system is described by a type I algebra (or simply put, by bounded operators in the respective finite-dimensional Hilbert spaces) is incorrect. Let us try to "solve" the paradox from a type III perspective.

The first observation is that one should be more explicit about the local nature of the algebras. That is, we denote by $\mathcal{O}_A$ the spacetime region where atom $A$ is located and by $\mathfrak{M}(\mathcal{O}_A)$ the local algebra associated with it. It is physically intuitive to actually include observables in this algebra which are associated with the electromagnetic field, previously described by the algebra $B(\mathcal{H}_\gamma)$ and was, in a sense, detached from the rest of the system.

Now, in order to have the spatial distance between atoms $A$ and $B$, we should write the regions $\mathcal{O}_A$ and $\mathcal{O}_B$ as

$$\mathcal{O}_A = \mathcal{R}_A \times \{0\}, \ \ \mathcal{O}_B = \mathcal{R}_B \times \{0\},$$

where $\mathcal{R}_A \subset \mathbb{R}^3$ is a ball containing the atom $A$ and is a distance $R$ apart from $\mathcal{R}_B \subset \mathbb{R}^3$, which has an analogous interpretation.

Concretely, in order to measure the effect that the decay of atom $B$ has on atom $A$, we must find a way to compare the states in the total system. In particular we should be able to compare the state $\omega_0$ which describes the atom $B$ in $\mathcal{R}_B$ at $t = 0$ and in an excited state and the state $\omega_0^{(0)}$, which describes either the absence of $B$ in $\mathcal{R}_B$ or the atom $B$ being in the ground state. In both states, atom $A$ is still in the ground state and localized at $\mathcal{R}_A$, hence they can't be distinguished *by operations conducted outside of* $\mathcal{R}_B$. This is the local nature of the algebras at work: properties of atom $B$ can only be assessed by measurements or observables contained in $\mathfrak{M}(\mathcal{O}_B)$.

Now, by taking the Schrödinger picture, we consider the states as evolving to $\omega_t$ and $\omega_t^{(0)}$, respectively. As long as atom $A$ is not excited, these states are still indistinguishable from one another, meaning that

$$\omega_t(E) = \omega_t^{(0)}(E), \ \forall E \in \mathfrak{M}(\mathcal{O}_A). \tag{5.6.7}$$

We then introduce the *deviation* of these states by

$$D(t) = \sup_{E \in \mathfrak{M}(\mathcal{O}_A)} |\omega_t(E) - \omega_t^{(0)}|. \tag{5.6.8}$$

Now, at $t = 0$, we have that $D(0) = 0$ following the definition of $\omega_0$ and $\omega_0^{(0)}$. If the theory satisfies the time-slice axiom, we know that the algebra $\mathfrak{M}(\mathcal{O}_A)$ is unchanged for $t > 0$. Hence, equation 5.6.7 holds for all values of $t$. In particular, it holds for $0 < t < R/c$ and hence we have $D(t) = 0$ for a non-identically zero state. This was not the case for the type I algebra, where we picked $E = P_A = \mathbf{1} - |\psi\rangle\langle\psi|$ which projects the vector state on the subset of linear combinations of states that are orthogonal to the ground state of $A$ (hence, *all of them* describe $A$ in an excited state). That choice implied that $\omega_t(E) = 0$ if, and only if, $\omega_t = 0$ for all $t$.

# 6 Locally Covariant Quantum Field Theory

> *First quantization is a mystery, but second quantization is a functor.*
>
> ———————————————
>
> Edward Nelson

In this chapter, we present the generalization of the lessons from AQFT to general globally hyperbolic spacetimes. This is done by means of Category Theory.

This chapter is divided in three sections: since we will be working in curved spacetimes, some notions of General Relativity will be presented in the first part, with an emphasis on *globally hyperbolic spacetimes*, which are understood as describing "physically admissible spacetimes". This is by no means a self-contained course in GR and some of the notions presented here have been mentioned or discussed in footnotes in the previous chapters. However, we include these notions here for completeness and clarity for the upcoming topics. Part of this section is based on the review paper [143]. For the good and standard references for General Relativity, we recommend [39, 135, 84], roughly in order of difficulty. For the mathematical background, we recommend [108] for a more Physics-oriented mind and [97] for an excellent introduction to smooth manifolds. A reader well acquainted with this literature or the definitions presented here can skip this section without loss.

We then give a brief introduction to Category Theory with emphasizes in the structures presented in this Thesis. Category Theory is a very rich field of study and research, but we need only some basic structures to understand the main themes presented here. Even in Physics there has been a lot more use of Category Theory than presented here: recently there has been a formulation of the HAK axioms in terms of what are called *stacks* [18] and there has been some use of concepts of *higher categories* and operads in AQFT [19]. We shall not comment here about these structures. For nice books on Category Theory, we recommend a classic and a more recent treatment, respectively [101, 114]. For a more pedagogical treatment and applications in Computer Science, we recommend [11, 98]. Finally, for applications in Physics we recommend the introductory paper [46] and its expansion [47]. A very nice source of information is the website by John Baez [13], one of the main advocates for Category Theory in Mathematics, Computer Science and Physics.

Finally, in the third part, we present the formulation of (A)QFT in terms of a functor following [30] and discuss the formulation of state spaces in terms of this language. This will allow us to formulate Bell's inequalities in terms of Category Theory and extend the results to general globally hyperbolic spacetimes. We do not mention aspects of dynamics or interactions in this framework in order to be concise.

## 6.1   Introduction

We want to generalize the Haag-Kastler axioms which associate an algebra of observables to causally convex subsets of Minkowski spacetime to include more general spacetimes.

There is a natural candidate to a language that formalizes such construction which essentially associates a algebraic structure to a geometrical construction. Such candidate is Category Theory, which allows to treat the quantum field theory as a functor. This generalization was first presented in the seminal paper [30] and is known in the literature as *Locally Covariant Quantum Field Theory.*

Category theory can be described as a "context" for the study of a particular class of objects. Alternatively, it can be understood as a way to relate different areas of mathematics. It can also be interpreted as a language: there are nouns and verbs and ways to connect them using *morphisms*.

The notion of a category is useful when we want to generalize a concept, a property or a notion common to different mathematical structures. For example, a transformation that maps the structure where it is defined to itself is called an *isomorphism* and it is realized in the context of group theory in the form of *homomorphisms* or in the context of topological spaces as *bijections*. The general property of mapping an object to another object of the same type is what is seen by the eyes of category theory.

The natural question that arises is if there is a formal way to categorize[1] different mathematical entities by means of these general common properties. The answer is affirmative and category theory provides us with the notion of a *functor* which can be intuitively understood as a map between different areas of math. Examples of such maps can be as simple as tangent spaces, which relate vector spaces to smooth manifolds or even canonical quantization, justifying the epigraph at the beggining of this chapter.

Since Physics is built upon different areas of math, it is interesting to find a common language that allows us to describe different areas and thus obtain a complete description of nature. It can also help us explore problems that are considered difficult by translating them to a easier or better understood structure. In fact, it has even been argued that Category Theory might be the natural language to describe Physics [47].

---

[1]   No pun intended.

## 6.2 Spacetimes, Causality and All That

This section is intended to be a "crash course" in General Relativity. We present here only the relevant elements to be used in the remainder of this Thesis. An emphasis is given to notions of causal structures.

### 6.2.1 Basic definitions

We will consider spacetime to have the structure of a real 4-dimensional Lorentzian smooth manifold. A *real manifold* $M$ is a topological space which is locally homeomorphic to an open set in $\mathbb{R}^n$. The value $n$ determines the dimension of our manifold, hence we will be interested in spaces that locally "looks like" $\mathbb{R}^4$. This is expected from our intuition since spacetime does resemble flat spacetime in a very good approximation at small distances.

To characterize such spaces, we consider the topology of $\mathbb{R}^n$, whose open sets $\mathcal{O} \subset \mathbb{R}^n$ are those for which to each point $x = (x^1, ..., x^n) \in \mathcal{O}$ we can obtain an *open ball* which contains points $y = (y^1, ..., y^n)$ that are at a distance $r$ of $x$:

$$|x - y| = \sqrt{\sum_{\mu=0}^{n} (x^\mu - y^\mu)^2} < r. \tag{6.2.1}$$

This is just the Euclidean metric for $n$ dimensions. Notice that we start our summation from $\mu = 0$, which is a common convention in General Relativity, since the component $x^0$ is usually reserved for the time direction (and similarly, the zeroth component of the four-momentum $p^0$ is taken to be the energy).

A *smooth manifold* has the additional properties

1. Each $p \in M$ lies in at least one open set $\mathcal{O}_\alpha$ and the set $\{\mathcal{O}_\alpha\}$ covers $M$;

2. For each $\alpha$, there is a bijective map $\psi_\alpha : \mathcal{O}_\alpha \to U_\alpha \subset \mathbb{R}^n$ called a *chart* or a *coordinate system*;

3. For any two sets $\mathcal{O}_1$, $\mathcal{O}_2$ such that $\mathcal{O}_1 \cap \mathcal{O}_2 \neq \varnothing$, the map $\psi_2 \circ \psi_1^{-1} : \mathbb{R}^n \supset U_1 \supset \psi_1(\mathcal{O}_1 \cap \mathcal{O}_2) \to \psi_2(\mathcal{O}_1 \cap \mathcal{O}_2) \subset U_2 \subset \mathbb{R}^n$ (see figure 2) is infinitely differentiable in the sense of ordinary Calculus.

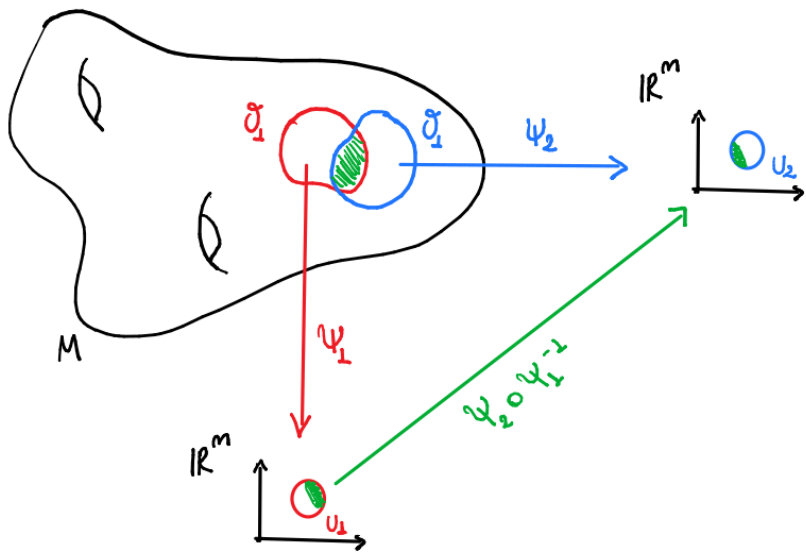

Figure 2 – An illustration of the smooth change of coordinates between different charts on a manifold $M$. We have depicted a 2-dimensional manifold for illustration but this idea should be generalized to higher dimensions.

An example of a smooth manifold is the 2-sphere $S^2 = \{(x^1, x^2, x^3) \in \mathbb{R}^3 : (x^1)^2 + (x^2)^2 + (x^3)^2 = 1\}$, which can be thought of six hemispherical open sets $\mathcal{O}_i^{\pm}$ for $i \in \{1, 2, 3\}$ sewn together:

$$\mathcal{O}_i^{\pm} = \{(x^1, x^2, x^3) \in S^2 : \pm x^i > 0\}.$$

We would like to add a vector space structure to our smooth manifold. This is done by defining the *tangent space* whose elements are *tangent vectors* or simply *vectors*. There are many ways to characterize this space [135, 56], but we appeal to the intuition of the reader and simply characterize the tangent space at a point $p$ as an assignment $p \mapsto TM_p$ where $TM_p$ has a vector space structure. One can show that this space has dimension $n$ (see [135], theorem 2.2.1) and its elements are written in terms of a basis

$$v = v^{\mu} X_{\mu}, \tag{6.2.2}$$

where the summation convention is implied and $X_{\mu}$ are the basis elements, sometimes denoted by $\partial / \partial x^{\mu}$. We will sometimes be loose and denote the tangent vectors by $X$, the same letter as the coordinate basis.

Finally, we need to introduce the notion of a *metric tensor*. This will give a causal structure to our manifold as we will discuss in the following section. This is a symmetric and nondegenerate tensor field of type $(0, 2)$ on the manifold $M$ and it can be written in the

coordinate basis as

$$g = g_{\mu\nu}dx^\mu \otimes dx^\nu. \tag{6.2.3}$$

Sometimes the notation

$$ds^2 = g_{\mu\nu}dx^\mu dx^\nu, \tag{6.2.4}$$

where the tensor product has been omitted is preferred since it conveys an idea of a line element. In fact, if we consider the matrix elements $g_{\mu\nu}$ to be the identity matrix, equation (6.2.4) reduces to the Euclidean metric in terms of the differentials of the tangent vectors. We can always find an orthonormal basis such that the matrix $g_{\mu\nu}\mathbf{1}$ is diagonal with diagonal elements being $\pm1$. The number of "plus signs" and "minus signs" on the diagonal is called the *signature* of the metric tensor and its independent of the choice of orthonormal basis (see [135], chapter 2 problem 7). When the signature comprises only of plus signs (such as the case for the Euclidean metric tensor) we have a *Riemannian metric tensor*, which is positive-definite. If at least one of the signs in the signature is flipped, we have a *Semi-Riemannian metric tensor* and the special case where only one sign is different for the rest is called a *Lorentzian metric tensor*. Hence, when we consider the pair $(M, g)$, we mean a smooth manifold with a Lorentzian metric tensor of signature $(-, +, +, +)$ or $(+, -, -, -)$ depending on the convention. When eqquiped with the metric tensor, we usually refer to the points on the manifold as being *events*.

A great deal of General Relativity is concerned in solving the *Einstein Field Equations*, written in components as

$$R_{\mu\nu} - \frac{1}{2}Rg_{\mu\nu} = \frac{8\pi G}{c^4}T_{\mu\nu}. \tag{6.2.5}$$

The left-hand-side is concerned with the curvature aspects while the right-hand-side is essentially dictated by the *energy-momentum tensor* $T_{\mu\nu}$, which describes the matter content ($G$ and $c$ are Newton's gravitational constant and the speed of light, respectively). The *Ricci curvature tensor* $R_{\mu\nu}$ and its trace $R$ (also called the *Ricci scalar*) are determined in terms of the metric tensor $g$. The goal is then to solve the equation for $g$ for a given energy-momentum tensor which can be constrained to some energy-conditions.

## 6.2.2  Causality

The Lorentzian metric tensor allows us to define a non-positive-definite inner product in the tangent space. It is given by

$$g(X, Y) = g_{\mu\nu}X^\mu Y^\nu, \tag{6.2.6}$$

where we have expressed the tangent vectors in local coordinates. Notice that this inner product is not positive definite if we choose, for instance, the Minkowski metric tensor:

$$g_{\mu\nu}\mathbf{1} = \begin{pmatrix} -1 & 0 & 0 & 0 \\ 0 & +1 & 0 & 0 \\ 0 & 0 & +1 & 0 \\ 0 & 0 & 0 & +1 \end{pmatrix}. \tag{6.2.7}$$

The *causal structure* can be defined using the non-positive-definiteness by looking at the norm of a vector defined by the inner product:

1. If $g(X, X) < 0$, then $X$ is said to be *timelike*;

2. If $g(X, X) = 0$, then $X$ is said to be *lightlike* or *null*;

3. If $g(X, X) > 0$, then $X$ is said to be *spacelike*.

With these definitions, we can divide spacetime into regions which we can attribute the quality of "past" and "future". This is typically done in terms of light-cones, such as in figure 3. Hence, for each point $p \in M$, we can assign a light cone which is a subset of the tangent space at that point. A spacetime for which a continuous choice of past and future can be made is called a *time-orientable spacetime*.

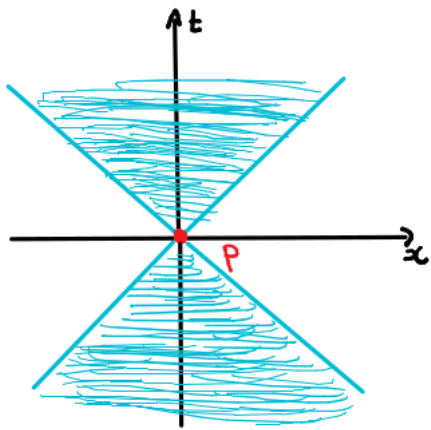

Figure 3 – The light-cone associated with a point $p \in M$. All points in the shaded area are causally influenced by $p$.

A *path* or a *curve* in a spacetime is an assignment $s \mapsto x^{\mu}(s)$, where the parameter $s \in [0, 1]$ is a real number and $x^{\mu}$ are local coordinates in spacetime. With that, the tangent vector can be defined as $dx^{\mu}/ds$. A path is then said to be *causal* if its tangent vector is everywhere timelike or null. A point $p \in M$ is said to be a *future endpoint* of a causal path $x^\mu(s)$ if for every neighborhood $\mathcal{O}$ of $p$ there is a $s_0$ such that $x^\mu(s) \in \mathcal{O}$ for all $s > s_0$. If the curve has no future endpoints, it is said to be *future inextendible*. An analgous construction can be made changing "future" for "past".

Given two points $p$ and $q$, the set of causal paths connecting them is called the *causal diamond*, and its denoted by $D^p_q = J^+(p) \cup J^-(q)$, where

$$J^+(p) = \left\{ q \in M : \exists x^\mu(s), x^\mu(0) = p, x^\mu(1) = q, \frac{dx^\mu}{ds} \leq 0 \right\}, \tag{6.2.8}$$

and similar for $J^-(q)$. That is, the causal diamond is the union of the regions which are causally connected to both $p$ and $q$ to their future and past respectively. A related construction is given by the sets $I^\pm(p)$, where we restrict to timelike curves, or $\frac{dx^\mu}{ds} < 0$. Naturally, one could replace the points $p$ and $q$ by a whole set $S$ and describe the causally connected regions to $S$, or $I^\pm(S)$, for instance.

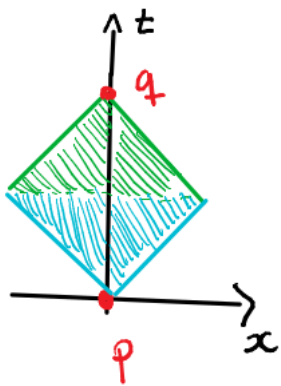

Figure 4 – The causal diamond between two points $p$ and $q$. The area in blue corresponds to the future lightcone of $p$ or $J^+(p)$ while the area in green corresponds to the past of $q$ or $J^-(q)$.

### 6.2.3   Domains of dependence and global hyperbolicity

We are interested in the collection of events which are influenced by a set $S$, that is, the sets $I^+(S)$ or $J^+(S)$, for example. If $S$ is a closed and achronal set (the latter meaning that $I^+(S) \cap S = \varnothing$), then the following theorem is valid:

**Theorem 6.2.1.** *Let $S \neq \varnothing$ be a closed, achronal set with empty edge. Then, $S$ is a three-dimensional, embedded, $C^0$ submanifold of $M$.*

*Proof.* See [135], theorem 8.3.1. ■

With this, we can define the *future domain of dependence of S* by the set of points $p \in M$ such all past inextendible causal curves through $p$ intersect $S$. This set is denoted by $D^+(S)$ and it conveys the idea that any signal sent from a point $p \in D^+(S)$ must intersect the surface $S$. Thus, if we are given some set of "initial conditions" on $S$, we can predict what happens to the point $p$. The *past domain of dependence of S*, $D^-(S)$, is obtained analogously and the full domain of dependence is just $D(S) = D^+(S) \cup D^-(S)$.

A closed achronal set $\Sigma$ such that $D(\Sigma) = M$ is called a *Cauchy surface* and a spacetime which possesses a Cauchy surface is called a *globally hyperbolic spacetime*. Thus, in a globally hyperbolic spacetime, one can predict the entire future and past history from the conditions at the instant of time represented by $\Sigma$. An equivalent definition is due to a theorem by Geroch:

**Theorem 6.2.2.** *Let $\Sigma$ be a Cauchy surface for a spacetime $(M, g)$. It follows then that $M$ is topologically equivalent to $\Sigma \times \mathbb{R}$.*

*Proof.* See [73], proposition 7. ■

Hence, a globally hyperbolic spacetime is one that admits a foliation in terms of Cauchy surfaces. This is particularly important in the context of *hyperbolic equations* such as the Klein-Gordon equation. Global hyperbolicity implies that the Cauchy problem is well-posed and that there are unique global solutions to such equations specified on a Cauchy surface [14].

## 6.3 Basic Category Theory

### 6.3.1 Categories

A *category* $\mathscr{C}$ is a collection (or class[2]) of *objects*, usually denoted by $\mathrm{ob}(\mathscr{C})$, which are related to each other by *morphisms*. Given two objects $X, Y \in \mathscr{C}$, the collection of morphisms $f : X \to Y$ is sometimes denoted by $\mathrm{Hom}(X, Y)$[3]. In this case, $X$ is called the *domain* of the morphism $f$ while $Y$ is called the *codomain*. Each object $X$ has an assigned morphism $1_X$ whose domain and codomain are $X$ itself. This is called the *identity morphism*.

---

[2]  We avoid using the term "set", since we will be working with the category whose objects are sets. Hence, any attempt to characterize the class of all sets as being itself a set leads to well-known paradoxes in set theory and mathematical logic.

[3]  This notation can be traced to the origins of Category Theory (more precisely, to the study of natural transformations) in the context of Group Theory, concretely in the study of group extensions. The morphisms in the category of groups are called homomorphisms, justifying the notation. For a brief historic note on the origin of category theory, see chapter 1 of [114].

Morphisms and objects are also required to obey the following laws:

1. **Composition**: For every pair of morphisms $f$, $g$ where the codomain of $f$ is the domain of $g$, we can construct a *composite morphism $gf$* (sometimes denoted by $g \circ f$) that takes an object from the domain of $f$ to the codomain of $g$, that is:

$$f : X \to Y, \, g : Y \to Z \implies gf : X \to Z$$

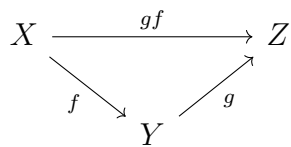

Figure 5 – Diagram representing the aforementioned composition law.

The analogy with set theory is quite obvious. It is common to employ *commutative diagrams* in category theory, especially when we are dealing with more than two morphisms.

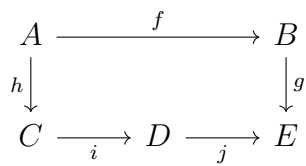

Figure 6 – Commutative diagram of a category with more than two objects. We notice that we can compose the morphisms $f$ and $g$ in order to obtain a morphism $gf : A \to E$, for example. The word "commutative" means that the morphism $gf$ coincides with the morphism $jih : A \to E$, and we denote this relation by $gf = jih$.

2. **Associativity**: For every triple of morphisms $f \in \mathrm{Hom}(A, B)$, $g \in \mathrm{Hom}(B, C)$ and $h \in \mathrm{Hom}(C, D)$, we have

$$h \circ (g \circ f) = (h \circ g) \circ f.$$

Because of this, sometimes we drop the parenthesis and simply write the composite morphism as $hgf$.

3. **Identity laws**: For every $f \in \mathrm{Hom}(A, B)$, we have

$$f \circ 1_A = f \circ 1_B = f,$$

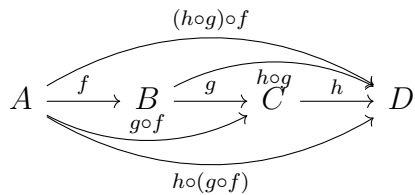

Figure 7 – Associativity law in practice.

where $1_A$ and $1_B$ are the identity morphisms of the objects $A$ and $B$

With the notion of identity morphisms and the law of composition, we can define a notion of "sameness".

**Definition 6.3.1.** A map $f : A \to B$ in a category $\mathscr{C}$ is called an *isomorphism* if there exists a map $g : B \to A$ such that the compositions with $f$ are $gf = 1_A$ and $fg = 1_B$, where $1_A$ and $1_B$ denote the identity morphisms in each object. We call $g$ the *inverse* of $f$ and denote it by $g = f^{-1}$.

In this case, we say that $A$ and $B$ are *isomorphic* (the "same" in the categorical sense) and write $A \cong B$.

An *endomorphism* is a morphism whose domain is equal to its codomain. If an endomorphism is also an isomorphism, it is called an *automorphism*.

We note the clear analogy between the adjectives defined here and those used elsewhere in this Thesis.

**Proposition 6.3.1.** For every morphism $f : A \to B$, there is, at most, a unique inverse $g = f^{-1}$ such that $gf = 1_A$ and $fg = 1_B$

*Proof.* Suppose there are two inverse morphisms, $g$ and $\bar{g}$. Then, we can write

$$\bar{g}f = 1_A \implies (\bar{g}f)g = (1_A)g,$$

where we have composed both sides of the equation with $g$. Applying the associativity rule (2) on the left side together with the identity law (3) on the right side of the equation, we arrive at

$$\bar{g}(fg) = g \implies \bar{g}(1_B) = g \implies \bar{g} = g,$$

where, in the last implication, we made use of the identity law again. ∎

We should now give some explicit examples of categories. Essentially any set equipped with some extra structure, such as a topology, a $\sigma$-algebra, a vector space or group operations is bound to be described in terms of categories. These are what are called *concrete categories*: the objects are simply sets with extra structure and the morphisms are structure-preserving arrows[4].

1. $\mathfrak{Grp}$, whose objects are groups and whose morphisms are group homomorphisms. The isomorphisms are bijective homomorphisms.

   This can be seen by the definition of a group homomorphism: it is a function between two groups that preserves the group structure. Representing the group operation by the juxtaposition of elements, we can define a homomorphism $\phi : G \to H$ as a function with the property

   $$\phi(ab) = \phi(a)\phi(b)$$

   for $a, b \in G$. Notice that the right-hand side of the equation represents the group operation in $H$.

   The composition of homomorphisms is a homomorphism. Let $\phi$ and $\psi$ be two homomorphisms:

   $$\phi(\psi(ab)) = \phi(\psi(a)\psi(b)) = \phi(\psi(a))\phi(\psi(b)) = (\phi\psi)(a)(\phi\psi)(b)$$

   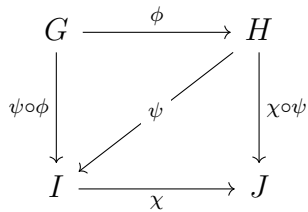

   Figure 8 – Diagram showing both the composition law and associativity (note that $\chi \circ (\psi \circ \phi) = (\chi \circ \psi) \circ \phi$) properties of homomorphisms between groups. Indeed, for any concrete category that is a set with extra structure, this diagram can be realized.

   The identity morphism is nothing but the identity element of the group which, by definition, commutes with all the elements of the group and with the morphisms:

   $$\phi(abe_G) = \phi(e_G ab) = e_H \phi(ab) = \phi(ab)e_H,$$

   where $e_G$ and $e_H$ are the identity elements in $G$ and $H$, respectively.

   A similar construction can be made for $\mathfrak{Field}$ and $\mathfrak{Rng}$, the categories of fields and rings, respectively.

---

4   We avoid using the names "transformation" or "maps" since these can be seen as examples of morphisms.

2. $\mathfrak{Set}$, whose objects are sets and whose morphisms are functions with specified domains and codomains. Naturally, the composition of two functions $f : X \to Y$, $g : Y \to Z$ is a new function $g \circ f : X \to Z$, where $X, Y, Z$ are sets. We can do even better: if both functions are injective, their composition is injective while if both are surjective, then the composition is also surjective.

   *Proof.* Let $f : X \to Y$ and $g : Y \to Z$ be injective functions between sets $X, Y, Z$. Suppose the composite function coincides for two point $a, \bar{a} \in X$, that is:

   $$g(f(a)) = g(f(\bar{a}))$$

   Since $g$ is injective, this means that $f(a) = f(\bar{a})$. Since $f$ is also injective, this means that $a = \bar{a}$, completing the proof.

   Let $j : Q \to R$ and $k : R \to S$ be surjective functions between sets $Q, R, S$. This means that every element in $R$ can be written as $j(q)$, with $q \in Q$, while every element in $S$ can be written as $k(r)$, with $r \in R$. Naturally, we can then write the element in $S$ as the composite $k(r) = k(j(q))$. Thus, any element of $S$ can be written as the image of an element in $Q$ under the composite function, making it surjective. ∎

   This idea will be generalized further ahead to morphisms that act in an injective or surjective-like way (mono and epimorphisms, respectively).

3. $\mathfrak{Top}$, whose objects are topological spaces and morphisms are continuous functions. Notice that isomorphisms in this category are homeomorphisms.

4. $\mathfrak{GlobHyp}$, whose objects are smooth manifolds and morphisms are smooth maps while isomorphisms are diffeomorphisms.

5. $\mathfrak{Vect}_\mathbb{K}$, whose objects are vector spaces over a field $\mathbb{K}$ and (iso)morphisms are linear maps (isomorphisms).

6. $\mathfrak{Alg}$, whose objects are ($*$-)algebras and whose morphisms are ($*$-)algebraic homomorphisms.

   We invite the reader to come up with new examples of concrete categories. We mention that there is also the important notion of an *abstract category*. This shows how powerful Category Theory can be: objects need not to look remotely like sets nor morphisms need to look like functions. Some examples include:

1. The empty set with no morphisms at all defines, trivially, a category. Accordingly, the set with a single object and a identity morphism also defines a category.

2. The elements of a set together with their respective identity morphisms define a category. Notice that this is different from $\mathfrak{Set}$, where we considered sets as the objects in the category and functions between sets as morphisms. This new more abstract category is sometimes dubbed the *discrete category*, because their elements appear isolated from one another.

3. A group (or more generally a monoid) $G$ defines a category B$G$ which has a single element and the morphisms are the elements (the composition law is nothing but the multiplication of the group elements). Again, notice how different this is from $\mathfrak{Grp}$ since now we consider a single group as the object and its *elements* as morphisms. A interesting way to picture this category is with the following diagram:

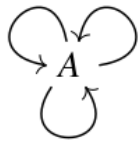

We notice that every morphism in this category is an isomorphism. Categories with this property are called **grupoids**, and B$G$ is defined as a one-element grupoid.

4. $\mathfrak{Htpy}$, whose objects are topological spaces but, unlike $\mathfrak{Top}$, the morphisms are homotopy classes of continuous functions, i.e., functions related by a homotopy, which defines an equivalence class [108]. This is an important example, so let's develop it a little bit.

**Definition 6.3.2.** Given two topological spaces $X$ and $Y$ and continuous maps $f$, $g$ : $X \to Y$, a *homotopy* between the maps $f$ and $g$ is a continuous function $H : X \times [0,1] \to Y$ such that

$$H(x,0) = f(x) \ \text{ and } \ H(x,1) = g(x), \forall x \in X$$

Intuitively, the existence of a homotopy between two maps in the topological space means that they can be deformed into each other continuously in the "time interval" $[0,1]$ (note that in "time" $t = 0$, $H(x,t)$ is just $f(x)$ while at $t = 1$, $H(x,t)$ becomes $g(x)$).

We finish with some considerations about arrows or morphisms. They can be divided into two classes.

**Definition 6.3.3.** Let $f : x \to y$ be a morphism in a category $C$. It can be

1. a **monomorphism** if for a pair of morphisms to the domain, i.e., $h, k : w \to x$, then $fh = fk \implies h = k$.

2. a **epimorphism** if for a pair of morphisms from the codomain, i.e., $h, k : y \to w$, then $hf = kf \implies h = k$.

We will sometimes use refer to a monomorphism by the adjectiv *monic* while reserving *epic* to epimorphisms..

**Proposition 6.3.2.** *A monomorphism in the category of* $\mathfrak{Set}$ *is an injective function while a epimorphism is a surjective function.*

*Proof.* Let $f : X \to Y$ be a monomorphism in the category of $\mathfrak{Set}$. The function is injective if $\forall x, y \in X, f(x), f(y) \in Y$ we have that $f(x) = f(y) \implies x = y$.

Let $h, k : W \to X$. Since $f$ is a monomorphism, we have that

$$f(h(w)) = f(k(w)), w \in W \implies h(w) = k(w)$$

Since both $h(w)$ and $k(w)$ are elements of $X$ by construction, that means that $f$ is injective.

Let $g : X \to Y$ be a epimorphism in $\mathfrak{Set}$. The function is surjective if $\forall y \in Y, \exists x \in X$ such that $g(x) = y$.

Let $h, k : Y \to Z$. Since $g$ is a epimorphism, we have that

$$h(g(x)) = k(g(x)) \implies h(y) = k(y) \forall y \in Y$$

Thus, $h$ and $k$ coincide in the image of $g$. The equality $h = k$ implies that the image of $g$ is the whole codomain $Y$, which means that every element of $Y$ has at least one partner in $X$. Therefore, $g$ is a surjective function. ∎

## 6.3.2 Functors

A natural question we may ask is whether or not we can define a analogous notion of a morphism between the categories themselves.

**Definition 6.3.4.** A (covariant) **functor** $F : C \to D$ is a morphism between categories $C$ and $D$ such that

- $\forall c \in C, \exists Fc \in D$;

- $\forall f : c \to c' \in C, \exists Ff : Fc \to Fc' \in D$

That is, a functor maps objects in one category to objects in the other while morphisms are mapped to other morphisms whose domain and codomain are the ones of the parent category under the action of the functor.

This data is required to satisfy the **functoriality axioms**:

1. $\forall f, g \in C, Fg \cdot Ff = F(g \cdot f)$ (that is, the functor preserves compositions);

2. $\forall c \in C, F(1_c) = 1_{F_c}$ (that is, the functor maps identities to identities).

Again, some examples are in order.

1. **The chain-rule:** the well-known chain-rule obeyed by derivatives in elementary calculus has an intrinsic functoriality.

$$D : \text{Euclid}_* \to \text{Mat}_{\mathbb{R}}$$

Where $\text{Euclid}_*$ is the category whose objects are pointed finite dimensional euclidean spaces and whose morphisms are point-wise differentiable functions and $\text{Mat}_{\mathbb{R}}$ is the category whose objects are positive integers and the morphisms are $n \times m$ real matrices.

The action of the functor $D$ is to send the euclidean spaces to its dimension (naturally, a positive integer) and to send the differentiable functions to the Jacobian matrices, i.e.:

$$f = (f_1, f_2, ..., f_m) \mapsto \frac{\partial f_i}{\partial x_j}(a)$$

where

$$f : \mathbb{R}^m \to \mathbb{R}$$

Functors must preserve composition, that is, for functions $f : \mathbb{R}^n \to \mathbb{R}^m$ and $g : \mathbb{R}^m \to \mathbb{R}^k$, we must have $D(g \cdot f) = D(g) \cdot D(f)$. Translating this to a more familiar notation, we get:

$$\frac{\partial (g \cdot f)_i(a)}{\partial x_j} = \frac{\partial g_i(f_i(a))}{\partial x_j} \frac{\partial f_i}{\partial x_j}(a)$$

which is nothing but the chain-rule in multivariable calculus.

2. **Dual space**: We can also define the notion of a *contravariant functor* which is the functor between the opposite category and the target category, i.e., $F_{\text{cont}} : C^{\text{op}} \to D$. The opposite category can be obtained by inverting the direction of the morphisms in $C$, which encapsulates the concept of *duality* in category theory.

In this spirit, we define the following contravariant functor:

$$(-)^* : \text{Vect}_{\mathbb{K}} \to \text{Vect}_{\mathbb{K}}^{\text{op}}$$

which takes the vector spaces to their respective duals (i.e., the vector space whose vectors are $\mathbb{K}$-linear functionals) and the linear transformations $\phi : V \to W$ to the linear transformations $\phi^* : W^* \to V^*$. Notice the direction of the morphisms in both categories.

This construction is contravariant in the categorical sense since we would like to compose maps $\phi : V \to W$ and $\omega : W \to \mathbb{K}$ in order to get a map from $V$ to the field $\mathbb{K}$. That is, the action of the functor is

$$\phi^*(\omega) = \omega\phi$$

If this functor were covariant, the composition wouldn't work because of the incompatibility of domains and codomains, so we would be forced to invert the arrows in one of the categories, yielding a opposite category.

3. **Forgetful functor:** this is the functor which assigns any concrete category to the category $\mathfrak{Set}$, sending each object to its underlying set and the morphisms to the corresponding functions between these sets.

4. **Fundamental group:** this is an example of a functor that associates an invariant to a mathematical object. The fundamental group can be thought as a covariant functor from $\mathfrak{Top}_*$, the category of pointed topological spaces and pointed continuous maps to $\mathfrak{Grp}$. It sends each tangent space to the group of equivalence classes under homotopy of the loops in that space and the continuous maps to the corresponding group homomorphisms.

5. **Tangent space:** this is the functor from $\mathfrak{Diff}_*$, the category of pointed smooth manifolds to $\mathfrak{Vect}_{\mathbb{R}}$, the category of real vector spaces. It assigns to each pointed manifold $(M, p)$ its tangent space $TM_p$ and each smooth map $F : (M, p) \to (M, F(p))$ to the differential $dF_p$. That this assignment is a covariant functor can be seen from the following properties of differentials for smooth maps $F : M \to N$ and $G : N \to P$ (see [97], proposition 3.6):

$$d(G \circ F)_p \equiv dG_{F(p)} \circ dF_p : TM_p \to T_{G \circ F(p)}P,$$
$$d(\text{id}_M)_p \equiv \text{id}_{TM_p} : TM_p \to TM_p.$$

6. **The action of a group:** Let $G$ be a group. We can describe it categorically as a one object category B$G$ whose morphisms are nothing but the group elements themselves. Let $C$ be another category. We can construct a functor $X : \text{B}G \to C$ that assigns the sole object of B$G$ to a unique oject $X \in \mathrm{ob}(C)$. The interesting part is to what morphisms we assign the group elements $g$, which are $g_* : X \to X$ such that:

a) $h_* g_* = (hg)_*, \forall h, g \in G$;

b) $e_* = 1_X$, where $e$ is the identity in $G$.

Now, recall that the *action of a group* on a set $X$ is a function $\alpha : G \times X \to X$ such that:

a) $\alpha(g, \alpha(h, x)) = \alpha(gh, x)$ or $g \cdot (h \cdot x) = (g \cdot h) \cdot x$;

b) $\alpha(e, x) = x$, or $e \cdot x = x$.

We see that both items 1 and 2 coincide in both definitions. Thus, the action of a group is a functorial construction and can be extended to any concrete category.

7. **Dual vector space:** there is a contravariant functor $(-)^* : \mathfrak{Vect}_{\mathbb{K}}^{\mathrm{op}} \to \mathfrak{Vect}_{\mathbb{K}}$ which assigns to a vector space $V$ its dual space of linear functionals over $V$. Linear maps $\phi : V \to W$ are then sent to to the linear map $\phi^* : W^* \to V^*$ which pre-composes with the linear functional $\omega : W^* \to \mathbb{K}$ so that the composition becomes a functional $\phi\omega : V \to \mathbb{K}$.

Example 6 is an important illustration of the following general lemma.

**Lemma 6.3.1.** *Functors preserve isomorphisms.*

*Proof.* Let $F : C \to D$ be a functor between categories $C$ and $D$. Let $f \in \mathrm{Mor}_C(c, c')$ be an isomorphism in $C$ with inverse $g : c' \to c$. So:

$$F(g \cdot f) = F(g) \cdot F(f)$$

by the preservation of composition. At the same time,

$$F(g \cdot f) = F(1_c) = 1_{Fc}$$

since $f$ and $g$ are inverses of one another and we used the preservation of identities property. Both these results yield

$$F(g) \cdot F(f) = 1_{F_c}$$

and, by an analogous reasoning,

$$F(f) \cdot F(g) = 1_{F_{c'}}.$$

Thus, $F(f)$ and $F(g)$ are inverses of one another, making $F(f)$ an isomorphism. ∎

In light of example 6, the previous lemma has the following important corollary.

**Corollary 6.3.1.** *When a group $G$ acts functorially on a object $X \in ob(C)$, then the elements $g \in G$ must act by automorphisms $g_* : X \to X$. Moreover, $(g_*)^{-1} = (g^{-1})_*$.*

### 6.3.3 Natural Transformations

We have defined morphisms between categories which are called functors. What about morphisms between morphisms?

**Definition 6.3.5.** A *natural transformation* $\alpha : F \implies G$ between functors $F, G : C \to D$ consists of an arrow $\alpha_c : Fc \to Gc$ in $D$ for each $c \in ob(C)$ such that the following diagram commutes:

$$
\begin{array}{ccc}
Fc & \xrightarrow{\alpha_c} & Gc \\
{\scriptstyle Ff}\downarrow & & \downarrow{\scriptstyle Gf} \\
Fc' & \xrightarrow{\alpha_{c'}} & Gc'
\end{array}
$$

Where $f$ is any $f : c \to c'$ in $C$. The morphisms $\alpha_c$ are called the components of the natural transformation $\alpha$. If all the components of the natural transformation are isomorphisms, then we say that $\alpha$ is a *natural isomorphism.*

The field of Category Theory was born as the "general study of natural equivalences" [62][5]. Some examples are in order.

1. The canonical example of a natural transformation is defined by the map $V \to V^{**}$, $V$ being a vector space and $V^{**}$ its double dual. Because the identification with the dual $V^*$ depends on the choice of basis, Eilenberg and Mac Lane were able to show in the original paper [62] that there is no natural isomorphism between $V$ and $V^*$. However, for each $v \in V$, the evaluation map given by $\text{eval}_v : V^* \to \mathbb{K}$ defined by $\text{eval}_v(f) = f(v)$ for $f \in V^*$ defines a natural transformation between the identity functor and the functor which assigns $V$ to its double dual $V^{**}$. The naturality square

$$
\begin{array}{ccc}
V & \xrightarrow{\text{eval}} & V^{**} \\
{\scriptstyle \phi}\downarrow & & \downarrow{\scriptstyle \phi^{**}} \\
W & \xrightarrow{\text{eval}} & W^{**}
\end{array}
$$

---

[5]  In fact, one of the authors of this seminal paper, Saunder MacLane, has been quoted as: *"I did not invent Category Theory to talk about functors. I invented it to talk about natural transformations."*

commutes once we recognize that $\text{eval}_{\phi v}(f) = f(\phi(v))$ and that by the action of the dual functor in example 7 we have that $\phi^{**}(\text{eval}_v)(f) = f(\phi(v))$.

2. We previously defined the action of a group (understood as a one object grupoid) as a functorial construction that assigns the one object of B$G$ to one object $X \in \text{ob}(C)$ and the group elements $g$ to automorphisms $g_* : X \to X$.

   On the other hand, suppose we have two functors $X, Y : \text{B}G \to C$. What is the natural transformation $\alpha : X \implies Y$?

   *Naturally*, since both functors take a single object (the sole element of B$G$) to a single object which is its image, then the natural transformation has a single component. Noting that the diagram

$$
\begin{array}{ccc}
X & \xrightarrow{\ \alpha\ } & Y \\
{\scriptstyle g_*}\big\downarrow & & \big\downarrow{\scriptstyle g_*} \\
X & \xrightarrow{\ \alpha\ } & Y
\end{array}
$$

commutes, we see that the natural transformation is nothing but a morphism between $X$ and $Y$ in the category $C$ which is said to be $G$-equivariant.

   Now, it is easy to see that there is a natural inverse morphism $\alpha^{-1}$ such that composition with $\alpha$ yields the identity morphism in either $X$ or $Y$. Thus, $\alpha$ cannot be any isomorphism between $X$ and $Y$, but an isomorphism. This tells us that the action of the group defines not only automorphisms in the target object in $C$ but also isomorphisms between the objects in $C$.

3. There is a "natural way" to see that complements and preimages commute. First, we define a contravariant functor $\mathcal{O} : \text{Top}^{\text{op}} \to \text{Poset}$ which assigns to each topological space its underlying set of open subsets. This functor is contravariant because the morphisms in Top $f : X \to Y$ (continuous maps) give rise to an inverse function $f^{-1} : \mathcal{O}(Y) \to \mathcal{O}(X)$. This comes from the definition of continuity in topological spaces: the preimage of a function is open in the domain.

   In a similar fashion, we can define a covariant functor $\mathcal{C} : \text{Top}^{\text{op}} \to \text{Poset}$ which maps the topological space to the set of closed subsets in the topology of the space.

   So: what is the natural transformation between $\mathcal{C}$ and $\mathcal{O}$? If we define $\alpha : \mathcal{O}(X) \to \mathcal{X}$ to be the "take the complement" map (a closed set is defined as the complement of an open set), then we can write down the following commuting diagram:

$$
\begin{array}{ccc}
\mathcal{O}(X) & \xrightarrow{\ \alpha\ } & \mathcal{C}(X) \\[4pt]
\scriptstyle{f^{-1}}\big\uparrow & & \big\uparrow\scriptstyle{f^{-1}} \\[4pt]
\mathcal{O}(Y) & \xrightarrow[\ \alpha\ ]{} & \mathcal{C}(Y)
\end{array}
$$

Commutativity of this diagram means that we can commute $f^{-1}$ and $\alpha$, which means that the operations of taking the preimage of a subste and taking it's complement commute!

The operation of taking the complement is an isomorphism, which means that $\alpha$ is an natural isomorphism and $\mathcal{O}(X) \cong \mathcal{C}(X)$.

## 6.4   The Categories for Locally Covariant Quantum Field Theory

We now proceed to define the categories which act as domain and codomain for the functor that defines our QFT.

**Definition 6.4.1.** Let $\mathfrak{GlobHyp}$ denote the category where the objects are four-dimension orientable and time-orientable globally hyperbolic spacetimes $(M, g)$ and the morphisms are isometric embeddings $\psi : (M_1, g_1) \to (M_2, g_2)$. We also demand that:

1. If $\gamma : [a, b] \to M_2$ is a causal curve and $\gamma(a), \gamma(b) \in \psi(M_1)$ then the whole curve $\gamma(t)$ is in $\psi(M_1)$;

2. $\psi$ preserves orientation and time-orientation.

Condition 1 is required for the induced causal structure of the embedded spacetime to coincide with that of $M_2$. Condition 2, on the other hand, may be relaxed to accommodate the discussion of the PCT-theorem.

We could also envisage a larger class of spacetimes which could define the category $\mathfrak{GlobHyp}$. An interesting question is how general can this category be? Can any category whose objects admit the interpretation of a spacetime be a valid domain for this functor? Some progress has been made in this direction [76] which indicate that this functorial formulation can be very general.

We verify that this construction is in fact a category. First, we identify that the identity map $\mathrm{id}_M : x \mapsto x$, $x \in M$ acts as the identity morphism for each globally hyperbolic spacetime. It suffices to show that the composition of isometric embeddings yields a new isometric embedding (associativity follows immediately). As we are embedding one spacetime into another, it is reasonable to demand that the morphisms are "injective" or monic.

**Definition 6.4.2.** Let $\mathfrak{Alg}$ denote the category whose objects are unital $C^*$-algebras and the morphisms are unit-preserving $*$-homomorphisms.

Since the morphisms in $\mathfrak{GlobHyp}$ are monic, we demand the morphisms in $\mathfrak{Alg}$ to be as well. We can also envisage other algebras to serve as the codomain to the functor. Examples of algebras that appear in the literature are von Neumann algebras, Borchers-algebras and even more general $*$-algebras. In the case of pertubative locally covariant QFT models, for example, one employs the category with locally convex topological unital $*$-algebras as objects. For our goals, considering $C^*$-algebras appears to be general enough, but we will specialize to von Neumann algebras as well.

Naturally, this construction in fact defines a category since the composition of two $*$-morphisms (which preserve the algebra structure and the involution operation) yields a new $*$-morphism and associativity is naturally implied. The identity morphism can be identified with the identical map $\mathrm{id}_a$, for all $a \in \mathfrak{A}$.

## 6.5 The Functor

We are now in the position to define the Quantum Field Theory as a functor.

**Definition 6.5.1.** A *locally covariant quantum field theory* is a covariant functor $\mathscr{A} : \mathfrak{GlobHyp} \to \mathfrak{Alg}$. In diagrammatic form:

$$
\begin{array}{ccc}
(M,g) & \xrightarrow{\ \psi\ } & (M',g') \\
\Big\downarrow{\scriptstyle \mathscr{A}} & & \Big\downarrow{\scriptstyle \mathscr{A}} \\
\mathscr{A}(M,g) & \xrightarrow{\ \alpha_\psi\ } & \mathscr{A}(M',g')
\end{array}
$$

where $\alpha_\psi \equiv \mathscr{A}(\psi)$, $\mathscr{A}(M,g) \equiv \mathfrak{A}(M,g)$ and we have the covariance properties $\alpha_{\psi'} \circ \alpha_\psi = \alpha_{\psi' \circ \psi}$ and $\alpha_{\mathrm{id}_M} = \mathrm{id}_{\mathscr{A}(M,g)}$.

The theory is said to be *causal* if for any pair of morphisms $\psi_1 \in \mathrm{Hom}((M_1,g_1),(M,g))$ and $\psi_2 \in \mathrm{Hom}((M_2,g_2),(M,g))$ such that the ranges $\psi_1(M_1)$ and $\psi_2(M_2)$ are causally separated in $(M,g)$, we have

$$
[\alpha_{\psi_1}(\mathscr{A}(M_1,g_1)), \alpha_{\psi_2}(\mathscr{A}(M_2,g_2))] = \{0\}
$$

We also say that the theory obeys the *time-slice axiom* if

$$\alpha_\psi(\mathscr{A}(M,g)) = \mathscr{A}(M',g')$$

provided that $\psi(M)$ contains a Cauchy surface for $(M',g')$, for all $\psi \in \mathrm{Hom}_{\mathfrak{GlobHyp}}((M,g),(M',g'))$.

### 6.5.1 Example: the Klein-Gordon field

We now discuss how the Klein-Gordon theory can be understood as a functor.

Let $(M,g) \in \mathfrak{GlobHyp}$. Global hyperbolicity tells us that the Cauchy problem for the scalar Klein-Gordon equation is well-posed on $(M,g)$:

$$(\nabla^\mu \nabla_\mu + m^2 + \xi R)\varphi = 0$$

Well-posedness tells us that there are uniquely determined advanced and retarded fundamental solutions of the equation, $E^{\mathrm{adv/ret}} : C_0^\infty(M,\mathbb{R}) \to C^\infty(M,\mathbb{R})$. The difference between these solutions is the causal propagator of the KG equation.

The way to obtain the Klein-Gordon theory as a functor is by defining a symplectic form on the range of the propagator solution. To this symplectic space, we associate its Weyl-algebra $\mathfrak{W}(\mathcal{R},\sigma)$, which is generated by a family of unitary elements satisfying the canonical commutation relations in the exponential form (see chapter 4)

$$W(\phi)W(\tilde{\phi}) = e^{-i\sigma(\phi,\tilde{\phi})/2}W(\phi + \tilde{\phi}).$$

The point is that both the symplectic space and the Weyl-algebra are completely determined by $(M,g)$, leading us to consider $\mathfrak{A}(M,g) = \mathfrak{W}(\mathcal{R}(M,g), \sigma_{(M,g)})$ as our functor. The choice is correct once we construct a $C^*$-monomorphism in a sutiable way and prove that it obeys the covariance property.

## 6.6 Haag-Kastler Net as a Functor

The Haag-Kastler axioms can be understood as a particular case of the construction presented up to this point. This was presented as a proposition in [30], where the proof can be found.

Before presenting this proposition, we introduce some definitions. Let $\mathcal{K}(M,g)$ be the set of compact open subsets of $M$ with causal curves between any pair of points. For a $\mathcal{O} \in \mathcal{K}(M,g)$, let $g_\mathcal{O}$ be the metric of $M$ restricted to $\mathcal{O}$ ($g_\mathcal{O} = g \restriction_\mathcal{O}$) such that $(\mathcal{O}, g_\mathcal{O}) \in \mathfrak{GlobHyp}$ and $i_{M,\mathcal{O}} : (\mathcal{O}, g_\mathcal{O}) \to (M,g) \in \mathrm{Hom}(\mathfrak{GlobHyp})$.

**Theorem 6.6.1.** *Let $\mathscr{A}$ be a LCQFT functor. Define a map $\mathcal{O} \mapsto \mathcal{A}(\mathcal{O}) \subset \mathscr{A}(M, g)$ by setting:*

$$\mathcal{A}(\mathcal{O}) = \alpha_{M,\mathcal{O}}(\mathscr{A}(\mathcal{O}), g_{\mathcal{O}})$$

*where $\alpha_{M,\mathcal{O}} \equiv \alpha_{i_{M,\mathcal{O}}}$. Then the following statements hold:*

1. **Isotony**

   $$\mathcal{O}_1 \subset \mathcal{O}_2 \implies \mathcal{A}(\mathcal{O}_1) \subset \mathcal{A}(\mathcal{O}_2), \forall \mathcal{O}_1, \mathcal{O}_2 \in \mathcal{K}(M, g)$$

2. **Representation**

   *If there is a group $G$ of isometric diffeomorphisms $\kappa : M \to M$ with $\kappa_* g = g$ preserving orientation and time-orientation, then there is a representation $\kappa \mapsto \tilde{\alpha}_\kappa$ of $G$ by $C^*$-automorphisms $\tilde{\alpha}_\kappa : \mathcal{A} \to \mathcal{A}$, where $\mathcal{A}$ is the minimal $C^*$-algebra generated by $\{\mathcal{A}(\mathcal{O}) : \mathcal{O} \in \mathcal{K}(M, g)\}$ such that*

   $$\tilde{\alpha}_\kappa(\mathcal{A}(\mathcal{O})) = \mathcal{A}(\kappa(\mathcal{O})), \mathcal{O} \in \mathcal{K}(M, g).$$

3. **Causality**

   *If $\mathscr{A}$ is causal, then*

   $$[\mathcal{A}(\mathcal{O}_1), \mathcal{A}(\mathcal{O}_2)] = \{0\}$$

   *for causally separated $\mathcal{O}_1$ and $\mathcal{O}_2$.*

4. *Suppose that $\mathscr{A}$ satisfies the time-slice axiom and let $\Sigma$ be a Cauchy surface in $(M, g)$ and let $S \subset \Sigma$ be open and connected. Then for each $\mathcal{O} \in \mathcal{K}(M, g)$ and $S \subset \mathcal{O}$:*

   $$\mathcal{A}(S^{\perp\perp}) \subset \mathcal{A}(\mathcal{O})$$

   *where $S^{\perp\perp}$ denotes the double causal complement of $S$ (causal closure) and $\mathcal{A}(S^{\perp\perp})$ is the smallest $C^*$-algebra formed by all $\mathcal{A}(\mathcal{O}_1)$, $\mathcal{O}_1 \subset S^{\perp\perp}$, $\mathcal{O}_1 \in \mathcal{K}(M, g)$.*

*Proof.* See [30], proposition 2.3.                                                                 ■

Interestingly, it was Dyson who first proposed the generalization of the Haag-Kastler axioms to the context of more general spacetimes in his famous "Missed opportunities" work [60].

### 6.6.1  Equivalence between QFTs

As mentioned previously, a natural isomorphism defines an equivalence relation between functors. We can use this construction to define the notion of equivalent QFTs. By "equivalent", we mean that the theories are physically indistinguishable, which is encapsulated by the fact that the algebras of observables are isomorphic to each other.

The natural transformation between different QFTs can be represented diagrammatically as:

$$
\begin{array}{ccc}
\mathscr{A}(M_1, g_1) & \xrightarrow{\ \beta_{(M_1, g_1)}\ } & \mathscr{A}'(M_1, g_1) \\
{\scriptstyle \alpha_\psi}\big\downarrow & & \big\downarrow{\scriptstyle \alpha'_\psi} \\
\mathscr{A}(M_2, g_2) & \xrightarrow[\ \beta_{(M_2, g_2)}\ ]{} & \mathscr{A}'(M_2, g_2)
\end{array}
$$

where $\beta_{(M_2, g_2)} \circ \alpha_\psi = \alpha'_\psi \circ \beta_{(M_1, g_1)}$. Notice that since this diagram is taking place in $\mathfrak{Alg}$, then the morphisms $\beta$ are nothing but the $C^*$-morphisms. If they are isomorphisms, in the $C^*$-algebra sense, then the algebras are isomorphic and the theories are equivalent.

#### 6.6.1.1  KG fields with different masses are not equivalent

We can slightly modify proposition 2.1 in order to show that Klein-Gordon theories with different masses do not equivalent in the sense that there is no natural isomorphism (concretely a $*$-isomorphism between the local algebras) between them.

Let $\mathcal{O} \mapsto \mathcal{A}_1(\mathcal{O})$ and $\mathcal{O} \mapsto \mathcal{A}_2(\mathcal{O})$ be the local $C^*$-algebras in Minkowski spacetime derived from the functors $\mathscr{A}_1$ and $\mathscr{A}_2$. Let $\tilde\alpha_L^1$ and $\tilde\alpha_L^2$ denote the automorphic actions of the Poincaré group on the algebras ($L \in \mathcal{P}_+^\uparrow$).

Suppose now that we have an isomorphism $\beta : \mathcal{A}_1 \to \mathcal{A}_2$ such that

$$
\beta(\mathcal{A}_1(\mathcal{O})) = \mathcal{A}_2(\mathcal{O}), \quad \beta \circ \tilde\alpha_L^1 = \tilde\alpha_L^2 \circ \beta
$$

Since this theory is formulated in Minkowski spacetime, we know there is a unique state $\omega_j$ for each $\mathcal{A}_j$ that is invariant under the action of the Poincaré group and a ground state with respect to the corresponding action of timelike translations. Thus, we would conclude that $\beta(\omega_1) = \omega_2$, which implies that the spectra of the generators of time-translations in the vacuum representations of the Klein-Gordon field for different masses are the same.

### 6.6.2   Quantum fields as natural transformations

In the same spirit that the one employed for the energy-momentum tensor discussed in section 1, we would like to impose covariance to the quantum fields. In order to do so, we will have to allow the codomain of the functor $\mathscr{A}$ to be more general than $\mathfrak{Alg}$.

The definition of the fields is given the following way: consider a family $\Phi \equiv \{\Phi_{(M,g)}\}$ of quantum fields indexed by all spacetimes. These fields are generalized (i.e., not necessarily linear) algebra-valued distributions, which means that there is a family $\{\mathcal{A}(m,g)\}$ of topological $*$-algebras indexed by the spacetimes and for each spacetime, there is a continuous map $\Phi_{(M,g)} : C_0^\infty(M) \to \mathcal{A}(M,g)$. Then we demand that there is a continuous monomorphism $\alpha_\psi : \mathcal{A}(M_1, g_1) \to \mathcal{A}(M_2, g_2)$ for any morphism $\psi \in \mathrm{Hom}_{\mathfrak{GlobHyp}}((M_1, g_1), (M_2, g_2))$ such that

$$\alpha_\psi(\Phi_{(M_1,g_1)}(f)) = \Phi_{(M_2,g_2)}(\psi_*(f))$$

where $f \in C_0^\infty(M_1)$ is any test function and $\psi_*(f) = f \circ \psi^{-1}$.

Let $\mathfrak{Test}$ denote the category of test function spaces on manifolds, whose objects are spaces $C_0^\infty(M)$ of smooth, compactly supported test-functions on $M$ and the morphisms are the pushforwards $\psi_*$ of (injective) embeddings $\psi : M_1 \to M_2$.

Lets consider $\mathscr{A}$ to be the functor between $\mathfrak{GlobHyp}$ and $\mathfrak{TAlg}$ of topological $*$-algebras. Moreover, let $\mathscr{D}$ denote the covariant functor between $\mathfrak{GlobHyp}$ and $\mathfrak{Test}$ such that $(M,g) \mapsto C_0^\infty(M)$ and $\mathscr{D}(\psi) = \psi_*$. Both these categories are subcategories of $\mathfrak{Top}$.

**Definition 6.6.1.** A locally covariant quantum field $\Psi$ is a natural transformation between the functors $\mathscr{D}$ and $\mathscr{A}$ such that the following diagram commutes:

$$\begin{array}{ccc}
\mathscr{D}(M_1, g_1) & \xrightarrow{\;\Phi_{(M_1,g_1)}\;} & \mathscr{A}(M_1, g_1) \\
\psi_* \downarrow & & \downarrow \alpha_\psi \\
\mathscr{D}(M_2, g_2) & \xrightarrow[\;\Phi_{(M_2,g_2)}\;]{} & \mathscr{A}(M_2, g_2)
\end{array}$$

where the commutativity of the diagrams satisfies the covariance for fields:

$$\alpha_\psi \circ \Phi_{(M_1,g_1)} = \Phi_{(M_2,g_2)} \circ \psi_*$$

## 6.7   States

In this section, we provide the formulation of the space of states as a functor from the category of globally hyperbolic spacetimes to the category of states over the algebra of observables of our theory. This was first proposed in [30] and we follow their notation.

We remind ourselves that a state is a linear functional $\omega : \mathfrak{A} \to \mathbb{C}$ having the property of being positive, i.e., $\omega(A^*A) \geq 0$, $\forall A \in \mathfrak{A}$ and normalized, i.e., $\omega(\mathbf{1}_\mathfrak{A}) = 1$, $\mathbf{1}_\mathfrak{A}$ being the identity in the $C^*$-algebra $\mathfrak{A}$. The collection of all states over an algebra $\mathfrak{A}$ is called a *state space*. This construction can be done in terms of a functor.

## 6.7.1  The Category of State Spaces

We define $\mathfrak{Sts}$ as the category whose objects $\mathbf{S}$ are state spaces of a $C^*$-algebra $\mathfrak{A}$ which are closed under taking finite convex combinations and operations $\omega(.) \mapsto \omega_A(.) = \omega(A^*.A)/\omega(A^*A)$, $A \in \mathcal{A}$.

The morphisms between objects $\mathbf{S}'$ and $\mathbf{S}$ are maps $\gamma^* : \mathbf{S}' \to \mathbf{S}$ which arise as dual maps from the corresponding $C^*$-monomorphisms from the underlying algebras, $\gamma : \mathfrak{A} \to \mathfrak{A}'$. This map is defined as

$$\gamma^* \omega'(A) = \omega'(\gamma(A)), \quad \omega' \in \mathbf{S}', \quad A \in \mathfrak{A}. \tag{6.7.1}$$

Notice the inverted order of the domain and codomain of $\gamma$ and $\gamma^*$.

## 6.7.2  The functor

As presented in the Introduction, one can define a QFT as a functor between a category of spacetimes and a category of algebras of observables (usually taken as $C^*$-algebras for definiteness).

In a similar fashion, one can define the functor which assigns to each spacetime the corresponding space of states.

**Definition 6.7.1.** Let $\mathscr{A}$ be a LCQFT. A *state space for $\mathscr{A}$* is a contravariant functor $\mathbf{S}$ between $\mathfrak{GlobHyp}$ and $\mathfrak{Sts}$:

$$
\begin{array}{ccc}
(M, \mathbf{g}) & \xrightarrow{\quad \psi \quad} & (M', \mathbf{g}') \\
\Big\downarrow{\scriptstyle \mathbf{s}} & & \Big\downarrow{\scriptstyle \mathbf{s}} \\
\mathbf{S}(M, \mathbf{g}) & \xleftarrow{\quad \alpha_\psi^* \quad} & \mathbf{S}(M', \mathbf{g}')
\end{array}
$$

where $\mathbf{S}(M, \mathbf{g})$ is a set of states on $\mathscr{A}(M, \mathbf{g})$ and $\alpha_\psi^*$ is the dual map of $\alpha_\psi$. The contravariance property is given by

$$\alpha_{\psi \circ \psi}^* = \alpha_\psi^* \circ \alpha_{\tilde{\psi}}^*. \tag{6.7.2}$$

We note that the inversion of the domain and codomain of each $C^*$-monomorphism, as observed in the previous section, is reflected in the contravariance nature of the state space functor.

# Part III

# Entanglement

# 7 Entanglement in Quantum Field Theory

> *Quantum carburetor? Jesus Morty, you can't just add a sci-fi word to a car word and hope it means something. Looks like something's wrong with the micro-verse battery.*
>
> Rick Sanchez, *Rick and Morty*, S02E06

In this chapter, we discuss quantum entanglement from an algebraic point of view, that is, making reference to the states over a local algebra. We then discuss the first manifestation of quantum entanglement in the form of what are called Bell inequalities and discuss their violation by the vacuum state of a QFT.

## 7.1 Introduction

Quantum entanglement has been the subject of intense research both in Fundamental Physics and in Applied Physics. In the latter, even though entanglement does not allow the passing of information, there has been a great progress in quantum computing and quantum optics as well as advancements in understanding physical phenomena such as superconductivity, super-radiance and disordered systems (see the comprehensive Horodecki[4] review [87] and references therein). In keeping the spirit of this Thesis, we will be concerned with the former aspects, that is, mathematical and some philosophical aspects of entanglement and in its realization in the form of Bell's inequalities, in particular.

Much of the folklore about quantum entanglement can be traced back to the famous EPR paper by Einstein, Podolsky and Rosen [63]. The paper poses that the quantum description of reality can't be considered complete since it seems to imply that quantum objects do not have intrinsic properties prior to measurement[1]. This conclusion was the starting hypothesis taken by John Bell, then working at CERN, who would publish a reply to the EPR paper in the short-lived *Physics, Physique, Fizika* [17].

---

[1] Well-known classical analogies to this situation include the question of whether the Moon exists when no-one is looking at it.

The analysis by Bell gave rise to the so-called *hidden variables theories*, which essentially sought to "complete" Quantum Mechanics thus eliminating the probabilistic interpretation of the theory. In order to achieve this, Bell assumed the following hypotheses:

1. **Realism:** the quantum objects have properties irrespective of their measurements, which is equivalent to assuming the EPR paper as correct;

2. **Locality:** experiments conducted at spacelike distances are independent[2];

3. **Free will:** the setting of the measurement apparatus is independent of the hidden variables which determine the local results.

From these assumptions, Bell was able to prove that the statistical correlations in experiments involving bipartite systems satisfy a set of inequalities which are nowadays known as *Bell inequalities*. He was also able to show that when these experiments involve entangled states, these inequalities are violated. Hence, the statistical correlations of quantum systems can't be obtained from a classical system, i.e., a hidden-variables theory. This conclusion alongside the inequalities is sometimes referred to as *Bell's theorem*. It is usually credited to Aspect et al. the first convincing experimental measurements of violations of Bell inequalities using entangled photon states [8, 9].

Even though Bell's theorem seems to rule out a classical alternative to Quantum Mechanics in terms of hidden variables, these theories still attract some interest from a foundational perspective and regarding different interpretations of Quantum Mechanics. For a review on such theories, see [72].

## 7.2  Separability of States

Entangled states are non-separable states. It is important to notice that this phrase is interpreted differently if one uses a physicist's convention or the one adopted by a mathematician. In keeping the spirit of this Thesis, we will adopt the mathematician's convention, but we will cite the physicist's convention for completeness. The mathematician's convention also has the advantage of being formulated directly in terms of von Neumann algebras, which allows a direct application of the definition of entanglement to QFT. We advise the reader to be careful in comparing different references.

---

[2]  We use the word "locality" here as it is common in the literature on Foundations of Quantum Mechanics and Quantum Information, but the meaning conveyed here is equivalent to the axiom of Einstein Causality in the context of AQFT

For physicists, a "state" is a vector in a Hilbert space representing a possible physical state the system may achieve. Consider the Hilbert space as being $\mathcal{H} = \mathcal{H}_1 \otimes \mathcal{H}_2$. A *product state* in $\mathcal{H}$ is then a vector state which can be written as

$$|\psi\rangle = |\psi_1\rangle \otimes |\psi_2\rangle, \tag{7.2.1}$$

where $|\psi_1\rangle \in \mathcal{H}_1$ and $|\psi_2\rangle \in \mathcal{H}_2$ and we have adopted Dirac's notation in keeping the spirit of using the physicist's notation. We say then that a state $|\phi\rangle$ is *separable* if it can be written as a convex combination of product states $|\psi\rangle$, i.e.:

$$|\phi\rangle = \sum_i \lambda_i |\psi\rangle, \tag{7.2.2}$$

where $\sum_i \lambda_i = 1$. Notice that we are considering bipartite systems, but this definition can be generalized for tensor products of more Hilbert spaces. Notice that a product state is, in particular, a separable state[3]. Hence, an *entangled state* is a non-separable state.

In the mathematician's convention, a "state" is a normalized functional over an algebra, which we will take as being given by a von Neumann algebra $\mathfrak{M}$.

**Definition 7.2.1.** Let $\mathfrak{M}$ be a von Neumann algebra over a Hilbert space $\mathcal{H}$ with $\mathfrak{M}_1$ and $\mathfrak{M}_2$ as subalgebras. A normal state $\omega(A) = \text{Tr}(\rho A)$ with $A, \rho \in B(\mathcal{H})$ and $\rho$ being a density matrix is said to be

1. a *product state* for the pair $(\mathfrak{M}_1, \mathfrak{M}_2)$ if there are density matrices $\rho_1$ and $\rho_2$ in $B(\mathcal{H})$ such that

$$\text{Tr}(\rho AB) = (\text{Tr}(\rho_1 A))\,(\text{Tr}(\rho_2 B)), \tag{7.2.3}$$

   for all $A \in \mathfrak{M}_1$ and $B \in \mathfrak{M}_2$;

2. a *separable state* for $(\mathfrak{M}_1, \mathfrak{M}_2)$ if there exists a weak $*$-limit (see discussion following definition 3.3.4) of convex combinations of product states such that

$$\text{Tr}(\rho AB) = \sum_i \lambda_i \,\text{Tr}(\rho_i AB), \tag{7.2.4}$$

   with $\sum_i \lambda_i = 1$ and all $A \in \mathfrak{M}_1$ and $B \in \mathfrak{M}_2$;

3. an *entangled state* if it is not separable.

---

[3] Many textbooks also mention that separable states are *pure states*, since they define a pure states as being normalized vectors in a Hilbert space. A lot of care needs to be taken here, since the GNS construction allows one to represent any state (in the mathematical convention) over a suitable algebra as a normalized vector in a Hilbert space, even mixed ones. For more details on this difference, see [16].

Notice that both definitions are the same in spirit and convey the intuitive meaning that separable states are those which are statistically independent from one another. The difference is in the mathematical objects to which the quality of being "separable" or "entangled" is attributed to.

## 7.3 Bell Inequalities

In this section, we will obtain the so-called first and second Bell inequalities approximately in their original form. For now, we will work with bounded operators acting on a Hilbert space $\mathcal{H}$. The abstraction to a generic $C^*$-algebra is immediate. This section is based on [95].

### 7.3.1 The first Bell inequality

We consider a set of four self-adjoint operators $A_i$ and $B_i$ with $i \in \{1, 2\}$ whose eigenvalues are $\pm 1$. We also assume that

1. $[A_i, B_j] = 0$, for all $i, j \in \{1, 2\}$;

2. $[A_1, A_2] \neq 0$ and $[B_1, B_2] \neq 0$.

Physically speaking, this set of operators describes a bipartite system whose subsystems don't interact with one another. Now, we define what is known as the *Bell operator* in the literature:

$$\mathcal{C} \equiv A_1(B_1 + B_2) + A_2(B_1 - B_2) = B_1(A_1 + A_2) + B_2(A_1 - A_2). \tag{7.3.1}$$

Since we are assuming that the $A$'s commute with the $B$'s, it is straightforward to prove that $\mathcal{C}$ is self-adjoint and hence may be interpreted as an "observable" whose possible eigenvalues are $\{-4, -2, +2, +4\}$. This statement is true both for Quantum Mechanics and hidden variable theories. Bell inequalities put certain bounds on the expectation value of the Bell operator $\langle \mathcal{C} \rangle = \langle \psi | \, \mathcal{C} \, | \psi \rangle = \omega(\mathcal{C})$, for some vector state $| \psi \rangle$ or, equivalently, a state $\omega$ (we use Dirac notation here only for historical purposes and continue the discussion employing the state $\omega$).

Recalling the Cauchy-Schwarz inequality for states

$$|\omega(A^*B)| \leq \sqrt{\omega(A^*A)\omega(B^*B)}, \tag{7.3.2}$$

which implies that for self-adjoint operators

$$|\omega(AB)| \leq \sqrt{\omega(A^2)\omega(B^2)}. \tag{7.3.3}$$

In particular, if we pick $A = \mathcal{C}$ and $B = \mathbf{1}$, we have that

$$\omega(\mathcal{C})^2 \leq \omega(\mathcal{C}^2). \tag{7.3.4}$$

By an elementary calculation, we find that

$$\begin{aligned}
\mathcal{C}^2 &= 4 - [A_1, A_2]\,[B_1, B_2] \\
&= 4 - (A_1 B_1)(A_2 B_2) + (A_1 B_2)(A_2 B_1) + (A_2 B_1)(A_1 B_2) - (A_2 B_2)(A_1 B_1),
\end{aligned}$$

where we omit the identity. This implies that

$$\begin{aligned}
\omega(\mathcal{C}^2) &= 4 + S \\
S &= -\omega\left((A_1 B_1)(A_2 B_2)\right) + \omega\left((A_1 B_2)(A_2 B_1)\right) + \omega\left((A_2 B_1)(A_1 B_2)\right) - \omega\left((A_2 B_2)(A_1 B_1)\right).
\end{aligned} \tag{7.3.5}$$

A closer look to the terms of $S$ together with the use of the Cauchy-Schwarz identity shows that

$$|\omega\left((A_a B_b)(A_c B_d)\right)|^2 \leq \omega\left((A_a B_b)^2\right)\omega\left((A_c B_d)^2\right) = 1,$$

where $a, b, c, d = 1, 2$ and we used the hypothesis that the $A$'s commute with the $B$'s. Since $\mathcal{C}$ is self-adjoint the expectation value must be a real number. Hence, we have that

$$\omega(\mathcal{C}^2) \leq 8 \implies -2\sqrt{2} \leq \langle \mathcal{C} \rangle \leq 2\sqrt{2}, \tag{7.3.6}$$

where we have used equation (7.3.4). This is the first Bell inequality and it is valid *both* for Quantum Mechanics and hidden variable theories and it has been exhaustively compared with experiment and simulations. The bound $2\sqrt{2}$ is sometimes called the *Cirel'son bound* [41].

If we can find some lower bound $|\beta| \leq 2\sqrt{2}$ such that a hidden variable theory must satisfy $|\langle \mathcal{C} \rangle| \leq \beta$, we would be able to compare hidden variable theories with Quantum Mechanics and speak of a *violation of a Bell inequality*. In fact, this is what is obtained in the second Bell inequality which we now show.

### 7.3.2 The second Bell inequality

We now assume that there are hidden variables and that the $A$'s and $B$'s commute with each other. This means that we are throwing away hypothesis 2 in the derivation of the first Bell inequality.

In terms of a hidden variables theory, the operators are interpreted as *random variables* which can take the values of $+1$ or $-1$. These operators are then defined in some sample space which includes the hidden variables, and are interpreted as *measurable functions*. Hence, they should *commute* in a theory with hidden variables. We then have the expectation value of the Bell operator given by

$$
\begin{aligned}
\langle \mathcal{C} \rangle &= \langle A_1 B_1 \rangle + \langle A_1 B_2 \rangle + \langle A_2 B_1 \rangle - \langle A_2 B_2 \rangle \\
&= \int_\Lambda d\mu(\lambda) A_1(\lambda) B_1(\lambda) + A_1(\lambda) B_2(\lambda) + A_2(\lambda) B_1(\lambda) - A_2(\lambda) B_2(\lambda) \\
&\equiv \int_\Lambda d\mu(\lambda) C_0(\lambda),
\end{aligned}
\tag{7.3.7}
$$

where we integrate over the sample space $\Lambda$ and consider the point-wise product of measurable functions. Again from the Cauchy-Schwarz inequality we have

$$
\left( \int_\Lambda d\mu(\lambda) C_0(\lambda) \right)^2 \leq \int_\Lambda d\mu(\lambda) C_0(\lambda)^2 = 4,
$$

where the last equality can be easily verified from the fact that the functions $A(\lambda)$ and $B(\lambda)$ commute and satisfy $A_i(\lambda)^2 = B_j(\lambda)^2 = 1$, for all $i, j \in \{1, 2\}$. Hence, we have that

$$
-2 \leq \langle \mathcal{C} \rangle \leq 2.
\tag{7.3.8}
$$

This is the second Bell inequality and from the elementary fact that $2 < 2\sqrt{2}$ we see that we have found a criterion for testing hidden variables against Quantum Mechanics. That is, for both theories, the expectation value is constrained by the Cirel'son bound but hidden variables are even more constrained by the bound of 2. The hard work of experimental physicists has shown that this bound of 2 is violated in experiments with objects described by Quantum Mechanics, thus strongly suggesting the absence of hidden classical variables in the theory. Therefore, hidden variable theories are incompatible with observations.

### 7.3.3 An example

As an example, we will consider a set of operators which satisfy the conditions of the first Bell inequality. This example is taken from [15]. Let us consider the Hilbert space $\mathcal{H} = \mathbb{C}^2 \otimes \mathbb{C}^2$, where the vector describing two electrons with total spin zero is given by

$$
\psi = \frac{1}{\sqrt{2}} \left[ \begin{pmatrix} 1 \\ 0 \end{pmatrix} \otimes \begin{pmatrix} 0 \\ 1 \end{pmatrix} - \begin{pmatrix} 0 \\ 1 \end{pmatrix} \otimes \begin{pmatrix} 1 \\ 0 \end{pmatrix} \right].
\tag{7.3.9}
$$

This vector is the solution for the equations

$$(s_3 + t_3)\psi = 0;$$
$$[(s_1 + t_1)^2 + (s_2 + t_2)^2 + (s_3 + t_3)^2]\psi = 0,$$
(7.3.10)

where $s_i = \sigma_i \otimes \mathbf{1}$ and $t_i = \mathbf{1} \otimes \sigma_i$ and $\sigma_i$ are the Pauli matrices for $i \in \{1, 2, 3\}$.

After constructing the Bell operator $\mathcal{C}$, we show that it satisfies the first Bell inequality while violating the second. To do so, we need to construct the dichotomic operators $A_i$ and $B_j$. We will consider the coordinate space for our problem as being $\mathbb{R}^3$. Consider the following normalized vectors

$$\vec{a}_1 = (\cos\alpha_1, 0, \sin\alpha_1);$$
$$\vec{a}_2 = (\cos\alpha_2, 0, \sin\alpha_2);$$
$$\vec{b}_1 = (\cos\beta_1, 0, \sin\beta_1);$$
$$\vec{b}_2 = (\cos\beta_2, 0, \sin\beta_2).$$
(7.3.11)

From these we construct the operators

$$A_1 = \vec{a}_1 \cdot \vec{s} = \cos\alpha_1 s_1 + \sin\alpha_1 s_2;$$
$$A_2 = \vec{a}_2 \cdot \vec{s} = \cos\alpha_2 s_1 + \sin\alpha_2 s_2;$$
$$B_1 = \vec{b}_1 \cdot \vec{t} = \cos\beta_1 t_1 + \sin\beta_1 t_2;$$
$$B_2 = \vec{b}_2 \cdot \vec{t} = \cos\beta_2 t_1 + \sin\beta_2 t_2.$$
(7.3.12)

It is straightforward to verify that $A_i^2 = B_j^2 = \mathbf{1}$ and that each operator has eigenvalues $\pm 1$. By combining these operators into the Bell operator and computing the expectation value with the vector $\psi$ in equation (7.3.9), we obtain

$$\langle \mathcal{C} \rangle = \langle \psi, \mathcal{C}\psi \rangle = -\cos(\alpha_1 - \beta_1) - \cos(\alpha_1 - \beta_2) - \cos(\alpha_2 - \beta_1) + \cos(\alpha_2 - \beta_2).$$

To illustrate our result, pick $\alpha_1 = 0$, $\alpha_2 = \pi/4$, $\beta = \pi$ and $\beta_2 = \pi/2$. This results in

$$\langle \mathcal{C} \rangle = 1 + \sqrt{2} > 2.$$
(7.3.13)

Hence, the second Bell inequality is violated. In this model, the operators defined in equation (7.3.12) do not commute with one another, and hence are part of the quantum mechanical paradigm. However, this is a textbook example of a real system describing two identical spin 1/2 particles, with the operators $s_i$ and $t_i$ measuring such spin.

## 7.4 Violation of Bell Inequalities in AQFT

The inequalities derived by Bell have the disadvantage that they assume perfect correlations exhibited by the singlet state. This is not true in experiments where a perfect correlation is nearly impossible.

However, an alternative inequality was derived by Clauser, Horne, Shimony and Holt [43], henceforth referred to as CHSH inequality, which allows to probe hidden variable theories independently of the quantum formalism. This is the inequality we will work with even though we might refer to it as being *the* Bell inequality.

A very interesting result exploring this inequality was achieved by Summers and Werner: the vacuum state of a AQFT violates *maximally* the CHSH inequality [127, 128]. In order to explore this result, we follow their formulation and conventions to describe this inequality in terms of a $C^*$-algebra (see also [12]).

**Definition 7.4.1.** A *correlation duality* consists of two mutually commuting $C^*$-algebras $\mathfrak{A}_1$ and $\mathfrak{A}_2$ together with a state $\omega$ over $\mathfrak{C} = \mathfrak{A}_1 \otimes \mathfrak{A}_2$ such that for positive elements $a \in \mathfrak{A}_+$ and $b \in \mathfrak{B}_+$ we have that $\omega(ab) \geq 0$.

Now, a measurement with a finite number of outcomes in this setting will be represented by a family $\{a_i\}_{i \in I}$ with $a_i \in \mathfrak{A}_+$ and $\sum_{i \in I} = \mathbf{1}$, with $\mathbf{1}$ being the identity in $\mathfrak{C}$. Physically speaking, the ordered set $I$ indexes the possible outcomes of the experiment, whose measuring device is represented by the positive element $a_i$, whose spectrum is a subset of the real line (following definition 2.3.2, thus justifying the choice of positive elements from a physical point of view). The state $\omega$ then represents the preparation device while the value $\omega(a_i)$ represents the probability of obtaining the $i$th result in the experiment.

**Definition 7.4.2.** Let $\mathfrak{A}$ be a $C^*$-algebras with subalgebras $\mathfrak{A}_1$ and $\mathfrak{A}_2$. If for all states $\varphi_1$ and $\varphi_2$ over $\mathfrak{A}_1$ and $\mathfrak{A}_2$ respectively there is a state $\varphi$ over $\mathfrak{A}$ such that $\varphi \restriction_{\mathfrak{A}_1} = \varphi_1$ and $\varphi \restriction_{\mathfrak{A}_2} = \varphi_2$, then we say that $\mathfrak{A}_1$ and $\mathfrak{A}_2$ are statistically independent.

An important theorem due to the works of Roos and Buchholz guarantees the existence of such state in the case of local algebras.

**Theorem 7.4.1.** *Let $\mathfrak{A}(\mathcal{O}_1)$ and $\mathfrak{A}(\mathcal{O}_2)$ be local algebras in spacelike separated regions $\mathcal{O}_1$ and $\mathcal{O}_2$. Then, these algebras are statistically independent and there exists a pure state $\varphi$ whose restrictions $\varphi \restriction_{\mathfrak{A}_1}$ and $\varphi \restriction_{\mathfrak{A}_2}$ are themselves pure states over the subalgebras.*

*Proof.* See [116][4] and [31]. ∎

We notice a crucial implication of this theorem. If we consider two families of measuring devices $\{a_i\}_{i \in I}$ and $\{a'_j\}_{j \in J}$, then the independence of the local algebras $\mathfrak{A}(\mathcal{O}_1)$ and $\mathfrak{A}(\mathcal{O}_2)$ implies that

$$\sum_i \omega(a_i b) = \sum_i \omega(a_i) \omega(b) = \mathbf{1} \omega(b) = \sum_j \omega(a_j) \omega(b) = \sum_j \omega(a_j b), \tag{7.4.1}$$

since by definition $\sum_i a_i = \sum_j a'_j = \mathbf{1}$ and $b \in \mathfrak{A}(\mathcal{O}_2)$. Hence, the choice of an element or measuring device in $\mathfrak{A}(\mathcal{O}_1)$ does not influence the choice in $\mathfrak{A}(\mathcal{O}_2)$. This is the locality assumption used to derive Bell's theorem. Hence, the statistical independence of the local algebras which stems from Einstein causality implies locality in the sense used by Bell.

In the same spirit of generic Bell inequalities, the CHSH was formulated in terms of systems with two outcomes. In the algebraic formulation, this is equivalent to consider positive elements $a_+$ and $a_-$ in $\mathfrak{A}$ such that $a_+ + a_- = \mathbf{1}$. One can map these elements to generic elements of $\mathfrak{A}$ by

$$a_\pm = \frac{1}{2}(\mathbf{1} \pm a), \tag{7.4.2}$$

*as long as* $-\mathbf{1} \leq a \leq \mathbf{1}$. We are then led to the following definition.

**Definition 7.4.3.** Given two $C^*$-algebras $\mathfrak{A}$ and $\mathfrak{B}$, an *admissible quadruple* is a set $\{a_1, a_2, b_1, b_2\}$ with $a_i \in \mathfrak{A}$, $b_i \in \mathfrak{B}$, $i \in \{1, 2\}$ and $-\mathbf{1} \leq A_i \leq \mathbf{1}$ and similar for $b_i$.

Given an admissible quadruple, we define the *Bell operator* for $(\mathfrak{A}, \mathfrak{B})$ by

$$\begin{aligned}
\mathcal{C} &\equiv |\omega(a_1 b_1) + \omega(a_1 b_2) + \omega(a_2 b_1) - \omega(a_2 b_2)| \\
&= |\omega(a_1(b_1 + b_2)) + \omega(a_2(b_1 - b_2))|
\end{aligned} \tag{7.4.3}$$

A Bell operator is said to satisfy *Bell's inequality* if

$$\mathcal{C} \leq 2. \tag{7.4.4}$$

Clearly this definition is equivalent to the second Bell inequality presented in section 7.3.2. The first inequality shows up in a theorem due to Summers and Werner [127] and Landau [95]. First, we present a technical lemma valid for general $C^*$-algebras.

**Lemma 7.4.1.** *Let $a$ be an element of a $C^*$-algebra $\mathfrak{A}$. Then, there exists a pure state $\omega$ over $\mathfrak{A}$ such that*

$$\omega(a^* a) = ||a||^2. \tag{7.4.5}$$

---

[4] In the original proof, it was also assumed that the algebras satisfy the *Schlieder property*, which tells us that if $a_1 \in \mathfrak{A}(\mathcal{O}_1)$ and $a_2 \in \mathfrak{A}(\mathcal{O}_2)$ then $a_1 a_2 = 0$ if and only if $a_1 = 0$ or $a_2 = 0$. However, it can be shown that this property follows from statistical independence, see the appendix of [116].

*Proof.* See [25], lemma 2.3.23.                                                            ∎

In particular, if $\omega$ is a product state and $a$ is self-adjoint, we have that $\omega(a) = \pm||a||$. Now, by considering the Bell operator in the form

$$\mathcal{C}^2 = 4 - [a_1, a_2] [b_1, b_2], \tag{7.4.6}$$

where we again omit the identity, we can use the Cauchy-Schwarz inequality for states to show that $\omega(\mathcal{C}^2) \leq 4 + ||[a_1, a_2] [b_1, b_2]||$. On the other hand, the norm is taken as the supreme over all states:

$$||4 - [a_1, a_2] [b_1, b_2]|| = \sup_{\varphi} ||\varphi(4 - [a_1, a_2] [b_1, b_2])|| \geq \omega(4 - [a_1, a_2] [b_1, b_2]).$$

Hence, combining both inequalities we have that $||\mathcal{C}^2|| = 4 + ||[a_1, a_2] [b_1, b_2]||$.

**Theorem 7.4.2 (Summers, Werner, Landau).** *Let $\mathfrak{M}$ be a von Neumann algebra with two independent subalgebras $\mathfrak{M}_1$ and $\mathfrak{M}_2$. Then, there exists an admissible quadruple $\{A_1, A_2, B_1, B_2\}$ such that $[A_1, A_2] \neq 0$ and $[B_1, B_2] \neq 0$ (hence satisfying the hypotheses for the first Bell inequality) and a pure state $\omega_0$ over $\mathfrak{M}$ such that we have a maximal violation of Bell's inequality:*

$$|\omega_0(\mathcal{C})| = 2\sqrt{2}. \tag{7.4.7}$$

*Proof.* We begin by showing by construction that we can find the admissible quadruple $\{A_1, A_2, B_1, B_2\}$ satisfying the hypotheses for the fist Bell inequality. Since any element of a von Neumann algebra can be written as a sum of self-adjoint elements (by equation (3.5.1)), we can always find orthogonal projections which do not commute with one another, which are the spectral projections associated with the self-adjoint elements. If the algebra is non-commutative, these projections must indeed commute with each other.

Hence, consider $E$ and $F$ as orthogonal projections which do not commute with one another. Define the element $T \in \mathfrak{M}$ by

$$T \equiv EF(\mathbf{1} - E). \tag{7.4.8}$$

This element is certainly non-vanishing, otherwise we would have that $EF = EFE \implies FE = EFE \implies EF = FE$, by self-adjointness, which is a contradiction. On the other hand, by an elementary computation, we see that $T^2 = 0$. This means that the range of $T$ is a subset of its kernel.

Let $T = V|T|$ be the polar decomposition of $T$, with partial isometry $V$, which is contained in the von Neumann algebra. This isometry satisfies $\ker(V) = \ker(T)$ and $\operatorname{ran}(V) = \overline{\operatorname{ran}(T)}$ and from $\operatorname{ran}(T) \subset \ker(T)$, we have that $V^2 = 0$. It is easy to see that the partial isometry satisfies

$$VV^*V = V, \ \ V^*VV^* = V^*, \tag{7.4.9}$$

since $V^*V$ is an orthogonal projection over $(\ker(V)^\perp$. Now, we define

$$X \equiv V^*V, \ \ Y \equiv VV^*, \ \ Z \equiv X + Y. \tag{7.4.10}$$

The operators $X$ and $Y$ are trivially projection operators and they are orthogonal to one another. From this, it follows that $Z$ is an orthogonal projection (by the orthogonal decomposition theorem, theorem B.4.4). Next, define

$$\overline{A}_1 \equiv V + V^*, \ \ \overline{A}_2 \equiv i(V^* - V). \tag{7.4.11}$$

Again, an elementary computation shows that $\overline{A}_1^2 = \overline{A}_2^2 = Z$ and $[\overline{A}_1, \overline{A}_2] = 2i(Y - Z)$. From the $C^*$-property (recalling that a von Neumann algebra is in particular a $C^*$-algebra), it follows that $||(Y - X)||^2 = ||(Y - X)^2|| = ||X + Y||^2 = ||Z||^2 = 1$, which implies

$$||[\overline{A}_1, \overline{A}_2]|| = 2. \tag{7.4.12}$$

Finally, define

$$A_1 = \overline{A}_1 + Z - \mathbf{1}, \ \ A_2 = \overline{A}_2 + Z - \mathbf{1}. \tag{7.4.13}$$

By a simple computation, using $\overline{A}_1^2 = \overline{A}_2^2 = Z$ and equation (7.4.12), it follows that $A_1^2 = A_2^2 = \mathbf{1}$ and that $||[A_1, A_2]|| = 2$. From the first result, it follows that the spectrum of $A_1$ and $A_2$ must be $\{-1, +1\}$ and from the second it follows that $A_1 \neq \mathbf{1}$ and similar for $A_2$. The proof for $B_1$ and $B_2$ is identical.

From equation (7.4.6), it follows that

$$||\mathcal{C}|| = 2\sqrt{1 + \frac{1}{4}||[A_1, A_2]\,[B_1, B_2]||} = 2\sqrt{2}, \tag{7.4.14}$$

with $A_1, A_2, B_1, B_2$ defined as the construction above. By lemma 7.4.1, there is a state $\omega_0$ such that

$$|\omega_0(\mathcal{C})| = 2\sqrt{2}, \tag{7.4.15}$$

thus completing the proof. $\blacksquare$

## 7.4.1   The vacuum violates Bell's inequality

The natural physical question that arises from theorem 7.4.2 is whether there are any physical states which violate the inequalities. The answer was found in the series of papers by Summers and Werner [126, 127, 128, 129] which poses that the vacuum state does indeed violate the inequality.

We denote the vacuum vector by $\Omega$, which is cyclic and separating for the local algebras by the Reeh-Schlieder theorem. The state defined by this vector is given by

$$\omega_0(A) = \langle \Omega, A\Omega \rangle, \tag{7.4.16}$$

for all $A \in \mathfrak{M}$. We will be considering the algebra for a free Boson, which is generated by the Weyl operators defined in equation (4.3.22) wit the vacuum state given by equation (4.5.1).

We will consider fields to be formal associations $f \mapsto \Phi(f)$, where $f$ are test functions (smooth functions of compact support) and $\Phi(f)$ are symmetric (unbounded) operators (see chapter 4). We will consider the inner product as given by

$$\langle f, g \rangle = \frac{1}{2} \langle \Omega, \Phi(f)\Phi(g)\Omega \rangle = \int d^4 p \, \delta(p^2 - m^2) \tilde{f}(p)^* \tilde{g}(p), \tag{7.4.17}$$

where we are integrating in the Fourier space (space of momenta) the delta in the integral enforces the *on-mass shell condition* of ordinary Quantum Field Theory (determined by the equation of motion, in this case the Klein-Gordon equation). We will assume that the test-function space $\mathcal{T}$ is complete with respect to the topology induced by this inner product and we will consider the von Neumann algebra generated by the weak closure of the Weyl algebra.

We can reduce the discussion of the algebras generated by the fields by considering subspaces of $\mathcal{T}$ and the test function algebra defined there. Hence, for some local region $\mathcal{O}$ with test function algebra $M(\mathcal{O})$, the commutant is given by

$$M(\mathcal{O})' = \{f \in \mathcal{T} : \sigma(f, g) = 0, \forall g \in \mathcal{T}, \text{supp}(g) \subset \mathcal{O}\}, \tag{7.4.18}$$

where we have considered the CCR for the fields defined in equation (4.3.21) and the symplectic form defined by equation (4.4.1).

We will now define $\mathfrak{M}(M(\mathcal{O}))$ as the von Neumann algebra generated by the Weyl operators $\{W(f) : f \in M(\mathcal{O})\}$. Now, if we assume Haag's duality as valid, we find that $\mathfrak{M}(M(\mathcal{O}))' = \mathfrak{M}(M(\mathcal{O}'))$, which is expected: physically speaking, we expect that the commutant of the algebra generated by the test functions $f$ with support in some region $\mathcal{O}$ to be the algebra generated by the test functions whose support is spacelike separated from the support of the functions $f$!

Since the Reeh-Schlieder theorem implies that the vacuum vector state, given by equations (7.4.16) and (4.5.1), is cyclic and separating for the local algebras, we can use the tools from Modular Theory to characterize the automorphisms of the algebra. In order to do so, we define the the modular involution on $\mathcal{T}$ by

$$s(f + ig) \equiv f - ig, \tag{7.4.19}$$

for $f, g \in M$, which implies that $s$ leaves $M$ invariant. The polar decomposition of $s$ is given by $s = j\delta^{1/2}$ with the unitary group $t \mapsto \delta^{it}$ serving as an automorphism group for $M$. The relationship of these operators with the ordinary modular operator and conjugation is given by

$$\begin{aligned} \Delta^{it} W(f) \Delta^{-it} &= W(\delta^{it} f) \\ J W(f) J &= W(jf). \end{aligned} \tag{7.4.20}$$

Now, since we are restricting our discussion to the subspaces $M$ of test functions, we present an equivalent condition of the Reeh-Schlieder property due to Rieffel and van Daele [113]. This condition is to say that $M(\mathcal{O})$ is *standard*, in the sense that $M \cap iM = \{0\}$ and $M \oplus iM$ is dense in $\mathcal{T}$.

The main result we sketch here is that when we consider the regions $\mathcal{O}$ to be the Rindler wedges of Minkowski spacetime (that is, the regions outside the lightcone depicted in figure 3) denoted by $\mathcal{W}$, then the result by Bisognano and Wichmann [21, 22] states that

$$\Delta^{it} = V(\pi t), \tag{7.4.21}$$

where $\mathbb{R} \ni t \mapsto V(t)$ is the unitary subgroup representing the Lorentz velocity transformations $(v(\pi t))$ which leaves both $\mathcal{W}$ and $\mathcal{W}'$ invariant. Then

$$\begin{aligned} W(v(\pi t)f) &= V(\pi t) W(f) V(\pi t)^* \\ &= \Delta^{it} W(f) \Delta^{-it} = W(\delta^{it} f), \end{aligned} \tag{7.4.22}$$

which implies that

$$v(\pi t)f = \delta^{it} f, \tag{7.4.23}$$

for all $f \in M(\mathcal{W})$. Because $M$ is standard, the operators must coincide! The importance of this resides in the following theorem

**Theorem 7.4.3.** *Let $M$ be a standard space with canonical involution $s = j\delta^{1/2}$. Suppose that for some $\lambda \in [0, 1]$, we have that $\lambda^2$ is in the spectrum of $\delta$. It follows then that for the Bell operator $\mathcal{C}(f) = (A_1 B_1 + A_1 B_2 + A_2 B_1 - A_2 B_2)(f)$ with $A_i(f) \in \mathfrak{M}(M(\mathcal{W}))$ and $B_j(f) \in \mathfrak{M}(M(\mathcal{W}'))$ for $i, j = 1, 2$, we have that*

$$|\omega_0(\mathcal{C})| \geq \sqrt{2} \left[ \frac{2\lambda}{1 + \lambda^2} \right]. \tag{7.4.24}$$

*In particular, if* 1 *is not an isolated eigenvalue of* $\delta$, *then*

$$|\omega_0(\mathcal{C})| = 2\sqrt{2}. \tag{7.4.25}$$

*Proof.* See [128], corollary 3.2. ∎

From the equality in equation (7.4.23), we conclude that the spectrum of $\delta$ must coincide with that of the generator of Lorentz boosts, which is the real line!

Therefore, our conclusion is that for the von Neumann algebras generated by the Weyl operators associated with bosonic fields, whose test functions are supported in the Rindler wedges of Minkowski spacetime, the vacuum state violates maximally Bell's inequalities! This has been made even more explicit in [129], where an analogue of the above theorem is proved for type III$_1$ factors, which as we saw describe all QFTs. The lesson to take away is that not only entanglement is a intrinsic feature of the mathematical structure of QFT but this feature is made very explicit in the algebraic approach.

## 7.5  Entanglement Measures

In this section, we briefly comment about using Bell inequalities as an entanglement measure. Even though the violations of Bell inequalities work are historically important to show that entanglement is a feature of Quantum Mechanics with no classical counterpart, it is not a good measurement of the *amount* of entanglement in a system. To understand this statement, we note that an entanglement measure $E(\omega)$ associated with some normal state $\omega$ on a $C^*$-algebra $\mathfrak{A}_A \otimes \mathfrak{A}_B$ should satisfy the following physically motivated properties:

1. **Symmetry:** The labels of the subalgebras $\mathfrak{A}_1$ and $\mathfrak{A}_2$ can be interchanged;

2. **Non-negative:** $E(\omega) \in [0, \infty]$ with $E(\omega) = 0$ if and only if $\omega$ is separable and $E(\omega) = \infty$ when $\mathfrak{A}_1$ and $\mathfrak{A}_2$ are not independent;

3. **Continuity:**[5] Let $\mathfrak{A}_{A_1} \subset \mathfrak{A}_{A_2} \subset ... \subset \mathfrak{A}_A$ be an increasing net of type I factors and similarly for $B$. Let $\omega_i$ and $\omega_{i'}$ be normal states on each pair $\mathfrak{A}_{A_i} \otimes \mathfrak{A}_{B_i}$ such that $\lim_{i \to \infty} ||\omega_i' - \omega_i|| = 0$. Then

$$\lim_{i \to \infty} \frac{E(\omega_i') - E(\omega_i)}{\ln n_i} = 0,$$

where $n_i$ is the dimension of the matrices isomorphic to each factor;

---

[5]  This is motivated by the fact that entanglement can't be increased by mixing states.

4. **Convexity:** If $\omega = \sum_j \lambda_j \omega_j$ is a convex combination of states $\omega_j$ then $\omega \mapsto E(\omega)$ is convex, i.e.

$$E(\omega) \leq \sum_j \lambda_j E(\omega_j).$$

Notice that if we define the entanglement measure in terms of the Bell operator by

$$E_B(\omega) = \sup \left\{ \frac{1}{2} \omega(A_1(B_1 + B_2) + A_2(B_1 - B_2)) \right\}, \tag{7.5.1}$$

then it does not satisfy property 2, since $1 \leq E_B(\omega) \leq \sqrt{2}$. One could try to change the normalization and require that $E(\omega) \in [1, \infty]$, but there are entangled states $\omega$ with $E_B(\omega) = 1$ [75].

There are many examples of entanglement measures which satisfy the above properties, including Modular Nuclearity, which makes use of the Tomita-Takesaki theory. Each one has its particularities and deserve a Thesis of their own. We refer the interested reader to chapter 3 of [86] for further information and references.

## 7.6   Bell Inequalities in LCQFT

In this final section, we present a formulation of the violation of Bell inequalities in the language of Category Theory and Locally Covariant Quantum Field Theory [30] and discuss some of its possible consequences. This section is in part based on [77].

### 7.6.1   The Category of Bell Observables

Let $\mathfrak{Bell}$ denote the subcategory of $\mathbf{S}(M, g)$ (see section 6.7) composed by the states over the algebras generated by admissible quadruples which satisfy Bell's inequalities, which we rewrite in terms of a state over a von Neumann algebra:

$$|\omega(A_1 B_1) + \omega(A_1 B_2) + \omega(A_2 B_1) - \omega(A_2 B_2)| \leq 2. \tag{7.6.1}$$

Since we constructed the category $\mathbf{S}(M, g)$ as being closed by convex combinations, this subcategory is well defined and it inherits the same morphisms between its objects.

Making use of (6.7.1), we obtain trivially

$$\omega(AB) = (\gamma^* \omega')(AB) = \omega'(\gamma(A)\gamma(B)), \tag{7.6.2}$$

where we have used the fact that $\gamma$ is a $*$-monomorphism. The first equality is justified by our restriction to the range of $\gamma$ in the algebra $\mathcal{B} \in \mathrm{ob}(\mathfrak{Bell})$, which defines a surjective map. Thus, we obtain the inequality

$$|\omega'(\gamma(A_1)\gamma(B_1)) + \omega'(\gamma(A_1)\gamma(B_2)) + \omega'(\gamma(A_2)\gamma(B_1)) - \omega'(\gamma(A_2)\gamma(B_2))| \leq 2\sqrt{2}. \quad (7.6.3)$$

In particular, if we take $\mathbf{S}(M, g)$ to be the state space generated over the Weyl algebra over the wedge regions and take the vacuum state, the maximal Bell correlation must be preserved, that is, the supremum is still atained.

### 7.6.2  Physical Interpretation

The previous result is illuminating in the following sense. Functoriality of the state space and of the formulation of quantum field theory in globally hyperbolic spacetimes asserts us that the inequalities must be preserved when we move from one state space to the other (or correspondingly, from one globally hyperbolic spacetime to the other). This indicates that *in any quantum field theory that admits a formulation in terms of an algebra of observables, there must be a set of observables which satisfy Bell's inequalities in the sense of equation 7.6.1*.

Furthermore, if a state violates Bell's inequalities maximally in one theory described by a $C^*$-algebra, then functoriality guarantees that the corresponding state in the target state space must also violate maximally the inequalities. In other words, this indicates that *maximal violations of Bell inequalities are a general feature of quantum field theories*.

These results seem to indicate that the formulation of quantum field theory both as a $C^*$-algebra and as a functor give rise naturally to both Bell's inequalities and to violations of such.

### 7.6.3  Speculations

In this section, we propose some paths that can be taken given the previous results. We mention that this simple result can be further explored by considering more modern formulations of Bell's inequalities in algebraic quantum field theory [129] using the Murray-von Neumann classification of factors, Connes classification of type III factors [130] and more general spacetime regions.

#### 7.6.3.1  Non-commutative spacetimes

There is a close relationship between non-commutative $C^*$-algebras and non-commutative spacetimes in the sense that any such algebra uniquely generates such a spacetime [49, 50, 7].

This equivalence may allow us to translate the algebraic form of Bell's inequalities (equation (7.6.1)) into a "geometric inequality" in the domain of the LCQFT functor. In other words, one could take the domain category of the LCQFT functor to be composed of non-commutative spacetimes as objects and associate them uniquely with the $C^*$-algebra of the codomain of such functor, meaning we define our domain category in terms of the codomain. A corresponding inequality for the spacetime might be an indication that, at least for this class of manifolds, Bell's inequalities are fundamentally related to the structure of such spacetimes.

This might be a step towards the research programs which treat spacetime as emergent from quantum correlations [123, 38] or towards the program known as the "It from Qubit" program [102, 134].

The main challenge with this idea is to prove that the category of non-commutative spacetimes (say, $\mathfrak{NonCommSp}$) is a suitable domain category for the LCQFT functor, in the sense proposed by Benini et. al. [20] and generalized by Grant-Stuart [76]. In particular, since the concept of global hyperbolicity is senseless in the context of non-commutative spacetimes, there is still a challenge to define the morphisms in this hypothetical category after choosing which structures should be preserved and are relevant in the context of quantum field theory.

### 7.6.3.2   Conformal invariance

Conformal invariance has played a significant role in theoretical physics [54]. In particular, its extensive use in quantum field theory has allowed insights into quantum gravity and strongly coupled systems [103, 141].

In two dimensions, the wedge regions and causal diamond regions given by $X^1 > 0$, $X^+ \geq 0$, $X^- \geq 0$ (with $X^{\pm} = X^1 \pm X^0$) and $|x| + |t| \leq R$ respectively (see figure 9) are related by a conformal transformation ([78], section V.4.2), namely

$$x^{\mu} = 2R \frac{X^{\mu} - b^{\mu} X^2}{1 - 2b \cdot X + b^2 X^2} + R b^{\mu},  \tag{7.6.4}$$

where $b^{\mu} = (0, -1)$.

The generalization of the work in [127] and [128] to causal diamond regions in [129], together with the functorial formulation presented in this paper suggest that it might be possible to extend the morphisms in the domain category of the LCQFT functor (i.e., the $\mathfrak{GlobHyp}$ category of globally hyperbolic spacetimes) to include conformal transformations.

This generalization would be useful in proving the consistency of the formulation presented here by showing that it leads to the same result obtained in [129]. It would perhaps also be useful for describing holography and phenomena in conformally related regions in terms of locally covariant quantum field theory.

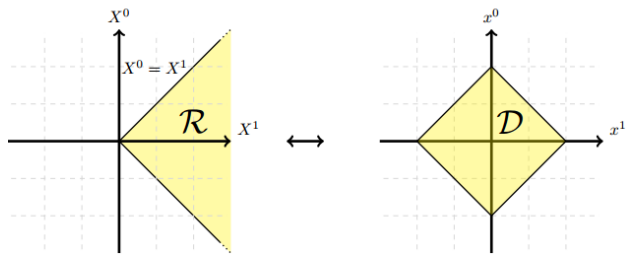

Figure 9 – The relation between the (Rindler) wedge region and a causal diamond region is given by a conformal transformation.

### 7.6.3.3 Cosmology

It is speculated in some inflationary models of the early universe that there was an initial singularity [120]. This would lead to a topological defect in the beginning of spacetime, making the globally hyperbolic assumption of LCQFT invalid for describing our world.

Nevertheless, violations of the inequalities have been successfully verified [43, 42, 44], leading the 2022 Nobel Prize in Physics to work in this field.

If we would expect that LCQFT is only feasible in globally hyperbolic spacetimes, then this would entail that there could be no topological defects in the universe. Furthermore, the experimental verification of the violation of Bell's inequalities together with the functorial formulation presented in this paper which assumes global hyperbolicity would be an indication against the hypothesis of an initial singularity.

# 8 Conclusions and Outlook

In this Thesis, we discussed many aspects of Algebraic Quantum Field Theory which are relevant to the problem of defining entanglement in a field-theoretic setting. Our analysis focused on the violation of Bell's inequalities which, as we have seen, is not an adequate measure of the amount of entanglement, even though it is the first indication of the non-classical behavior of such phenomenon.

Our extension of the results on the violations of Bell's inequalities to more general spacetimes opens up interesting but challenging routes for future research. One can even envision treating the *bona fide* measures of entanglement to such spaces, in particular Modular Nuclearity which is based on the Tomita-Takesaki Theory.

The deep connection between the local algebras of observables and von Neumann algebras deserves a separate Thesis on its own. Even though we merely sketched the results on the universal structure of the local algebras, we hope to explore this richness in the future. In particular, the relationship with non-commutatative spacetimes seems to be promising since the LCQFT functor would be trivial provided that one can treat such spacetimes in this setting.

We also hope to explore the functorial nature of LCQFT. The interplay between Category Theory and QFT is much older than the works of Brunetti, Fredenhagen and Verch, having its origins in the axiomatization of *Topological Quantum Field Theories* by Atiyah [10] based on work by Witten [140]. In particular, algebraic structures arise in a different number of dimensions, with the particular case of dimension 2 arising as a *Frobenius algebra* [92]. These algebras admit matrix representations, which means that one might hope to make a connection to $C^*$-algebras via the GNS construction.

We hope to have convinced the reader that the algebraic approach in not simply a "way to make QFT mathematically rigorous". Rather, it provides new physical insights, such as the case of entanglement, where treating space states is more direct, and more general theorems and aspects which are model independent. A deeper understanding of Nature requires a deeper understanding of the language that She speaks in.

# A  Elementary Topology and Measure Theory

## A.1  Topological spaces

**Definition A.1.1.** A *topological space* is a ordered pair $(X, \tau)$ where $X$ is some set and $\tau$ is a subset of $\mathbb{P}(X)$ (the power set of $X$, which consists of all the possible subsets of $X$). The sets of $\tau$ are called *open sets* and they satisfy:

1. $X \in \tau$;

2. $\varnothing \in \tau$;

3. $\forall U, V \in \tau \implies U \cap V \in \tau$;

4. $\forall U_\lambda \in \tau, \lambda \in \Lambda \implies \bigcup_\lambda U_\lambda \in \tau$;

where, in the last property, $\Lambda$ is some set that labels the elements $U$ of the topology.

An example of a topological space of great importance is the real line $\mathbb{R}$ together with the set of open sets in the usual sense, i.e., all intervals $(a, b)$. This is what is usually referred as the *real line with the usual topology*.

One could have also defined the real line with a topology given by the *closed* intervals in the usual sense, $[a, b]$. In this case, we would now, by declaration, consider these sets as being our open sets. We define *closed sets* in the more general sense as being the sets whose complement is open, i.e., $U \subset (X, \tau)$ is closed if $U^c = X \setminus U$ is open.

There are many ways to classify a topological space regarding its properties. A particularly useful classification in the context of this work is that of *compactness*. To introduce this, we need to understand the meaning of a *covering*:

**Definition A.1.2.** A *covering* $\mathcal{C}$ of a topological space $(X, \tau)$ is a set of open sets $U_\lambda$ such that $\bigcup_\lambda U_\lambda = X$. A *subcovering* is any subset of $\mathcal{C}$ that is also a covering of $(X, \tau)$.

**Definition A.1.3.** A topological space is called a *compact set* if every covering admits a finite subcovering.

One of the simplest examples of a compact set is given by the famous *Heine-Borel theorem*

**Theorem A.1.1** (**Heine-Borel**). *For a subset $S$ of an Euclidean space $\mathbb{R}^n$, the following statements are equivalent:*

    *1. $S$ is closed and bounded;*

    *2. $S$ is compact.*

That is, the compact sets in the real line are exactly the closed sets in the usual topology! This theorem has actual physical relevance in computations of momentum integrals in perturbative quantum field theory, as was shown by Weinberg in [137].

In particular, we are interested in vector spaces in general, which motivates us to define the following:

**Definition A.1.4.** A *topological vector space (TVS)* is a topological space $(X, \tau)$ in which $X$ is a vector space such that:

1. Every point of $X$ defines a closed set with respect to $\tau$;

2. The vector space operations (sum of vectors and multiplication by a scalar) are continuous functions with respect to $\tau$.

## A.1.1 Local Base

We will denote a TVS only by its underlying vector space, thus writing $X$ only. There are many ways to characterize the sets that constitute the topology on a TVS (see [118] for a complete list), but it suffices to consider the following for a locally convex topology:

**Definition A.1.5.** Let $X$ be a TVS over a field $\mathbb{K} = \mathbb{R}$ or $\mathbb{C}$ and let $V \in \tau$ be a subset of $X$. We say that

1. $E$ is *balanced* if $\forall \lambda \in \mathbb{K}, |\lambda| = 1 : \lambda V \subseteq V$;

2. $E$ is *convex* if $\forall \lambda \in \mathbb{R}, \lambda \geq 0$ and $x_1, x_2 \in V$ implies $\lambda x_1 + (1 - \lambda)x_2 \in V$.

An important way to characterize a topological space in general is via the notion of a *base*.

**Definition A.1.6.** A *local base* of a TVS $X$ is a collection $\mathscr{B}$ of neighborhoods of the origin $0$ such that if $\mathcal{O}$ is a neighborhood of $0$ then there exists a $\mathcal{B} \in \mathscr{B}$ such that $\mathcal{B} \subseteq \mathcal{O}$. The open sets of $X$ are then the unions of translates of members of $\mathscr{B}$.

This definition might sound confusing and redundant, so an example is in order. Consider the following open sets in the standard topology of $\mathbb{R}$:

$$\mathscr{B} = \left\{ \left( -\frac{1}{n}, \frac{1}{n} \right), n \in \mathbb{N} \right\}.$$

Clearly, these are all open neighborhoods of the origin. Because the real line is complete, every neighborhood $\mathcal{O}$ of 0 contains some ball of radius $r > 0$ centered at 0. Hence, $(-r, r) \subseteq \mathcal{O}$. The Archimedean property of the real numbers [117] tells us we can find an $k \in \mathbb{N}$ such that $\frac{1}{k} < r$, hence we can take $\left( -\frac{1}{k}, \frac{1}{k} \right) \in \mathscr{B}$ and $\left( -\frac{1}{k}, \frac{1}{k} \right) \subseteq (-r, r) \subseteq \mathcal{O}$.

## A.2 Measurable spaces

**Definition A.2.1.** A collection $\mathcal{M}$ of subsets of $X$ is said to be a $\sigma$-algebra on $X$ if

1. $\varnothing \in \mathcal{M}$ and $X \in \mathcal{M}$;

2. If $A \in \mathcal{M}$, then $A^c \equiv X \setminus A \in \mathcal{M}$;

3. If $\{A_n, n \in \mathbb{N}\}$ is a countable collection of elements of $\mathcal{M}$, then $\bigcup_{n \in \mathbb{N}} A_n \in \mathcal{M}$.

The elements of $\mathcal{M}$ are called *measurable sets*. The pair $(X, \mathcal{M})$ is called a *measurable space*.

An important notion is that of *Borelian sets*. Given a topology $\tau$, the Borelian sets are defined as the smallest $\sigma$-algebra that contains that topology. In the case of the real line, the Borelian sets include the open intervals, closed intervals and semi-open intervals.

On these spaces, one can define a function which gives a notion of "length" of each measurable set.

**Definition A.2.2.** A *measure* on a measurable space $(X, \mathcal{M})$ is a function $\mu : \mathcal{M} \to \mathbb{R}_+ \cup \{+\infty\}$ such that

1. $\mu(\varnothing) = 0$;

2. If $A_i$ with $i \in \mathbb{N}$ is a countable and disjoint collection of measurable sets in $\mathcal{M}$, then

$$\mu \left( \bigcup_{n \in \mathbb{N}} A_n \right) = \sum_{n \in \mathbb{N}} \mu(A_n).$$

The advantage of measure theory is that it allows us to define an abstract notion of integration. This is done by summing over the "lengths" of the sets in a clever way.

**Definition A.2.3.** Let $M$ be a non-empty set. A function $s : M \to \mathbb{R}$ or $s : M \to \mathbb{C}$ is called a *simple function* if it assumes a finite number of values. The preimage of these functions are defined as

$$A_k = \{x \in M : s(x) = s_k\},$$

where $s_k$ are the values assumed by the function. Hence, the function has the following representation

$$s(x) = \sum_{k=1}^{n} s_k \chi_{A_k}(x),$$

where $\chi$ is the characteristic function, which is equal to 1 if $x \in A_k$ and vanishes if not.

We can now define the abstract notion of an integral of simple functions.

**Definition A.2.4.** If $s$ is a simple function defined on a measurable space $(X, \mathcal{M})$ with measure $\mu$, then the integral of $s$ with respect to $\mu$ is

$$\int_X s d\mu \equiv \int_X s(x) d\mu(x) = \sum_{k=1}^{n} s_k \mu(A_k).$$

One can verify that the ordinary properties of Riemann integrals are satisfied in this construction. For measurable functions $f$ (that is, those for which the preimage of a measurable set is a measurable set) we can define the Lebesgue integral on a measurable set $E$ as

$$\int_E f d\mu = \sup_{s \in S(f)} \int_E s d\mu, \tag{A.2.1}$$

where $S(f)$ is the set of simple functions on $(X, \mathcal{M})$ such that $0 \le s(x) \le f(x)$, for all $x \in X$.

# B  Elements of Functional Analysis

This chapter is dedicated to present some important notions in Functional Analysis which are needed to understand the mathematical structure and physical content of Quantum Physics. We assume an understanding of the reader of basic concepts from Linear Algebra such as Vector Spaces and Linear Transformations. For the readers who are unfamiliar with basic concepts from Topology such as continuity and Cauchy sequences, we provide a quick review of these topics in appendix A.

For the more introductory topics, we follow references [111], [118] and [93]. For notions about $C^*$-algebras, von Neumann algebras, the GNS construction and other physically relevant tools, we follow closely references [25] and [26].

## B.1  Normed Spaces

A vector space can be viewed as a set with added structure: in particular, we define a sum between vectors and the product by a scalar. A normed space is what we get when we add a structure of a *norm*, intuitively understood as measuring the "size" of a vector. We remind the reader that although such intuitions may be helpful in the beginning of the learning process, relying too much on it can be harmful when learning more abstract Mathematics and Physics. We can define a norm rigorously in the following definition.

**Definition B.1.1.** Let $X$ be a vector space. A function $||\cdot|| : X \to \mathbb{K}$, where $\mathbb{K}$ denotes the field of real or complex numbers, is called a *norm* if it satisfies the following properties for all $x, y \in X$ and $\alpha \in \mathbb{K}$:

1. $||x|| \geq 0$;

2. $||x|| = 0 \iff x = 0$;

3. $||\alpha x|| = |\alpha| \, ||x||$;

4. $||x + y|| \leq ||x|| + ||y||$,

where the last item is called the *Triangle Inequality*.

Throughout this Thesis, we will specialize ourselves in real norms. One can see that the defining properties of a norm are a natural generalization of the idea of the length of a vector in Euclidean space. In fact, properties 1 and 2 indicate that such "length" is always a positive real number (again, we are only considering real norms) except in the case of the zero vector, whose length is of course zero. Property 3 generalizes the idea that if one stretches or shrinks each component of the vector by the same number $\alpha$, then the length of the vector gets multiplied by that same $\alpha$. Finally, property 4, as the name suggests, comes from the fact that in Euclidean geometry, the sum of two sides of a triangle is always larger than the remaining side[1].

**Definition B.1.2.** A *normed space* $(X, ||\cdot||)$ is a pair defined by a vector space and a norm defined on it.

We note that a norm defines a metric $d$ on $X$ (called the *metric induced by the norm*), which is given by

$$d(x, y) = ||x - y||. \tag{B.1.1}$$

Thus, one can define a (metrizable) topology and, consequently, notions of continuity and convergence. This is not a mere technicality: in Quantum Physics, one is sometimes interested in spaces of infinite dimension. Naively, one could write an generic element of such a space in terms of some countable "basis":

$$x = \sum_{n=1}^{\infty} \alpha_n e_n.$$

In this case, one has to work with a vector space with added structure: it is not possible to define convergence of an infinite sum on a infinite dimensional "naked" vector space. One needs at least a metric function on such spaces.

We mention *en passant* that there is also interest in defining *pseudo* or *semi* norms, which are the same as a norm except for property 2. In this case, we define *non-metrizable topologies*, which are arise in the study of Distributions, for example.

To finish this section and for completeness, we also mention that *not all metrics come from a norm*. This is the content of the following simple lemma:

**Lemma B.1.1.** *A metric $d$ induced by a norm $||\cdot||$ on a normed space $X$ if it is* translation invariant, *i.e.:*

1. $d(x + a, y + a) = d(x, y)$;

2. $d(\alpha x, \alpha y) = |\alpha|\, d(x, y)$,

*for all $x$, $y$, $a \in X$ and $\alpha \in \mathbb{K}$.*

---

[1] This fact was realized by Euclid and the proof can be found as Proposition 20 in Book One of *Elements*.

*Proof.* By a straightforward computation using equation B.1.1, we can verify each of the items:

1. $d(x + a, y + a) = ||(x + a) - (y + a)|| = ||x - y|| = d(x, y)$.

2. $d(\alpha x, \alpha y) = ||\alpha x - \alpha y|| = |\alpha| \, ||x - y|| = |\alpha| \, d(x, y)$.

∎

### B.1.1 Banach Spaces

A Banach space is a normed vector space with the additional property that it is *complete* in the metric induced by the norm. We give the following definition for this property which is standard in Real and Complex Analysis.

**Definition B.1.3.** A sequence $(x_n)_{n \in \mathbb{N}}$ in a metric space $(X, d)$ is called a *Cauchy sequence* if for every $\epsilon > 0$, there exists a natural number $N$ (which may depend on $\epsilon$) such that for all $m, n > N$,

$$d(x_m, x_n) < \epsilon.$$

The space $(X, d)$ is said to be *complete* if every Cauchy sequence on $(X, d)$ converges to an element of $(X, d)$.

Intuitively, the definition of a Cauchy sequence simply tells us that no matter how small the distance (or tolerance) $\epsilon > 0$ is, we can always find point in the sequence such that from that point forward the distance between two neighbors in the sequence is smaller than this $\epsilon$. In other words, *the members of the sequence get arbitrarily close to one another as the sequence goes on.*

The astute reader may have noticed that we choose $\epsilon$ strictly greater than zero. This means that the distance gets smaller and smaller, but it doesn't vanish. If it does, we say that the sequence *converges* (for a review of the concept of convergence, see Appendix A). This implies the simple but important theorem:

**Theorem B.1.1.** *Every convergent sequence in a metric space is a Cauchy sequence.*

*Proof.* If $x_n \to x$, then for every $\epsilon > 0$, there exists an $N$ such that

$$d(x_n, x) < \frac{\epsilon}{2}.$$

By the triangle inequality, we obtain for $m, n > N$:

$$d(x_m, x_n) \leq d(x_m, x) + d(x, x_n) < \frac{\epsilon}{2} + \frac{\epsilon}{2} = \epsilon,$$

which shows that the sequence $(x_n)_{n \in \mathbb{N}}$ is a Cauchy sequence.                    ∎

Thus, in a Banach space, every Cauchy sequence converges in the metric/topology induced by the norm. The simplest example of such spaces are the Euclidean spaces $\mathbb{R}^n$ and complex spaces $\mathbb{C}^n$, which are both complete in the norm defined by

$$||x|| = \left( \sum_{j=1}^{n} |x_j|^2 \right)^{1/2} = \sqrt{|x_1|^2 + ... + |x_n|^2}, \tag{B.1.2}$$

where $x_j$ denotes the components of the vectors in $\mathbb{R}^n$ or $\mathbb{C}^n$ and $|\cdot|$ denotes the absolute value in the real numbers or complex numbers, respectively.

Another important example is the case of the so-called $\ell^p$ spaces, with $p \geq 1$ (the case where $p = 2$ is particularly important in Quantum Mechanics, as we shall see briefly). These are the infinite sequences of numbers $x = (x_n)_{n \in \mathbb{N}}$ such that the sum $|x_1|^p + |x_2|^p + ...$ converges, i.e.:

$$\sum_{j=1}^{\infty} |x_j|^p < \infty. \tag{B.1.3}$$

The $\ell^p$ spaces form a Banach space when they are equipped with the norm given by

$$||x|| = \left( \sum_{j=1}^{\infty} |x_j|^p \right)^{1/p}.$$

A last, but important example is that of the space $C[a, b]$ of continuous functions on the closed real interval $[a, b]$, together with the norm

$$||x|| = \max_{t \in [a,b]} |x(t)|.$$

As mentioned in the previous section, the presence of a norm allows us to define notions of convergence of infinite series. This allows us to construct a basis for elements in general normed spaces. This can be done if the space $(X, ||\cdot||)$ contains a sequence $(e_n)_{n \in \mathbb{N}}$ such that for every $x \in X$ there is a unique sequence of scalars $(\alpha_n)_{n \in \mathbb{N}}$ such that

$$\left\| x - \left( \sum_{j=1}^{n} \alpha_j e_j \right) \right\| \to 0.$$

The sequence $(e_n)_{n \in \mathbb{N}}$ is then called a *Schauder basis* for $X$ (as opposed to the usual notion of a *Hamel basis* in ordinary vector spaces). In this case, we can write an expansion for an element of $X$ as

$$x = \sum_{j=1}^{\infty} \alpha_j e_j. \qquad (B.1.4)$$

For example, the space $\ell^p$ has a Schauder basis given by $e_n = (\delta_{nj})$, that is, the sequence whose $n$th term is 1 and all others are zero.

A classic result in Functional Analysis tells us that if $X$ has a Schauder basis, then $X$ is separable (Appendix A). Even though the proof for this statement is simple, the converse question, i.e., *if every separable Banach space has a Schauder basis*, was an open problem in Mathematics for many years. The answer was surprisingly *no* and was given by Enflo in 1973 [64] by constructing a counterexample.

We finish this section with an important result about *completion*. Completeness is a nice property for a space to have and, luckily, even if our space is not complete, we can guarantee that the space is *isometric* to a subspace of a complete space. Two spaces are said to be isometric if there is a bijective function between them which preserves distances (isometry), i.e., $T : (X, d) \to (\tilde{X}, \tilde{d})$ is an isometry if $\tilde{d}(Tx, Ty) = d(x, y)$, for all $x, y \in X$. Since the only difference between metric (normed) spaces to a simple set is the added structure of distances (sizes of elements), an isometry defines an equivalence relation between two spaces, i.e., they are, in the eyes of metric (normed) spaces, the same except for a relabling of the elements of the underlying set.

**Theorem B.1.2.** *Let $(X, || \cdot ||)$ be a normed space. Then, there is a Banach space $\hat{X}$ and an isometry $A : X \to W$, where $W$ is a dense subspace of $\hat{X}$. The space $\hat{X}$ is unique except for isometries.*

The reader interested in the proof of this theorem is referred to reference [93].

## B.2 Operators

In this section, we begin to study one the main subjects of this Thesis which are the mappings between vector spaces and, in particular, between normed spaces (and later, inner product spaces). These maps are called *operators* and are a generalization of the notion of functions from the study of Calculus.

## B.2.1   Linear Operators

It so happens that in Physics, the more naturally occurring operators are the so-called *linear operators*. We give a precise definition of these maps:

**Definition B.2.1.** A *linear operator* $T$ is an operator such that

- the domain $\mathcal{D}(T)$ of $T$ is a vector space and the range $\mathcal{R}(T)$ lies in a vector space over the same field $\mathbb{K}$;

- for all $x$, $y \in \mathcal{D}(T)$ and scalars $\alpha$, $\beta \in \mathbb{K}$,

$$T(\alpha x + \beta y) = \alpha T x + \beta T y.$$

Linearity has many consequences in both physical and mathematical viewpoint. It expresses the fact, for example, that a linear operator is a *homomorphism* of a vector space (its domain) into another vector space, that is, it is a *structure preserving* map. These kinds of maps will be of great value later in this Thesis in the study of Category Theory. In Functional Analysis, linearity and finite dimensional spaces are ubiquitous and they are sufficient conditions to prove many statements and we shall see a few of these. In fact, linearity is such a strong requirement that it sparked interest in studying the consequences of *non-linearity*, thus giving birth to the subject of *Non-Linear Functional Analysis* (for a reference, see [121]).

We mention the following theorem whose immediate consequence is that *linear operators preserve linear dependence* (hence, supporting the discussion above).

**Theorem B.2.1.** *Let $T$ be a linear operator. Then:*

1. *The range $\mathcal{R}(T)$ is a vector space;*

2. *If $dim\mathcal{D}(T) = n < \infty$, then $dim\mathcal{R}(T) \leq n$;*

3. *the null space (the subset of the domain of vectors that are mapped to zero) $\mathcal{N}(T)$ is a vector space.*

## B.2.2   The Inverse of an Operator

Recall from the study of Calculus or Analysis that a function is said to be *injective* or *one-to-one* if different points in the domain have different images. This is generalized in the obvious way to operators: an operator is thus injective or one-to-one if for any $x_1$, $x_2 \in \mathcal{D}(T)$ such that $x_1 \neq x_2$ implies $Tx_1 \neq Tx_2$, or, equivalently:

$$Tx_1 = Tx_2 \implies x_1 = x_2.$$

Injectiveness is a sufficient condition to guarantee the existence of an inverse between $\mathcal{D}(T)$ and $\mathcal{R}(T)$. This inverse is denoted

$$T^{-1} : \mathcal{R}(T) \to \mathcal{D}(T)$$
$$y_0 \mapsto x_0$$

(B.2.1)

where $y_0$ is such that $Tx_0 = y_0$. From equation B.2.1, we have for all $x \in \mathcal{D}(T)$ and $y \in \mathcal{R}(T)$

$$T^{-1}Tx = x$$
$$TT^{-1}y = y.$$

The following theorem stated without proof gives sufficient conditions for the existence of the inverse and some of its properties.

**Theorem B.2.2.** *Let $X$, $Y$ be vector spaces over the same field $\mathbb{K}$. Let $T : \mathcal{D}(T) \to \mathcal{R}(T)$ be a linear operator with domain $\mathcal{D}(T) \subset X$ and range $\mathcal{R}(T) \subset Y$. Then:*

1. *The inverse $T^{-1} : \mathcal{R}(T) \to \mathcal{D}(T)$ exists if, and only if*

$$Tx = 0 \implies x = 0;$$

2. *If $T^{-1}$ exists, it is a linear operator;*

3. *If dim $\mathcal{D}(T) = n < \infty$ and $T^{-1}$ exists, then dim $\mathcal{R}(T) =$ dim $\mathcal{D}(T)$.*

We finish this section with an important technical lemma which provides an important property of the inverse.

First, given two operators $T : X \to Y$ and $S : Y \to Z$, we can define the *product* of such operators as being the *composition* $ST : X \to Z$. If both $T$ and $S$ are linear, then the composition $ST$ is easily seen as linear and the same happens if both $T$ and $S$ are bijective. Hence, we can define a inverse for $ST$ which mimics the known property of the inverse of the product of matrices (this relation will be discussed further ahead).

**Lemma B.2.1.** *Let $T : X \to Y$ and $S : Y \to Z$ be bijective linear operators acting on the vector spaces $X$, $Y$ and $Z$. Then, the inverse of the product $ST$, written as $(ST)^{-1} : Z \to X$ exists and is given by*

$$(ST)^{-1} = T^{-1}S^{-1}.$$

*Proof.* Because both $T$ and $S$ are bijective, the product $ST$ is also bijective, so $(ST)^{-1}$ exists. We thus have

$$ST(ST)^{-1} = I_Z,$$

where $I_Z$ is the identity operator on the vector space $Z$. Applying $S^{-1}$ and using $S^{-1}S = I_Y$, we obtain

$$S^{-1}ST(ST)^{-1} = T(ST)^{-1} = S^{-1}I_Z = S^{-1}.$$

Now, applying $T^{-1}$ and using $T^{-1}T = I_X$, we obtain

$$T^{-1}T(ST)^{-1} = (ST)^{-1} = T^{-1}S^{-1},$$

thus completing the proof. ∎

### B.2.3   Bounded Linear Operators

The notion of bounded operators is ubiquitous in Physics. Most of the time, physicists are inadvertently working with the related notion of *unbounded operators* as if they were bounded. This leads to apparent "paradoxes" as we shall discuss further in this Thesis. The main issue is that in the unbounded case, there must be an extra care in working in the domain of the operator. Bounded operators have the nice property due to a theorem that their domain can be extended to the whole normed space $(X, ||\cdot||)$, where their domain $\mathcal{D}(T)$ is a subset.

Physically speaking, we can get away in many cases with the use of bounded operators because experiments take place within some energy or length scale which allows us to put some bounds on the quantities of interest. This argument is not to be taken lightly: in fact, one can use this same reasoning to justify not using the real numbers to characterize a measurement, since experimental apparatuses can only measure rational numbers (you will *never* measure an energy of $\sqrt{2}$ eV in a lab).

However, in systems of finite degrees of freedom, bounded operators are adequate for the description of physical quantities and measurements. Care must be taken when discussing this in the context of Quantum Field Theory.

**Definition B.2.2.** Let $X$ and $Y$ be normed spaces and $T : \mathcal{D}(T) \to Y$ be a linear operator, where $\mathcal{D}(T) \subset X$. The operator $T$ is said to be *bounded* if there is a real number $c$ such that for all $x \in \mathcal{D}(T)$

$$||Tx|| \leq c||x||, \tag{B.2.2}$$

where the norm on the left is the norm on $Y$ and the norm on the right is the norm on $X$.

The reader familiar with the concept of a bounded *function* might find this definition strange. Indeed, a bounded function is one whose range $A$ is a bounded set (there is a maximum finite distance between the members, i.e., $\delta(A) \equiv \sup_{x,y \in A} d(x, y) < \infty$, where $\delta(A)$ is called the *diameter* of the set $A$). Both these concepts are standard and they are *not* the same.

Equation B.2.2 allows us to define a norm for a bounded operator. This comes from the fact that, considering $x \neq 0$, the smallest possible value that $c$ can hold is trivially determined by

$$\frac{||Tx||}{||x||} \leq c,$$

which means that $c$ must be the *least upper bound* or the *supremum* of the expression. One can verify that the quantity

$$||T|| \equiv \sup_{x \in \mathcal{D}(T)} \frac{||Tx||}{||x||}, \tag{B.2.3}$$

satisfy the requirements of a norm, and thus receives the name *operator norm*. An equivalent way of defining this norm is by

$$||T|| = \sup_{\substack{x \in \mathcal{D}(T) \\ ||x||=1}} ||Tx|| \tag{B.2.4}$$

Trivial examples of bounded operators are the identity operator $I : X \to X$ which has norm $||I|| = 1$ and the zero operator, which has norm zero. A not so trivial example is the *integral operator* $T : C[0,1] \to C[0,1]$ defined by

$$y = Tx, \quad y(t) = \int_0^1 k(t, \tau) x(\tau) d\tau, \tag{B.2.5}$$

where the function $k$ is called the *kernel* of $T$ and is assumed to be continuous in the $[0,1] \times [0,1]$ closed square. Boundedness of this operator comes from the fact that the kernel is continuous on the closed square, and thus it is bounded: $|k(t, \tau)| \leq k_0$ for all $(t, \tau) \in [0,1] \times [0,1]$, where $k_0$ is a real number. We also have that, considering $C[0,1]$ as a Banach space with norm given by $||x|| = \max_{t \in [0,1]} |x(t)|$ that

$$|x(t)| \leq ||x||,$$

and hence

$$||y|| = ||Tx|| = \max_{t \in [0,1]} \left| \int_0^1 k(t, \tau) x(\tau) d\tau \right|$$

$$\leq \max_{t \in [0,1]} \int_0^1 |k(t, \tau)| \, |x(\tau)| d\tau \leq k_0 ||x||,$$

proving that $T$ is bounded.

This is an important example in Functional Analysis and in the study of *integral equations*. In fact, the main motivation for Hilbert to develop the concept of what is now known as a *Hilbert space* was to study integral equations of the form

$$(T - \lambda I)x(t) = y(t), \quad Tx(t) = \int_a^b k(t, \tau)x(\tau)d\tau,$$

where $\lambda \in \mathbb{C}$. This is what is called a *Fredholm integral equation of the first kind*. The fact that essential results on the solvability of this type of equation does not depend on the integral representation but on whether or not $T$ is a *compact operator* led to the subsequent study of these type of operators (which we will not delve into here, but the interested reader is referred to [111] and [52]).

An important example of an operator which is *not* bounded is given by the *differentiation operator*: let $X$ be the normed space of all polynomials on $[0, 1]$ with the norm given by $||x|| = \max_{t \in [0,1]} |x(t)|$. The differentiation operator is defined as

$$Tx(t) = \frac{d}{dt}x(t) \equiv x'(t).$$

This operator is linear, but not bounded. Indeed, if we let $x_n(t) = t^n$, with $n \in \mathbb{N}$, then we have that $||x_n|| = 1$ (we take the $t$ to be the maximum value in the closed interval $[0, 1]$ which is of course 1) and

$$Tx_n(t) = nt^{n-1} \implies ||Tx_n|| = n \implies \frac{||Tx_n||}{||x_n||} = n.$$

Now, since $n \in \mathbb{N}$ is arbitrary, this shows that we can't have a fixed real number $c > 0$ such that $||Tx_n||/||x_n|| \leq 0$. Hence, $T$ is not bounded.

### B.2.3.1 Finite Dimension and Continuity

As mentioned elsewhere, the fact that the space is finite dimensional gives some constraints. In the case of operators, it implies the following important theorem:

**Theorem B.2.3.** *If a normed space $(X, ||\cdot||)$ is finite dimensional, then every linear operator $T$ on $X$ is bounded.*

*Proof.* Since $X$ is finite dimensional, we can ascribe a basis without worrying about convergence. Let $\{e_1, ..., e_n\}$ be such a basis and consider an element $x = \sum_{j=1}^n x_j e_j$. Since $T$ is linear, we have

$$||Tx|| = \left|\left|\sum x_j T e_j\right|\right| \leq \sum |x_j| \, ||T e_j|| \leq \max_k ||T e_k|| \sum |x_j|.$$

One can show ([93], lemma 2.4-1) that for a linearly independent set of vectors $\{y_j\}$ in a normed space, there is a number $c > 0$ such that for every choice of scalars $\alpha_1, ..., \alpha_n$ we have

$$\left\| \sum_{j=1}^{n} \alpha_j y_j \right\| \geq c \left( \sum_{j}^{n} |\alpha_j| \right). \tag{B.2.6}$$

Putting $\alpha_j = x_j$ and $y_j = e_j$ in equation B.2.6, we get

$$\sum_{j}^{n} |x_j| \leq \frac{1}{c} \left\| \sum_{j}^{n} x_j e_j \right\| = \frac{1}{c} ||x||.$$

These in turn imply

$$||Tx|| \leq \gamma ||x||, \quad \gamma = \frac{1}{c} \max_{k} ||Te_k||,$$

which tells us that $T$ is bounded. ∎

Linear operators are mappings, and thus we can ascribe a definition of *continuity* to them. It turns out that *boundedness and continuity are equivalent concepts in the study of linear operators*. This is the content of the last theorem of this section, which we state without proof and we will proceed to use the terms bounded and continuous operator interchangeably.

**Theorem B.2.4.** *Let $T : \mathcal{D}(T) \to Y$ be a linear operator. Then:*

1. *$T$ is continuous if, and only if it is bounded;*

2. *If $T$ is continuous at a single point, it is continuous everywhere.*

### B.2.3.2 Extensions

Building upon the notion of an operator as a generalization of a function in Calculus, we can define two operators $T_1$ and $T_2$ being *equal* if they have the same domain $\mathcal{D}(T_1) = \mathcal{D}(T_2)$ and if $T_1 x = T_2 x$ for all $x \in \mathcal{D}(T_1) = \mathcal{D}(T_2)$.

We may also be interested in the action of an operator in a subset of the domain, rather than the whole domain. In this sense, we define the *restriction* of an operator $T : \mathcal{D}(T) \to Y$ to a subset $B \subset \mathcal{D}(T)$ as being the operator defined by

$$T|_B : B \to Y, \quad T|_B x = Tx, \forall x \in B.$$

The converse of this definition occurs when we are interested in extending the operator to a larger set than its domain. In this case, we refer to the *extension* of $T$ to a set $M \supset \mathcal{D}(T)$, defined by the operator

$$\tilde{T} : M \to Y, \quad \tilde{T}|_{\mathcal{D}(T)} = T,$$

which simply means that $\tilde{T}$ coincides with $T$ in the sense presented above for all $x \in \mathcal{D}(T) \subset M$. In other words, $T$ is the restriction of $\tilde{T}$ in $\mathcal{D}(T)$.

Naturally, one can think of many extensions given a linear operator. For instance, one could define to an operator defined on a closed subset $[a, b]$ of the real line an extension defined on the whole line by demanding that the image vanishes for all $x$ not in $[a, b]$. Thus, some criterion for the extension is warranted.

This is the motivation behind the following theorem on bounded linear extensions. This theorem is sometimes called *Bounded Linear Transformation theorem* or, amusingly, the *BLT theorem*[2].

**Theorem B.2.5** (BLT Theorem). *Let $T : \mathcal{D}(T) \to Y$ be a bounded linear operator, where $\mathcal{D}(T)$ lies in a normed space $X$ and $Y$ is a Banach space. Then, $T$ has an extension*

$$\tilde{T} : \overline{\mathcal{D}(T)} \to Y$$

*where the bar denotes the topological closure of the set $\mathcal{D}(T)$ and $\tilde{T}$ is bounded, linear and satisfies $||\tilde{T}|| = ||T||$.*

### B.2.3.3 Matrix Representation of Bounded Operators

There is a close relationship between bounded linear operator on *finite* dimensional spaces and matrices in the sense that there is one-to-one relation between them. This reduces the problem in finite dimensional spaces to the familiar matrix algebra.

We first note that a $m \times n$ real matrix $A = (\alpha_{jk})$ defines an operator $T : \mathbb{R}^n \to \mathbb{R}^m$ by

$$y = Ax, \tag{B.2.7}$$

where $x = (x_j)$ and $y = (y_j)$ are column vectors with $n$ and $m$ components, respectively. In terms of the components, equation B.2.7 becomes

$$y_j = \sum_{k=1}^{n} \alpha_{jk} x_k. \tag{B.2.8}$$

Linearity of $T$ comes from the linearity of matrix multiplication. To prove that $T$ is bounded, we consider the norm on $\mathbb{R}^n$ given by equation B.1.2 together with equation B.2.8

---

2   This notation is commonly used by researches who learned Functional Analysis through reference [111] and makes, of course, reference to the traditional sandwich whose ingredients are Bacon, Lettuce and Tomato.

and the Cauchy-Schwarz inequality (appendix A):

$$\begin{aligned}
||Tx||^2 = \sum_{j=1}^{m} y_j^2 &= \sum_{j=1}^{m} \left( \sum_{k=1}^{n} \alpha_{jk} x_k \right)^2 \\
&\leq \sum_{j=1}^{m} \left[ \left( \sum_{k=1}^{n} \alpha_{jk}^2 \right)^{1/2} \left( \sum_{l=1}^{n} x_l^2 \right)^{1/2} \right]^2 \\
&= ||x||^2 \sum_{j=1}^{m} \sum_{k=1}^{n} \alpha_{jk}^2.
\end{aligned}$$

Since the last double sum does not depend on $x$, we can write

$$||Tx||^2 \leq c^2 ||x||^2, \quad c^2 \sum_{j=1}^{m} \sum_{k=1}^{n} \alpha_{jk}^2,$$

which proves that $T$ is bounded.

Now we make the relation between matrices and bounded operators precise. Let $X$ and $Y$ be finite dimensional vector spaces over the same field $\mathbb{K}$, and $T : X \to Y$ a linear operator. We choose a basis $E = \{e_1, ..., e_n\}$ for $X$ and $B = \{b_1, ..., b_m\}$ for $Y$. Then, each $x \in X$ has a unique representation

$$x = \sum_{j=1}^{n} x_j e_j, \tag{B.2.9}$$

and linearity of $T$ allows us to write

$$y = Tx = T \left( \sum_{j=1}^{n} x_j e_j \right) = \sum_{j=1}^{n} x_j T e_j. \tag{B.2.10}$$

This allows us to conclude that since representation B.2.9 is unique, than $T$ is uniquely determined by the images $Te_j$ of the basis $E$.

We can now represent $y$ and $Te_k$ (we change the index to avoid confusion) both in terms of the basis $B$:

$$\begin{aligned}
y &= \sum_{j=1}^{m} y_j b_j \\
Te_k &= \sum_{j=1}^{m} \tau_{jk} b_j,
\end{aligned} \tag{B.2.11}$$

which we plug in equation B.2.10 to obtain

$$y = \sum_{j=1}^{m} \left( \sum_{k=1}^{n} \tau_{jk} x_k \right) b_j \implies y_j = \sum_{k=1}^{n} \tau_{jk} x_k, \tag{B.2.12}$$

where the last implication comes from the fact that the basis $B$ is comprised of a linearly independent set of vectors and must be equal on both sides of the equality.

The coefficients $\tau_{jk}$ in equation B.2.12 form a matrix $T_{EB}$ with $m$ rows and $n$ columns. Given the basis $E$ and $B$ with elements arranged in a fixed order, the matrix $T_{EB}$ is uniquely determined by the operator $T$, and it is called the *matrix representation.*

This discussion essentially formalizes the use of matrices to treat finite dimensional problems in quantum mechanics. For example, the *Pauli matrices* are a representation of the spin operators in systems with spin. We can also represent angular momentum operators with matrices. What we *can't* do is assign matrices to systems with infinite degrees of freedom.

### B.2.3.4 Normed Spaces of Operators

In this final subsection, we mention some notation and results which will become important in working with bounded linear operators between vector (normed) spaces.

First, we denote by $\mathcal{B}(X, Y)$ the set of all bounded linear operators from $X$ into $Y$[3]. This set forms a vector space under the natural operations between the operators:

$$
\begin{aligned}
(T_1 + T_2)x &= T_1 x + T_2 x \\
(\alpha T)x &= \alpha T x.
\end{aligned}
\tag{B.2.13}
$$

**Theorem B.2.6.** *The vector space $\mathcal{B}(X, Y)$ is itself a normed space with the norm defined by B.2.3.*

The more interesting result, which we won't prove here, is the condition where $\mathcal{B}(X, Y)$ forms a Banach space.

**Theorem B.2.7.** *If $Y$ is a Banach space, then $\mathcal{B}(X, Y)$ is a Banach space.*

## B.3 Linear Functionals

We will now discuss an important class of bounded linear operators on normed spaces which are the ones whose range lies on the field $\mathbb{R}$ or $\mathbb{C}$. These operators are called *functionals* and their importance in Quantum Physics cannot be overstated: they will give rise to what we understand as being *states* (not to be confused with *vector states* which are the elements of a Hilbert space, as we shall see further) and have the physical interpretation of being expectation values of measurable quantities.

---

[3]   Some authors denote this space by $\mathcal{L}(X, Y)$.

**Definition B.3.1.** A *linear functional* $f$ is a linear operator with domain in a vector space $X$ and range in the scalar field of $X$, $\mathbb{K}$:

$$f : \mathcal{D}(f) \to \mathbb{K}.$$

A *bounded linear functional* satisfies the additional property that there exists a real number $c$ such that for all $x \in \mathcal{D}(f)$

$$|f(x)| \le c\,||x||.$$

This allows us to define the norm of $f$ in an analogous way to equation B.2.3:

$$||f|| \equiv \sup_{x \in \mathcal{D}(f)} \frac{|f(x)|}{||x||}. \tag{B.3.1}$$

There is also an analogous theorem to theorem B.2.4 for linear functionals which we state without proof.

**Theorem B.3.1.** *A linear functional $f$ with domain $\mathcal{D}(f)$ in a normed space is continuous if, and only if, it is bounded.*

Hence, for linear functionals (as for the general case of linear operators), boundedness and continuity are equivalent properties.

It is important to notice that the set of all linear functionals on a vector space $X$ defines, itself a vector space, which we denote $X^{\#}$, called the *algebraic dual space* of $X$. The vector operations are defined in the natural way

$$\begin{aligned}
(f_1 + f_2)(x) &= f_1(x) + f_2(x) \\
(\alpha f)(x) &= \alpha f(x),
\end{aligned} \tag{B.3.2}$$

which are the same as B.2.13.

The set of all *bounded* linear functionals on $X$ is denoted by $X^*$ and is called the *dual space of $X$*.[4]

As in the case of bounded linear operators, linear functionals also constitute a normed space with norm given by equation B.3.1 and, unlike the operator case, is *always* a Banach space. This is a consequence of theorem B.2.7 for $Y = \mathbb{R}$ or $Y = \mathbb{C}$, both of which are Banach spaces themselves. For clarity, we enunciate the following theorem.

**Theorem B.3.2.** *The dual space $X^*$ of a normed space $X$ is a Banach space whether or not $X$ is.*

---

[4] Different authors have different conventions both for notation and name of the spaces. Sometimes the dual space space is denoted by $X'$ and called the *topological dual*, but we reserve the notation for the commutant in the context of von Neumann algebras. We follow the notation from [105].

## B.4 Inner Product Spaces

We now present the concept of inner products and inner product spaces which will eventually lead us to the concept of a Hilbert space. Hilbert spaces are an important ingredient in Quantum Physics as the phase space is to Classical Mechanics or a smooth manifold is to General Relativity.

The inner product, like the norm, is an added structure to vector spaces which allows us to take combinations of vectors and producing a number. Inner product spaces turn out to be a special case of normed spaces and Hilbert spaces are a particular case of Banach spaces, as we shall see. So what do we gain by restricting ourselves to this special case? To answer this, recall from ordinary geometry in three dimensions the concept of a *dot product*:

$$\vec{v} \cdot \vec{w} = v_1 w_1 + v_2 w_2 + v_2 w_3. \tag{B.4.1}$$

This formula allows us to define two concepts: the length of a vector, obtained by taking the dot product of a vector with itself,

$$|\vec{v}| = \sqrt{\vec{v} \cdot \vec{v}}, \tag{B.4.2}$$

and the important concept of *orthogonality*, defined by

$$\vec{v} \cdot \vec{w} = 0. \tag{B.4.3}$$

As we will see, orthogonality is a key feature which separates Hilbert spaces from general Banach spaces. And it is one of the reasons why we use Hilbert spaces instead of Banach spaces in Quantum Mechanics: suppose we are working in a two-level spin system which can be either in a state called "up" or in a state called "down". We would like to describe a general state of this system as a linear combination of both these states (vector space structure). However, in Quantum Mechanics we can only measure probabilities, so a state which is in the "up" state needs to have zero probability to be measured in the "down" state. In this case, orthogonality suggests a way to quantify this by constructing a way to combine vectors and producing a number that tells us "how much" of one vector is in another.

In many Quantum Mechanics textbooks, one is then introduced to the so-called "Braket notation" due to Dirac:

$$\langle \psi | \phi \rangle = \int d^3 x \, \overline{\psi}(x) \phi(x). \tag{B.4.4}$$

This is nothing but an inner product, even though most textbooks fail to mention so. In this notation, one usually regards the "bra" as the "out" state (the one that is measured) and the "ket" as the "in" state (the one that is prepared). The braket (or rather, the modulus squared) is then taken to be the probability of getting the out state given an in state.

We will not be using Dirac's notation in the remainder of this Thesis. We believe that a serious student of Physics should become acquainted with the underlying mathematical structure of a physical theory and understand it well. We think that Dirac's notation camouflages some of the mathematical nuances behind Quantum Physics. Eventually, after becoming familiar with the "true" concepts behind the notation, the student or researcher might use the notation they find to be the most convenient[5].

We now introduce the proper concept of an inner product and the "correct" notation.

**Definition B.4.1.** An *inner product space* or a *pre-Hilbert space* is a vector space $X$ over a field $\mathbb{K}$ with an *inner product*, which is a mapping $\langle \cdot, \cdot \rangle : X \times X \to \mathbb{K}$ that satisfies:

1. $\langle x + y, x \rangle = \langle x, z \rangle + \langle y, z \rangle$;

2. $\langle \alpha x, y \rangle = \alpha \langle x, y \rangle$;

3. $\langle x, y \rangle = \overline{\langle y, x \rangle}$;

4. $\langle x, x \rangle \geq 0$, $\langle x, x \rangle = 0 \iff x = 0$.

We note that properties 1 to 3 imply the following formula:

$$\langle x, \alpha y + \beta z \rangle = \overline{\alpha} \langle x, y \rangle + \overline{\beta} \langle x, z \rangle. \tag{B.4.5}$$

The reader should keep in mind that the Physics and Mathematics conventions diverge in this with the former adopting anti-linearity (the complex conjugate of the scalars) in the first entry of the inner product while the later, as evidenced in equation B.4.5, take anti-linearity on the second entry of the inner product.

We also note that one can construct a norm from an inner product by

$$||x|| = \sqrt{\langle x, x \rangle}, \tag{B.4.6}$$

which states that inner-product spaces are special cases of normed spaces. We note that not all norms can be obtained from an inner product: one can show that the norm has to satisfy the equality

$$||x + y||^2 + ||x - y||^2 = 2(||x||^2 + ||y||^2), \tag{B.4.7}$$

[5] In the experience of the author, even after understanding the mathematics behind the notation, the notation used here is still the clearest one and there is but a fez justification to actually use Dirac's notation.

better known as the *parellelogram equality*. This shows us that while all inner product spaces are normed spaces, the converse is not in general true. An interesting fact is that we can obtain the inner product that gives rise to the norm which satisfies the parallelogram equality. This is given by the so-called *polarization identity*, which can be written, for real inner product spaces, as

$$\langle x, y \rangle = \frac{1}{4}(||x + y||^2 - ||x - y||^2), \tag{B.4.8}$$

and for complex inner product spaces as

$$\mathrm{Re}\langle x, y \rangle = \frac{1}{4}(||x + y||^2 - ||x - y||^2)$$
$$\mathrm{Im}\langle x, y \rangle = \frac{1}{4}(||x + iy||^2 - ||x - iy||^2). \tag{B.4.9}$$

We now formalize the concept of orthogonality in inner product spaces.

**Definition B.4.2.** An element $x$ of an inner product space $(X, \langle \cdot, \cdot \rangle)$ is said to be *orthogonal* to an element $y$ in that same space if

$$\langle x, y \rangle = 0.$$

We then write $x \perp y$. Similarly, for subsets $A$, $B \subset X$, we write $x \perp A$ if $x \perp a$ for all $a \in A$ and $A \perp B$ if $a \perp b$ for all $a \in A$ and $b \in B$. In this last case, it is common to write $B = A^\perp$.

### B.4.1  Hilbert Spaces

We now arrive at the concept of a Hilbert space. Similarly to a Banach space, a Hilbert space is a *complete* space, in the sense that every Cauchy sequence in it converges in the topology defined by the inner product. To make this clear, we note one can obtain a metric from an inner product by

$$d(x, y) = ||x - y|| = \sqrt{\langle x - y, x - y \rangle}, \tag{B.4.10}$$

and proceed to an analogous discussion to that of a Banach space.

Some examples are in order. The Euclidean space $\mathbb{R}^n$ is a Hilbert space when paired with the inner product

$$\langle x, y \rangle = \sum_{j=1}^{n} x_j y_j, \tag{B.4.11}$$

which of course produces the norm in equation B.1.2 when we take the square-root. In the case where $n = 3$, we recover the familiar dot product from equation B.4.1. The same happens for $\mathbb{C}^n$ when paired with the inner product given by

$$\langle x, y \rangle = \sum_{j=1}^{n} x_j \overline{y_j}. \tag{B.4.12}$$

A very important example in the context of Quantum Mechanics is the $L^2[a,b]$ space of square-integrable functions in the closed interval $[a,b]$ of the real line. Considering complex-valued functions, we can make this space a Hilbert space when we pair it with the inner product given by

$$\langle x, y \rangle = \int_a^b x(t)\overline{y(t)}dt \tag{B.4.13}$$

with the real-valued functions case being recovered by dropping the complex conjugate in the second entry. Notice that this is the general case for the "representation of the Dirac braket in position space", from equation B.4.4. However, there is another reason why $L^2$ spaces are important in the context of Quantum Mechanics: recall that we defined the dual space of a Banach space as the vector space made of all bounded functionals acting on that space. The same definition applies to a Hilbert space (since it is a Banach space itself, but we will dive more into functionals on Hilbert spaces). It turns out that the dual space for $L^p$ spaces are the $L^q$ spaces such that $\frac{1}{p} + \frac{1}{q} = 1$. In particular, taking $p = 2$, we see that the $L^2$ space is *dual to itself*! This means that this space is dual to the space of functionals acting on it or, in Physics lingo, *the space of kets is isomorphic to the space of bras*!

An interesting example of something that is not even an inner product space is the Banach space $C[a,b]$ with norm given by equation B.1.1. In fact, one can show that this norm does *not* satisfy the parallelogram equality.

A very important inequality which relates an inner product to the norm it produces is the *Cauchy-Schwarz inequality*:

$$|\langle x, y \rangle| \leq ||x||\,||y||. \tag{B.4.14}$$

The proof of this inequality goes as follows. We assume that both $x$ and $y$ are non-zero since the converse is trivial. Thus, we take $\langle x, y \rangle \neq 0$. Using the fact that the norm is always a non-negative number, we can write for all $\alpha \in \mathbb{C}$

$$\begin{aligned}
0 &\leq \left|\left| \frac{\alpha}{||y||}y - \frac{1}{||x||}x \right|\right|^2 \\
&= \left\langle \frac{\alpha}{||y||}y - \frac{1}{||x||}x, \frac{\alpha}{||y||}y - \frac{1}{||x||}x \right\rangle \\
&= |\alpha|^2 + 1 - \frac{2}{||x||\,||y||}\text{Re}(\alpha\langle x, y \rangle) \implies 2\text{Re}(\alpha\langle x, y \rangle) \leq (|\alpha|^2 + 1)||x||\,||y||.
\end{aligned}$$

Since the last implication is valid *for all* $\alpha \in \mathbb{C}$, then it is valid, in particular, for $\alpha = \frac{|\langle x, y \rangle|}{\langle x, y \rangle}$, which yields $|\alpha| = 1$ and the last implication becomes $|\langle x, y \rangle| \leq ||x||\,||y||$, as we wanted.

It is important to notice that this inequality becomes an *equality* only when $x$ and $y$ are linearly dependent vectors. For a proof of this statement, see any introductory book on Functional Analysis.

Another important property of the inner product is that it is a continuous operation, in the following sense.

**Lemma B.4.1.** *If in an inner product space we have $x_n \to x$ and $y_n \to y$, then we have $\langle x_n, y_n \rangle \to \langle x, y \rangle$.*

*Proof.* This can be proven by the following simple calculation:

$$
\begin{aligned}
|\langle x_n, y_n \rangle - \langle x, y \rangle| &= |\langle x_n, y_n \rangle - \langle x_n, y \rangle + \langle x_n, y \rangle - \langle x, y \rangle| \\
&\leq |\langle x_n, y_n - y \rangle| + |\langle x_n - x, y \rangle| \\
&\leq ||x_n|| \, ||y_n - y|| + ||x_n - x|| \, ||y|| \to 0
\end{aligned}
$$

where the last inequality goes to zero since $y_n - y \to 0$ and $x_n - x \to 0$ as $n \to \infty$ and we have made use of the triangle inequality for numbers and the Cauchy-Schwarz inequality. ■

The last theorem of this section is analogous to theorem B.1.2 but for Hilbert spaces.

**Theorem B.4.1.** *For any inner product space $(X, \langle \cdot, \cdot \rangle)$ there exists a Hilbert space $\mathcal{H}$ and an isomorphism $A : X \to \mathcal{W}$ where $\mathcal{W}$ is a dense subspace of $\mathcal{H}$. This Hilbert space is unique up to isomorphisms.*

Note that the concept of an isomorphism here is adapted to inner products, i.e., it is a bijective linear operator $A : X \to \tilde{X}$ that preserves the inner product:

$$
\langle Tx, Ty \rangle = \langle x, y \rangle, \quad \forall x, y \in X.
$$

## B.4.2 Subspaces

We briefly mention the concept of a subspace in the context of inner product spaces. There are some interesting results regarding subspaces of normed spaces, especially of Banach spaces, but we restrict ourselves to inner product and Hilbert spaces.

**Definition B.4.3.** A *subspace* $Y$ of an inner product space $(X, \langle \cdot, \cdot \rangle)$ is a vector subspace of $X$ taken with the inner product restricted to $Y \times Y$. If $X$ is a Hilbert space, then $Y$ is defined in the same way.

Note that $Y$ need not be complete when taken as a subspace of a Hilbert space. The conditions where $Y$ is itself complete are stated in the following theorem.

**Theorem B.4.2.** *Let $Y$ be a subspace of a Hilbert space $\mathcal{H}$. Then:*

1. *$Y$ is complete if, and only if, $Y$ is closed in $\mathcal{H}$;*

2. *If $Y$ is finite dimensional, then $Y$ is complete.*

Conditions 1 and 2 imply a result which can be proven independently but we mention here for completeness.

**Corollary B.4.1.** *Every finite dimensional subspace $Y$ of a normed space $(X, || \cdot ||)$ is complete. In particular, every finite dimensional normed space is complete.*

### B.4.3   On the importance of completeness in Physics

We illustrate the importance of *completeness* in defining a Hilbert space with some basic results of geometrical nature.

Most traditional Physics textbooks on Quantum Mechanics fail to recognize this important feature of these spaces, either treating it as a mere technicality or even not mentioning it at all![6] Thus, the student might be led to believe that a Hilbert space is merely a vector space equipped with a inner product, i.e., that it is synonymous to a inner product space. We hope this brief explanation encourages the reader to familiarize themselves with the proper definition of a Hilbert space.

The property we will show is that one can decompose a Hilbert space $\mathcal{H}$ into two mutually orthogonal subspaces. As a consequence, a general vector $\psi$ of $\mathcal{H}$ can be written as a *unique* combination of two vectors $\varphi$ and $\chi$ which are orthogonal to one another. We have already seen that the concept of orthogonality arises naturally in inner product spaces, but the fact that we can write an element of $\mathcal{H}$ as a sum of elements of some "basis" is not straightforward, and the proof for the theorem that guarantees this makes explicit use of the concept of completeness, as we shall see.

The decomposition of $\mathcal{H}$ into two closed orthogonal subspaces makes the treatment of a Hilbert space similar to that of some Euclidean space $\mathbb{R}^n$, in the sense that one can think of a basis of orthonormal vectors. This is a feature of Hilbert spaces which working physicists rely on all the time to write, for example, a generic spin state as a combination of an "up" and "down" states, in the case of a two-level quantum system.

---

6   Some even treat a Hilbert space as just a letter $\mathcal{H}$ without providing *any* insight on what such space is.

### B.4.3.1  Convex sets

We provide a definition which we will make use of again later in this Thesis, in the discussion of states.

**Definition B.4.4.** Let $V$ be a vector space over the field of real or complex numbers. A linear combination of two vectors $x, y \in V$ which can be written as

$$\lambda x + (1 - \lambda)y, \quad \lambda \in [0, 1] \tag{B.4.15}$$

is said to be a *convex linear combination* of $x$ and $y$.

A set $A \subset V$ is said to be a *convex set* if, for all $x, y \in A$ and all $\lambda \in [0, 1]$, then $\lambda x + (1 - \lambda)y \in A$.

We note that, trivially, every sub*space* of a vector space is convex.

### B.4.3.2  The Best Approximation Theorem for Hilbert Spaces

We present a version of the Best Approximation Theorem, which holds in general for normed spaces. Here, we prove its validity for Hilbert spaces. This theorem shows that convex and closed subsets of Hilbert spaces admit an unambiguous notion of "distance to a set".

**Theorem B.4.3.** *Let $A$ be a convex and closed subset of a Hilbert space $\mathcal{H}$. Then, for all $x \in \mathcal{H}$, there exists a unique vector $y \in A$ such that the distance between $x$ and $y$ is equal to the minimum possible distance between $x$ and $A$, i.e.:*

$$||x - y|| = \inf_{y' \in A} ||x - y'||.$$

*The vector $y$ is usually called* the best approximant of $x$ in $A$.

*Proof.* Let $D \leq 0$ be defined as

$$D = \inf_{y' \in A} ||x - y'||.$$

For each $n \in \mathbb{N}$, let $y_n \in A$ be a vector with the following property:

$$||x - y_n||^2 < D^2 + \frac{1}{n}.$$

Such vectors always exist: otherwise, we would have that $||x - y'||^2 \geq D^2 + \frac{1}{n}$ for all $y' \in A$, which contradicts the hypothesis of $D$ being the infimum of $||x - y'||$, with $y' \in A$.

We now show that the sequence $(y_n)_{n \in \mathbb{N}}$ with each member $y_n$ as defined above is a Cauchy sequence in $\mathcal{H}$. Recalling the parallelogram identity B.4.7 for all $a$, $b \in \mathcal{H}$. We take $a = x - y_n$ and $b = x - y_m$. Thus:

$$||2x - (y_m + y_n)||^2 + ||y_m - y_n||^2 = 2||x - y_n||^2 + 2||x - y_m||^2,$$

which can in turn be rewritten as

$$||y_m - y_n||^2 = 2||x - y_n||^2 + 2||x - y_m||^2 - 4\left|\left|x - \frac{y_m + y_n}{2}\right|\right|^2.$$

We now use the fact that $||x - y_n||^2 < D^2 + \frac{1}{n}$ for all $n$:

$$||y_m - y_n||^2 \leq 4D^2 + 2\left(\frac{1}{n} + \frac{1}{m}\right) - 4\left|\left|x - \frac{y_m + y_n}{2}\right|\right|^2.$$

Since $A$ is a convex set by hypothesis, then $\frac{y_m + y_n}{2} \in A$. By definition of $D$,

$$\left|\left|x - \frac{y_m + y_n}{2}\right|\right|^2 \geq D^2.$$

Hence, we have that

$$||y_m - y_n||^2 \leq 2\left(\frac{1}{n} + \frac{1}{m}\right),$$

thus making the distance between $y_m$ and $y_n$ arbitrarily small for large values of $m$ and $n$. Therefore, $(y_n)_{n \in \mathbb{N}}$ is a Cauchy sequence.

Since $\mathcal{H}$ is complete, then $(y_n)_{n \in \mathbb{N}}$ converges to some $y \in \mathcal{H}$. Even further, we can assume that $y \in A$, because of the hypothesis that $A$ is a closed subset of $\mathcal{H}$.

For all $n$, we have that

$$||x - y|| = ||(x - y_n) - (y - y_n)|| \leq ||x - y_n|| + ||y - y_n|| < \sqrt{D^2 + \frac{1}{n}} + ||y - y_n||.$$

By taking $n \to \infty$, we conclude that $||x - y|| \leq D$. On the other hand, by definition of $D$, we have that $||x - y|| \geq D$. Hence, $||x - y|| = D$.

We now prove that $y$ is the only element of $A$ with this property. Suppose there is a $y' \in A$ such that $||x - y'|| = D$. By employing again the paralelogram identity with $a = x - y$ and $b = x - y'$, we have that

$$||2x - (y + y')||^2 + ||y - y'||^2 = 2||x - y||^2 + 2||x - y'||^2 = 4D^2,$$

which can be rewritten as

$$||y - y'||^2 = 4D^2 - ||2x - (y + y')||^2 = 4D^2 - 4\left|\left|x - \frac{y + y'}{2}\right|\right|^2.$$

Since $\frac{y+y'}{2} \in A$ (convex combination), we have that

$$\left\|x - \frac{y+y'}{2}\right\|^2 \geq D^2 \implies ||y-y'||^2 \leq 0 \implies y = y',$$

which completes the proof. ∎

### B.4.3.3  The Orthogonal Decomposition Theorem

Now we prove the aforementioned property of Hilbert spaces, namely, that it admits a decomposition into orthogonal subspaces. This is a direct consequence of the previous theorem.

The following definition is in order.

**Definition B.4.5.** Let $E$ be a subset of a Hilbert space $\mathcal{H}$. We define the *orthogonal complement* of $E$ as the set

$$E^\perp \equiv \{y \in \mathcal{H} | \langle y, x \rangle = 0, \forall x \in E\}.$$

It can be shown that $E^\perp$ is a closed linear subspace of $\mathcal{H}$. We also mention a simple lemma:

**Lemma B.4.2.** *If $A$ and $B$ are subsets of a Hilbert space $\mathcal{H}$ and $A \subset B$, then $B^\perp \subset A^\perp$*

*Proof.* By definition, if $y \in B^\perp$, then $y$ is orthogonal to every element of $B$. Since $A$ is a subset of $B$, then $y$ is also orthogonal to every element of $A$, thus $y \in A^\perp$. ∎

We now proceed to the main theorem, sometimes dubbed the *Orthogonal Decomposition Theorem*.

**Theorem B.4.4.** *Let $\mathcal{M}$ be a closed linear subspace of a Hilbert space $\mathcal{H}$. Then, every $x \in \mathcal{H}$ can be uniquely written as $x = y + z$, with $y \in \mathcal{M}$ and $z \in \mathcal{M}^\perp$. The vector $y$ is such that $||x - y|| = \inf_{y' \in \mathcal{M}} ||x - y'||$ (hence, it is the best approximant of $x$ in $\mathcal{M}$).*

*Proof.* Theorem B.4.3 guarantees to us the existence and uniqueness of an element $y \in \mathcal{M}$ such that $||x - y|| = \inf_{y' \in \mathcal{M}} ||x - y'||$. We now define $z = x - y$ and prove that $z \in \mathcal{M}^\perp$ and that $y$ and $z$ are unique.

Proving that $z \in \mathcal{M}^\perp$ is equivalent to showing that $\langle z, y' \rangle = 0$ for all $y' \in \mathcal{M}$. By definition of $y$, we have that

$$||x - y||^2 \leq ||x - y - \lambda y'||^2,$$

for all $\lambda \in \mathbb{C}$ and $y' \in \mathcal{M}$. We note that the fact that $\mathcal{M}$ is a subspace guarantees us that $y + \lambda y' \in \mathcal{M}$. From this last relation, we obtain, by definition of $z$:

$$||z||^2 \leq ||z - \lambda y'||^2 = \langle z - \lambda y', z - \lambda y' \rangle.$$

Expanding on the last inequality using the linearity of the inner product results in

$$||z||^2 \leq ||z||^2 - 2\mathrm{Re}(\lambda \langle z, y' \rangle) + |\lambda|^2 ||y'||^2 \implies 2\mathrm{Re}(\lambda \langle z, y' \rangle) \leq |\lambda|^2 ||y'||^2.$$

Writing the complex number $\langle z, y' \rangle$ in its polar form $|\langle z, y' \rangle| e^{i\alpha}$ for some $\alpha \in \mathbb{R}$ and noting that the last implication is valid for all $\lambda \in \mathbb{C}$ (including $\lambda = te^{-i\alpha}$, for some $t > 0$) we obtain

$$2t|\langle z, y' \rangle| \leq t^2 ||y'||^2 \implies |\langle z, y' \rangle| \leq \frac{t}{2}||y'||^2.$$

The last inequality is valid for all $t > 0$, which can only happen if we have $|\langle z, y' \rangle| = 0$. Since $y'$ is arbitrary, this means that $z \in \mathcal{M}^\perp$.

The last step is to show that both $y$ and $z$ are unique. Suppose we could write $x = y_0 + z_0$ with $y_0 \in \mathcal{M}$ and $z_0 \in \mathcal{M}^\perp$. We would then have $y + z = y_0 + z_0$, that is, $y - y_0 = z_0 - z$. Hence, we have an element of $\mathcal{M}$ which is equal to an element in $\mathcal{M}^\perp$. This can only mean that $y - y_0 = z_0 - z = 0$, proving uniqueness. ∎

## B.5 The Riesz Representation Theorem

Hilbert spaces are considerably simpler than Banach spaces when it comes to defining functionals in them. This will become clear with the next theorem due to Riesz [115].

A word about nomenclature is in order. The following theorem is usually known as the *Riesz Representation Theorem*, but sometimes this name is reserved to a theorem regarding the representation of sesquilinear forms in terms of the inner product (see [93], theorem 3.8-4), with the theorem below receiving the name of *Riesz's theorem*. Since these names are not universal in the literature, the reader should be careful when comparing different references.

**Theorem B.5.1** (Riesz's Theorem). *Let $f$ be a bounded linear functional on a Hilbert space $\mathcal{H}$. Then, there exists a unique $\phi \in \mathcal{H}$ such that*

$$f(x) = \langle \phi, x \rangle, \quad \forall x \in \mathcal{H} \tag{B.5.1}$$

*Proof.* Let $\mathcal{K} \subset \mathcal{H}$ be the kernel of $f$, that is, the set of all vectors in $\mathcal{H}$ which vanish under $f$:

$$\mathcal{K} = \{y \in \mathcal{H} | f(y) = 0\}. \tag{B.5.2}$$

We now prove that $\mathcal{K}$ is a closed linear subspace of $\mathcal{H}$ (hence, by theorem B.4.2, it is complete).

The fact that $\mathcal{K}$ is a linear subspace comes from the linearity of linear bounded functionals: if $x, y \in \mathcal{K}$, then $f(\alpha x + \beta y) = \alpha f(x) + \beta f(y) = 0$ and thus $f(\alpha x + \beta y) \in \mathcal{K}$.

$\mathcal{K}$ is closed since we can characterize it as being the preimage of $0 \in \mathbb{C}$ by $f$: $\mathcal{K} = f^{-1}(\{0\})$ and since a point is a closed set in the usual topology of $\mathbb{C}$, continuity of $f$ asserts us that the preimage is closed.

We now suppose $\mathcal{K} \neq \mathcal{H}$, with the converse being the trivial case where $f(x) = 0$, for all $x \in \mathcal{H}$ in which case we need only to take $\phi = 0$ in equation B.5.1. Since $\mathcal{K}$ is closed, the Orthogonal Decomposition Theorem B.4.4 tells us that every $x \in \mathcal{H}$ is of the form $x = y + z$ with $y \in \mathcal{K}$ and $z \in \mathcal{K}^{\perp}$.

Let $z_0$ be a non-zero element of $\mathcal{K}^{\perp}$. For any vector $u \in \mathcal{H}$, we have that $f(z_0)u - f(u)z_0$ is an element of $\mathcal{K}$ since

$$f(f(z_0)u - f(u)z_0) = f(z_0)f(u) - f(u)f(z_0) = 0,$$

by linearity of $f$ and commutativity of the field. Since $f(z_0)u - f(u)z_0$ is an element of $\mathcal{K}$, it must be orthogonal to $z_0$, which by hypothesis is in $\mathcal{K}^{\perp}$. Hence:

$$\langle z_0, f(z_0)u - f(u)z_0 \rangle = 0 \implies f(z_0)\langle z_0, u \rangle - f(u)||z_0||^2.$$

The last implication can be rewritten as

$$f(u) = \frac{f(z_0)}{||z_0||^2}\langle z_0, u \rangle = \left\langle \frac{\overline{f(z_0)}}{||z_0||^2}z_0, u \right\rangle,$$

which proves the theorem by defining

$$\phi = \frac{\overline{f(z_0)}}{||z_0||^2}z_0. \tag{B.5.3}$$

The last step in the proof is to show that the vector $\phi$ defined in equation B.5.3 is unique. In order to do so, suppose there were a $\phi'$ which satisfies $f(x) = \langle \phi', x \rangle$ for all $x \in \mathcal{H}$. Thus, we have that $\langle \phi', x \rangle = \langle \phi, x \rangle \implies \langle \phi - \phi', x \rangle = 0$. Since this relation is valid for all $x \in \mathcal{H}$, it is also valid for $x = \phi - \phi'$. Hence, we have $\langle \phi - \phi', \phi - \phi' \rangle = ||\phi - \phi'||^2 = 0$ which implies $\phi - \phi' = 0 \iff \phi = \phi'$, thus proving unicity. ∎

## B.6   Bounded Linear Operators in Hilbert Spaces

We now begin the main mathematical concept of this Thesis. The main properties and definitions for bounded linear operators in Hilbert spaces are essentially the same as those for Banach spaces, including the definition of the operator norm (considering of course the inner product). In Hilbert spaces, however, the presence of the inner product allow us to add more structure and to give a classify different types of operators. This is but a small part in what is called Operator Theory in Mathematics (for a classic text devoted to this area, see [89] and [90]).

Bounded linear operators on Hilbert spaces (henceforth denoted by $\mathcal{B}(\mathcal{H})$) are the first example of what we call a $C^*$-algebra, which we will define in the following sections. This type of algebra is the fundamental mathematical structure behind Quantum Mechanics and to Quantum Field Theory (where, in fact, an important subset of the bounded linear operators turns out to be the more important concept rather than the more abstract $C^*$-algebra). The importance of $\mathcal{B}(\mathcal{H})$ among the more abstract $C^*$-algebras is that we can obtain concrete information about the abstract algebras by representing them in a concrete $\mathcal{B}(\mathcal{H})$ via a process known as the *GNS construction*, to be discussed later.

In Physics, particularly Quantum Mechanics, the choice of Hilbert space to describe the problem is of great importance, but not always obvious. In the context of Quantum Field Theory in Curved Spacetimes [136], there is even a bigger *aggravant* which is the lack of symmetries in general spacetimes, which turns the problem of selecting a vacuum state (hence, a Hilbert space with such vacuum) a highly non-trivial problem. In this context, it was realized early on that $C^*$-algebras could supply valuable information about such theories and actual calculations could be carried out once a representation and Hilbert space are chosen based on other physical principles pertaining the problem.

It is important to note that, physically, operators on Hilbert spaces act as a measurement of a certain quantity. It is common to read the phrase that "observables are described by a Hermitian operator on a Hilbert space" which sometimes is inadequate (either because the observable is *not* Hermitian or because this might give the impression that any Hermitian operator is associated to a physical quantity, which is not true).

### B.6.1   The Adjoint Operator

In this section, we introduce the general properties of what is called the *adjoint* or *Hilbert-adjoint* operator. In the following section we will classify these operators in three classes according to their properties.

**Definition B.6.1.** Let $T : \mathcal{H}_1 \to \mathcal{H}_2$ be a bounded linear operator, where $\mathcal{H}_1$ and $\mathcal{H}_2$ are Hilbert spaces. The *adjoint* or *Hilbert-adjoint* operator of $T$ is the operator

$$T^* : \mathcal{H}_2 \to \mathcal{H}_1$$

such that, for all $x \in \mathcal{H}_1$ and $y \in \mathcal{H}_2$

$$\langle Tx, y \rangle = \langle x, T^*y \rangle.$$

The existence of such operator is guaranteed by the following theorem, which also enumerates some of the properties of such operator.

**Theorem B.6.1.** *The adjoint operator $T^*$ exists, is unique and is a bounded linear operator with norm*

$$||T^*|| = ||T||.$$

*In addition, the following properties are satisfied for operators $S, T : \mathcal{H}_1 \to \mathcal{H}_2$ and their respective adjoints:*

1. *$\langle T^*y, x \rangle = \langle y, Tx \rangle$, for all $x \in \mathcal{H}_1$ and $y \in \mathcal{H}_2$;*

2. *$(S + T)^* = S^* + T^*$;*

3. *$(\alpha T)^* = \overline{\alpha} T^*$;*

4. *$(T^*)^* = T$;*

5. *$||T^*T|| = ||TT^*|| = ||T||^2$;*

6. *$T^*T = 0 \iff T = 0$;*

7. *$(ST)^* = T^*S^*$, when $\mathcal{H}_1 = \mathcal{H}_2$.*

For a proof, see [93], theorems 3.9-2 and 3.9-4.

## B.6.2 Self-Adjoint, Unitary and Normal Operators

We now classify bounded linear operators on Hilbert spaces with respect to their adjoints. The names are frequently (mis-) used in Physics and can lead to some apparent inconsistencies in the theory (see [74]). One of the goals of this section is to familiarize the student with the proper definition of these terms.

**Definition B.6.2.** A bounded linear operator $T : \mathcal{H} \to \mathcal{H}$ on a Hilbert space $\mathcal{H}$ is said to be

- *self-adjoint* or *Hermitian* if $T^* = T$, i.e., $\langle x, Ty \rangle = \langle Tx, y \rangle$, for all $x, y \in \mathcal{H}$;

- *unitary* if $T$ is a bijection and $T^* = T^{-1}$;

- *normal* if $T^*T = TT^*$

We first note that if either $T$ is self-adjoint or unitary, it is normal. An example of an operator which is normal but neither self-adjoint nor unitary is the operator defined by $T = 2iI$ where $I$ is the identity operator $I : \mathcal{H} \to \mathcal{H}$.

An important property concerning self-adjoint operators is that *any* bounded operator $B$ acting on $\mathcal{H}$ can be written as a combination of two self-adjoint operators $B = \mathrm{Re}(B) + i\mathrm{Im}(B)$, where

$$\mathrm{Re}(B) = \frac{1}{2}(B + B^*)$$
$$\mathrm{Im}(B) = \frac{1}{2i}(B - B^*).$$

A criterion for self-adjointness is given by the following theorem.

**Theorem B.6.2.** *Let $T : \mathcal{H} \to \mathcal{H}$ be a bounded linear operator on a Hilbert space $\mathcal{H}$. Then:*

1. *If $T$ is self-adjoint, $\langle Tx, x \rangle$ is real for all $x \in \mathcal{H}$.*

2. *If $\mathcal{H}$ is complex and $\langle Tx, x \rangle$ is real for all $x \in \mathcal{H}$, the operator $T$ is self-adjoint.*

*Proof.* 1. If $T$ is self-adjoint, then for all $x$, we have the following

$$\overline{\langle Tx, x \rangle} = \langle x, Tx \rangle = \langle Tx, x \rangle,$$

making $\langle Tx, x \rangle$ real.

2. Assuming $\langle Tx, x \rangle$ real for all $x$,

$$\langle Tx, x \rangle = \overline{\langle Tx, x \rangle} = \overline{\langle x, T^*x \rangle} = \langle T^*x, x \rangle,$$

and hence

$$\langle Tx, x \rangle - \langle T^*x, x \rangle = \langle (T - T^*)x, x \rangle = 0 \implies T = T^*,$$

where we use the same reasoning as in theorem B.5.1

■

Since the compostition of bounded linear operators is important (and in fact is what is going to give rise to the *algebra* of bounded linear operators), we mention the following simple theorem.

**Theorem B.6.3.** *The product of two bounded self-adjoint linear operators $S$ and $T$ on a Hilbert space $\mathcal{H}$ is self-adjoint if, and only if, the operators commute, i.e.,*

$$ST = TS,$$

*also denoted by*

$$[S, T] = ST - TS = 0.$$

*Proof.* The proof follows easily from the property $(ST)^* = T^*S^*$:

$$(ST)^* = T^*S^* = TS,$$

since the operators are self-adjoint, by assumption. Hence

$$ST = (ST)^* \iff ST = TS.$$

■

Unitary operators are very important in the context of Quantum Physics because they are associated to symmetry transformations and because they preserve the inner product:

$$\langle U\psi, U\phi \rangle_{\mathcal{H}_2} = \langle \psi, \phi \rangle_{\mathcal{H}_2}$$
$$\langle U^*\psi, U^*\phi \rangle_{\mathcal{H}_1} = \langle \psi, \phi \rangle_{\mathcal{H}_2}. \tag{B.6.1}$$

Since the inner product is associated with the probability in the context of Quantum Mechanics, transformations described by unitary transformations preserve probability. Other properties of unitary operators are described in the following theorem.

**Theorem B.6.4.** *Let the operators $U : \mathcal{H} \to \mathcal{H}$ and $V : \mathcal{H} \to \mathcal{H}$ be unitary. Then:*

1. *$||Ux|| = ||x||$, for all $x \in \mathcal{H}$ (hence, $U$ is an* isometry*);*

2. *$||U|| = 1$;*

3. *$U^{-1} = U^*$ is unitary;*

4. *$UV$ is unitary.*

### B.6.2.1 Self-Adjoint and Hermitian Operators

In this brief section, we clarify a important difference between self-adjoint and Hermitian operators. Even though we took the terms to be synonymous in definition B.6.2, this is only valid for *bounded* operators. The reason for this is given by the following theorem, which is known as the *Hellinger-Toeplitz theorem*:

**Theorem B.6.5.** *Let $\mathcal{H}$ be a Hilbert space and $A$ a linear operator such that $\mathcal{D}(A) = \mathcal{H}$ and*

$$\langle x, Ay \rangle = \langle Ax, y \rangle, \quad \forall x, y \in \mathcal{H}. \tag{B.6.2}$$

*Then, $A$ is bounded.*

The proof of this theorem makes explicit use of the *Closed Graph Theorem*, one of the fundamental theorems of Functional Analysis (alongside the *Open Mapping Theorem*, the *Uniform Boundedness Theorem* and the *Hahn-Banach Theorem*). Even though of great importance, we will not present these theorems in this Thesis and instead, we refer to the bibliography (in particular, for a simple proof of the Hellinger-Toeplitz theorem, we recomend [111]).

The takeaway lesson from the Hellinger-Toeplitz theorem is that if we are working in the context of *unbounded* operators, we cannot have equation B.6.2 being valid *for all* $x, y \in \mathcal{H}$. In other words, the operator *fails* to be self-adjoint in all the Hilbert space. In this discussion, the domain of the operator becomes a crucial part of the theory, and we need to make a difference between *self-adjoint operators* (which satisfy both $A = A^*$ *and* $\mathcal{D}(A) = \mathcal{D}(A^*)$) and *Hermitian* or *symmetric* operators (which only satisfy $A = A^*$).

Even though we will not be explicitly considering unbounded operators in this Thesis, we feel it to be necessary to warn the reader about these differences, as the misuse of the notation can lead to unpleasant surprises.

## B.6.3 Orthogonal Projectors and Partial Isometries

In this brief section, we introduce important concepts for the discussion of the Spectral Theorem. We will eventually generalize the concept of projectors, understood as being the precise mathematical description of the *act of measurement* (thus forcing a quantum vector state to be in a subspace of the full Hilbert space), to the more abstract entities known as *projection-valued measures* which become relevant in Quantum Field Theory.

### B.6.4  Projectors and Orthogonal Projectors

**Definition B.6.3.** A linear operator $P$ on a Hilbert space $\mathcal{H}$ is said to be a *projector* if it satisfies

$$P^2 = P \cdot P = P.$$

A projector is said to be an *orthogonal projector* if it is self-adjoint, that is $P^* = P$.

It is important to note that *not all projectors are orthogonal*. A counter-example is given by the operator with matrix representation given by

$$P = \begin{pmatrix} 1 & 0 \\ 1 & 0 \end{pmatrix}$$

acting on the Hilbert space $\mathbb{C}^2$.

**Proposition B.6.1.** *Every orthogonal projector $P$ on a Hilbert space is bounded and satisfies* $||P|| = 1$.

*Proof.* Orthogonality and self-adjointness of $P$ implies that $P^2 = P = P^*$. Hence, for every $x \in \mathcal{H}$:

$$||Px||^2 = \langle Px, Px \rangle = \langle P^*Px, x \rangle$$
$$= \langle P^2x, x \rangle = \langle Px, x \rangle \leq ||x|| \, ||Px||,$$

where the last inequality comes from the Cauchy-Schwarz inequality. We then have $||Px|| \leq ||x||$, for all $x \in \mathcal{H}$, which implies $||P|| \leq 1$.

On the other hand, assuming $P \neq 0$, there exists some $y \in \mathcal{H}$ such that $Py \neq 0$. Denoting $x = Py$, we have $Px = P^2y = Py = x$, which implies

$$||P|| = \sup_{\substack{\psi \in \mathcal{H} \\ \psi \neq 0}} \frac{||P\psi||}{||\psi||} \geq \frac{||Px||}{||x||} = \frac{||x||}{||x||} = 1 \implies ||P|| \geq 1.$$

Combining both results, we get $||P|| = 1$. ∎

If $P$ is a orthogonal projector, then $\mathbf{1} - P$ is also an orthogonal projector as the reader can easily verify.

An important example of an orthogonal projector are the so-called *one-dimensional projectors on subspaces spanned by a vector*. We take the vectors to have $||v|| = 1$ and define the projector as

$$P_v u = \langle v, u \rangle \, v, \tag{B.6.3}$$

where $u$ is some vector on the Hilbert space. These are the type of projection operators that are commonly presented in textbooks on Quantum Mechanics. Note that if $u$ and $v$ are orthogonal to one another, then we have $P_u P_v = P_v P_u = 0$.

### B.6.5   Isometries and Partial Isometries

Even though we have already mentioned the concept of an isometry, we repeat it here for clarity and to illuminate the following definition of a *partial isometry*.

**Definition B.6.4.** Let $\mathcal{H}_1$ and $\mathcal{H}_2$ be two Hilbert spaces. A bounded linear operator $U \in \mathcal{B}(\mathcal{H}_1, \mathcal{H}_2)$ is said to be an *isometry* if

$$||Ux||_{\mathcal{H}_2} = ||x||_{\mathcal{H}_1},$$

for all $x \in \mathcal{H}_1$. This is equivalent to

$$\langle Ux, Ux \rangle_{\mathcal{H}_2} = \langle x, x \rangle_{\mathcal{H}_1}.$$

By means of the polarization identity in equation B.4.8, we can show that $U$ is an isometry if, and only if $\langle Ux, Uy \rangle_{\mathcal{H}_2} = \langle x, y \rangle_{\mathcal{H}_1}$ for all $x, y \in \mathcal{H}_1$.

**Definition B.6.5.** A bounded linear operator $U \in \mathcal{B}(\mathcal{H}_1, \mathcal{H}_2)$ is said to be a *partial isometry* if it is an isometry when restricted to $\text{Ker}(U)^{\perp}$.

By use of the polarization identity B.4.8, we find that $U$ is a partial isometry if, and only if $\langle Ux, Uy \rangle_{\mathcal{H}_2} = \langle x, y \rangle_{\mathcal{H}_1}$ for all $x, y \in \text{Ker}(U)^{\perp}$.

## B.7   Elements of Spectral Theory

In this section, we will present some of the main concepts and results of Spectral Theory which we will make use in the remaining of the Thesis. Spectral Theory by itself is a very rich field of Mathematics and our approach here barely scratches the surface of it. For a brief introduction to Spectral Theory, we recommend reference [6] or any good book on Functional Analysis such as the ones already cited (however, the extent to which each reference treats this subject might vary, with references such as [105] treating only Spectral Theory in Banach spaces).

## B.7.1   Basic concepts

We assume the student is familiar with the process of diagonalization of a square matrix to find its eigenvalues and eigenvectors using the characteristic equation. Here, we present an abstract version of that process and present an adequate definition for *spectrum*, which is sometimes misinterpreted as being the set of eigenvalues of an operator.

Let $(X, ||\cdot||)$ be a complex normed space and $T : \mathcal{D}(T) \to X$ be a linear operator with domain $\mathcal{D}(T) \subset X$. To $T$, we associate the operator

$$T_\lambda = T - \lambda I, \tag{B.7.1}$$

where $\lambda \in \mathbb{C}$ and $I$ is the identity operator. If $T_\lambda$ is invertible, then we denote it by $R_\lambda(T)$:

$$R_\lambda(T) = T_\lambda^{-1} = (T - \lambda I)^{-1}, \tag{B.7.2}$$

which is called the *resolvent* of $T$. The name indicates the fact that the existence of this operator is related to solving the problem given by

$$(T - \lambda I)x = y,$$

which is the operator form of the "eigenvalue equation", as commonly presented in Quantum Mechanics textbooks. The properties of $R_\lambda(T)$ are key to understanding both the operator $T$ itself, but also its spectrum.

We now proceed to the key definitions of this section.

**Definition B.7.1.** Let $(X, ||\cdot||)$ be a complex normed space and $T : \mathcal{D}(T) \to X$ a linear operator such that $\mathcal{D}(T) \subset X$. A *regular value* $\lambda$ of $T$ is a complex number such that

1. $R_\lambda(T)$ exists;

2. $R_\lambda(T)$ is a bounded operator;

3. $R_\lambda(T)$ is defined on a set which is dense in $X$.

The set of all regular values of $T$ is called the *resolvent set* of $T$ and is denoted by $\rho(T)$. The complement set $\sigma(T) = \mathbb{C} \setminus \rho(T)$ is called the *spectrum* of $T$, with each element $\lambda \in \sigma(T)$ being called a *spectral value* of $T$.

One can partition the spectrum $\sigma(T)$ of the operator $T$ into three disjoint sets, which allows us to classify the spectral values of this operator.

1. The *point spectrum* or *discrete spectrum* $\sigma_p(T)$ is the set where $R_\lambda(T)$ does not exist. A complex number $\lambda \in \sigma_p(T)$ is called an *eigenvalue* of T.

2. The *continuous spectrum* $\sigma_c(T)$ is the set where $R_\lambda(T)$ exists, is defined on a dense subset of $X$ but is not bounded.

3. The *residual spectrum* $\sigma_r(T)$ is the set where $R_\lambda(T)$ exists, it may or not be bounded but its domain is not dense in $X$.

For the point spectrum, we note that the condition for $R_\lambda(T)$ to exist (as for any inverse of an operator) is that $T_\lambda x = 0 \implies x = 0$. Hence, $R_\lambda(T)$ does not exist if we have $(T - \lambda I)x = 0$ for $x \neq 0$. The vectors $x$ which satisfy this equation are called the *eigenvectors* of $T$ (or *eigenfunctions* if $X$ happens to be a function space). It is important to note that eigenvectors are *non-zero* vectors.

Note that the spectrum is more than just the point spectrum. In fact, we present an example of an operator with a spectral value which is *not* an eigenvalue.

Let $\mathcal{H} = \ell^2$, i.e., the Hilbert space of square-summable sequences $(x_n)_{n \in \mathbb{N}}$ with inner product $\langle x, y \rangle = \sum_{j=1}^{\infty} x_j \overline{y_j}$. Define a linear operator $T : \ell^2 \to \ell^2$ such that

$$(x_1, x_2, ...) \mapsto (0, x_1, x_2, ...).$$

This operator is called the *right-shift operator*, for obvious reasons. Because the sequence is square-summable, $T$ is bounded:

$$||Tx||^2 = \sum_{j=1}^{\infty} |x_j|^2 = ||x||^2.$$

The inverse $T^{-1} = R_0(T)$ exists and is given by the *left-shift operator*:

$$R_0(T) : T(\ell^2) \to \ell^2$$
$$(x_1, x_2, ...) \mapsto (x_2, x_3, ...).$$

So, the operator $R_0(T)$ always exists, so $\sigma_p = \varnothing$. On the other hand, it is *not* defined on a dense subset of $\ell^2$, since the set $T(\ell^2)$ is composed of square-summable sequences with the first entry being zero. Therefore, a sequence of elements of $T(\ell^2)$ can only converge to a sequence of $T(\ell^2)$, never to a sequence with the first entry being non-zero. Thus, not all points of $\ell^2$ are accumulation points of $T(\ell^2)$, so this subset is not dense. This also implies that $\lambda = 0$ is a spectral value of $T$, which can't be an eigenvalue since it would imply $Tx = 0 \implies x = 0$, and, as we stated before, eigenvectors are non-zero vectors.

## B.7.2 Spectral Theorem

Here we present the important Spectral Theorem for bounded self-adjoint operators. There are more general forms of this theorem (see for instance [15, 111, 118]) but we will content ourselves with this version. It should be known that this theorem is analogous in the case of matrices since there is a one-to-one correspondence between bounded operators in finite dimensional spaces and matrices.

We will be interested in the Borelian sets of the spectrum of an operator $A$, $\mathfrak{B}(\sigma(A))$ and we define a measure in these sets (see appendix A).

**Theorem B.7.1 (Spectral Theorem).** *Let $\mathcal{H}$ be a Hilbert space and $A \in B(\mathcal{H})$ self-adjoint. Then there is a unique projection-valued measure $\mathfrak{B}(\sigma(A)) \ni B \mapsto P_B = \chi_B(A) \in B(\mathcal{H})$ such that*

$$A = \int_{\sigma(A)} \lambda \, dP_\lambda. \tag{B.7.3}$$

*Moreover, for any borelian function $g$, we have that*

$$g(A) = \int_{\sigma(A)} g(\lambda) \, dP_\lambda. \tag{B.7.4}$$